\documentclass[dvips]{article}
\usepackage{amsmath,amssymb,amsfonts,amscd,theorem}
\usepackage[latin1]{inputenc}
\def\simarrow{\mathrel{\raise -0.5mm\hbox{$\sim$}}\hspace{-1.8mm}{\rightarrow} } 

\def\bsimarrow{\leftarrow\hspace{-0.7mm}\mathrel{\raise -0.5mm\hbox{$\backsim$}} }

\newcommand{\tim}[1]{\;\; \mbox{#1} \;\;}

\def\bt{\begin{tabular}}
\def\te{\end{tabular}}

\def\lettrine#1#2#3{\noindent\hangindent#1\hangafter-#2
\hskip-#1\smash{\hbox to #1{#3\hfill}}\ignorespaces}

\newcommand{\To}{\begin{CD}@>>>\end{CD}}

\def\BM{\begin{pmatrix}}
\def\EM{\end{pmatrix}}

\def\txt{\textstyle}
\def\ds{\displaystyle}

\def\d=f{\buildrel\hbox{\scriptsize d\'{e}f}\over \Longleftrightarrow}

\def\cit{\text{\it I\hskip -6ptC\/}}
\def\ptcit{\hbox{${\scriptstyle I\hskip -4ptC\/}$}}

\def\square{\hfill\hbox{\vrule height .9ex width .8ex depth -.1ex}}
\def\rit{\text{\it I\hskip -2pt  R}}

\def\zit{\text{\it Z\hskip -4pt  Z}}

\def\nit{\text{\it I\hskip -2pt  N}}

\def\rl {\rit^{\hskip 1pt\ell}}

\def\Bd{{\text B}}

\def\Cs{{\cal C}}
\def\Ds{{\cal D}}
\def\Ed{{\text E}}

\def\Hs{{\cal H}}

\def\Ls{{\cal L}}

\def\ung{\hbox{1\hskip -4.2pt \rm 1}}

\def\be{\begin{equation}}
\def\ee{\end{equation}}
\def\beqn{\begin{eqnarray}}
\def\eeqn{\end{eqnarray}}
\def\nobeqn{\begin{eqnarray*}}
\def\noeeqn{\end{eqnarray*}}
\def\ba{\left(\begin{array}}
\def\ea{\end{array} \right) }

\def\bpr{\paragraph{Proof.}}

\def\epr{\square\vskip 6pt}
\def\eop{\hbox{\vrule height .9ex width .8ex depth -.1ex}}

\def\o{\overline}
\def\and{\; \mbox{and} \;}

\newcommand{\half}{\frac{1}{2}}

\def\hfl#1#2{\smash{\mathop{\hbox to 12mm{\rightarrowfill}}
\limits^{\scriptstyle #1}_{\scriptstyle #2}}}

\def\vfl#1#2{\llap{$\scriptstyle #1$}\left\downarrow
\vbox to 6mm{}\right. \rlap{$\scriptstyle #2$}}

\def\vupfl#1#2{\llap{$\scriptstyle #1$}\left\uparrow
\vbox to 6mm{}\right. \rlap{$\scriptstyle #2$}}

\def\Ker{\mathop{\rm Ker}\nolimits}

\def\mod{\mathop{\rm mod}\nolimits}

\def\Be{\begin{enumerate}}
\def\Ee{\end{enumerate}}

\def\Bena{\begin{enumerate}
\def\labelenumi{\theenumi)}
\def\theenumi{\arabic{enumi}}
\def\labelenumii{\theenumii)}
\def\theenumii{\alph{enumii}}}

\def\Bean{\begin{enumerate}
\def\labelenumii{\theenumii)}
\def\theenumii{\arabic{enumii}}
\def\labelenumi{\theenumi)}
\def\theenumi{\alph{enumi}}}

\def\Bero{\begin{enumerate}
\def\labelenumii{\theenumii)}
\def\theenumii{\arabic{enumii}}
\def\labelenumi{(\theenumi)}
\def\theenumi{\roman{enumi}}}

\def\BeRo{\begin{enumerate}
\def\labelenumii{\theenumii)}
\def\theenumii{\arabic{enumii}}
\def\labelenumi{(\theenumi)}
\def\theenumi{\Roman{enumi}}}

\def\Bi{\begin{itemize}\itemsep=11pt}
\def\Ei{\end{itemize}}

\def\Bd{\begin{description}}
\def\Ed{\end{description}}

\def\R{\right}
\def\L{\left}
\def\F{\frac}

\def\Hfl#1#2{\smash{\mathop{\hbox to 14mm{$\Longleftarrow\joinrel\Longrightarrow$}}
\limits^{\scriptstyle #1}_{\scriptstyle #2}}}

\def\aa{{\mathbb{A}\,}}
\def\Aa{{\mathbb{A}\,}}
\def\ZZ{{\mathbb{Z}\,}}
\def\FF{{\mathbb{F}\,}}
\def\cit{{\mathbb{C}\,}}
\def\NN{{\mathbb{N}\,}}
\def\QQ{{\mathbb{Q}\,}}
\def\pZ{{\rm{Z}}\,}
\def\ptaa{{\mathbb{A}\,}}

\def\labelenumi{\alph{enumi})}
\def\theenumi{\alph{enumi}}
\def\labelenumii{\arabic{enumii}.}
\def\theenumii{\arabic{enumii}}

\def\Aut{\operatorname{Aut}}
\def\Hom{\operatorname{Hom}}
\def\Ind{\operatorname{Ind}}
\def\End{\operatorname{End}}
\def\trace{\operatorname{trace}}
\def\Frob{\operatorname{Frob}}
\def\cent{\operatorname{cent}}
\def\Gal{\operatorname{Gal}}
\def\Im{\operatorname{Im}}
\def\Rep{\operatorname{Rep}}
\def\Repsp{\operatorname{Repsp}}
\def\AnRepsp{\operatorname{An\,Repsp}}
\def\RL{_{R,L}}

\def\FRep{\operatorname{FRep}}

\def\GL{$GL$}

\def\ELLIP{\operatorname{ELLIP}}
\def\Irr{\operatorname{Irr}}
\def\Int{\operatorname{Int}}
\def\Hilb{\operatorname{Hilb}}

\def\mit{{\mathbb{M}\,}}
\def\MM{{\mathbb{M}\,}}
\def\ptmit{{\mathbb{M}\,}}

\def\EE{E}

{\theorembodyfont{\rmfamily}
  \newtheorem{defi}{Definition}[subsection]}
{\theorembodyfont{\rmfamily}
  \newtheorem{defis}[defi]{Definitions}}
\newtheorem{lm}[defi]{Lemma}
\newtheorem{propo}[defi]{Proposition}
\newtheorem{coro}[defi]{Corollary}
{\theorembodyfont{\rmfamily}
  \newtheorem{rem}[defi]{Remark}}
{\theorembodyfont{\rmfamily}
  \newtheorem{rms}[defi]{Remarks}}
\newtheorem{axiom}[defi]{Axiom I}
\newtheorem{Axiom}[defi]{Axiom II}
{\theorembodyfont{\rmfamily}
  \newtheorem{sub}[defi]{}}

\def\lr{left (resp. right) }
\def\rl{right (resp. left) }
\def\gl{{\mathfrak{g}}\ell}

\def\To{\begin{CD} @>>>\end{CD}}

\begin{document}

\pagestyle{myheadings}

\def\thepage{\arabic{page}}

\title{Algebraic quantum theory}
\author{Christian Pierre (Université de Louvain)}
\date{}

\thispagestyle{empty}

\maketitle

{\begin{abstract}{The main objective consists in endowing the elementary particles with an algebraic space-time structure in the perspective of unifying quantum field theory and general relativity: this is realized in the frame of the Langlands global program based on the infinite dimensional representations of algebraic groups over adele rings. In this context, algebraic quanta, strings and fields of particles are introduced.}\end{abstract}}
\vskip 11pt

\tableofcontents

\section*{Introduction}

\addcontentsline{toc}{section}{Introduction}

This work is a first attempt for endowing the elementary particles with an
algebraic space-time structure. The quantum essence of the quantum (field)
theory is then of algebraic nature and the most adequate mathematical frame
envisaged to carry out this project is the Langlands program which sets up bijections between  the set of equivalence classes of representations of the Weil-Deligne group and the equivalence classes of cuspidal representations of the general linear group. \vskip 11pt

The algebraic part of the Langlands program is realized by the Galois cohomology and more particularly by the Eisenstein cohomology which, being in one-to-one correspondence with the representation of  the general linear group, constitutes a representation of the Weil-Deligne group while the analytic part of the Langlands program is given by the cuspidal representations of the general linear group. The cuspidal representation of a general linear group is constituted by the sum of its irreducible representations inflated from the corresponding unitary irreducible representations of the additive group of $\rit$ or $\cit$~.
\vskip 11pt

But, instead of working with a linear mathematical frame as commonly envisaged
in quantum theories, a bilinear mathematical frame will be considered for
describing the structure of the elementary particles.  
This is justified mathematically in the sense that  the enveloping algebra of a
given algebra allows to define a representation of this algebra.  This leads us to
consider that if an algebra is supposed to be tractable physically, for example by
a system of observations and measures, its mathematical management will only be
reached by considering its enveloping algebra, i.e. by tensoring the given algebra
by its opposite algebra.  Generally, if a given algebra can become ``measurable'',
its opposite algebra will not be and its enveloping algebra will constitute the
manageable algebra corresponding to an objectivable reality.  But then, the quantum
theories should be of bilinear nature as the invariants of the group theory and,
more particularly, as the invariants of special relativity.  If bilinearity is
taken into account, then the quantum theories can be merged fashionably with the
special relativity: this allows
 to find a simple solution to the problem of the interactions between
elementary particles in the sense that the inextricable $N$-body problem
\cite{Der} becomes easily solvable if it is replaced by a $N$-bibody problem. 
Owing to that, a new light is brought to the gravitational force which then
results from diagonal interactions between bibodies.
\vskip 11pt

An elementary particle must then be considered as a biobject, called a
bisemiparticle: it is composed of the union of a left and a right semiparticle
localized respectively in the upper  and in the lower half space: this
constitutes the basis of the model worked out by the author since the end of
the seventies.  The fundamental algebraic (space-)time structure of a
bisemiparticle is then given by a ``physical field'' consisting of a sheaf of rings on
a bisemimodule defined by the tensor product of
a right and a left semimodule respectively over a right and a left adele (semi)ring.
 \vskip 11pt

Bisemiparticles find a physical conceptual basis in the fact that every action
implies and needs a reaction.  This concept of action-reaction was worked out
by several authors in the frame of quantum mechanics: let us mention the
famous paper of R.P. Feynman and J.A. Wheeler \cite{F-W} on the absorber theory
of radiation and other papers \cite{Whe} such as, for example, the paper of
C.W. Rietdyk \cite{Rie} conjecturing a retroactive influence based on the EPR
paradox.
\vskip 11pt

The vacuum implicit in the Dirac theory by negative energy levels \cite{Dir1},
\cite{Dir2} corresponds in this bisemiparticle model to the existence of a right
semiparticle associated to a left semiparticle.  It is likely that P.A.M. Dirac
had several times the presentiment of this hidden reality, especially in his
celebrated paper ``the quantum theory of the electrons'' \cite{Dir1} and in a
more recent paper \cite{Dir6} on the epistemology of relativity and quantum
mechanics.  Notice that the set of dual right semiparticles associated to the left semiparticles in the bisemiparticles could be a candidate for the dark matter.
\vskip 11pt

On the other hand, it is conceptually acceptable to think that elementary particles have
an internal structure which looks pointlike to the observer but which must be
very complex in order to explain their transformations and decays.  Furthermore, a spinning particle cannot be pointlike.
\vskip 11pt

The idea then consists in endowing elementary (semi)particles with an internal algebraic space-time structure from which their ``mass'' shell could be generated.  Indeed, a way of bridging the gap between quantum field theory and general relativity is to consider that the expanding space-time, to which the cosmological constant of the general relativity equations can correspond, could constitute the fundamental structure of the vacuum of quantum field theory \cite{Pie1}.  As, this vacuum of QFT is generator of matter, it is natural to admit that its space-time structure will generate matter, i.e. massive elementary particles due to the fluctuations of these elementary vacua associated to local strong curvatures of the space-time at the origin of degenerated singularities.  So, the vacuum of the QFT becomes peopled of massless bisemiparticles potentially able to generate their mass shells due to the fluctuations of these bisemiparticles internal vacua which could contribute to the dark energy.  If the geometry of general relativity is envisaged at the elementary particle level and if elementary particle internal vacua are taken into account, then the vacuum quantum fields correspond to them in the perspective of the Langlands program.  Indeed, the generation of the ``discontinued'' algebraic space-time obtained by the (representation of) algebraic groups resulting from the Eisenstein cohomology of Shimura varieties is in bijection with its analytic ``continued'' counterpart given by global elliptic modules (which are truncated Fourier series) included in the corresponding automorphic forms.  So, the ``discontinued'' behavior of the space-time of quantum field theory is in bijection with the ``continued'' geometry of space-time of classical general relativity by means of the Langlands correspondences. \vskip 11pt

More concretely, the internal vacuum space-time structure of an elementary bisemiparticle will originate from its internal time structure which will be localized in the orthogonal complement space with respect to its space structure.  So, the internal time structure of a bisemiparticle, which is its vacuum time field, will be assumed to be of algebraic nature and will correspond to a bisemisheaf of rings over
a general bilinear algebraic semigroup.

The ``time'' bilinear algebraic semigroup will decompose into conjugacy ``bi''classes which are in one-to-one correspondence with the ``bi''places of the considered algebraic extension (bisemi)field.  The functional representation space of the ``time'' bilinear algebraic semigroup is given by the bilinear Eisenstein cohomology which decomposes following the bicosets of the Shimura bisemivariety such that the conjugacy ``bi''classes of the bilinear algebraic semigroup correspond to the Shimura bisemivariety bicosets which are $G_t(\aa_R\times \aa_L)$-subbisemimodules decomposing into one-dimensional irreducible (bi)components.  Note that the conjugacy classes of $G_t(\aa_R\times \aa_L)$ are defined with respect to its smallest ramified  normal bilinear subsemigroup which 
implies that the equivalent representatives of the $\mu $-th conjugacy class of $G_t(\Aa_R\times \Aa_L)$ can be cut into $(p)+\mu $ equivalent conjugacy subclass representatives having a rank equal to $N^2$ and
interpreted as time biquanta, i.e. the products of  right quanta by  left quanta.  So, the representations of the one-dimensional components of the right and left conjugacy classes of $G_t(\aa_R\times \aa_L)$ are one-dimensional subsemimodules whose rank is a multiple of the rank $N$ of a time quantum constituting also the representation of the global inertia subgroup of the considered conjugacy class.  These one-dimensional subsemimodules are isomorphic to one-dimensional (semi)tori constituting the irreducible analytic representations of $G_t(\aa_R\times \aa_L)$ associated to the considered right or left places of the algebraic real number semifield and are interpreted physically as elementary ``time'' waves and strings. \vskip 11pt

So the vacuum algebraic time
structure of a left and of a right semiparticle will be given by a set of correlated
left and right waves represented by a  left and a right
time semisheaf of rings generated by Eisenstein cohomology from a
$1D$-time symmetric splitting semifield $L^{\mp}$~.  The union of these left and right 
semisheaves of rings is the vacuum time ``physical field'' of the particle.

But, the Eisenstein cohomology needs a cuspidal automorphic representation allowing to give an analytic representation to the bilinear algebraic semigroup $G_t(\Aa_R\times \Aa_L)$~.  In this purpose, it is assumed that the space of global elliptic semimodules is included into the space of cusp forms so that the ring of endomorphisms acting on global elliptic bisemimodules is generated by the tensor product of Hecke operators whose coset representatives are given in function of the decomposition group associated to the split Cartan subgroup.

These global elliptic bisemimodules are expanded in formal power series whose coefficients can be obtained from the eigenvalues of the coset representatives of the tensor product of Hecke operators.  Each term of a global  elliptic semimodule is a one-dimensional irreducible torus whose radius can be obtained from the coefficient mentioned above.

In fact, only global elliptic bisemimodules have a real meaning and decompose into a sum of pairs of one-dimensional tori constituting irreducible analytic representations on pairs of places of the considered algebraic extension (bisemi)field.

The  algebraic space structure of a semiparticle can also be constructed as the functional representation of a bilinear algebraic semigroup in the context of the Langlands program or can be 
 generated from
its $1D$-time wave structure by a $(\gamma _{t\to r}\circ E)$ morphim where:

\Be
\item $E$ is an endomorphism based upon a Galois antiautomorphism which
transforms the Eisenstein cohomology in Eisenstein homology and in a
complementary Eisenstein cohomology associated to the generation of a
disconnected complementary $1D$-time wave structure.
\vskip 11pt

\item $\gamma _{t\to r}$ is a morphism transforming partially the complementary
$1D$-time wave structure which is a $1D$-time semisheaf into a complementary
$3D$-space semisheaf representing the structure of a spatial wave.
\Ee
\vskip 11pt

If the smooth endomorphism $E_t$ is such that the complementary Eisenstein
cohomology is associated to the generation of a complementary connected
semisheaf which is proved to be three dimensional, then the resulting
complementary $3D$-semisheaf will constitute the basic time structure of the
three semiquarks of a semibaryon.

The fundamental $1D$-time structure of a semibaryon is thus composed of a
$1D$-core time semisheaf generated by Eisenstein homology on the basis of the
smooth endomorphism $E_t$ and of three $1D$-time complementary semisheaves
generated by the complementary Eisenstein cohomology and constituting the time
structure of the three semiquarks.  The space structure of the semiquarks is
then obtained by $(\gamma _{t_i\to r_i}\circ E_i)$ morphisms, $1\le i\le 3$~,
as explained above.
\vskip 11pt

Semiphotons result from the nearly complete transformation of $1D$-time wave
semisheaves into $1D$-complementary space wave semisheaves by the $(\gamma
_{t\to r}\circ E)$ morphism.

The vacuum space-time structure of semiparticles is thus assumed to be given by:
\Be
\item the number of sections of the space-time semisheaves representing their
structure;
\item the set of ranks of these sections and especially the set of
parameters $c_{t\to r}(\rho )_{R,L}$ measuring the generation of the
complementary $3D$-space semisheaf with respect to the reduced $1D$-time
semisheaf.\Ee
\vskip 11pt

Each one-dimensional section representative $\mu$ of the semiparticle time or space semisheaf of rings has thus a rank $n_\mu=(p+\mu)N$ and is composed of $(p+\mu)$ quanta having a rank $N$~.  And, a one-dimensional spatial section representative $\mu$~, which is a string, is interpreted as the internal vacuum structure of a semiphoton at $(p+\mu)$ quanta.  By this way, the vacuum time and space structure of semiparticles is quantified.
\vskip 11pt

As the vacuum fundamental space-time structure of semiparticles is strongly perturbed
because it is assumed to have likely a spatial extension of the order of the
Planck length, singularities are generated on the sections of the space-time
semisheaves $\theta ^{1-3}_{R,L}(t,r)_{ST}$~.  Consequently, these sections
are  submitted to versal deformations and spreading-out isomorphisms. Recall
that a spreading-out isomorphism constitutes an algebraic extension of the
quotient algebra of the corresponding versal deformation such that the base
sheaves  of the quotient algebra can be pulled out partially or
completely by a blowing-up morphism, called spreading-out map which depends on
a smooth endomorphism based on a Galois antiautomorphism.
\vskip 11pt

As the base sheaves  of the versal deformation do not necessarily cover compactly the
fundamental space-time semisheaf, a gluing-up of these base sheaves  
  is envisaged so that they cover it by patches.
\vskip 11pt

Taking into account the codimension of the singularities on the fundamental
space-time semisheaf $\theta ^{1-3}_{R,L}(t,r)_{ST}$~, it is proved that a
maximum of two successive spreading-out isomorphisms consecutive to versal
deformations can occur leading to the generation of two embedded semisheaves
covering the fundamental space-time semisheaf of a semiparticle.
\vskip 11pt

This allows to give a new light on the nature of the quantum field theory
vacuum which is a state of zero-energy from which elementary particles are
created by pairs.  Indeed, one of the objectives of the present algebraic
quantum theory is to consider that the vacuum of QFT must be viewed as being part of the internal structure of bisemiparticles in the
sense that the fundamental $4D$-space-time semisheaf $\theta
^{1-3}_{R,L}(t,r)_{ST}$ of a semiparticle and the first covering semisheaf
$\theta ^{1-3}_{R,L}(t,r)_{MG}$~, obtained from $\theta ^{1-3}_{R,L}(t,r)_{ST}$
by versal deformation and spreading-out isomorphism and
called the ``middle-ground'' structure, constitute the ``vacuum physical semifield'' of the
semiparticle from which the second covering semisheaf $\theta
^{1-3}_{R,L}(t,r)_{M}$~, obtained from $\theta
^{1-3}_{R,L}(t,r)_{MG}$ by versal deformation and spreading-out isomorphism,
constitute the mass physical semifield of the semiparticle.  Note that this way of generating the mass of a (semi)particle replaces advantageously the Higgs mechanism.
\vskip 11pt

These two space-time and middle-ground structures have likely a spatial
extension of the order of the Planck length to which it is well known that
there is a breakdown of the standard quantum field theory \cite{Pen}.

Matter is then created from space-time.  This reflects as aspiration of A.
Einstein (June 9, 1952): ``I wish to show that space-time is not necessarily
something to which one can ascribe a separate existence independently of the
actual objects of physical reality.  Physical objects are not in space, but
these objects are spatially extended.  In this way, the concept of ``empty
space'' loses its meaning''.
\vskip 11pt

The quantification of the $4D$-semisheaves ``~$ST$~'', ``~$MG$~'' and ``~$M$~'', given
by their algebraic structure, involves that the frequencies of vibration of the
semisheaves ``~$ST$~'', ``~$MG$~'' and ``~$M$~'' are quantified.  This allows to
demonstrate that the
$4D$-semisheaf ``~$M$~'' of a semiparticle is observable while the $4D$-semisheaves
``~$ST$~'' and ``~$MG$~''  are unobservable because the vibration frequency of the
semisheaf ``~$M$~'' is inferior to the vibration frequencies of the semisheaves
``~$ST$~'' and ``~$MG$~''.
\vskip 11pt

The semisheaves  $\theta
^{1-3}_{R,L}(t,r)_{ST}$~,  $\theta
^{1-3}_{R,L}(t,r)_{MG}$ and  $\theta
^{1-3}_{R,L}(t,r)_{M}$ constitute commutative algebras while one of these
algebras extended by versal deformation(s) and spreading-out isomorphism(s)
becomes noncommutative.

\vskip 11pt

As it was briefly taken up above, the algebraic structure of the ``~$ST$~'',
``~$MG$~'' and ``~$M$~'' levels of an elementary particle is crudely given by bisemisheaves over
$10D$-bisemimodules of the corresponding bisemiparticle composed of the union
of a left and of a right semiparticle.  It is then demonstrated that a
``~$ST$~'', ``~$MG$~'' or ``~$M$~'' bisemimodule, noted $(M_{R;ST,MG,M}\otimes
M_{L;ST,MG,M})$ can break down under a blowing-up morphism into:
\Be
\item a diagonal bisemimodule  $(M_{R;ST,MG,M}\otimes_{D}
M_{L;ST,MG,M})$ of dimension 4, characterized by a flat geometry and a diagonal
orthogonal basis: it gives the diagonal central bistructure of the ``~$ST$~'',
``~$MG$~'' or ``~$M$~'' level of the bisemiparticle;
\item a magnetic bisemimodule  $(M^S_{R;ST,MG,M}\otimes_{m}
M^S_{L;ST,MG,M})$ of dimension 3, characterized by a non\-orthogonal basis and a
metric called magnetic.  It is composed of  ``~$ST$~'',
``~$MG$~'' or ``~$M$~'' magnetic biquanta and constitutes the magnetic moment of
the corresponding level of the bisemiparticle;  this magnetic bisemimodule results from the off-diagonal spatial interactions between a left and a right semiparticle;
\item an electric bisemimodule $(M^{T-(S)}_{R;ST,MG,M}\otimes_{e}
M^{S-(T)}_{L;ST,MG,M})$ of dimension 3, characterized by 
 a metric called electric. It is composed of time-space or space-time
biquanta which constitute the electric charge of the corresponding level of the
bisemiparticle. This electric bisemimodule results from off-diagonal interactions between the time (resp. space) components of the right semiparticle and the space (resp. time) components of the associated left semiparticle.

\Ee
\vskip 11pt

The diagonal, magnetic and electric bisemimodules are defined respectively
by the ``diagonal'', ``magnetic'' and ``electric'' tensor products between the right
and left semimodules $M_R$ and $M_L$ (resp. $M^S_R$ and $M^S_L$ or $M^S_R$ and
$M^T_L$~,
\ldots).
\vskip 11pt

If we consider the projection of the right (resp. left) semimodule on the left
(resp. right) semimodule, then the right (resp. left) semimodule becomes the
dual semimodule of the left (resp. right) semimodule.  

Furthermore, a bijective
linear isometric map from the projected right (resp. left) semimodule to the
left (resp. right) semimodule transforms each covariant element into a
contravariant element and gives rise to a left (resp. right) diagonal, magnetic
or electric bisemimodule.

Associated with the appropriate internal bilinear form, the left (resp. right)
diagonal, magnetic or electric bisemimodule allows to define a left (resp.
right) bilinear internal Hilbert, magnetic or electric space.
\vskip 11pt

The algebras of operators acting on bilinear Hilbert, magnetic and electric
spaces are bialgebras of bioperators owing to the bilinearity of these spaces. 
We have thus to consider bialgebras of von Neumann of bounded bioperators
acting on these bilinear spaces.  As the representation space of a given algebra is
isomorphic to its enveloping algebra, the extended bilinear Hilbert spaces
characterized by a nonorthogonal Riemanian metric will be taken as natural
representation spaces for the bialgebras of bounded operators.

In this context, an (elliptic differential) bioperator (~$T_R\otimes T_L$~) maps 
the bisemisheaf $\widetilde M_R\otimes \widetilde M_L$ on the 
$GL_n(\aa_R\times \aa_L)$-bisemimodule $M_R\otimes M_L$ into
the bisemisheaf $\widetilde M_R^a\otimes \widetilde M_L^a$ on
 the shifted $GL_{n[m]}((\aa_R\otimes\cit)\times (
\aa_L\otimes \cit))$-bisemimodule $(M^a_R\otimes M_L^a)$~.   $ \widetilde M^a_R\otimes  \widetilde M_L^a$ is a perverse bisemisheaf whose sections are defined over the conjugacy classes $\mu$ with multipliticies $m_\mu$
of $GL_{n[m]}((\aa_R\otimes\cit)\times (
\aa_L\otimes \cit))$.  Now, the Langlands 
program, setting up bijections between the algebraic $GL_n(\aa_R\times \aa_L)$-bisemimodule 
$(M_R\otimes M_L)$ and the corresponding analytic global elliptic bisemimodule $((\phi _R(s_R)
\otimes \phi _L(s_L))$~, can be extended, under the action of a bioperator, into a shifted 
Langlands program establishing bijections between the shifted bisemimodule $(M_R^a\times M_L^a)$ and the corresponding analytic shifted global elliptic bisemimodule $(\phi _R^a(s_R)\otimes\phi _L^a(s_L))$~.  An eigenbivalue equation directly follows from the shifted global elliptic bisemimodule such that its eigenbivalues form an embedded sequence in one-to-one correspondence with the embedded eigenbifunctions of $T_R\otimes T_L$ which form an increasing sequence of truncated global elliptic bisemimodules, i.e. products, right by left, of truncated Fourier series.  In this perspective, a bisemiparticle spatial wave bifunction will be given by the  analytical representation of the $GL_2(\aa_R\times \aa_L)$-bisemimodule 
and will consist in an increasing set of products, right by left, of truncated Fourier series whose number of terms corresponds to the number of considered conjugacy class representatives of $GL_2(\aa_R\times \aa_L)$~: so, a bisemiparticle wave bifunction is described in terms of its spectral representation which corresponds to the classical assertion saying that the spectral theorem is equivalent to consider that any unitary representation of a compact Lie group is a direct sum of irreducible representations.

As the consequence of the algebraic spectral representation of a wave bifunction, the group of automorphisms of an analytic
von Neumann algebra is isomorphic to the group of ``shifted'' automorphisms of Galois which
has for consequence that the entire dimensions of the von Neumann algebras are in fact the integers labelling the classes of degrees of Galois extensions.

It is then proved that the discrete spectrum of a bioperator is obtained by means
of an isomorphism from the bialgebra of von Neumann on an extended bilinear Hilbert
space to the corresponding bialgebra of von Neumann on a bilinear diagonal Hilbert
space.
\vskip 11pt

The bilinear structure of this quantum theory involves that:
\Bena
\item the traditional calculus with the amplitudes of probability of quantum
field theories is replaced by a calculus with intensities of probability;
\item the rotation of the sections of the semisheaves of a right (resp. left)
semiparticle with respect to the sections of the semisheaves of its associated
left (resp. right) semiparticle allows to define the internal angular momentum
of the right (resp. left) semiparticle from which it results that a right and a
left semiparticles rotate in opposite senses and have only two possible spin
states.
\Ee
\vskip 11pt

To each $1D$- and $3D$-``~$ST$~'',
``~$MG$~'' and ``~$M$~'' semisheaf corresponds a phase space  which has
the structure of a $F$-Steenrod
bundle whose basis is given by the considered semisheaf.

The bisections of the bisemisheaves are tensor products of differentiable functions for which wave
equations are studied: they are degenerated second order elliptic differential
bilinear equations.
\vskip 11pt

The right (resp. left) wave function, solution of a wave equation, is proved to have a spectral decomposition in terms of eigenfunctions having an algebraic representation as mentioned above.  The statistical interpretation of the wave function is the same as in quantum theories.  This allows to reconcile the Bohr and Einstein points of view about quantum theories.
\vskip 11pt

Every elementary bisemiparticle has a ``mass'' central algebraic structure composed of pairs of right and left one-dimensional sections (which are in fact one-dimensional waves or open strings) behaving like (damped) harmonic oscillators.  Thus, a ``field'' (in the physical sense, but having an algebraic structure) can be associated to the mass structure of each elementary bisemiparticle.  And, as each one-dimensional subsection is an open string, this algebraic quantum model has also a ``string'' aspect.
\vskip 11pt

The mass equation for a bisection  ``~$\mu$~'' of the bisemielectron is especially
considered: it is the equation of a damped harmonic oscillator whose general
solution consists in the superposition of two damped waves in phase opposition
with frequencies given by $E_\mu=\frac \hbar c\nu_\mu s$ and whose general
motion corresponds to a damped sinusoidal motion whose dephasage is
proportional to the linear momentum of the considered left (resp. right)
section of the left (resp. right) semielectron.

It is proved that the energy of a section $s_\mu$ at $\mu_p=p+\mu$ quanta can be given 
in function of the energy $E^I_\mu$ of a quantum on this section following $E_\mu=\mu_pE^I_\mu$~.  And, 
the energy $E^I_\mu$ of a quantum $\widetilde M^I_\mu\in s_\mu$ can be calculated from the analytic development of the corresponding 
nontrivial zero of the Riemann Zeta function $\zeta (s)$~.
\vskip 11pt

The internal machinery of a bisemiparticle allows to justify the absorption and
the emission of right and left quanta.  In fact, each spatial one-dimensional bisection of the  
``~$ST$~'',
``~$MG$~'' and ``~$M$~'' bisemisheaf behaves globally like two adjacent
gyroscopes having opposite torques which allows to understand that diagonal
biquanta are emitted under the action of a diagonal centrifugal biforce
represented mathematically by a diagonal smooth biendomorphism corresponding to an inverse deformation and that
magnetic biquanta are emitted under the action of a Coriolis biforce
represented mathematically by a magnetic smooth biendomorphism.

The emission and reabsorption of left and right magnetic quanta by left and
right semisheaves having different magnitudes of rotational velocities generate by
reaction a global movement of translation of the bisemiparticle.
\vskip 11pt

The structure of bisemiparticles is given by bisemisheaves so that an
action-reaction process is generated by the interactions between the right
semisheaves of the right semiparticle and the left semisheaves of the left
semiparticle.  Generalizing this concept to a set of bisemiparticles, it can be
easily demonstrated that the interactions between a set of bisemiparticles
result from the interactions between the right and left semisheaves belonging
to different bisemiparticles, leading to a set of mixed action-reaction
processes of bilinear nature associated to interferences.  This allows to get rid of the inextricable
problems of Physics originated from linearity as A. Einstein outlined in
\cite{Ein5}: ``Linear laws have solutions which satisfy the superposition
principle but they do not describe the interactions between elementary
particles''.

The general mathematical frame allowing to describe the interactions between a set of $N$ bisemiparticles is the Langlands reducible program as developed in \cite{Pie9}.  In this context, the spatial mass structure (i.e. the ``mass'' field) of $N$ interacting bisemiparticles is given by the 
nonorthogonal completely reducible functional representation space of $GL_{2N}(\aa_R\times \aa_L)$ as introduced in chapter 5.
\vskip 11pt

The bilinear interactions generate gravitational, magnetic and electric
biquanta giving rise to a gravito-electro-magnetic field such that the
gravitation results from diagonal interactions between bisemiparticles while the
electromagnetism originates from off-diagonal interactions.

It is then proved that:
\Be\item a set of bisemifermions interact by means of a
gravito-electro-magnetic field;
\item a set of bisemiphotons interact by means of a gravito-magnetic field;
\item a set of bisemifermions and of bisemiphotons interact by means of a
gravito-electro-magnetic field.
\Ee

\vskip 11pt

The biwave equation of $N$ interacting bisemiparticles separates automatically
into $N_q$ biwave equations of the $N_q$ bisections of the $N$ bisemiparticles
and into $((N_q)^2-N_q)$ biwave equations referring to the interactions between
the right and left sections of these $N$ bisemiparticles.

In this context, the antisymmetric electromagnetic field tensor is replaced by
a gravito-electro-magnetic tensor whose diagonal components are the components
of a gravitational field.  This leads to a new conceptual approach of the electromagnetism and of the quantum gravity.
\vskip 11pt

In this algebraic quantum model, the strong interactions and the cause of the
confinement of the semiquarks result from the new structure proposed for the
semibaryons.  Indeed, the confinement of the semiquarks originates from the
generation of the three semiquarks from the core time semisheaf of the
semibaryon by a smooth endomorphism $E_t$~.  The core time structure of a
semibaryon is physically justified by the fact that the quarks contribute only
to about $15\%$ of the spin of the nucleon \cite{Ash}.  

We then have that a
right and a left semibaryon of a given bisemibaryon interact by means of:
\Be
\item the electric charges and the magnetic moments of the three bisemiquarks;
\item a gravito-electro-magnetic field resulting from the bilinear interactions
between the right and the left semiquarks of different bisemiquarks;
\item a strong gravitational and electric fields resulting from the bilinear
interactions between the central core structures of the left and right
semibaryons and the right and left semiquarks.
\Ee
\vskip 11pt

The leptonic decay of a bisemibaryon results essentially from the diagonal
emission of a bisemilepton throughout a diagonal biendomorphism.  The emitted
bisemineutrino allows to take into account the bilinear interactions between
the emitting bisemiquark and the emitted bisemilepton.

The nonleptonic decay of a bisemibaryon consists essentially in the emission
of a meson by a bisemiquark throughout a nonorthogonal biendomorphism.
\vskip 11pt

Finally, it is shown that the EPR paradox receives a new lighting because the
linear frame of quantum theories is replaced by a bilinear frame so that two
elementary bisemiparticles interact through the space by means of a
gravito-magneto-(electric) field and nonlocally through the time by means of a $1D$-time
gravitational field. \vskip 11pt

Let us also make the following last remarks:

\Bena
\item This algebraic quantum theory is an algebraic quantum field theory describing the structure of elementary particles in terms of bisemiparticles from their internal algebraic space-time structures interpreted as elementary internal vacua whose union corresponds to the essential part of the vacuum of QFT.

This algebraic quantum theory lies on the Langlands mathematical program and does not proceed from Lagrangian methods of classical mechanisms as the quantum field theories.  This theory is thus not a priori directed towards the description of the trajectories of particles. However, the study of the structure of bisemiparticles corresponds to bringing up to light the existence of an internal dynamics with respect to internal variables of which the most popular are the proper time and the proper mass.  But, this elementary ``internal dynamics'' also evolves with respect to an external time variable throughout the equivalent of the Stone theorem and leads to an ``external dynamics'' corresponding to the classical or quantum dynamics.
\vskip 11pt

\item In this AQT, all observables are quantified due to the algebraic nature of the theory: thus, the internal time, the internal space, the mass, the energy, the linear momentum, the charge, the electromagnetism and the gravitation are quantified.
\vskip 11pt

\item The internal structure of a massive bisemiparticle is composed of biwave packets localized into the $1D$-time and $3D$-spatial orthogonal spaces: the $3D$-spatial structure of a bisemiparticle thus has a wave aspect which becomes evident when it interacts with other bisemiparticles by interference process.  The corpuscular aspect of the $3D$-spatial structure of a bisemiparticle can become apparent when its $3D$-spatial biwave packet is flattened into two dimensions as resulting from a collision.
\vskip 11pt

\item The internal time structures of semiparticles are perhaps not localized in a traditional one-dimensional time space but in a three-dimensional space.  Then, a magnetic moment and/or field related to $3D$-time structures ought to be envisaged as resulting from off-diagonal interactions between time right-semisheaves and time-left semisheaves.
\vskip 11pt

\item Some of the difficulties of the standard model seem to have been solved in this algebraic quantum model, as for example the origin of the  mass, the nature of the dark matter and energy and the existence of three families of elementary particles.
\Ee

\vskip 11pt

\setcounter{defi}{0}
\section{\boldmath Algebraic representation of the fundamental $4D$-space-time structure
of semiparticles}

\subsection[Generation of $1D$-semisheaves of rings by Eisenstein cohomology]{\boldmath Generation of $1D$-semisheaves of rings by Eisenstein cohomology}

The aim of this section consists in the generation of two symmetric right and left $1D$-time
semisheaves of rings $\theta ^1_R(-t)$ and $\theta ^1_L(+t)$ whose $q$ right
and left sections are continuous functions over the completions of finite Galois extensions of global number field $K$ of characteristic zero.  The finite Galois extensions of $K$ are the splitting field over $K$ of the polynomial ring $K[t]$ in the time indeterminate ``~$t$~''. 

The dimension corresponding to the time variable will be called the
{\em generative dimension\/}.
\vskip 11pt

\paragraph{Notation} : ``~$R,L$~'' means ``~$R$ (respectively $L$~)''.
\vskip 11pt

\begin{defi}[Symmetric polynomial ring] { The polynomial ring $K[t]$ is
assumed to have for elements the polynomials $P_{\mu_\nu } (t)$ and $P_{\mu_\nu } (-t)$ which
are such that~:

\begin{enumerate}
\item All the polynomials $P_{\mu_\nu } (t)$~, $1\le \mu \le q$~, $1\le \nu \le\infty $~, have a same number
of positive simple (real) roots ``~$N^+_{\mu  L}$~'' and a same number of negative
simple (real) roots ``~$N^+_{\mu  R}$~'', i.e. $N^+_{1  L} = \cdots=N^+_{\mu  L}
= \cdots =N^+_{q L}$ and
$N^+_{1  R} = \cdots=N^+_{\mu  R}
= \cdots =N^+_{q R}$ with $N^+_{\mu  L}$ in general not equal to $N^+_{\mu 
R}$ and $N^{\pm}_{\mu  R,L}\in \nit$~.
\vskip 11pt

\item The polynomials $P_{\mu_\nu } (-t)$~, $1\le \mu \le q$~, $1\le\nu \le\infty $~, have a number of
positive simple (real) roots $N^-_{\mu  L}$ equal to the number of negative
simple (real) roots $N^+_{\mu  R}$ of the polynomials $P_\mu (t)$ and vice versa.

So we have that
\begin{enumerate}
\item $N^+_{\mu  L}=N^-_{\mu R}$~, $N^+_{\mu  R}=N^-_{\mu L}$~, $\forall\ \mu
$~, $1\le \mu \le q$~;
\item $N^+_{\mu  L}+N^-_{\mu L}=N^-_{\mu  R}+N^+_{\mu R}$~.
\end{enumerate}
\end{enumerate}} \end{defi} \vskip 11pt

Remark that, when the polynomials $P^c_{\mu_\nu }(t)$ and $P_{\mu_\nu }^c(-t)$ of the polynomial ring
$K[t]$ have simple complex roots, $K[t]$ is manifestly a symmetric polynomial ring if all
the polynomials $P^c_{\mu_\nu} (t)$ and $P^c_{\mu_\nu} (-t)$~,  $1\le\mu \le q$~, $1\le\nu \le\infty $~, have a same number of simple complex roots.
\vskip 11pt

\begin{defi}[Symmetric splitting semifield]  { This polynomial ring $K[t]$
is then composed of a set of pairs of polynomials $\{ P_{\mu_\nu }(t),P_{\mu_\nu }(-t)\}_{\mu
,\nu }$~.  Each set of pairs of polynomials for the index $\mu $ generates the symmetric splitting subfield $L_\mu
$ which is composed of the set of positive simple roots, noted $L^+_\mu $~, and
of the symmetric set of negative simple roots, noted $L^-_\mu $~.  $L_\mu $ is
thus characterized by the properties:
\begin{enumerate}
\item $L_\mu =L^-_\mu \cup L^+_\mu $~.

\item $L^-_\mu \cap L^+_\mu  =\emptyset$~.

\item To each positive simple root $\alpha _{\mu +}\in L^+_\mu $ corresponds
the symmetric negative simple root $\alpha _{\mu -}\in L^-_\mu $~.
\end{enumerate}

$L^+_\mu $ and $L^-_\mu $ are respectively a left and a right algebraic
extension semisubfields.  They are semisubfields because they are commutative
division semisubrings.  They are ``{\em semisubrings\/}'' because $(L^+_\mu ,+)$ and
$(L^-_\mu ,+)$ are abelian semisubgroups $[H-N]$ lacking for inverses with respect
to the addition and endowed with associative multiplication and distributive
laws.
\vskip 11pt

Similarly, each set of pairs of polynomials $\{ P^c_{\mu_\nu } (t),P^c_{\mu_\nu } (-t)\}_{\mu ,\nu }\in K[t]$
generates a complex symmetric splitting subfield $L^c_\mu $ composed of the set of complex simple roots,
noted $L^{c+}_\mu $~, and of the symmetric set of complex conjugate simple roots, noted
$L^{c-}_\mu $~.  $L^{c+}_\mu $ and $L^{c-}_\mu $ are also respectively a left and a right algebraic
extension semisubfields.}
\end{defi}
\vskip 11pt

\begin{defi}[Right and left specializations]  { We consider the right and
left specializations \cite{Wei1} of the right and left semisubrings $A_{\mu  R}$ and
$A_{\mu  L}$ (included respectively in the semirings $A_R$ and $A_L$~) from the
polynomial subring $\{P_{\mu_\nu } (t),P_{\mu_\nu } (-t)\}\in K[t]$~.  

The right (resp. left)
specialization of $A_{\mu  R}$ (resp. $A_{\mu  L}$~) is completely determined by
$p_{\mu  R}$ (resp. $p_{\mu  L}$~) which is a nonzero prime right (resp. left)
specialization ideal of $A_{\mu  R}$ (resp. $A_{\mu  L}$~), i.e. the set of all
negative (resp. positive) nonunits of $A_{\mu  R}$ (resp. $A_{\mu  L}$~), such
that $p_{\mu  R}\cap p_{\mu  L}=\emptyset$~.
\vskip 11pt

We denote by $B_{\mu  R}$ (resp. $B_{\mu  L}$~) the integral closure of
$A_{\mu  R}$ (resp. $A_{\mu  L}$) in $L^-_\mu $ (resp. $L^+_\mu $~) (i.e. the
set of elements of $L^-_\mu $ (resp. $L^+_\mu $~) which are integral over
$A_{\mu  R}$ (resp. $A_{\mu  L}$~)).  Then,  the
right (resp. left) semisubring $B_{\mu  R}$ (resp. $B_{\mu  L}$~) is a finitely
generated $A_{\mu  R}$-right semimodule (resp. $A_{\mu  L}$-left semimodule)
\cite{Ser3}.
\vskip 11pt

Let $b_{\mu  R_{1_\mu} }\subset \cdots\subset b_{\mu R_{n_\mu }}$ (resp.
$b_{\mu  L_{1_\mu}}\subset \cdots\subset b_{\mu L_{n_\mu }}$~) be a chain of
distinct prime right (resp. left) ideals of $B_{\mu  R}$ (resp.
$B_{\mu  L}$~) obtained under the right (resp. left) action of the right (resp.
left) Galois group $\Gamma_{\mu  R} =\Aut_KL^-_\mu $ (resp. 
$\Gamma_{\mu  L} =\Aut_KL^+_\mu $~).

If $p_{\mu  R}=b_{\mu  R_{i_\mu }}\cap A_{\mu  R}$ (resp.
$p_{\mu  L}=b_{\mu  L_{i_\mu} }\cap A_{\mu  L}$~), $1\le i_\mu \le n_\mu $~, then
$b_{\mu  R_{i_\mu}}$ (resp. $b_{\mu  L_{i_\mu }}$~) divides (or is above) $p_{\mu  R}$
(resp. $p_{\mu  L}$~).

Then, $B_{\mu  R}/b_{\mu R_{n_\mu }}$ (resp.  $B_{\mu  L}/b_{\mu L_{n_\mu }}$~) is an extension
of $A_{\mu  R}/p_{\mu R}$ (resp. $A_{\mu  L}/p_{\mu L}$~) of finite degree,
called the right (resp. left) global residue degree of $b_{\mu  R}\equiv b_{\mu R_{n_\mu }}$ (resp. $b_{\mu 
L}\equiv b_{\mu L_{n_\mu }} $~) and noted $f_{b_{\mu R}}$ (resp. $f_{b_{\mu L}}$~).
}\end{defi}
\vskip 11pt

\begin{sub}{\bf Inertia subgroups and adele semirings}
\end{sub}

If the \rl ideal $b_{\mu _R}$ (resp. $b_{\mu _L}$~) is assumed to be unramified, we have more precisely that:
\[ [L^{-(nr)}_\mu :K]=f_{b_{\mu _R}} \qquad \text{(resp.}\quad [L^{+(nr)}_\mu :K]=f_{b_{\mu _L}}\ )\]
where $L^{-(nr)}_\mu $ (resp. $L_\mu^{+(nr)}$~) is a \rl unramified algebraic extension.

Let $\Gal(L^{-(nr)}_\mu /K)$ (resp. $\Gal(L^{+(nr)}_\mu /K)$~) denote the Galois subgroup of the unramified \rl extension $L_\mu ^{-(nr)}$ (resp. $L_\mu ^{+(nr)}$~) of $K $ and let $\Gal(L^{-}_\mu /K)$ (resp. $\Gal(L^{+}_\mu /K)$~) be the Galois subgroup of the corresponding ramified \rl extension $L^-_\mu $ (resp. $L^+_\mu $~).

If $I_{L^-_\mu }$ (resp. $I_{L^+_\mu }$~) denotes the global inertia subgroup of $\Gal(L^-_\mu /K)$ (resp. $\Gal(L^+_\mu /K)$~), then the equalities follow:
\begin{align*}
\Gal(L^-_\mu /K) \big/ I_{L^-_\mu } &= \Gal (L^{-(nr)}_\mu /K)\;, \\
\text{(resp.} \quad 
\Gal(L^+_\mu /K) \big/ I_{L^+_\mu } &= \Gal (L^{+(nr)}_\mu /K)\;), \end{align*}
leading to the exact sequences:
\begin{alignat*}{5}
1 & \To I_{L^-_\mu } & \To \Gal(L^-_\mu /K) & \To \Gal(L^{-(nr)}_\mu /K) &\To& 1\\
\text{(resp.} \quad
1 & \To I_{L^+_\mu } & \To \Gal(L^+_\mu /K) & \To \Gal(L^{+(nr)}_\mu /K) &\To& 1\;).\end{alignat*}
On the other hand, it was seen in \cite{Pie9} that
\begin{align*} [L^{-(nr)}_\mu :K ]&=f_{b_{\mu _R}}=kp+\mu '=p+\mu \;, \quad 1\le\mu \le q\le \infty \;, \\
\noalign{\noindent where $k\le p-1$ is an integer referring to congruence classes modulo $p$ such that $kp+\mu '=p+\mu $}
\text{(resp.} \quad  [L^{+(nr)}_\mu :K]&=f_{b_{\mu _L}}=kp+\mu '=p+\mu \;).\end{align*}
If the global residue degree $f_{b_{\mu _R}}$ (resp. $f_{b_{\mu _L}}$~) is an integer and not an integer modulo $p$~, then $p=0$ and $f_{b_{\mu _R}}=f_{b_{\mu _L}}=\mu $~.

If $N$ denotes the order of the global inertia subgroups $I_{L^-_\mu }$ (resp. $I_{L^+_\mu }$~), 
$1\le \mu \le q\le \infty $~, then the degrees of the right (resp. left) ramified extensions ${L^-_\mu }$ 
(resp. ${L^+_\mu }$~) are given by integers modulo $N$~:
\begin{align*}
n_{\mu _R} &= [L^-_\mu:K ]
=*+f_{b_{\mu _R}}\cdot N
\approx f_{b_{\mu _R}}\cdot N = (p+\mu )N\\
\text{(resp.} \quad 
n_{\mu _L} &= [L^+_\mu:K]
=*+f_{b_{\mu _L}}\cdot N
\approx f_{b_{\mu _L}}\cdot N = (p+\mu )N\;)\end{align*}
where $*$ denotes an integer inferior to $N$~.

Let $L_{\o v_\mu }$ (resp. $L_{v_\mu }$~) be the $\mu $-th completion corresponding to the \rl ramified algebraic extension $L^-_\mu $ (resp. $L^+_\mu $~) and associated to the place $\o v_\mu $ (resp. $v_\mu $~).

The completion $L_{\o v_\mu }$ (resp. $L_{v_\mu }$~), which is a one-dimensional $K$-semimodule, is assumed to be generated from an irreducible (central) $K$-semimodule $L_{\o v^1_\mu }$ (resp. $L_{v^1_\mu }$~) of rank (or degree) $N$ such that $L_{\o v^1_\mu }\simeq p_{\mu _R}$ (resp. $L_{v^1_\mu }\simeq p_{\mu _L}$~).

As a result, $L_{\o v_\mu }$ (resp. $L_{v_\mu }$~) is cut into a set of $(p+\mu )$ equivalent real subcompletions $L_{\o v^{\mu '}_\mu }$ (resp. $L_{v^{\mu '}_\mu }$~), $1\le \mu '\le \mu $~, of rank $N$~: since the rank of
$L_{\o v_\mu }$ (resp. $L_{v_\mu }$~) is also given by:
\begin{align*}
n_{\mu _R} &= [L_{\o v_\mu }:K] \simeq f_{b_{\mu _R}}\cdot N=(p+\mu )\ N\\
\text{(resp.} \quad 
n_{\mu _L} &= [L_{v_\mu }:K] \simeq f_{b_{\mu _L}}\cdot N=(p+\mu )\ N\ ).\end{align*}
So, the ranks or degrees of the real completions $L_{\o v_\mu }$ (resp. $L_{v_\mu }$~),  $1\le \mu \le q$~, are integers of $\ZZ\big/p\ N\ \ZZ$~, noted in condensed form $\o\ZZ_{p_q}$~.

On the other hand, as a place is an equivalence class of completions, we have to consider at a place $\o v_\mu $ (resp. $v_\mu $~) a set of real completions $\{L_{\o v_\mu ,m_\mu }\}$ (resp. $\{L_{v_\mu ,m_\mu }\}$~), $m_\mu \in\NN$~, equivalent to the basic completion $L_{\o v_\mu }$ (resp. $L_{v_\mu }$~) and having the same rank $n_{\mu _R}$ (resp. $n_{\mu _L}$~) as $L_{\o v_\mu }$ (resp. $L_{v_\mu }$~); the integer $m^{(\mu )}=\sup(m_\mu )$ is interpreted as the multiplicity of
$L_{\o v_\mu }$ and $L_{v_\mu }$~.

Then, a \rl ``ramified'' adele semiring $\Aa_{L_{\o v}}$ (resp. $\Aa_{L_v}$~) can be introduced by:
\begin{align*}
\Aa_{L_{\o v}} &= \txt\prod\limits _\mu  L_{\o v_\mu } \txt\prod\limits _{m_\mu } L_{\o v_\mu ,m_\mu }\\
\text{(resp.} \quad 
\Aa_{L_{v}} &= \txt\prod\limits _\mu  L_{v_\mu } \txt\prod\limits _{m_\mu } L_{v_\mu ,m_\mu }\ ).\end{align*}
\pagebreak

\begin{sub}{\bf Representations of the algebraic bilinear general semigroup}
\end{sub}

Let $T_2^t(\Aa_{L_{\o v}})$ (resp. $T_2(\Aa_{L_{v}})$~) denote the matrix algebra of lower (resp. upper)
 triangular matrices of order $2$ over the adele semiring $\Aa_{L_{\o v}}$ (resp. $\Aa_{L_{v}}$~)~. Then, according to \cite{Pie10}, 
an algebraic bilinear general semigroup over the product of $\Aa_{L_{\o v}}$ by $\Aa_{L_{v}}$ can be introduced by:
\[ GL_2(\Aa_{L_{\o v}}\times \Aa_{L_v})=T_2^t(\Aa_{L_{\o v}})\times T_2(\Aa_{L_v})\]
such that:
\Bena
\item $GL_2(\Aa_{L_{\o v}}\times \Aa_{L_v})$ has a bilinear Gauss decomposition:
\[ GL_2(\Aa_{L_{\o v}}\times \Aa_{L_v})=[(D_2(\Aa_{L_{\o v}})\times D_2(\Aa_{L_v})][UT_2(\Aa_{L_v})\times UT^t_2(\Aa_{L_{\o v}})]\]
where
\Bi
\item $D_2(\cdot)$ is a subgroup of diagonal matrices,
\item $UT_2(\cdot)$ is a subgroup of unitriangular matrices;
\Ei
\item $GL_2(\Aa_{L_{\o v}}\times \Aa_{L_v})$ has for modular representation space $\Repsp(GL_2(\Aa_{L_{\o v}}\times \Aa_{L_v}))$ given by the tensor product $M_R\otimes M_L$ of a right $T_2^t(\Aa_{L_{\o v}})$-semimodule $M_R$ by a left $T_2(\Aa_{L_v})$-semimodule $M_L$~.
\Ee

$M_R$ (resp. $M_L$~) decomposes into $T^t_2(L_{\o v_{\mu }})$-subsemimodules $M_{\o v_{\mu ,m_\mu }}$ 
(resp. $T_2(L_{v_{\mu}})$-subse\-mi\-modules $M_{v_{\mu ,m_\mu }}$ ~) following:
\[ M_R=\txt\bigoplus\limits^q_{\mu =1}\txt\bigoplus\limits_{m_\mu }M_{\o v_{\mu,m_\mu } }\qquad
\text{(resp.} \quad
M_L=\txt\bigoplus\limits^q_{\mu =1}\txt\bigoplus\limits_{m_\mu }M_{v_{\mu ,m_\mu }}\;).\]

Each $T_2^t(L_{\o v_\mu })$-subsemimodule $M_{\o v_{\mu,m_\mu } }$ (resp. $T_2(L_{v_\mu })$-subsemimodule $M_{v_{\mu,m_\mu } }$~) constitutes an equivalent representative of the $\mu $-th conjugacy class of $T_2^t(\Aa_{L_{\o v}})$
 (resp. $T_2(\Aa_{L_{v}})$~) with respect to the fixed global inertia subgroup $I_{L_\mu }$ and has a rank given by $n_{\mu _R}=(p+\mu )\cdot N$ (resp. $n_{\mu _L}=(p+\mu )\cdot N$~). So, the $T^t_2(\Aa_{L_{\o v}})$-semimodule $M_R$ (resp. $T_2(\Aa_{L_{v}})$-semimodule $M_L$~) has a rank:
\begin{align*}
n_R &= \txt\bigoplus\limits^q_{\mu =1} \txt\bigoplus\limits_{m_\mu} n_{\mu _R}=\txt\bigoplus\limits_\mu  \txt\bigoplus\limits_{m_\mu} (p+\mu )\cdot N\\
\text{(resp.} \quad
n_L &= \txt\bigoplus\limits^q_{\mu =1} \txt\bigoplus\limits_{m_\mu} n_{\mu _L}=\txt\bigoplus\limits_\mu  \txt\bigoplus\limits_{m_\mu} (p+\mu )\cdot N\;).\end{align*}

On the other hand, the \rl global inertia subgroup $I_{L_{\o v_\mu }}$ (resp. 
$I_{L_{v_\mu }}$~) has a representation space given by $\Repsp(T_2^t(L_{\o v_{\mu}^1}))$ (resp. 
$\Repsp(T_2^t(L_{v_{\mu}^1}))$~) where $L_{\o v_\mu ^1}$ (resp. $L_{v_\mu ^1}$~) is an irreducible completion of rank $N$ as introduced in section 1.1.4.

\begin{sub}{\bf Quanta, strings and field are introduced}
\end{sub}

Consequently, each representative $M_{\o v_{\mu ,m_\mu }}$ (resp. $M_{v_{\mu ,m_\mu }}$~) of the 
$\mu $-th conjugacy class of $T^t_2(\Aa_{L_{\o v}})$ (resp. $T_2(\Aa_{L_{v}})$~) is cut into $(p+\mu )$ equivalent 
conjugacy subclass representatives $M_{\o v^{\mu '}_{\mu ,m_\mu }}$ (resp. $M_{v^{\mu '}_{\mu ,m_\mu }}$~), 
$1\le \mu '\le \mu $~, having a rank equal to $N$ and being in one-to-one correspondence with the $(p+\mu )$ 
equivalent subcompletions $L_{\o v^{\mu '}_{\mu ,m_\mu }}$ (resp. $L_{v^{\mu '}_{\mu ,m_\mu }}$~) of 
$L_{\o v_\mu }$  (resp. $L_{v_\mu }$~).

These conjugacy subclass representatives $M_{\o v^{\mu '}_{\mu ,m_\mu }}$ (resp. $M_{v^{\mu '}_{\mu ,m_\mu }}$~) are interpreted as  \rl time quanta which are thus closed irreducible $1D$ algebraic sets of degree $N$~.

Each representative $M_{\o v_{\mu ,m_\mu }}$ (resp. $M_{v_{\mu ,m_\mu }}$~), being a one-dimensional 
$T_2^t(L_{\o v_\mu })$-subsemi\-module (resp. $T_2(L_{v_\mu })$-subsemimodule), is a string localized in the lower (resp. upper) half space.  So, each string $M_{\o v_{\mu ,m_\mu }}$ (resp. $M_{v_{\mu ,m_\mu }}$~) is composed of $(p+\mu )$ quanta, where $(p+\mu )$ is the global residue degree $f_{b_{\mu_R} }$ (resp. $f_{b_{\mu _L}}$~) referring to the dimension of a quantum class representative.

On the other hand, we want to introduce the set of smooth continuous (bi)functions on the representation space $M_R\otimes M_L=\Repsp (\GL_2(\Aa_{L_{\o v}}\times \Aa_{L_v}))$ of the algebraic bilinear semigroup $\GL_2(\Aa_{L_{\o v}}\times \Aa_{L_v})$~.  Due to the bilinear Gauss decomposition of $\GL_2(\Aa_{L_{\o v}}\times \Aa_{L_v})$~, we have to envisage the set of smooth continuous functions $\phi _{G_R}(x_{g_R})$~, $x_{g_R}\in T^t_2(\Aa_{L_{\o v}})$~, on $M_R
=\Repsp (T^t_2(\Aa_{L_{\o v}}))$ and localized in the lower half space as well as the corresponding symmetric set of smooth continuous functions $\phi _{G_L}(x_{g_L})$~, $x_{g_L}\in T_2(\Aa_{L_{v}})$~, on  $M_L
=\Repsp (T_2(\Aa_{L_{v}}))$ and localized in the upper half space.

On $M_R\otimes M_L$~, the tensor products   $\phi _{G_R}(x_{g_R})\otimes  \phi _{G_L}(x_{g_L})$ of smooth continuous functions have to be considered: the are called bifunctions.

But, as $\GL_2(\Aa_{L_{\o v}}\times \Aa_{L_v})$ is partitioned into conjugacy classes, we have to take into account the bifunctions $\phi _{G_{\mu ,m_\mu ,R}}(x_{\mu _R})\otimes \phi _{G_{\mu ,m_\mu ,L}}(x_{\mu _L})$ on the conjugacy class representatives $M_{\o v_{\mu ,m_\mu }}\otimes M_{v_{\mu ,m_\mu }}$~.

The set of smooth continuous bifunctions $\{\phi _{G_{\mu ,m_\mu ,R}}(x_{\mu _R})\otimes \phi _{G_{\mu ,m_\mu ,L}}(x_{\mu _L})\}_{{}^\mu_{m_\mu }} $ on the $\GL_2(\Aa_{L_{\o v}}\times \Aa_{L_v})$-bisemimodule $M_R\otimes M_L$ is a bisemisheaf of rings, noted $\Cs_{M_R}\otimes \Cs_{M_L}$ or $\widetilde M_R\otimes \widetilde M_L$~, in such a way that the set of continuous bifunctions are the (bi)sections of $\Cs_{M_R}\otimes \Cs_{M_L}$~.

Note that $\Cs_{M_R}$ (resp. $\Cs_{M_L}$~), having as sections the smooth continuous functions 
$\phi _{G_{\mu ,m_\mu ,R}}(x_{\mu _R})$ (resp. $\phi _{G_{\mu ,m_\mu ,L}}(x_{\mu _L})$~), is a semisheaf of rings because it is a sheaf of abelian semigroups $\Cs_{M_R}(x_{\mu _R})$ (resp. $\Cs_{M_L}(x_{\mu _L})$~) for every \rl point $x_{\mu _R}$ (resp. $x_{\mu _L}$~) of the topological semispace $M_R=\Repsp(T^t_2(\Aa_{L_{\o v}}))$
(resp. $M_L=\Repsp(T_2(\Aa_{L_{v}}))$~) where $\Cs_{M_R}(x_{\mu _R})$ (resp. $\Cs_{M_L(x_{\mu _L}}$~) has the structure of a semiring \cite{Ser1}, \cite{G-D}.

Remark that the pair $\{\Cs_{M_R},\Cs_{M_L}\}$ of semisheaves of ring or their product $\Cs_{M_R}\otimes \Cs_{M_L}$ is what the physicists call a field because each pair  
$\{\phi _{G_{\mu ,m_\mu ,R}}(x_{\mu _R}),\phi _{G_{\mu ,m_\mu ,L}}(x_{\mu _L})\}$ of smooth continuous symmetric functions behaves like a harmonic oscillator as it will be seen in the following.

As each representative $M_{\o v_{\mu ,m_\mu }}$ (resp. $M_{v_{\mu ,m_\mu }}$~) of the $\mu $-th conjugacy class of $T^t_2(\Aa_{L_{\o v}})$ (resp. $T_2(\Aa_{L_{v}})$~) has a rank equal to $n_{\mu _R}=(p+\mu )N$ (resp. $n_{\mu _L}=(p+\mu )N$~), we will say by abuse of language that the function 
$\phi _{G_{\mu ,m_\mu ,R}}(x_{\mu _R})$ (resp. $\phi _{G_{\mu ,m_\mu ,L}}(x_{\mu _L})$~) on $M_{\o v_{\mu,m_\mu }}$ (resp. $M_{v_{\mu,m_\mu }}$~) is characterized by a rank $n_{\mu _R}$ (resp. $n_{\mu _L}$~).

If $\phi _{G^{\mu '}_{\mu ,m_\mu ,R}}(x_{G^{\mu '}_{\mu _R}})$ 
(resp. $\phi _{G^{\mu '}_{\mu ,m_\mu ,L}}(x_{G^{\mu '}_{\mu _L}})$~) 
denotes  of the smooth continuous function on the $\mu '$-th equivalent conjugacy 
subclass representative $M_{\o v^{\mu '}_{\mu ,m_\mu }}$ 
(resp. $M_{v^{\mu '}_{\mu ,m_\mu }}$~), then 
$(M_{\o v^{\mu '}_{\mu ,m_\mu }} ,\phi _{G^{\mu '}_{\mu ,m_\mu ,R}}(x_{G^{\mu '}_{\mu _R}}))$ 
(resp. $(M_{v^{\mu '}_{\mu ,m_\mu }} ,
\phi _{G^{\mu '}_{\mu ,m_\mu ,L}}(x_{G^{\mu '}_{\mu _L}}))$~) 
is a closed irreducible one-dimensional subscheme of rank $N$ associated to the \rl quantum 
$M_{\o v^{\mu '}_{\mu ,m_\mu }} $ (resp. $M_{v^{\mu '}_{\mu ,m_\mu }} $~) and noted 
$\widetilde M_{\o v^{\mu '}_{\mu ,m_\mu }} $ 
(resp. $\widetilde M_{v^{\mu '}_{\mu ,m_\mu }} $~).
\vskip 11pt

\begin{sub}{\bf Emergent projection and Borel-Serre compactification}
\end{sub}

\Bena
\item As the \rl subsemimodules $M_{\o v_{\mu,m_\mu}}$ (resp. $M_{v_{\mu,m_\mu}}$~) are not necessarily closed strings, the emergent toroidal projective isomorphisms:
\[ \gamma _{\mu _R}: \quad M_{\o v_{\mu,m_\mu}}\To T^1_{\mu _R} \qquad
\text{(resp.} \quad \gamma _{\mu _L}: \quad M_{\mu,m_\mu}\To T^1_{\mu _L}\;)\]
are introduced such that \cite{Pie3}:
\Be
\item the geometric points of $M_{\o v_{\mu,m_\mu}}$ (resp. $M_{v_{\mu,m_\mu}}$~) are mapped onto the origin, called the emergence point which can be viewed as the point at infinity of the resulting projective variety;
\item these geometric points are then projected symmetrically from the origin into the affine 
connected compact algebraic varieties $\widetilde M^T_{\o v_{\mu,m_\mu}}$ (resp. $\widetilde M^T_{v_{\mu,m_\mu}}$~) \cite{C-S} 
which ``are'' $1D$-(semi)tori $T^1_{\mu _R}$ (resp. $T^1_{\mu _L}$~) \cite{B-T} characterized by a 
radius of ejection $r_{\mu _{R,L}}$ and such that:
\Bi
\item $\widetilde M^T_{\o v_{\mu,m_\mu}}$ (resp. $\widetilde M^T_{v_{\mu,m_\mu}}$~) are localized in the lower (resp. upper) half space with respect to the time variable ``~$t$~'';
\item each time quantum $M^T_{\o v^{\mu '}_{\mu,m_\mu}}$ is localized on a closed affine subset of $M^T_{\o v_{\mu,m_\mu}}$~, taking into account the $\gamma _\mu : M_{\o v^{\mu '}_{\mu,m_\mu}}\to M^T_{v^{\mu '}_{\mu,m_\mu}}$ morphism.

Remark that it will also be considered in the following that $\widetilde M^T_{\o v^{\mu '}_{\mu,m_\mu}}$ (resp. $\widetilde 
M^T_{v^{\mu '}_{\mu,m_\mu}}$~) are isomorphic to $1D$-(semi)tori, the distinction between the two  cases being in general evident.
\Ei\Ee

\item the space $X=GL_2(\rit)\big/GL_2(\ZZ)$ corresponds to the set of lattices of $\rit$~.  In this perspective, we have introduced in \cite{Pie9} a lattice bisemispace $X_{S_{R\times L}}=GL_2(\aa_{L_{\o \omega }}\times \aa_{L_\omega })\big/GL_2(\o\ZZ^2_{p_q})$~, where $\aa_{L_\omega }$ is a ramified adele semiring over a complex semifield $L_\omega $~, such that the boundary $\partial \o X_{S_{R\times L}}$ of the compactified bisemispace $\o X_{S_{R\times L}}$ corresponds to the boundary of the Borel-Serre compactification \cite{B-S} and is given by:
\[\partial \o X_{S_{R\times L}}=GL_2(\aa_{L^T_{\o v}}\times \aa_{L^T_v})\big/GL_2(\o\ZZ^2_{p_q})\] where $\aa_{L^T_{\o v}}$ (resp. $\aa_{L^T_{v}}$~) is the \rl ramified adele semiring with respect to the ``toroidal'' completions of the $L^T_{\o v_{\mu,m_\mu}}$ (resp. $L^T_{v_{\mu,m_\mu}}$~)~: $\Aa_{L^T_{\o v}}=\prod\limits_\mu L^T_{\o v_\mu }\prod\limits_{m_\mu }L^T_{\o v_{\mu ,m_\mu }}$~.

Let us note that there exits an isomorphism $\gamma _{R\times L}:\quad 
\o X_{S_{R\times L}}\To \partial \o X_{S_{R\times L}}$ between the compactified lattice 
bisemispace $\o X_{S_{R\times L}}$ and its boundary $\partial \o X_{S_{R\times L}}$ 
such that a one-to-one correspondence exists between the complex ``bipoints'' of 
$\o X_{S_{R\times L}}$ and the real ``bipoints'' of $\partial \o X_{S_{R\times L}}$ (a bipoint being defined as the product of a right point localized in the lower half space by a left point localized in the upper half space).

\item The double coset decomposition $\partial \o S_{K^D_{R\times L}}$ of the boundary $\partial \o X_{S_{R\times L}}$ of the compactified lattice bisemispace corresponds to a Shimura bisemivariety and is given by:
\[\partial \o S_{K^D_{R\times L}}=P_2(\Aa_{L^T_{\o v^1}}\times \Aa_{L^T_{v ^1}})\setminus GL_2(\aa_{L^T_{\o v}}\times \aa_{L^T_v})\big/GL_2(\o\ZZ^2_{p_q})\]
where
\Bi
\item $P_2(\Aa_{L^T_{v^1}})$ is the standard parabolic subgroup over the adele subsemiring 
$\Aa_{L^T_{v^1}}=\prod\limits_\mu L^T_{v^1_\mu}\prod\limits_{m_\mu} L^T_{v^1_{\mu ,m_\mu }}$ 
where 
$L^T_{v^1_{\mu ,m_\mu }}$ denotes the $\mu $-th irreducible toroidal central 
subcompletion of $L^T_{v_{\mu ,m_\mu }}$ 
having a rank equal to  $N$~.

$P_2(\Aa_{L^T_{\o v^1}}\times \Aa_{L^T_{v^1}})$ is a bilinear parabolic subgroup and is considered as the 
smallest normal ramified bilinear subsemigroup of the bilinear algebraic semigroup $GL_2(\Aa_{L^T_{\o v}}\times 
\Aa_{L^T_v})$~.  The bilinear quotient semigroup $P_2(\Aa_{L^T_{\o v^1}}\times {\Aa_{L^T_{v^1}}})\big/
GL_2(\Aa_{L^T_{\o v}}\times \Aa_{L^T_v})$ has its (bi)cosets which are in one-to-one correspondence
 with the modular conjugacy classes of $GL_2(\Aa_{L^T_{\o v}}\times \Aa_{L^T_v})$ with respect 
to  fixed bielements which correspond to the product $L^T_{\o v^1_\mu }\times L^T_{v^1_\mu }$ of irreducible subcompletions.

\item the general bilinear semigroup $GL_2(\aa_{L^T_{\o v}}\times \aa_{L^T_v})$ is a bilinear algebraic semigroup \cite{Che1}, also noted $G_{t_{R\times L}}(\aa_{R\times L})$ in  abbreviated form, to which corresponds the bilinear semigroup of modular automorphisms of $G_{t_{R\times L}}(\aa_{R\times L})$~, such that the set of products, right by left, of orbits of $G_{t_{R\times L}}(\aa_{R\times L})$ coincide with its modular conjugacy (bi)classes.

The fixed bielements of a modular conjugacy class of $GL_2(\aa_{L^T_{\o v}}\times \aa_{L^T_v})$ 
are the elements the bilinear parabolic subgroup $P_2(\Aa_{L^T_{\o v^1}}\times \Aa_{L^T_{v^1}})$~, representing the product of global inertia subgroups \cite{Pie9} $(I_{L_{\o v_\mu }}\times I_{L_{v_\mu }})$~.

\item the modular conjugacy classes of $G_{t_{R\times L}}(\aa_R\times \aa_L)$ correspond to the (bi)cosets of $G_{t_{R\times L}}(\aa_R\times \aa_L)\big/GL_2(\o\ZZ^2_{p_q})$ since the subgroup $GL_2(\o\ZZ^2_{p_q})$ constitutes the representation of the (bi)cosets of the tensor product of Hecke operators as it will be seen in definition 1.1.18: it is also noted $K^D_{R\times L}(\o\ZZ^2_{p_q})$~.  The bilinear quotient semigroup $GL_2(\aa_{L^T_{\o v}}\times \aa_{L^T_v})\big/GL_2(\o\ZZ^2_{p_q})$ consists in a double symmetric tower of conjugacy class representatives characterized by increasing ranks, i.e. by increasing numbers of quanta or strings.
\Ei

\item The double coset decomposition $\partial \o S_{K^D_{R\times L}}$ restricted to the lower (resp. upper) half space then becomes:
\begin{align*}
\partial \o S_{K_{t_R}}&=P_2(\Aa_{L^T_{\o v^1}})\setminus T_2^t(\Aa_{L^T_{\o v}})\big/T_2^t(\o\ZZ_{p_q})
\\
\text{(resp.} \quad
\partial \o S_{K_{t_L}}&=P_2(\Aa_{L^T_{v^1}})\setminus T_2(\aa_{L^T_{v}})\big/T_2(\o\ZZ_{p_q})\;).\end{align*}
It will also be noted:
\begin{align*}
\partial \o S_{K_{t_R}}&=P_{t_R}(\Aa_{L^T_{\o v^1}})\setminus G_{t_R}(\aa_R)\big/K_R(\o\ZZ_{p_q})
\\
\text{(resp.} \quad
\partial \o S_{K_{t_L}}&=P_{t_L}(\Aa_{L^T_{v^1}})\setminus G_{t_L}(\aa_L)\big/K_L(\o\ZZ_{p_q})
 \;).\end{align*}
\Ee \vskip 11pt

\begin{sub}{\bf Right and left semisheaves of rings}
\end{sub}

{The set of products, right by left, of toroidal projective isomorphisms:
\[\{\gamma _{\mu _R}\times \gamma _{\mu _L}: M_{\o v_{\mu ,m_\mu }}\otimes M_{v_{\mu ,m_\mu }}\To M^T_{\o v_{\mu ,m_\mu }}\otimes M^T_{v_{\mu ,m_\mu }}
\equiv T^1_{\mu _R}\otimes T^1_{\mu _L}\}\]
transforms the $\GL_2(\Aa_{L_{\o v}}\times \Aa_{L_v})$-bisemimodule $M_R\otimes M_L=\bigoplus\limits_\mu \bigoplus\limits_{m_\mu }(M_{\o v_{\mu ,m_\mu }}\otimes M_{v_{\mu ,m_\mu }})$ into the 
 $\GL_2(\Aa_{L^T_{\o v}}\times \Aa_{L^T_v})$-bisemimodule $M^T_R\otimes M^T_L=\bigoplus_\mu \bigoplus_{m_\mu }( M^T_{\o v_{\mu ,m_\mu }} \otimes M^T_{v_{\mu ,m_\mu }})$~.

Each representative   $ M^T_{\o v_{\mu ,m_\mu }} $ (resp.   $ M^T_{v_{\mu ,m_\mu }} $~) of $M^T_R$ (resp. $M^T_L$~) is a semitorus localized in the lower (resp. upper) half space. In fact, we shall be essentially interested in \rl one-dimensional tori: so, we have to double the representatives $M^T_{\o v_{\mu ,m_\mu }} $ (resp. $M^T_{v_{\mu ,m_\mu }} $~), i.e. to consider representatives $M^T_{2\o v_{\mu ,m_\mu }} $ (resp. $M^T_{2v_{\mu ,m_\mu }} $~) characterized by double ranks and by double quanta in such a way that $M^T_{2\o v_{\mu ,m_\mu }} $ (resp. $M^T_{2v_{\mu ,m_\mu }} $~) be closed strings \cite{DelToWit}.  But, in the following, we shall maintain the condensed notation $M^T_{\o v_{\mu ,m_\mu }} $ (resp. $M^T_{v_{\mu ,m_\mu }} $~) for the two cases, the distinction being evident by itself. On the representation space
$M^T_R=\Repsp (T^t_2(\Aa_{L^T_{\o v}}))$ (resp. 
$M^T_L=\Repsp (T_2(\Aa_{L^T_{v}}))$~) of 
$T^t_2(\Aa_{L^T_{\o v}})$ (resp. $T_2(\Aa_{L^T_{v}})$~) 
$\subset \GL_2(\Aa_{L^T_{\o v}}\times \Aa_{L^T_v})$~, we can consider the set of 
differentiable smooth functions $\phi _{G^T_{\mu ,m_{\mu _R}}}(x_{\mu _R})$ 
(resp. $\phi _{G^T_{\mu ,m_{\mu _L}}}(x_{\mu _L})$~) on the representatives 
$M^T_{\o v_{\mu ,m_\mu }} $ (resp. $M^T_{v_{\mu ,m_\mu }} $~) of $M^T_R$ 
(resp. $M^T_L$~): it is a semisheaf of rings noted $\theta ^1_R$ 
(resp. $\theta ^1_L$~) or $\widetilde M^T_R$ (resp. $\widetilde M^T_L$~).

The differentiable functions $\phi _{G^T_{\mu ,m_{\mu _R}}}(x_{\mu _R})$ 
(resp. $\phi _{G^T_{\mu ,m_{\mu _L}}}(x_{\mu _L})$~) 
are the sections of the semisheaf of rings $\theta ^1_R$ 
(resp. $\theta ^1_L$~): they are noted in condensed notation  $s_{\mu _R}$ 
(resp. $s_{\mu _L}$~).

Remark that the following developments will essentially deal with the semisheaf of rings 
$\theta ^1_R$ (resp. $\theta ^1_L$~) because they naturally lead to automorphic 
representations (see sections 1.15 to 1.23) and to Langlands global correspondences 
(similar developments can be envisaged on the semisheaves of rings $\Cs_{M_R}$ 
(resp. $\Cs_{M_L}$~) or $\widetilde M_R$ and $\widetilde M_L$ (see section 1.1.6)).

It is then possible to define a graded algebra on the set of right (resp. left)
sections $s_{\mu R,L}$ of the right (resp. left) semisheaf of rings $\theta
^1_{R,L}$~.}
\vskip 11pt

\begin{propo} Let $n_{\mu R,L}$ be the right (resp. left) rank of the
right (resp. left) section $s_{\mu R,L}$ and $n_{(\mu +1)R,L}$ the right (resp.
left) rank of the right (resp. left) section $s_{(\mu +1)R,L}$~. 
Then, the inequality $n_{(\mu +1)R,L}>n_{\mu R,L}$ leads to the topological
embedding $s_{\mu R,L}\subset s_{(\mu +1)R,L}$ between the $\mu $-th and the
$(\mu +1)$-th section.\end{propo}
\vskip 11pt

\bpr If the inequality $n_{(\mu +1)R,L}> n_{\mu R,L}$ holds, then $s_{(\mu
+1)R,L} \supset s_{\mu R,L} $~. Indeed, if $r(x_{\mu +1})_{R,L}$ and $r(x_\mu
)_{R,L}$ denote respectively the radii of ejection of the points $x_{(\mu
+1)R,L}\in s_{(\mu
+1)R,L} $ and $x_{\mu R,L}\in s_{\mu R,L} $~, it is evident that $r(x_{\mu +1})_{R,L}>r(x_\mu
)_{R,L}$~.\epr
\vskip 11pt

\begin{coro} Let $s_{1R,L}\subset\cdots\subset s_{qR,L}$ be the increasing
filtration of the $q$ sets of sections of the semisheaf of rings $\theta ^1_{R,L}$ to
which is associated the sequence of ranks
\[ n_{\theta ^1_{R,L}}=\{ n_{1R,L},\cdots,n_{\mu R,L},\cdots,n_{qR,L}\}\;.\]
Then, the right (resp. left) semisheaf of rings $\theta ^1_{R,L}$ is characterized by the global rank given by
the set $n_{\theta ^1_{R,L}}$~.   \end{coro}
\vskip 11pt

We are now concerned with the cohomology of the boundary of the Borel-Serre
compactification [B-S], \cite{Sch2}  of semispace $\partial
\ \o S_{K_{t R,L}}$~: it is the Eisenstein cohomology, as nicely developed
by G. Harder \cite{Har1}, J. Schwermer \cite{Sch1} and others, which becomes
the so-called right (resp. left) Eisenstein cohomology when it leads to the
generation of a right (resp. left) semisheaf of rings $\theta ^1_{R,L}$ on the $G_{t_{R,L}}(\aa_{R,L})$-semimodule.
\vskip 11pt

\begin{defi}{ {\bf (Nilpotent fibration on the \rl Shimura semi\-variety)\/}\    If we take into account:
\Bi
\item the Gauss decomposition of the bilinear algebraic semigroup
\begin{align*}
G_{t_{R\times L}}(\aa_R\times \aa_L) \equiv GL_2(\aa_R\times \aa_L)&=T^t_2(\aa_R)\times T_2(\aa_L)\\
&= [D_2(\aa_R)\times UT^t_2(\aa_R)][D_2(\aa_L)\times UT_2(\aa_L)]\;, \end{align*}
as developed in 1.1.5. where $\Aa_R\equiv\Aa_{L^T_{\o v}}$ and $\Aa_L\equiv\Aa_{L^T_{v}}$~;
\item the Levi decomposition of the \rl parabolic subgroup
\begin{align*}
P_{t_R}(\Aa_{L^T_{\o v^1}}) 
&\equiv P_2(\Aa_{L^T_{\o v^1}})
= D_2(\Aa_{L^T_{\o v^1}})\cdot UT^t_2
(\Aa_{L^T_{\o v^1}})\\
\text{(resp.} \quad
P_{t_L}(\Aa_{L^T_{v^1}}) 
&\equiv P_2(\Aa_{L^T_{v^1}})
= D_2(\Aa_{L^T_{v^1}})\cdot UT^t_2
(\Aa_{L^T_{v^1}})\;)\end{align*}
\item and the similar decomposition of 
\begin{align*}
K_R(\o\ZZ_{p_q}) &\equiv    T_2^t( \o\ZZ_{p_q} )= D_2( \o\ZZ_{p_q} )\cdot UT^t_2
( \o\ZZ_{p_q} )\\
\text{(resp.} \quad
K_L(\o\ZZ_{p_q}) &\equiv    T_2( \o\ZZ_{p_q} )= D_2( \o\ZZ_{p_q} )\cdot UT_2
( \o\ZZ_{p_q} ) \;),\end{align*}
introduced in 1.1.7.,
\Ei
into product of unitriangular matrices of nilpotent subsemigroups by diagonal matrices of centralizers $\pZ(\cdot)$~, noted here $M(\cdot)$ to respect the notations of \cite{Sch1} and \cite{Har2}, we are led to define, following G. Harder and J. Schwermer, the fibration:
\begin{align*}
\partial \o S_{K_{t_R}} = P_2(\Aa_{L^T_{\o v^1}})\setminus T^t_2(\aa_R)\Big/T^t_2(\o\ZZ_{p_q})&\\
\To S^{M_R}_{K^{M_R}} &=  M(\Aa_{L^T_{\o v^1}})\setminus M(\aa_R)\Big/K^{M_R}_R(\o \ZZ_{p_q})\\
&\equiv  D_2(\Aa_{L^T_{\o v^1}})\setminus D_2(\aa_R)\Big/D_2(\o\ZZ_{p_q})\\[11pt]
\text{(resp.} \quad
\partial \o S_{K_{t_L}} = P_2(\Aa_{L^T_{v^1}})\setminus T_2(\aa_L)\Big/T_2(\o\ZZ_{p_q})&\\
\To S^{M_L}_{K^{M_L}} &=  M(\Aa_{L^T_{v^1}})\setminus M(\aa_L)\Big/K^{M_L}_L(\o \ZZ_{p_q})\\
&\equiv  D_2(\Aa_{L^T_{v^1}})\setminus D_2(\aa_L)\Big/D_2(\o\ZZ_{p_q}\;)\end{align*}
having as \rl fiber, the \rl nilpotent fiber
\begin{align*}
N(\Aa_{L^T_{\o v^1}})\setminus N(\aa_R)\Big/ K_R^N &\equiv UT^t_2(\Aa_{L^T_{\o v^1}})\setminus UT^t_2(\aa_R)\Big/UT^t_2(\o\ZZ_{p_q})\\
\text{(resp.} \quad
N(\Aa_{L^T_{v^1}})\setminus N(\aa_L)\Big/ K^N_L &\equiv UT_2(\Aa_{L^T_{v^1}})\setminus UT_2(\aa_L)\Big/UT_2(\o\ZZ_{p_q})\;).\end{align*}
}
\end{defi}
\vskip 11pt

\begin{propo}
The right (resp. left) Eisenstein cohomology associated with the generation of a
right (resp. left) semisheaf of rings $\theta ^1_{R,L}$ decomposes into:
\nobeqn
\lefteqn{H^*_{R,L}(\partial \ \o S_{K_{tR,L}},\theta ^1_{R,L})}\\
&=& H^*(P_{tR,L}(\Aa_{L^T_{v^1}})\setminus G_{tR,L}(\aa_{R,L})/K_{R,L}(\o\ZZ_{p_q}),\theta ^1_{R,L})\\
&\simeq& \txt\bigoplus\limits_{\xi_{R,L}\in \Xi_{K_{R,L}}}
H^*(S^{M_{R,L}}_{K^{M_{L,R}}(\xi_{R,L})},H^*(\tilde u_{R,L},\theta ^1_{R,L}))\noeeqn
where
 $S^{M_{R,L}}_{K^{M_{R,L}}}=M(\Aa_{L^T_{v^1}})\setminus M(\aa_{R,L})/K ^{M_{R,L}} _{R,L}(\o\ZZ_{p_q})$~.
\end{propo}
\vskip 11pt

\bpr The right (resp. left) Eisenstein cohomology
$H^*_{R,L}(\partial \ \o S_{K_{tR,L}},\theta ^1_{R,L})$ decomposes into the
direct sum of right (resp. left) cohomology classes referring to right (resp.
left) cosets $\xi_{R,L}$ of 
$(G_{tR,L}(\aa_{R,L})/ K_{R,L})$ such that the right (resp. left) coefficient
system be given by the right (resp. left) semisimple Lie algebra cohomology
$H^*_{R,L}(\tilde u_{R,L},\theta ^1_{R,L})$ which is a right (resp. left)
semimodule for the right (resp. left) algebraic semigroup $M_{R,L}(\aa_{R,L})$~.

If $u_{P_{R,L}}$ is the  right (resp. left) unipotent algebraic
semigroup, then $\tilde
u_{R,L} = \mbox{Lie\ } (u_{P_{R,L}})$ is its right (resp. left) nilpotent Lie algebra.

Note that this decomposition of the right (resp. left) Eisenstein cohomology is
an adaptation of the developments of G. Harder \cite{Har2}.\epr
\vskip 11pt

\begin{defi}[Algebraic Hecke characters]  { If $\Gamma  _{\mu R,L}$ denotes the
right (resp. left) Galois subgroup $\Gal(L^{\mp}_\mu /K)$~, let
\[ \lambda _{\mu R,L}= \{\lambda _{1R,L},\cdots,\lambda _{\mu R,L},\cdots,
\lambda _{qR,L}\}_{\Gamma _{\mu R,L}:L^T_{\o v^1}\to L^T_{\o v_\mu }} \]
be given by
\[X(T_{ R,L})=\Hom (T_{
R,L}\times_K\ L^T_v,G_{m})=\txt\bigoplus\limits_{\Gamma    _{\mu R,L}:L^T_{v^1 }\to L^T_{v_\mu}}X(T^1_{{\mu R,L}})\;,\]
where $G_{m}\equiv GL_{1}$~.

Then, the set $\lambda _{R,L}=\{\lambda _{1R,L},\cdots,\lambda _{\mu 
R,L},\cdots,\lambda _{q R,L}\}$ is the sequential set of weights in $X(\theta
^1_{R,L})$ referring to the $q$ basic right (resp. left) sections of the right (resp.
left) semisheaf of rings $\theta ^1_{R,L}$~.

Let $\omega _{R,L}=\{\omega _{1R,L},\cdots,\omega _{qR,L}\}$ be the set of $q$
right (resp. left) actions of the Weyl groups on $\lambda _{R,L}\in X(\theta
^1_{R,L})$~.

Remark that $\omega _{R,L}$ is a set of Weyl subgroups because this set acts
on the set of right (resp. left) characters $\lambda _{\mu R,L}$~.

Consequently, the maximal convex right (resp. left) subsets of $X(\theta
^1_{R,L})$ will be in negative (resp. positive) Weyl chambers.

Let finally $\phi_{R,L}=\omega _{R,L}\cdot \lambda _{R,L}$ be the set of right
(resp. left) algebraic Hecke characters \cite{Clo} on $\theta ^1_{R,L}$~.}
\end{defi}
\vskip 11pt

\begin{propo}  Let $B_{tR,L}(\aa_{R,L})$ be the right (resp. left) Borel subgroup
of upper (resp. lower) triangular matrices of the right (resp. left) algebraic semigroup
$G_{tR,L}(\aa_{R,L})$~.  Then, the right (resp. left) Eisenstein cohomology
$H^*(\partial \ \o S_{K_{t_{R,L}}},\theta ^1_{R,L})$ decomposes into
one-dimensional eigenspaces:
\nobeqn
\lefteqn{H_{R,L}^*(\partial \ \o S_{K_{t_{R,L}}},\,\theta ^1_{R,L}) }\\
\noalign{\vskip 6pt}
&\simeq& \txt\bigoplus\limits_{\omega _{R,L}} \bigoplus\limits_{\phi_{R,L}}
\Ind^{\pi _0(G_{R,L}(\ptaa_{R,L})} 
_{\pi _0(B_{R,L}(\ptaa_{R,L})}\
H^*_{R,L}(S^{M_{R,L}},H^*(\tilde u_{B_{tR,L}},\theta ^1_{R,L})(\omega
_{R,L}\cdot \lambda _{R,L}))\noeeqn
where $S^{M_{R,L}}=\lim\limits_{\overrightarrow{K_{R,L}}}
S^{M_{R,L}}_{K^{M_{R,L}}}$~.\end{propo}
\vskip 11pt

\bpr Indeed, the cohomology $H^*(\tilde u_{B_{tR,L}},\theta ^1_{R,L})$ is a
right (resp. left) semimodule for the set of tori $T^1_{R,L} 
=\{T^1_{1R,L}  ,\cdots,T^1_{\mu R,L}  ,\cdots,T^1_{qR,L}  \}$~.

In this context, Kostant's theorem says that the cohomology decomposes into
one-dimensional eigenspaces under $T^1_{R,L}$~.  The right (resp. left)
Eisenstein cohomology then decomposes into one-dimensional eigenspaces with
respect to $\omega _{R,L}$ and the type of algebraic Hecke characters
$\phi_{R,L}$ according to the considered induced representation $\pi _0$ of the Borel
right (resp. left) stratum of $B_{R,L}(\Aa_{R,L})\equiv P_{tR,L}$~.\epr
\vskip 11pt

In correlation with Kostant's theorem, it appears necessary to develop a bit further the problem of the representation of Eisenstein cohomology into irreducible one-dimensional components.  Taking into account that Eisenstein series are eigenfunctions of Hecke operators and that the decomposition of Eisenstein cohomology into irreducible submodules characterized by some weights needs a cuspical automorphic representation of the algebraic semigroups $G_{t_{R,L}}(\aa _{R,L})$~, we have to envisage the action of the Hecke operators in the space of cusp forms.

Note that cusp forms are directly related to the branes of ``string physicists''.
\vskip 11pt

\begin{defi}[Algebra of cusp forms] Let $H$ denote the Poincare upper half plane in $\cit$~.  Assume that $f_L$ is a normalized eigenform, holomorphic in $H$ and defined in $\{\Im (z_L)>0\}$ with respect to $z_L\in\cit$ of $q_L=e^{2\pi iz_L}$~.  The normalized eigenform $f_L$~, expanded in formal power series $f_L=\sum\limits_na_{n_L}q^n_L$~, are cusp forms of the space $S_L(N)$ and are eigenvectors of the Hecke operators $T_{\ell_L}$~, for $\ell\nmid N$~, and $U_{\ell_L}$~, for $\ell\mid N$ where $N$ is a positive integer.  Then, Fourier coefficients of $f_L$ and eigenvalues of the Hecke operator coincide: $a_1=1$ and $a_n=c(n,f_L)$ so that the $c(n,f_L)$ generate the ring of integers $\theta_L$ of the number field $L^+$ over $\QQ$~.  The space $S_L(N)$ can then be considered as a $\theta_L$-algebra over $\theta_L$~.
\vskip 11pt

As we are concerned with the endomorphisms of the algebra of cusp forms $S_L(N)$~, it is the bialgebra $S_L^e=S_L(N)\otimes_\theta S_R(N)$  which must be considered in the developments such that tensor products of Hecke operators acting on tensor products of cusp forms defined respectively in the upper and in the lower half plane will be envisaged.  The coalgebra $S_R(N)$ of cusp forms is defined in the Poincare lower half plane $H^*$ and has for elements the eigenforms $f_R=\sum\limits_na_{n_R} q^n_R$ with $q^n_R=e^{-2\pi inz_R}$ where $z_R$ is the complex conjugate of $z_L$~.  These eigenforms $f_R$ are eigenfunctions of Hecke operators $T_{\ell_R}$~, for $\ell\nmid N$ and $U_{\ell_R}$ for $\ell\mid N$~.
\end{defi}
\vskip 11pt

\begin{defi}[Global elliptic $\aa_{R,L}$-semimodule]  In order to get an automorphic irreducible representation of the algebra of cusp forms, we shall consider that the one-dimensional semisheaf of rings $\theta^1_{R,L}$ define a global  elliptic semimodule whose space is included in the space of cusp forms. Let $s_{R,L}=\Gamma (\theta^1_{R,L})$ denote the set of sections of $\theta^1_{R,L}$~.  For each section $s_{\mu _{R,L}}\in s_{R,L}$~, let $ \End(G_{s_{R,L}})$ be the Frobenius endomorphism of the group $G_{s_{R,L}}$ of the elements $s_{\mu _{R,L}}$ and let $q^{\pm p}\to q^{\pm(p+\mu)}\in \End_{\FF_p}(G_{s_{R,L}})$ be the corresponding Frobenius substitution with $q^{\pm (p+\mu)}=e^{\pm2\pi i(p+\mu )x}$~, $x\in \rit$~.

A global  elliptic right (resp. left) $s_{R,L}$-semimodule $\phi_{R,L}(s_{R,L})$ in the sense of Drinfeld \cite{Drin} is a ring homomorphism \cite{And}: $\phi_{R,L}:s_{R,L}\to \End(G_{s_{R,L}})$ given by $\phi_{R,L}(s_{R,L})=\sum\limits_\mu \sum\limits_{m_\mu }\phi(s_{q_{R,L}})_{\mu ,m_\mu }q^{\pm(p+\mu)} /\QQ_{R,L}$ where $\sum\limits_\mu $ runs over the sections $T^1_{\mu _{R,L}}$ of $\theta^1_{R,L}$ having ranks $n_\mu $ and where $\sum\limits_{m_\mu }$ runs over the number of ideals of the decomposition group $D_{\mu ^2}$ introduced in section 1.1.8 and  corresponding to the multiplicity $m^{(\mu)} $ of the $\mu $-th section.
\end{defi}
\vskip 11pt

\begin{lm} The space $S_{R,L}(\phi_{R,L})$ of global elliptic $s_{R,L}$-semimodules $\phi_{R,L}(s_{R,L})$ is included into the space $S_{R,L}(N)$ of cusp forms $f_{R,L}:S_{R,L}(\phi_{R,L})\hookrightarrow S_{R,L}(N)$ such that $f_{R,L}\simeq\phi_{R,L}(s_{R,L})$~.
\end{lm}
\vskip 11pt

\begin{defi}[The decomposition group] The ring of endomorphisms acting on the global elliptic 
$s_{R,L}$-semimodules included into weight two cusp forms is generated over 
$\o\zit_{p_q}$ by the Hecke operators $T_{q_{R,L}}$ for $N\nmid q_N$ and $U_{q_{R,L}}$ for 
$N\mid q_N$ \cite{M-W}, \cite{Lan3}.  The coset representatives of $U_{q_L}$ can be chosen to 
be upper triangular and given by integral matrices 
$\left(\begin{smallmatrix} 1&b_N \\ 0&q_N\end{smallmatrix}\right)$ 
while the coset representatives of $U{q_R}$ are lower triangular and are given by matrices 
$\left(\begin{smallmatrix} 1&0 \\ b_N&q_N\end{smallmatrix}\right)$~.  
For general $n=a\cdot d$~, we would have respectively the integral matrices 
$\left(\begin{smallmatrix} a&b_N \\ 0&d_N\end{smallmatrix}\right)$ and
$\left(\begin{smallmatrix} a_N&0 \\ b_N&d_N\end{smallmatrix}\right)$ of determinant 
$n\cdot N=ad\cdot N\equiv a_N\cdot d_N$ such that 
$q_N \begin{array}[t]{l}
=*\mod N\simeq q\cdot N\\
=q\cdot N
\mbox{\ (case $q_N=0\mod N$~)}\end{array}$ and $b_N=*\mod N$~.

But, as noticed in definition 1.1.15, we have to consider tensor products of Hecke operators.  So, taking into account that the group of matrices $u(b)= \left(\begin{smallmatrix} 1&b \\ 0&1\end{smallmatrix}\right)$ and $u(b)^t= \left(\begin{smallmatrix} 1&0 \\ b&1\end{smallmatrix}\right)$ generate $\FF_q$ \cite{Lan3}, the following coset representatives
\[ k^D_{R\times L}(\o\zit^2_{p_q})=\left[\BM 1&b_N\\ 0&1 \EM\BM 1&0\\ b_N&1\EM \right]\BM 1&0 \\ 0&q_N^2\EM\] 
will be adopted for $U_{q_R}\otimes U_{q_L}$ where $\alpha_{q^2_N}= \left(\begin{smallmatrix} 1&0 \\ 0&q_N^2\end{smallmatrix}\right)$ is the split Cartan subgroup matrix and where $D_{q_N^2,b_N}= \left(\begin{smallmatrix} 1&b_N \\ 0&1\end{smallmatrix}\right)  \left(\begin{smallmatrix} 1&0 \\ b_N&1\end{smallmatrix}\right)$ is the element of the decomposition group associated to $\alpha_{q_N^2}$~. Indeed, the semisimplicial form $D_{q_N^2,b_N}$ is unimodular.
\end{defi}
\vskip 11pt

\begin{propo} The eigenvalues $\lambda_{\pm}(q_N^2,b_N^2)$ of $k^D_{R\times L}(\o\zit^2_{p_q})$ of $U_{q_R}\otimes U_{q_L}$ are such that:
\Bena
\item $\lambda_+(q_N^2,b_N^2)$ being equivalent to $\lambda_-(q_N^2,b_N^2)$ is an algebraic Hecke character noted $\phi_{R,L}$ in definition 1.1.13.
\item they are the coefficients of the elliptic $s_{R,L}$-semimodule $\phi_{R,L}(s_{R,L})$~:  $ \phi(s_{q_{R,L}})_{q,b}=\lambda_{\pm}(q_N^2,b_N^2)$~.
\item they allow to define the radius of the torus $T^1_{q_{R,L}}$ by 
\[ r(q_N^2,b_N^2)=(\lambda_+(q_N^2,b_N^2)-\lambda_-(q_N^2,b_N^2))/2\;.\]
\Ee
\end{propo}\vskip 11pt

\bpr The eigenvalues of $k ^D_{R,L}(\o\zit^2_{p_q})$ are
\[ \lambda_{\pm}(q_N^2,b_N^2)=\F{(1+b_N^2+q_N^2)\pm [(1+b_N^2+q_N^2)^2-4q_N^2]^{\half}}2
\]
and verify 
\[ \begin{cases}
\trace ( k^D_{R\times L}(\o\ZZ^2_{p_q} )) = 1+b_N^2+q_N^2\; , \\
\det ( k^D_{R\times L}(\o\ZZ^2_{p_q})) = \lambda_+(q_N^2, b_N^2)\cdot \lambda_-(q_N^2,b_N^2)\;.
\end{cases}\]
Assume that there exists a global  elliptic $\aa_{R\times_D L}$-(bi)semimodule
\[ \phi_R(s_R)\otimes_D \phi_L(s_L) =
\txt\sum\limits_{\mu ,m_\mu }\phi(s_R)_{\mu ,b}q^{-(p+\mu )} \otimes_D \sum\limits_{\mu ,m_\mu }\phi(s_L)_{\mu ,b}q^{p+\mu }\;,\]
{where} $q^{p+\mu }=e^{2\pi i(p+\mu )x}$~, 
included into a diagonal tensor product of weight two cusp forms $f_R\otimes_D f_L$~, then the coefficients $\phi(s_{R,L})_{\mu ,b}$ are given by $\phi(s_{R,L})_{\mu ,b}=\lambda_{\pm}(\mu _N^2,b_N^2)$ according to definition 1.1.16.  

Notice that a diagonal tensor product, written $\otimes_D$~, is a tensor product whose only diagonal terms with respect to a basis $\{e_{\mu ,m_\mu }\otimes e_{\mu ,m_\mu }\}$ are different from zero.

Let $i_R\otimes_D i_L$ be the (bi)isomorphism:
\[ i_R\otimes_D i_L:\phi_R(s_R)\otimes_D \phi_L(s_L) \to \widehat \phi_R(s_R)\otimes_D\phi_L(s_L)\]
where
\[ \widehat \phi_R(s_R)\otimes_D\widehat \phi_L(s_L) = \txt\sum\limits_{\mu}\sum\limits_{m_\mu }r(\mu _N^2,b_N^2)q^{-(p+\mu )} \otimes_D \sum\limits_\mu \sum\limits_{m_\mu } r(\mu _N^2,b_N^2)q^{p+\mu }\;,\]
which maps the  eigenvalues $\lambda_{\pm}(\mu _N^2,b_N^2)$ to
\[ r(\mu _N^2,b_N^2)=(\lambda_+(\mu _N^2,b_N^2)-\lambda_-(\mu _N^2,b_N^2))/2\;.\]

Then, $\widehat \phi_R(s_R)\otimes_D\widehat \phi_L(s_L)$ decomposes into a sum of tensor products of irreducible (semi)tori $T^1_{\mu ,b_{R,L}}$ localized respectively in the upper and in the lower half space, corresponding between themselves by pairs of same ranks $n_\mu $ and same values of $b_N$ and such that each pair of (semi)tori be characterized by a radius $r(\mu _N^2,b_N^2)$ and a center at the origin.  Notice that the radius $r(\mu _N^2,b_N^2)$ is the radius of ejection $r(x_{\mu_{R,L}})$ considered in proposition 1.1.9.  The isomorphism $i_{R,L}$ translates the centers of the tori from $\cent(\mu _N^2,b_N^2)$  $=(\trace(\Frob \mu _N^2))/2$ to the origin.  The result is that the eigenvalues $\lambda_+(\mu _N^2,b_N^2)$ and $\lambda_-(\mu _N^2,b_N^2)$ are equivalent.\epr
\vskip 11pt

\begin{rms} {\bf 1)\/} A cuspidal automorphic representation of Eisenstein series has thus been given in terms of global elliptic $s                                                                             _{R,L}$-semimodules as developed in proposition 1.1.19: this constitutes a first step in the direction of Weil's conjectures suggesting a deep connection between the arithmetic of algebraic varieties defined over finite fields and the topology of algebraic varieties defined over $\cit$~.
\vskip 11pt

\noindent {\bf 2)\/} It has thus been proved from the developments of 1.1.15 to 1.1.19 that the analytic representation of the \rl Eisenstein cohomology $H^*(\partial \o{S}_{K_{t_{R,L}}},\theta^1_{R,L})$ is given by a global elliptic $s_{R,L}$-semimodule noted $\ELLIP_{R,L}(1,\mu ,m_\mu )$ where ``~1~'' refers to the dimension: this constitutes a central challenge in the Langlands program as developed in \cite{Pie1}.  The bilinear version of the Langlands program is only really relevant and will be introduced in 1.1.23. 
\end{rms}
\vskip 11pt

\begin{propo}  Each \lr exponential $U_{v_\mu }=e^{2\pi i(p+\mu )x}$ (resp. $U_{\o v_\mu }=e^{-2\pi i(p+\mu )x}$~) of the $(\mu ,m_\mu )$-term $\phi(s_L)_{\mu ,b}e^{2\pi i(p+\mu )x}$ (resp. $\phi(s_R)_{\mu ,b}e^{-2\pi i(p+\mu )x}$~) of $\ELLIP_L(1,\mu ,m_\mu )$ (resp.\linebreak $\ELLIP_R(1,\mu ,m_\mu )$~) constitutes a unitary irreducible representation $v_\mu \to U_{v_\mu }$ (resp. $\o v_\mu \to U_{\o v_\mu }$~) associated to the \lr place $v_\mu $ (resp. $\o v_\mu $~) of the algebraic extension semifield $L^+$ (resp. $L^-$~) with respect to the coset representative $k^D_{R\times L}(\o\ZZ^2_{p_q})$ of the tensor product $T_{\mu _R}\otimes T_{\mu _L}$ of Hecke operators.

So, each \lr $(\mu ,m_\mu )$-term of $\ELLIP_L(1,\mu ,m_\mu )$ (resp. $\ELLIP_R(1,\mu ,m_\mu )$~) 
forms an irreducible representation of $L^+$ (resp. $L^-$~) inflated from the corresponding 
unitary irreducible representation $U_{v_\mu }$ (resp. $U_{\o v_\mu }$~) 
by a value $r(\mu ^2_N,b^2_N)$ which is the radius of the considered (semi)torus
$T^1_{\mu _L}=r(\mu ^2_N,b_N^2)\cdot e^{2\pi i(p+\mu )x}$ (resp. $T^1_{\mu _R}=
r(\mu ^2_N,b_N^2)\cdot e^{-2\pi i(p+\mu )x}$~).\end{propo}
\vskip 11pt

As  the coset representatives $k^D_{R\times L}(\o\zit^2{p_q})= \alpha_{\mu _N^2}\cdot D_{\mu _N^2,b_N}$ of tensor products of Hecke operators have a real meaning, we are constrained to work in the context of an Eisenstein (bi)cohomology as follows:

\vskip 11pt

\begin{propo} Let the product of the semigroups $K_R(\o\ZZ_{p_q})\times K_L(\o\ZZ_{p_q})$ be given by $K^D_{R\times L}(\o\zit^2_{p_q})$~.  Then, the Eisenstein bicohomology $H^*_{R\times_{(D)} L}( (\partial\o{S}_R\times_{(D)} \partial \o{S}_L))_{K^D_{R\times L}(\o\zit^2_{p_q})} , \theta^1_R\otimes_{(D)} \theta^1_L)$ decomposes under the decomposition group $D_{\mu _N^2,b_N}$ into products of pairs of one-dimensional eigenspaces:
\begin{align*}
& H^*_{R\times_{(D)} L}((\partial \o S_R\times_{(D)}\partial \o S_L)_{K^D_{R\times L}(\o\zit^2_{p_q})},\theta^1_R\otimes_{(D)}\theta^1_L)\\
&\qquad \simeq \txt\bigoplus\limits_\mu  \bigoplus\limits_{m_\mu } \Ind^{(G_{R\times_{(D)} L}(\aa _R\times_{(D)}\aa _L))}_{(K^D_{R\times L}(\o\zit^2_{p_q}))} H^*_{R\times_{(D)} L}(S^{M_{R\times_{(D)} L}},H^*(\tilde u_{K^D_{R\times L}(\o\zit^2_{p_q})} ,\theta^1_R\otimes_{(D)} \theta^1_L))\end{align*}
where the sums $\bigoplus\limits_\mu  \bigoplus\limits_{m_\mu }$ run over the cosets of
$ G_{t_R}(\aa _R)\times_{(D)} G_{t_L}(\aa _L)/K^D_{R\times L}(\o\zit^2_{p_q})$ having multiplicities $m^{(\mu )}=\sup(m_\mu )$~.
\end{propo}
\vskip 11pt

\bpr This proposition reduced to the right or left case is clearly equivalent to  proposition 1.1.14.  The coefficient system given by the semisimple Lie algebra (bi)-  cohomology $H^*(\tilde u_{K^D_{R\times L}(\o\zit^2_{p_q})},\theta^1_R\otimes_{(D)}\theta^1_L)$ decomposes into sum of products of pairs of one-dimensional sections $(T^1_{\mu ,b_R}\times T^1_{\mu ,b_L})$ of $\theta^1_R\otimes_{(D)}\theta^1_L$ characterized by the (bi)weights  $ \lambda_+(\mu _N^2,b_N^2) \times \lambda_-(\mu _N^2,b_N^2) $~.\epr
\vskip 11pt

\begin{sub} {\bf Langlands bilinear global program\/} \end{sub}

According to the developments from 1.1.15 to 1.1.20, the $GL_2(\Aa_{L^T_{\o v}}\times 
\Aa_{ L^T_v})$-bisemimodule $M^T_R\otimes M^T_L$ has an analytic development given by the global 
elliptic $s_R\otimes_Ds_L$-bisemimodule $\phi _R(s_R)\otimes_D\phi _L(s_L)$ which is a 
product, right by left, of truncated Fourier series.

As the $\GL_2(\Aa_{L_{\o v}}\times \Aa_{L_v})$-bisemimodule $M_R\otimes M_L$ 
constitutes an irreducible representation
$\Irr_W^{(1)}(W^{ab}_{L^-}\times W^{ab}_{L^+})$ of the bilinear global Weil group 
$W^{ab}_{L^-}\times W^{ab}_{L^+}$ 
\cite{Pie9} and 
as the global elliptic bisemimodule $\phi _R(s_R)\otimes_D\phi _L(s_L)$~, also noted 
$\ELLIP_R(1,\cdot,\cdot)\otimes \ELLIP_L(1,\cdot,\cdot)$~, constitutes an irreducible cuspidal representation 
$\Irr\ELLIP(GL_2(\Aa_{ L^T_{\o v}}\times \Aa_{L^T_v})$) of $GL_2(\aa_{L_{\o v}}\times \aa_{L_v})$~, we have on the Shimura 
bisemivariety $\partial \o S_{K^D_{R\times L}}$~, the Langlands irreducible global 
correspondence, i.e. the bijection:
\[\Irr_W^{(1)}(W^{ab}_{L^-}\times W^{ab}_{L^+})\To \Irr\ELLIP(GL_2(\Aa_{ L^T_{\o v}}
\times \Aa_{L^T_v }))\]
according to \cite{Pie9}.

Let us recall that if we fix:
\begin{align*}
\Gal(L^-/ K) &= \txt\bigoplus\limits_\mu\bigoplus\limits_{m_\mu } (\Gal L^-_{\mu }/K)\;, \\
\Gal(L^+/ K) &= \txt\bigoplus\limits_\mu\bigoplus\limits_{m_\mu } (\Gal L^+_{\mu }/K)\;, \end{align*}
the \rl global Weil group
$W^{ab}_{L^-}$ (resp. $W^{ab}_{L^+}$~) is the Galois subgroup of $\Gal(L^-/ K)$ (resp. $\Gal(L^+/ K)$~) of the extensions $L^-_\mu $ (resp. $L^+_\mu $~) characterized by degrees:
\begin{align*}
n_{\mu _R} &= [L^-_\mu :K]=0\mod N=(p+\mu )\ N\;, \\
\mbox{(resp.} \quad n_{\mu _L} &= [L^+_\mu :K]=0\mod N=(p+\mu )\ N\ )\;. \end{align*}
\vskip 11pt

\begin{defi} {\bf The notion of quantum\/} on the time semisheaf of rings $\theta^1_{R,L}(t)$ can be introduced as follows: let $s_{1_{R,L}}\subset \cdots \subset s_{q_{R,L}}$ be the set of sections of $\theta^1_{R,L}(t)$ and $n_1<\cdots <n_\mu< \cdots<n_q$ be the corresponding set of ranks.  According to the preceding developments, it corresponds to the section $s_{\mu_{R,L}}$ a set of equivalent sections $\{ s_{\mu,1_{R,L}},\cdots,s_{\mu,b_{R,L}}\}$ relative to the decomposition group $D_{\mu^2,b}$ where all the $s_{\mu,b_{R,L}}$ have the same rank $n_\mu$~.

A section $s_{\mu ,b_{R,L}}$ has a rank $n_{\mu _{R,L}}=(p+\mu )\cdot N$~, following 1.1.6, 
where $N$ is the order of the inertia subgroup $I_{L_{\o v_{\mu }}}$ (resp. $I_{L_{v_{\mu }}}$~) 
having as representation space the $T^t_2(L_{\o v^1_\mu })$-subsemimodule $M^I_{\o v_\mu }$ (resp. $T_2(L_{v^1_\mu })$-subsemimodule $M^I_{v_\mu }$~) which was interpreted in 1.1.6 as a \rl quantum.  
Thus, the section $s_{\mu ,b_{R,L}}$ is composed of $(p+\mu )$ \rl time quanta, noted $\widetilde M^I_\mu (t)_{R,L}$~, or $\widetilde M^T_{v^{\mu '}_{\mu ,m_\mu }}=\AnRepsp(T^{(t)}_2(L^T_{v^{\mu '}_{\mu ,m_\mu }}))$ where $\AnRepsp(T^{(t)}_2(L^T_{v^{\mu '}_{\mu ,m_\mu }}))$~, denoting  the analytic representation space of the algebraic subgroup $T^{(t)}_2(L^T_{v^{\mu '}_{\mu ,m_\mu }})$ over the irreducible subcompletion $L^T_{v^{\mu '}_{\mu ,m_\mu }}$~, is a ``class of germ'' of continuous (differentiable) function over a big point centered on $T^{(t)}_2(L^T_{v^{\mu '}_{\mu ,m_\mu }})$~.

Notice that the writing ``~$s_{\mu_{R,L}}$~'' will mean, in the following developments, either a basic section $ s_{\mu_{R,L}} $ alone, i.e. for a value of $b=0$~, or a subset of equivalent sections $\{ s_{\mu_{1_{R,L}}} ,\cdots, s_{\mu_{b_{R,L}}}\}$ corresponding to all the ideals of the decomposition group $D_{\mu^2,b}$~, the distinction between the two cases being in general evident.
\end{defi}
\vskip 11pt

\subsection[Generation of $4D$-semisheaves of rings by Eisenstein homology and
$(\gamma _{t\to r}\circ E)$ morphism]{\boldmath Generation of $4D$-semisheaves of rings by Eisenstein homology and
$(\gamma _{t\to r}\circ E)$ morphism}

\begin{defis}[1. Galois antiautomorphism] { From the right (resp.
left) Galois automorphic group $\Gamma_{\mu R,L}=\Aut_K L^{\mp}_\mu $~, it is
possible to define a Galois antiautomorphic group $\Gamma^*_{\mu
R,L}=\widetilde{\Aut}_K L^{\mp}_\mu $ acting transitively on the left on the
set of right (resp. left) prime ideals $b_{\mu R,L}$ of the right (resp. left)
specialization semiring $B_{\mu R,L}$~. We thus have a descending chain of right (resp.
left) specialization ideals:
\[ b_{\mu_{n  \mu R,L}}\supset\cdots\supset b_{\mu _{(n _\mu -\rho_\mu
)R,L}}\;,\quad \rho <n  \;,\]
where $(n  _\mu -\rho _\mu )$ is a decreasing rank.
\vskip 11pt

\noindent {\bf (2. Reduced algebraic semigroups)}  From the right (resp. left)
boundary of the compactified semispace $\partial \ \o S_{K_{t R,L}}$ (see
definition 1.1.7), it is possible to introduce the reduced compactified semispace:
\[ \partial \ \o S_{K^*_{t R,L}}=P^*_{t R,L}\setminus G^*_{t
R,L}/K^*_{t R,L}\;,\]
where\begin{enumerate}
\item $G^*_{t R,L}$ is a reduced algebraic semigroup, i.e. an algebraic
semigroup submitted to  Galois antiautomorphisms, and having the following decomposition:
\[G^*_{t_L}(\aa^*_L) \equiv T_2(\aa^*_L)=D_2
(\aa^*_L)\times UT^{-1}_2(\aa^*_L)\]
where
\Bi
\item $\aa^*_L$ is a reduced adele semiring given by $\aa^*_L\equiv \Aa^*_{L^T_v}=\prod\limits_\nu L^T_{v_\nu} \prod\limits_{m_\nu}L^T_{v_{\nu ,m_\nu }} $~, $\nu\le \mu$~, and coming from $\aa_L= \prod\limits_\mu L^T_{v_\mu} \prod\limits_{m_\mu} L^T_{v_{\mu,m_\mu }}  $ (see 1.1.4 and 1.1.10);
\item $UT^{-1}_2(\aa^*_L)$ is the inverse of $UT_2(\Aa_L)$~.
\Ei
\item $P^*_{t R,L}$ is a reduced parabolic semisubgroup;

\item $K^*_{tR,L}$ is a reduced arithmetic semisubgroup of $G^*_{t R,L}$~.
\end{enumerate}}
 \end{defis}
\vskip 11pt

\begin{propo} The right (resp. left) Eisenstein homology, defined from the
action of a right (resp. left) Galois antiautomorphic group, is associated to
the generation of a right (resp. left) reduced semisheaf of rings $\theta
^{*1}_{R,L}$ and decomposes into \cite{Pie3}:
\nobeqn
\lefteqn{H_*(\partial\o S_{K^*_{R,L}},\theta ^{*1}_{R,L})}\\
&=& H_*(P^*_{tR,L}\setminus G^*_{tR,L}(\aa^*_{R,L})/K^*_{
R,L}(\o\ZZ_{p_q}),\theta ^{*1}_{R,L})\\
&=& \txt\bigoplus\limits_{\xi_{R,L}\in \Xi_{K^*_{R,L}}}
H_*(S^{M^*_{R,L}}_{K^{*M_{R,L}}(\xi_{R,L})},H_*(\tilde u^*_{R,L},\theta
^{*1}_{R,L}))\;.\noeeqn
\end{propo}
\vskip 11pt

\bpr This proposition is the homological version of proposition 1.1.12.\epr
\vskip 11pt

\begin{defi}[Reduced algebraic Hecke characters] \   { If $\aa^*_{R,L}$
 denotes a reduced adele semiring, let
\[ \lambda ^*_{ R,L}=(\cdots,\lambda ^*_{\nu_{R,L}},\cdots) _{\aa_{R,L} \to \aa^*_{R,L}} \]
be a sequence of decreasing weights.

Then, $\lambda ^*_{R,L}=\{\lambda ^*_{1R,L},\cdots,\lambda ^*_{\nu
R,L},\cdots,\lambda ^*_{qR,L}\}$ is the sequential set of decreasing weights in
$X(\theta ^{*1}_{R,L})$ referring to the $q$ set of right (resp. left) sections of the
right (resp. left) reduced semisheaf of rings $\theta ^{*1}_{R,L}$~.

Let $\omega ^*_{R,L}=\{\cdots,\omega ^*_{\mu R,L},\cdots\}$ be the set of right
(resp. left) inverse actions of the Weyl groups on $\lambda ^*_{R,L}$~.  Then,
$\phi^*_{R,L}=\omega ^*_{R,L}\cdot\lambda ^*_{R,L}$ will denote the set of
right (resp. left) reduced algebraic Hecke characters on $\theta ^{*1}_{R,L}$~.}
\end{defi}
\vskip 11pt

\begin{defis} {\bf (1.} { Every {\bf smooth endomorphism)\/} $E[G_{\mu
R,L}]$ of the algebraic semigroup $G_{\mu R,L}$~, representing the Galois
subgroup $\Gamma_{\mu R,L}=\Aut_KL^{\mp}_\mu $~, can decompose
into the direct sum of the two nonconnected algebraic semigroups \cite{Pie3}:
\begin{enumerate}
\item the reduced algebraic semigroup $G^*_{\mu R,L}$~, submitted to the Galois
antiautomorphic subgroup $\Gamma^*_{\mu R,L}=\widetilde{\Aut}_KL^{\mp}_\mu $~;

\item the complementary algebraic semigroup $G^I_{\mu R,L}$~, submitted to the
complementary Galois automorphic subgroup $\Gamma^I_{\mu
R,L}=\Aut^{(I)}_KL^{\mp}_\mu $~, such that $G^I_{\mu R,L}$ be a semisubgroup of
$G_{\mu R,L}$~.
\end{enumerate}

We then have
\[ G_{\mu R,L} =G^*_{\mu R,L}\oplus G^I_{\mu R,L}\;.\]
Recall that the semisubgroup $K_{\mu R,L}\subset G_{\mu R,L}$ can be defined following \cite{Har1}:
\[K_{\mu R,L}=SO(m,L^{T\mp}_\mu )\cdot Z^0_{\mu R,L}(L^{T\mp}_\mu )\;.\]
The nonconnectivity of $G^*_{\mu R,L}$ and $G^I_{\mu R,L}$ is a necessary
condition to avoid triviality if the groups\linebreak $SO(m,L_\mu^{T\mp})^*\in K^*_{\mu R,L}$ and
$SO(m,L_\mu^{T\mp})^I\in K^I_{\mu R,L}$ had the same Witt index and the same  order
``~$m$~''.
\vskip 11pt

\noindent {\bf (2.\/} The {\bf complementary Galois automorphic group)\/}
$\Gamma^I_{\mu R,L}=\Aut^I_KL^{\mp}_\mu $ can be defined by its transitive right
action on the set of prime ideals $b^I_{\mu  R,L}$ of the complementary
specialization semiring $B^I_{\mu R,L}$ leading to an ascending chain of complementary
specialization ideals $b^I_{\mu _{1R,L}}\subset\cdots\subset b^I_{\mu_{\rho ^I_\mu}
R,L}$ such that the maximal rank $\rho ^I_\mu $ be equal to the
integer $n _\mu $ when the decreasing rank is $(n _\mu -\rho _\mu )$
(see definition 1.2.1).
\vskip 11pt

\noindent {\bf (3.  Complementary Eisenstein cohomology)\/}  From the
compactified complementary semispace $\ \o S_{K^I_{ R,L}}$~, we define its
boundary by
\[ \partial \ \o S_{K^I_{ R,L}}=P^I_{ R,L}\setminus G^I_{
R,L}/K^I_{ R,L}\;.\]
 The right
(resp. left) complementary Eisenstein cohomology can then be introduced:
\[ H^*_I (\partial\o S_{K^I_{R,L}},\theta ^1_{I_{R,L}}) =
H^*_I(P^I_{tR,L}  \setminus G^I_{tR,L}(\aa^I_{R,L})/K^I_{
R,L}(\o\ZZ_{p_q}),\theta ^1_{IR,L})\;.\]
It is associated to the generation of a right (resp. left) complementary
semisheaf of rings $\theta ^1_{I_{R,L}}$~.
}
\end{defis}
\vskip 11pt

\begin{defi} { Let $\gamma _{t\to r}$ be the {\bf emergent morphism\/},
introduced in \cite{Pie3} and mapping the
complementary semisheaf of rings $\theta ^1_{I_{R,L}}(t)$ from the
complementary semispace $\partial \ \o S_{K^I_{R,L}}(t)$ into its orthogonal
complementary semispace $\partial \ \o S^{\perp}_{K^I_{R,L}}(r)$ where
$r=\{x,y,z\}$ is the triple of spatial variables:
\[ \gamma _{t\to r} : \begin{array}[t]{ccc}
\partial \ \o S_{K^I_{R,L}}(t) &\to & \partial
\ \o S^{\perp}_{K^I_{R,L}}(r)\;,\\
\theta ^1_{I_{R,L}}(t) &\to & 
\theta ^3_{I_{R,L}}(r)\;.\end{array}\]
}\end{defi}
\vskip 11pt

\begin{propo} Let $\theta ^{*1}_{R,L}(t)$ be the reduced semisheaf of rings
generated under the smooth endomorphism $E$ by the right (resp. left)
Eisenstein homology.  Then, the morphism $(\gamma _{t\to r}\circ E)$
transforms the semisheaf of rings $\theta ^1_{R,L}(t)$ into:
\[ \gamma _{t\to r}\circ E : \begin{array}[t]{ccc}
\partial \ \o S_{K_{R,L}}(t) &\to & \partial \ \o{\!
S}_{K^*_{R,L}}(t)\oplus \partial \ \o S^{\perp}_{K^I_{R,L}}(r)\;,\\
\theta ^1_{{R,L}}(t) &\to & \theta ^{*1}_{R,L}(t)\oplus
\theta ^3_{I_{R,L}}(r)\;,\end{array}\]
such that each section $s^*_{\mu R,L}\oplus s_{I_{\mu R,L}}\in \theta
^{*1}_{R,L}(t)\oplus \theta ^3_{I_{R,L}}(r)$ be $T^{*1}_\mu (t)_{R,L}\oplus
T^1_{I_\mu }(r)_{R,L}$~, called a right (resp. left) elementon of space-time
and noted $T^{1-1}_\mu (t,r)_{R,L}$~, where $T^{*1}_\mu (t)_{R,L}$ is a set of
$1D$-tori and $T^1_{I_\mu }(r)_{R,L}$ is also a set of  $1D$-tori.
\end{propo}
\vskip 11pt

\bpr \begin{enumerate}
\item The complementary semisheaf of rings $\theta ^3_{I_{R,L}}(r)$ is
three-dimensional because the groups $SO(2p,L^{T\mp})\in K^*_{R,L}$ and
$SO(2p+1,L^{T\mp})\in K^{I\perp}_{R,L}$ must have the same Witt index $p=1$ in
order that the endomorphism $E$ be smooth \cite{Pie3} but their order ``~$m$~''
may be different: consequently, $m=2$ for $K^*_{R,L}$ and $m'=3$ for
$K^{I\perp}_{R,L}$ \cite{Bum}.
\vskip 11pt

\item The fact that the section $s_{I_{\mu R,L}}$ is a set of $1D$-tori results from:
\Bi
\item the morphism $\gamma_{t\to r}\circ E$ where $\gamma_{t\to r}$ corresponds to the projective map :
\[GL_2(\aa_R\times \aa_L)\To P\ GL_2(\aa_R\times \aa_L)\hookrightarrow GL_3(\aa_R\times \aa_L)\] as developed by S. Gelbart \cite{Gel2};
\item the decomposition (or degeneration) of the representation space $\Repsp (GL_3(\Aa_R\times \Aa_L))$ into one-dimensional components.
\epr\Ei
\end{enumerate}
\vskip 11pt

\begin{defi}[The quantum of space]   { 
Assume that the section $T^1_{I_\mu}(r)_{R,L}\in \theta^3_{I_{R,L}}(r)$~, 
generated under the $(\gamma_{t\to r}\circ E)$-morphism from the $1D$-section $T^1_\mu(t)$ 
composed in fact of a set of $m_\mu $ equivalent sections $\{T^1_{\mu,b}\}$ under the decomposition group 
$D_{\mu^2,b}$~, is partitioned into $m_\mu $ corresponding $1D$-fibers,  having each one a rank $\rho_\mu=(p+\mu )\cdot N$~.  
Then, each $1D$ equivalent section $T^{1}_{I_{\mu,m_\mu }}(r)_{R,L}$ has $\mu_p=(p+\mu )$  spatial quanta, 
noted $\widetilde M^I_\mu (r)_{R,L}$~, which are functions on subsemimodules of rank $N$~.  And the section 
$T^1_{I_\mu}(r)_{R,L}=\{T^{1}_{I_{\mu_1}},\cdots,T^{1}_{I_{\mu,m_\mu }}\}$ counts $m_\mu (p+\mu )$ 
space quanta.}
\end{defi}
\vskip 11pt

\begin{coro} There exists an inverse morphism $(\gamma _{r\to t}\circ E')$
transforming gradually and sequentially the $3D$-complementary semisheaf of
rings $\theta ^3_{I_{R,L}}(r)$ into the $1D$-semisheaf of rings $\theta
^1_{R,L}(t)$~.
\end{coro}
\vskip 11pt

\bpr Let $n$ and $(n-\rho )$ be the set of $q$ graded ranks
referring to the $q$ sections respectively of $\theta ^1_{R,L}(t)$ and $\theta
^{*1}_{R,L}(t)$ according to corollary 1.1.10.

(Note that the proof is valid for the right and left cases but the indices
$R,L$ will be dropped for facility).

Then the morphism $(\gamma _{r\to t}\circ E')$ is such that
\begin{enumerate}
\item $E':\theta ^3_I(r)_\rho \to \theta ^{*3}_I(r)_{\rho -\rho '}\oplus \theta
^3_{I(I)}(r)_{\rho '}$ where $\theta ^3_{I(I)}(r)_{\rho '}$ is the complementary
semisheaf of $\theta ^{*3}_I(r)_{\rho -\rho '}$ obtained under the smooth
endomorphism $E'$~.

\item $\gamma _{r\to t}:\theta ^3_{I(I)}(r)_{\rho '}\to \theta ^1(t)_{\rho
'}$ where $\gamma _{r\to t}$ maps $\theta ^3_{I(I)}(r)_{\rho '}$~, ideal by
ideal, into its $1D$-time orthogonal complementary space giving rise to $\theta
^1(t)_{\rho '}$~.

\item $(\gamma _{r\to t}\circ E') : \theta ^{*1}(t)_{n -\rho }\oplus \theta
^3_I(t)_\rho \to \theta ^{*1}(t)_{n -(\rho -\rho ')}+\theta ^{*3}_I(r)_{(\rho
-\rho ')}$~.

If $\rho '=\rho $~, then under $(\gamma _{r\to t}\circ E')$~, $\theta
^3_I(r)_\rho $ has been totally transformed into $\theta ^1(t)_{\rho '}$~.\epr
\end{enumerate}
\vskip 11pt

\begin{defi}[Algebraic  Hecke parameters]  { Let $\phi^*_{t;(n -\rho
)_{R,L}}$ bet the set of algebraic Hecke characters referring to the generation
of the reduced semisheaf of rings $\theta ^{*1}_{R,L}(t)$ by Eisenstein homology
and let $\phi_{r;\rho_{R,L}}$ bet the set of algebraic Hecke characters
referring to the generation of the complementary semisheaf of rings $\theta
^3_{I_{R,L}}(r)$ by Eisenstein cohomology.

We then have the following equality between these two sets of algebraic Hecke
characters:
\[ \phi^*_{t;(n-\rho )_{R,L}}=c_{t\to r}(\rho )_{R,L}\cdot\phi_{r;\rho
_{R,L}}\]
where $c_{t\to r}(\rho )_{R,L}=\{c_1(\rho _1)_{R,L},\cdots,c_q(\rho
_q)_{R,L}\}$ is a set of parameters referring to the $q$ sections of the
semisheaf of rings $\theta ^3_{I_{R,L}}(r)$ and depending on the set of
sequential ranks ``~$\rho $~''.

$c_{t\to r}(\rho )_{R,L}$ can be considered as an algebraic measure giving the
ratio of the generation of the complementary semisheaf of rings $\theta
^3_{I_{R,L}}(r)$ with respect to the reduced semisheaf of rings $\theta
^{*1}_{R,L}(t)$~.

Consequently, $c_{t\to r}(\rho )$ is the most closed to the unity when $(n
-\rho )=\rho $~.}
\end{defi}
\vskip 11pt

\begin{propo} \qquad  Each right and left $4D$-elementon of space-time $(T^{*1}_\mu
(t)_{R,L}\oplus T^1_{I_\mu }(r)_{R,L}) \in \theta ^{*1}_{R,L}(t)\oplus \theta
^3_{I_{R,L}}(r)$~, $1\le \mu \le q$~, is composed of elementary subtori
$\tau^{1-1}_\mu (t,r)_{R,L}$~, characterized by a rank $2N$, which are sums of a time and of a space quantum.
\end{propo}
\vskip 11pt

\bpr (The indices $R,L$ will be dropped in this proof). Let $(n_\mu -\rho
_\mu )$ be the rank of the section $T^{*1}_\mu (t)$ and let
$\rho _\mu $ be the rank of the section $T^1_{I_\mu }(r)$~, taking
into account that the complementary section $T^1_{I_\mu }(r)$ is generated from
$T^1_\mu (t)$ by the morphism $(\gamma _{t\to r}\circ E)$~.

Considering the algebraic generation of $T^1_\mu (t)$ under the action of the
Galois automorphic group $\Gamma_\mu =\Aut_KL_\mu $ and envisaging the
$(\gamma _{t\to r}\circ E)$ morphism, we then have that the elementary time
prime ideal $\tau^{*1}_\mu (t)\in T^{*1}_\mu (t)$ has a rank $N$ and  the elementary space prime ideal $\tau^1_\mu (r)\in T^1_{I_\mu }(r)$
is characterized by a rank $N$~.\epr
\vskip 11pt

\begin{coro} Consider the morphism:
\[ \gamma _{t\to r}\circ E : T^{1-1}_\mu (t,r)_{R,L}\to T^1_\mu
(r)_{R,L}\;,\qquad\forall\ \mu \;, \quad 1\le \mu \le q\;,\]
such that the reduced section $T^{*1}_\mu (t)_{R,L}$ of the semisheaf of rings
$\theta ^{*1}_{R,L}(t)$ be completely transformed into the complementary space
section $T^1_\mu (r)_{R,L}$~.  Then, every elementary subtorus $\tau^1_\mu
(r)_{R,L}\in T^1_\mu (r)_{R,L}$ is also characterized by a rank $N$~.\end{coro}
\vskip 11pt

\bpr This is obvious by taking into account the proposition 1.2.10.\epr
\vskip 11pt

\begin{sub} {\bf Space-time structure of semiparticles\/}
\end{sub}
{\noindent \bf 1)}  The mathematical and physical
reasons given in the introduction and in the following developments lead us to
admit that elementary particles must be composed of two symmetric objects,
called a right and a left semiparticle.  The basic ``algebraic'' space-time structure
of a right and a left semiparticle (or, more exactly, of a right and a left
semilepton, or semiquark, as it will be developed in section 4) will be
assumed to be respectively a right and a left sequential semisheaf of rings
$(\theta ^{*1}_{R,L}(t)\oplus \theta ^3_{I_{R,L}}(r))$ of which
$\theta^3_{I_{R,L}}(r)$ can be regarded as the algebraic representation of a space
physical wave packet.

The right and left $1D$-semisheaves or rings $\theta ^{*1}_R(t)$ and $\theta
^{*1}_L(t)$ must be viewed as the basic time structure of the right and left
semiparticles while the right and left $3D$-semisheaves of rings $\theta
^3_{I_R}(r)$ and $\theta ^3_{I_L}(r)$ must be regarded as the basic space
structure of the respective semiparticles.
\vskip 11pt

{\noindent \bf 2)}  Indeed, the fact of endowing elementary (semi)particles with an internal 
space-time structure from which the ``mass'' shell could be generated results from an attempt of the 
author \cite{Pie1} to bridge the gap between general relativity and quantum field theory.  The 
problem is that general relativity is a ``classical'' theory describing the mutual interaction 
between the geometry of space-time and the matter without explaining how matter could be 
generated.  Now, quantum field theory asserts precisely that matter must be created from the 
vacuum to which the cosmological constant of the general relativity equations could correspond 
if it was associated to it an expanding space-time which could then constitute the fundamental 
structure of the vacuum of QFT.  On the basis of these considerations, I have developed, in an 
unpublished preprint \cite{Pie1}, equations in differential geometry rather close by the equations 
of general relativity but referring to the quantum structure of bisemiparticles such that their 
most internal structures, which are space-time structures, be the fundamental structures of their 
own vacua from which their matter shells could be generated due to the fluctuations of these 
internal vacua.  So, the vacuum of QFT becomes peopled to massless (bisemi)particles being 
potentially able to generate their mass shells due to the fluctuations of these (bisemi)particle 
internal vacua. \vskip 11pt

{\noindent \bf 3)}  To the internal ``space'' structure of an elementary semiparticle then corresponds 
its linear momentum $\vec p$ on its ``mass'' shell; and, to the internal ``time'' structure of a 
semiparticle would corresponds its rest mass $m_0$~.  The fact of considering the internal time 
of a semiparticle as corresponding to a topological structure can be justified by the annihilation 
of a pair of leptons into (pair(s)) of photons and by 2) of 1.2.10. \vskip 11pt

{\noindent \bf 4)}  The internal space structure of a semiparticle is thus given by the semisheaf of 
rings $\theta^3_{I_{R,L}}(r)$ on a $T_3(\Aa_L)$ (resp. $T_3(\Aa_R)$-semimodule 
$M_L^{(T)}$ (resp. $M_R^{(T)}$~) restricted to the upper (resp. lower) half space.  
Indeed, following the Langlands program briefly developed in 1.1.22, $\theta^3_{I_{R,L}}(r)$ has 
an analytic representation given by the global elliptic semimodule $\ELLIP_{R,L}(1,q,b)$ which 
corresponds to an eigenfunction of the spectral representation of an operator $T_{R,L}$ (see 
chapter 3) on the space structure.  On the other hand, each term 
$\phi (s_{\mu _{R,L}})_{\mu ,b}e^{\pm2\pi i(p+\mu )x}\in\ELLIP_{R,L}(1,q,b)$ will be interpreted 
as the ``space'' structure of a semiphoton at $(p+\mu )$ quanta, giving then a (semi)photonic 
spatial structure to the semiparticle. \vskip 11pt

{\noindent \bf 5)}  If we consider the space structure of a \rl semiparticle as given by the 
three-dimensional semisheaf of rings $\theta^3_{I_{R,L}}(r)$~, then this semiparticle will be 
interpreted as having a wave (packet) aspect.  But, we have seen in proposition 1.2.6 that we can 
consider the projective map: $P_{G_{3\to 2}}: T_3( \Aa_R)\to T_2( \Aa_R)$ (resp. $ T_3( \Aa_L)\to 
T_2( \Aa_L)$~) to which corresponds the projective map: $P_{\theta_{3\to 2}}: \theta^3_{I_{R,L}}(r)
\to \theta^2_{I_{R,L}}(r)$ mapping the three-dimensional semisheaf of rings $\theta^3_{I_{R,L}}(r)$ 
into its two-dimensional analogue $\theta^2_{I_{R,L}}(r)$~, giving then to the space structure of 
a \rl semiparticle a ``particle'' aspect. \vskip 11pt

We can then formulate the first axiom referring to the generation of the
``wave'' space-time structure of elementary right and left semiparticles.
\vskip 11pt

\begin{axiom} The basic space-time structure of elementary right and left
semiparticles is of algebraic nature.
\end{axiom}
\vskip 11pt

\bpr Indeed, the space-time structure of elementary right and left
semiparticles is assumed to be given by $4D$-space-time right and left
sequential semisheaves of rings $(\theta ^{*1}_{R,L}(t)\oplus \theta
^3_{I_{R,L}}(r))$ whose $q$ sets of sections are ``~$4D$~''-elementons $T^{1-1}_\mu
(t,r)_{R,L}$ generated from $1D$-symmetric splitting semifield(s) by Eisenstein
cohomology and homology and by the morphism $(\gamma _{t\to r}\circ E)$~.\epr
\vskip 11pt

\subsection[Algebraic representation of bisemiparticles by bilinear Hilbert
schemes ]{Algebraic representation of bisemiparticles by bilinear Hilbert
schemes}

\begin{defi}[Tensor product of semisheaves of rings]  { The right and left
semi\-sheaves of rings $(\theta ^{*1}_{R,L}(t)\oplus \theta ^3_{I_{R,L}}(r))$ are defined 
respectively on a $G_R(\aa_R)$-right semimodule, noted 
$M^{ST}_R$~, and on a $G_L(\aa_L)$-left semimodule, noted 
$M^{ST}_L$~.  The $G_R(\aa_R)$-right semimodule $M^{ST}_R$ and the $G_L(\aa_L)$-left
semimodule $M^{ST}_L$ represent the basic internal space-time structures of the right
and left semiparticles (essentially leptons) which act conjointly in order to form a bisemiparticle
localized inside a $4D$-openball centered on the emergence point.  
As the $G_{R,L}(\aa_{R,L})$-right (resp. left) semimodule
$M^{ST}_{R,L}$ is also a unitary right (resp. left) $\aa_{R,L}$-semimodule, it is an
$\aa_{R,L}$-right (resp. left) semialgebra $M^{ST}_{R,L}$~.  By construction, $M^{ST}_R$ is the
opposite semialgebra of $M^{ST}_L$~.  So,  the tensor product $M^{ST}_R\otimes _{\aa_R\times \aa_L}M^{ST}_L$
will be the enveloping semialgebra of $M^{ST}_L$ and will be assumed to constitute the
space-time structure of a bisemiparticle. $(M^{ST}_R\otimes M^{ST}_L)$ will be written for $(M^{ST}_R\otimes _{\aa_R\times\aa_L}M^{ST}_L)$~.

The space-time structure of a bisemiparticle will thus be given by the tensor
product $(M^{ST}_R\otimes M^{ST}_L)$ of 
$M^{ST}_R$ and $M^{ST}_L$ such that the right semimodule $M^{ST}_R$ be flat on
the left semimodule $M^{ST}_L$~, i.e. that for every left semimodule $M^{'ST}$
and for every injective homomorphism $\nu :M^{'ST}_L\to M^{ST}_L$~, the
homomorphism $\ung_{M^{ST}_R}\otimes \nu :M^{ST}_R\otimes M^{'ST}_L\to M^{ST}_R
\otimes M^{ST}_L$ is injective \cite{Bou1}.

If the right and left semisheaves of rings $\widetilde M^{ST}_R$ and $\widetilde M^{ST}_L$ are
defined respectively on the right and left semispaces $M^{ST}_R$ and
$M^{ST}_L$~,  we then get a right and a left ringed
semispace $(M^{ST}_R,\widetilde M^{ST}_R)$ and $(M^{ST}_L,\widetilde M^{ST}_L)$~.

Similarly, we can define the tensor product between the right and left ringed
semispaces:
\[ \otimes :\{ (M^{ST}_R,\widetilde M^{ST}_R),(M^{ST}_L,\widetilde M^{ST}_L)\} \to (M^{ST}_R\times M^{ST}_L,
\widetilde M^{ST}_R\otimes \widetilde M^{ST}_L)\]
where $ M^{ST}_R\otimes  M^{ST}_L$ represents the space-time
structure of a bisemiparticle.

As $M^{ST}_R$ and $M^{ST}_L$ are semisimple, then $(M^{ST}_R\otimes M^{ST}_L)$
is also semisimple according to C. Chevalley \cite{Che2} and J.P. Serre
\cite{Ser5}, \cite{Ser6}.

The tensor product $(M^{ST}_R\otimes M^{ST}_L)$~, called a bisemimodule, is
characterized by a $10$-dimensional noneuclidean geometry, 
reflecting its degree of compactness and of instability.  Consequently, a
blowing-up morphism will be considered in the following proposition.}
\end{defi}
\vskip 11pt

\begin{propo} There exists a blowing-up isomorphism
\[ S_L : M^{ST}_R\otimes M^{ST}_L\to (M^{ST}_R\otimes_D M^{ST}_L)\oplus
(M^{S}_R\otimes_{magn} M^{S}_L)\oplus (M^{S-(T)}_R\otimes_{elec} M^{(S)-T}_L)\]
transforming the bisemimodule $(M^{ST}_R\otimes M^{ST}_L)$ of dimension $10$ (lepton case)
into a set of disconnected bisemimodules which are:
\begin{enumerate}
\item the diagonal bisemimodule $(M^{ST}_R\otimes_D M^{ST}_L)$ of dimension $4$
characterized by  a diagonal orthogonal $4D$-basis
$\{e^\alpha \otimes f_\alpha \}^3_{\alpha =0}$~, $\forall\ e^\alpha \in
M^{ST}_R$ and $f_\alpha \in M^{ST}_L$~;

\item the magnetic bisemimodule $(M^{S}_R\otimes_{magn} M^{S}_L)$ characterized
by a $3D$-nonorthogonal basis $(e^\alpha \otimes f_\beta )^3_{\alpha \neq
\beta =1}$~, where $M^S_{R,L}=\theta ^3_{I_{R,L}}(r)$~;

\item the electric bisemimodule $(M^{S}_R\otimes_{elec} M^{T}_L)$ or
$(M^{T}_R\otimes_{elec} M^{S}_L)$ characterized respectively by a
$3D$-nonorthogonal basis $(e^\alpha \otimes f_0)^3_{\alpha =1}$ or $(e^0
\otimes f_\alpha )^3_{\alpha =1}$ where $M^T_{R,L}=\theta ^{*1}_{R,L}(t)$~.
\end{enumerate}
\end{propo}
\vskip 11pt

\bpr The blowing-up isomorphism can be understood algebraically by considering that
right and left quanta are taken away respectively from the
right and left semimodules $M^{ST}_R$ and $M^{ST}_L$ by the smooth endomorphism
$E$~, recalled \cite{Pie3} in definition 1.2.4, in such a way that the complete
bisemimodule $(M^{ST}_R\otimes M^{ST}_L)$ be transformed into the diagonal bisemimodule $(M^{ST}_R\otimes_{D} M^{ST}_L)$~.  Consequently, the
disconnected right and left quanta will generate two off-diagonal
bisemimodules having a magnetic and an electrical metric to keep a trace of the
off-diagonal metric of $(M^{ST}_R\otimes M^{ST}_L)$ (see also 4.3.4 and 4.3.5).  The magnetic metric is
given by $g^\alpha _\beta =(e^\alpha ,f_\beta )^3_{\alpha \neq \beta =1}$ and
the electric metric is given by $g^\alpha _0=(e^\alpha ,f_0)^3_{\alpha =1}$ or
$g^0_\alpha =(e^0,f_\alpha )^3_{\alpha =1}$ where $(\cdot,\cdot)$ is a scalar
product.\epr
\vskip 11pt

\begin{defi}[Diagonal tensor product] { Let the right ringed semispace  
$(M^{ST}_R,\widetilde M^{ST}_R)$ of the right semiparticle define locally the affine right
semischeme $S^{ST}_R$ and the left ringed semispace $(M^{ST}_L,\widetilde M^{ST}_L)$ of the left
semiparticle define locally the affine left semischeme $S^{ST}_L$ \cite{Hart}.

Let $(S^{ST}_R\otimes_D S^{ST}_L)$ be the diagonal tensor product between the
right and left semischemes $S^{ST}_R$ and $S^{ST}_L$ characterized by a
diagonal metric.

Consider the projective morphisms $p_L$ and $p_R$~:
\nobeqn p_L &:& S^{ST}_R\otimes_D S^{ST}_L \to S^{ST}_{R(P)/L}\;,\\
p_R &:& S^{ST}_R\otimes_D S^{ST}_L \to S^{ST}_{L(P)/R}\;,\noeeqn
such that:
\begin{enumerate}
\item the right semischeme $S^{ST}_R$ be projected under $p_L$ on the left
semischeme $S^{ST}_L$ giving rise to the bisemischeme $S^{ST}_{R(P)/L}$~;

\item the left semischeme $S^{ST}_L$ be projected under $p_R$ on the right
semischeme $S^{ST}_R$ giving rise to the bisemischeme $S^{ST}_{L(P)/R}$~.
\end{enumerate}}\end{defi}
\vskip 11pt

\begin{propo} The diagonal tensor product $S^{ST}_{R(P)/L}$ of the right and left semischemes $S^{ST}_R$
and $S^{ST}_L$ such that $S^{ST}_R$ is projected on $S^{ST}_L$ is a covariant
functor of $S^{ST}_L$ representable by the bilinear Hilbert scheme ${\rm
Hilb}_{S^{ST}_{R(P)}/S^{ST}_L}$ where $S^{ST}_{R(P)}$ is dual of $S^{ST}_L$~.
\end{propo}
\vskip 11pt

\bpr Let $S^{ST}_{R(P)}$ be a projective scheme on $S^{ST}_L$ \cite{G-R1}.

Let $c$ be the category of locally noetherian $S^{ST}_L$ preschemes.  If
$T^{ST}_L\in {\rm obj}(c)$~, consider $S_{T^{ST}_L}=S^{ST}_{R(P)}\otimes_D
T^{ST}_L$ and let $F(T^{ST}_L)$ be the set of closed subpreschemes of
$S_{T^{ST}_L}$ which are flat on $T^{ST}_L$~: it is a covariant functor of
$T^{ST}_L$ representable by the Hilbert scheme ${\rm
Hilb}_{S^{ST}_{R(P)}/S^{ST}_L}$ \cite{Gro1}, \cite{Gro2}, \cite{Got}.

There is bilinearity on ${\rm Hilb}_{S^{ST}_{R(P)}/S^{ST}_L}$ with linearity on
the left semischeme $S^{ST}_L$ and antilinearity on the right semischeme
$S^{ST}_{R(P)}$ if we take into account that the associated right and left
ringed semispaces $(M^{ST}_R,\widetilde M^{ST}_R)$ and $(M^{ST}_L,\widetilde M^{ST}_L)$ are defined
respectively on the lower half space $M^{ST}_R$ and on
the upper half space $M^{ST}_L$~.

The projective right semischeme $S^{ST}_{R(P)}$ is flat on $S^{ST}_L$~. 
Furthermore, the right semischeme $S^{ST}_{R(P)}$ is dual of the left
semischeme $S^{ST}_L$~.\epr
\vskip 11pt

\begin{coro} The Hilbert scheme ${\rm Hilb}_{S^{ST}_{R(P)}/S^{ST}_L}$ is
endowed with a diagonal metric $g^\alpha _\alpha $ of type $(1,1)$ 
{\em \cite{Pie4}}.
\end{coro}
\vskip 11pt

\bpr Indeed, the components $g^\alpha _\alpha =(e^\alpha ,f_\alpha )$ of the
metric tensor at each point of ${\rm Hilb}_{S^{ST}_{R(P)}/S^{ST}_L}$ are
external scalar products with respect to the basis vectors $\{(e^\alpha
)^*\}^3_{\alpha =0}\in S^{ST}_{R(P)}$ and $\{f_\alpha \}^3_{\alpha =0}\in
S^{ST}_L$~.\epr
\vskip 11pt

\begin{propo} If we consider a bijective linear isometric map
$B_L:S^{ST}_{R(P)}\to S^{ST}_L$ mapping each covariant element of
$S^{ST}_{R(P)}$ into the corresponding contravariant element of $S^{ST}_L$~,
then the Hilbert scheme ${\rm Hilb}_{S^{ST}_{R(P)}/S^{ST}_L}$ is transformed
into the internal Hilbert scheme ${\rm Hilb}_{S^{ST}_{L_R}/S^{ST}_L}$ characterized
by a diagonal metric $g_{\alpha \alpha }$ of type $(0,2)$~.\end{propo}
\vskip 11pt

\bpr Indeed, under the $B_L$ map, the covariant basis vectors $\{(e^\alpha)^*\}^3_{\alpha =0}$ are transformed into the contravariant basis vectors
$\{(e_\alpha )^*\}^3_{\alpha =0}$ and the components $g^\alpha _\alpha $ of the
metric tensor then become internal scalar products $g_{\alpha \alpha
}=(e_\alpha ,f_\alpha )$~.\epr
\vskip 11pt

\begin{coro} The diagonal bisemischeme
$S^{ST}_{L(P)/R}$ is a covariant functor
of $S^{ST}_R$ representable by the bilinear Hilbert scheme ${\rm
Hilb}_{S^{ST}_{L(P)}/S^{ST}_R}$ endowed with a metric $g^\alpha  _\alpha $ of
type $(1,1)$~.\end{coro}
\vskip 11pt

\begin{coro} By the bijective linear isometric map $B_R:S^{ST}_{L(P)}\to
S^{ST}_R$~, the Hilbert scheme ${\rm Hilb}_{S^{ST}_{L(P)}/S^{ST}_R}$ is
transformed into the internal Hilbert scheme 
${\rm Hilb}_{S^{ST}_{R_L}/S^{ST}_R}$ characterized by a metric $g^{\alpha \alpha
}$ of type $(2,0)$~.
\end{coro}
\vskip 11pt

The presentation of bilinear Hilbert schemes leads us to formulate the
\vskip 11pt

\begin{Axiom}  Nature is composed of bisemiparticles whose fundamental diagonal
space-time structure is given locally by bilinear diagonal Hilbert schemes.  This axiom
is a multiplicative axiom {\em \cite{Ati4}}.\end{Axiom}
\vskip 11pt

\bpr Due to the fact that the right semischeme $S^{ST}_R$ of the right
semiparticle is topologically very close to the left semischeme $S^{ST}_L$
of the left semiparticle such that these two semischemes $S^{ST}_R$ and
$S^{ST}_L$ be localized in the same openball centered on the emergence point,
only the left semischeme of the left semiparticle will be commonly observable
in the frame of bilinear Hilbert schemes with the right semischeme of the right
semiparticle generally unobservable because it is projected on the left
semischeme of the left semiparticle: this corresponds to the existence of the bilinear Hilbert scheme $\Hilb_{S^{ST}_{R(P)}/S^{ST}_L}$~.\epr

\vskip 11pt

\begin{rem}[Twin bisemiparticles]  {But, there also exists a bilinear Hilbert scheme $\Hilb_{S^{ST}_{L(P)}/S^{ST}_R}$~, as introduced in corollary 1.3.8, and resulting from the projective morphism $p_R$ (given  in definition 1.3.3) which maps the left semischeme $S_L^{ST}$ on the right semischeme $S_R^{ST}$~.\newline
Thus, next to the common world in which we live and described by the bilinear Hilbert scheme $\Hilb_{S^{ST}_{R(P)}/S^{ST}_L}$ at the level of bisemiparticles, there is also the possibility of the existence of a twin world described by the bilinear Hilbert scheme $\Hilb_{S^{ST}_{L(P)}/S^{ST}_R}$ at the level of ``twin bisemiparticles''.}
\end{rem} \vskip 11pt

\subsection[Fundamental algebraic space-time structure of semileptons, semibaryons and semiphotons ]{Fundamental algebraic space-time structure of  semileptons, semibaryons and semiphotons}

In sections 1 and 2, the basic algebraic space-time structure of the right and
left semiparticles was assumed to be given by right and left semisheaves of
rings $(\theta ^{*1}_{R,L}(t)\oplus \theta ^3_{I_{R,L}}(r))$ generated by
Eisenstein cohomology and by the $(\gamma _{t\to r}\circ E)$ morphism. 
However, as it was noticed in  1.2.12, this basic space-time structure
corresponds essentially to the algebraic space-time structure of the
semilepton of the first family, i.e. the semielectron.
\vskip 11pt

It will be seen in the first part of this section how the algebraic time
structure of the semiquarks can be generated from the central algebraic time
structure of a semibaryon.

\vskip 11pt

\begin{defi}[Smooth endomorphism $E_t$] {\rm 1.} { Instead of
considering as in definitions 1.2.4 a smooth endomorphism $E[G_{\mu R,L}]$ of
the algebraic semigroup $G_{\mu  R,L}$ decomposing it into the direct sum of
two nonconnected algebraic semigroups, we can envisage the following smooth
endomorphism \cite{Pie3}: $E_t[G_{\mu  R,L}]=G^*_{(c)_{\mu  R,L}}\oplus
G^I_{(c)_{\mu  R,L}}$ of the algebraic semigroup $G_{\mu  R,L}$ decomposing it
into the two connected algebraic semigroups $G^*_{(c)_{\mu  R,L}}$ and
$G^I_{(c)_{\mu  R,L}}$ where $G^*_{(c)_{\mu  R,L}}$ is the reduced algebraic
semigroup, submitted to a Galois antiautomorphic subgroup, and where $G^I_{(c)_{\mu 
R,L}}$ is the complementary algebraic semigroup resulting from a Galois
automorphic subgroup.

The smooth endomorphism $E_t$ is such that the subgroups $SO(m_1,L_\mu ^{T\mp})\in
K^*_{\mu  R,L}\subset G^*_{(c)_{\mu R,L}}$ and $SO(m_2,L_\mu ^{T\mp})\in K^I_{\mu
R,L}\subset G^I_{(c)_{\mu  R,L}}$ must have the same rank but different
orders, i.e. that $m_1=2t$ and $m_2=2t+1$~, $t$ being an odd integer taking the value $t=1$ here.}
\vskip 11pt

\noindent {\rm 2.} { Let $\theta ^1_{R,L}(t)$ be the semisheaf of rings generated by
Eisenstein cohomology on the boundary of the Borel-Serre compactification  $\partial \ \o S_{K_{t
R,L}}=P_{t R,L}\setminus G_{t R,L}/K_{t R,L}$~.

Then, the smooth endomorphism $E_t$ applied to the semisheaf $\theta
^1_{R,L}(t)$ gives the following decomposition:
\[ E_t[\theta ^1_{R,L}(t)]=\theta ^{*1}_{R,L}(t)\oplus \theta
^3_{I_{R,L}}(t_1,t_2,t_3)\]
where $\theta ^3_{I_{R,L}}(t_1,t_2,t_3)$ is a $3D$-complementary semisheaf of
rings connected to the reduced semisheaf of rings $\theta ^{*1}_{R,L}(t)$~.}
\end{defi}
\vskip 11pt

\begin{lm} The $3D$-complementary semisheaf of rings $\theta
^3_{I_{R,L}}(t_1,t_2,t_3)$ can only be expressed as the direct sum of 3
connected $1D$-time semisheaves of rings
\[ \theta ^3_{I_{R,L}}(t_1,t_2,t_3)=\theta ^1_{I_{R,L}}(t_1)\oplus \theta
^1_{I_{R,L}}(t_2)\oplus \theta ^1_{I_{R,L}}(t_3)\;.\]
\end{lm}
\vskip 11pt

\bpr Indeed, according to definition 1.4.1, the complementary semisheaf of
rings\\  $\theta ^3_{I_{R,L}}(t_1,t_2,t_3)$ generated from $\theta
^{*1}_{R,L}(t)$ by the smooth endomorphism $E_t$ must be three dimensional. 
But, considering that:
\begin{enumerate}
\item $\theta ^3_{I_{R,L}}(t_1,t_2,t_3)$ is defined on a $1D$ semispace
$\partial \ \o S_{K_{R,L}}(t)$~,

\item $\theta ^3_{I_{R,L}}(t_1,t_2,t_3)$ is localized in the orthogonal
complement space of the $3D$-space \cite{Sco} on which semisheaves
$\theta ^3_{I_{R,L}}(r_i)$ are defined and generated by the $\gamma _{t_i\to r_i}$
morphisms,
\end{enumerate}

\noindent the semisheaf $\theta ^3_{I_{R,L}}(t_1,t_2,t_3)$ can only be
composed of three orthogonal $1D$-semisheaves of rings $\theta
^1_{I_{R,L}}(t_i)$~, $1\le i\le 3$ (see also \cite{Pie9}, section 4.1).

Consequently, $\theta ^1_{I_{R,L}}(t_i)$ is a $1D$-time semisheaf whose
sections are given by $1D$-tori $T^1_{\mu R,L}(t_i)$~.\epr
\vskip 11pt

\begin{propo}  The algebraic time structure of a semibaryon is given by \[ \theta ^{\rm Bar}_{R,L}(t)=\theta ^{*1}_{R,L}(t_c)\txt\bigoplus\limits^3_{i=1} \theta
^1_{I_{R,L}}(t_i)\]
where $\theta ^{*1}_{R,L}(t_c)$ is its core time structure and where $\theta
^1_{I_{R,L}}(t_i)$ is the time structure of a semiquark.
\end{propo}
\vskip 11pt

\bpr The semisheaf or rings $\theta ^{\rm Bar}_{R,L}(t)$ results directly from
definition 1.4.1 and lemma 1.4.2 such that the reduced semisheaf or rings
$\theta ^{*1}_{R,L}(t_c)$ is connected to the complementary semisheaves
$\theta ^1_{I_{R,L}}(t_i)$~.

The interpretation of $\theta ^{\rm Bar}_{R,L}(t)$ as the time structure of a
semibaryon is justified by the ``bag'' model of the baryons \cite{C-J-J-T-W}
and the confinement of the three quarks \cite{Bjo}, \cite{C-R}.\epr
\vskip 11pt

\begin{propo} The algebraic space-time structure of a semibaryon is generated
from $\theta ^{\rm Bar}_{R,L}(t)$ by $\gamma _{t_i\to r_i}$ morphisms following:
\[ \gamma _{t_i\to r_i}\circ E_i : \theta ^{\rm Bar}_{R,L}(t)\to \theta ^{\rm
Bar}_{R,L}(t,r)\] where $\theta ^{\rm Bar}_{R,L}(t,r)$ is given by
\[ \theta ^{\rm Bar}_{R,L}(t,r)=\theta ^{*1}_{R,L}(t_c)\txt\bigoplus\limits^3_{i=1} \theta
^{1-3}_{R,L}(t_i,r_i)\;.\]
\end{propo}
\vskip 11pt

\bpr \begin{enumerate}\item The morphism $(\gamma _{t\to r}\circ E)$ does not
apply on $\theta ^{*1}_{R,L}(t_c)$ because it is a reduced semisheaf of rings
resulting from the smooth endomorphism $E_t$ on which
$\bigoplus\limits^3_{i=1} \theta ^1_{I_{R,L}}(t_i)_{q_i}$ are connected.  
\item The space structure of the three semiquarks is generated by considering
the $(\gamma _{t_i\to r_i}\circ E_i)$ morphisms on $\theta
^1_{I_{R,L}}(t_i)_{q_i}$~:
\vskip 6pt

\mbox{} \hfill $\gamma _{t_i\to r_i}\circ E_i :\theta ^1_{I_{R,L}}(t_i)_{q_i}\to \theta
^{*1}_{I_{R,L}}(t_i)_{q_i}\oplus \theta ^3_{I_{R,L}}(r_i)_{q_i}\quad\forall\
i\ , \; 1\le i\le 3\;.$\epr
\end{enumerate}
\vskip 11pt

\begin{defi}[Constant of the strong interaction] { Let $n_B$ denote the set
of $q_B$ ranks of the $q_B$ sections of $\theta ^{1}_{R,L}(t_c)$ and let $(n_B-\rho_B)$ be
the set of $q_B$ decreasing ranks of $\theta ^{*1}_{R,L}(t_c)$~.

Then, $\phi^*_{t_c;(n_B-\rho _B)_{R,L}}$ will be the set of algebraic Hecke
characters related to the generation of the reduced semisheaf or rings $\theta^{*1}_{R,L}(t_c)$ by Eisenstein homology and $\phi_{[t_1,t_2,t_3];\rho
_{B_{R,L}}}$
 will be the set of algebraic Hecke characters related to the generation of
the complementary semisheaf or rings $\theta
^3_{I_{R,L}}(t_1,t_2,t_3)=\bigoplus\limits^3_{i=3}\theta
^1_{I_{R,L}}(t_i)_{q_i}$ by Eisenstein cohomology.

As introduced in definition 1.2.9, there is the following equality between
these two sets of algebraic Heche characters:
\[ \phi^*_{t_c;(n_B-\rho _B)_{R,L}}=G(\rho _B)_{t_c\to
[t_1,t_2,t_3]}\phi_{[t_1,t_2,t_3];\rho _{B_{R,L}}}\]
where 
\[G(\rho _B)_{t_c\to
[t_1,t_2,t_3]} =\{G_1(\rho _{B_1}),\cdots,G_\mu (\rho _{B_\mu
}),\cdots,G_{q_B}(\rho _{B_q})\}\]
is the set of $q_B$ parameters measuring the generation of the complementary
semisheaf $\theta ^3_{I_{I_{R,L}}}(t_1,t_2,t_3)$ from $\theta
^{*1}_{R,L}(t_c)$~.}
\vskip 11pt

\begin{propo} The  parameter $\langle G(\rho _B)_{t_c\to
[t_1,t_2,t_3]}\rangle = \sum\limits^{q_B}_{\mu =1}G_\mu (\rho _{B_\mu)
}$ must correspond to the strong constant of the strong interaction.
\end{propo}
\vskip 11pt

\bpr Indeed, $G(\rho _B)_{t_c\to[t_1,t_2,t_3]}$ measures the generation of the
time structure of the three semiquarks $\theta ^3_{I_{R,L}}(t_1,t_2,t_3)$
from the core time structure $\theta ^{*1}_{R,L}(t_c)$ of the envisaged
semibaryon.

If $(n_B-\rho _B)\to 0$~, then $\phi^*_{t_c;(n_B-\rho _B)}\to 0$ and we have
asymptotic freedom \cite{G-W}, \cite{Pol}, \cite{Wein1}, corresponding to the
fact that the semiquarks become free since $\bigoplus\limits^3_{i=1}\theta
^1_{I_{R,L}}(t_i)$ are no more connected to $\theta ^{*1}_{R,L}(t_c)\to 0$~: this
is reflected by $\langle G(\rho _B)_{t_c\to[t_1,t_2,t_3]}\rangle \to 0$~.

\hspace{-3mm} On the other hand, $\!$ if $\phi_{[t_1,t_2,t_3]\rho _{B_{R,L}}\!}\!$ is small,$\!$ then
$\langle G(\rho _B)_{t_c \to[t_1,t_2,t_3]\!}\rangle$
 will take high values.\epr
\vskip 11pt

\begin{defi}[Algebraic Hecke characters] { Consider the morphism
\[ (\gamma _{t\to r}\circ E):\theta ^1_{R,L}(t)\to \theta
^{*1}_{R,L}(t)\oplus \theta ^3_{I_{R,L}}(r)\]
transforming sequentially and gradually the $1D$-time semisheaf of rings
$\theta ^1_{R,L}(t)$ into the $3D$-spatial semisheaf of rings $\theta 
^3_{I_{R,L}}(r)$~.  Let $(n-\rho )_{R,L}$ be the set of ranks of
the semisheaf $\theta ^{*1}_{R,L}(t)$ and $\phi^*_{t;(n-\rho )_{R,L}}$ the
corresponding set of algebraic Hecke characters.

Similarly, let $\rho _{R,L}$ be the set of ranks of the semisheaf
$\theta ^3_{R,L}(r)$ and $\phi_{r;\rho _{R,L}}$ the corresponding set of
algebraic Hecke characters.

If $(n-\rho )_{R,L}\to 0$~, then we have that
\[ \gamma _{t\to r}\circ E :\theta
^1_{R,L}(t)\stackrel{\sim}{\longrightarrow} \theta ^3_{I_{R,L}}(r)_P\]
which means that the semisheaf $\theta ^1_{R,L}(t)$ has been nearly
transformed into the $3D$-spatial semisheaf $\theta ^3_{I_{R,L}}(r)_P$~.

Let $\phi^*_{t;(n-\rho )_{R,L}}=c_{t\to r}(\rho )\phi_{r;\rho _{R,L}}$ be
the equality between the corresponding sets of algebraic Hecke characters. 
Then, 
\[c^{-1}_{t\to r}(\rho )=\{c^{-1}_1(\rho _1),\cdots,c^{-1}_\mu (\rho _\mu
),\cdots,c^{-1}_q(\rho _q)\}\]
is the set of $q$ inverse parameters measuring the generation of the semisheaf
$\theta ^3_{R,L}(r)_P $ from the semisheaf $\theta
^{*1}_{R,L}(t)$~.}\end{defi}
\vskip 11pt

\begin{propo} If $(n-\rho )_{R,L}\to 0$~, then
\begin{description}
\item{\rm 1.}  the average parameter $\langle c^{-1}_{t\to r}(\rho
)\rangle_{\max}=\left(\sum\limits^q_{\mu =1}c^{-1}_\mu (\rho _\mu )\right)/q$
is proportional to the velocity of the light ``~$c$~''.

\item{\rm 2.} the semisheaf of rings $\theta ^3_{I_{R,L}}(r)_P$ resulting from the
morphism:
\[(\gamma _{t\to r}\circ E)_{\max}:\theta
^1_{R,L}(t)\stackrel{\sim}{\longrightarrow} \theta ^3_{I_{R,L}}(r)_P\]
gives the structure of a set of right (resp. left) semiphotons.
\end{description}
\end{propo}
\vskip 11pt

\bpr If $(n-\rho )_{R,L}\to 0$~, the $1D$-time structure $\theta ^1_{R,L}(t)$ has
been nearly completely transformed into the $3D$-spatial structure
$\theta ^3_{I_{R,L}}(r)_P$~; consequently, $\langle c^{-1}_{t\to r}(\rho
)\rangle_{\max}$~, giving a measure of the ratio of the spatial structure
with respect to the time structure, must be proportional to the velocity of
the light: indeed, as the proper time of the semiphotons tends to zero, their
space(-time) structure is crudely given by $\theta ^3_{I_{R,L}}(r)_P$~.\epr
\vskip 11pt

\begin{defi}[A right (resp. left) semiphoton] { Considering that a photon with momentum $k$ corresponds to a plane wave 
and that to each normal mode $k$ is associated  $(p+\mu)_k$ quanta, we shall assume that the internal (vacuum) space structure of a right(resp. left) semiphoton with momentum $\vec{p}=\hbar\ \vec k$ will be described by a spatial section $T^1_{R,L}(r_k)$ (which is a $1D$-real torus according to definition 1.2.7) having $(p+\mu)$ quanta of momenta $\vec k_\tau $ and composed of $(p+\mu)\cdot N$ prime ideals  corresponding to $(p+\mu)\cdot N$ Galois automorphisms.

Then, $\vec k=(p+\mu )\vec k_\tau$ and $\vec p = \hbar_{ST}\ \vec k=(p+\mu) \hbar_{ST}\ \vec k_\tau$~, where the equivalent of the Planck constant $h_{ST}$  corresponds to the integer $N$ in the internal (vacuum) space time unit system (see proposition 2.2.13).}\end{defi}
\vskip 11pt

\begin{defi} {\bf The (vacuum) space-time structure\/} { of elementary right and left
semiparticles is assumed to be given at the fundamental level by:
\begin{description}
\item{1.} the number of sections of  the space-time semisheaf of rings
representing their structure;
\item{2.} the set of ranks of these sections and especially the set of
parameters $c_{t\to r}(\rho) _{R,L}$ (see definition 1.2.9) measuring the
generation of the complementary space semisheaf of rings $\theta
^3_{I_{R,L}}(r)$ with respect to the reduced time semisheaf of rings $\theta
^{*1}_{R,L}(t)$~.\end{description}}
\end{defi}
\vskip 11pt

More precisely, the fundamental algebraic structure of:
\begin{enumerate}
\item  semileptons will be characterized by:
\begin{description}
\item{\rm 1.} $g_\ell$ sections with $g_\ell\in \nit^+$ (~$\ell$ for leptons); 
\item{\rm 2.} a set of $g_\ell$ ranks $n_\ell$ referring to these sections
in such a way that the set of ranks $(n_\ell-\rho _\ell)$ refers to
the reduced $1D$-time semisheaf of rings $\theta ^{*1}_{R,L}(t)_\ell$ and
$\rho _\ell$ refers to the complementary $3D$ semisheaf of rings $\theta
^3_{I_{R,L}}(r)_\ell$~.  Then, the set of parameters
\[ c_{t\to r}(\rho _\ell)_{R,L}=\phi^*_{t;(n_\ell-\rho _\ell)_{R,L}}\ /\ \phi_{r;(\rho
_\ell)_{R,L}}\]
gives a measure of the generation of $\theta ^3_{I_{R,L}}(r)_\ell$ with
respect to $\theta ^{*1}_{R,L}(t)_\ell$~.
\end{description}
\vskip 11pt

\item semibaryons will be characterized by:
\begin{description}
\item{1.} $q_B$ sections;

\item{2.} a set of $q_B$ ranks $n_B$~;

\item{3.} a set of $q_B$ parameters $G(\rho _B)_{t_c\to[t_1,t_2,t_3]}$~, as
introduced in definition 1.4.5 and measuring the generation of the three
complementary $1D$-time semisheaves of rings $\theta ^1_{I_{R,L}}(t_i)_{q_i}$
from the reduced $1D$-time semisheaf of rings $\theta ^{*1}_{R,L}(t_c)$ of the
baryonic core.\\
Parallely, we have a set of $q_B$ ranks $(n_B-\rho _B)$
referring to $\theta ^{*1}_{R,L}(t_c)$ and a set of $q_B$ complementary
ranks $\rho _B$ referring to $\bigoplus\limits^3_{i=1}\theta
^1_{I_{R,L}}(t_i)_{q_i}$~.

\item{4.} three sets of $q_{B_i}$ parameters $c_{t\to r}(\rho _{q_i})$~, $1\le i\le
3$~, referring to the generation of the $3D$-spatial semisheaves of rings of
the three semiquarks $\theta ^3_{I_{R,L}}(r_i)_{q_i}$ from $\theta
^1_{I_{R,L}}(t_i)_{q_i}$~.
\end{description}\end{enumerate}
\vskip 11pt

\begin{propo} The sets of parameters $c_{t\to r}(\rho _\ell)$ of semileptons
and $c_{t\to r}(\rho _{q_i})$ of the semiquarks are obstruction parameters
with respect to the stability of these semiparticles.\end{propo}
\vskip 11pt

\bpr Indeed, these sets of parameters fix the space structures of these
semiparticles with respect to their time structures according to definition
1.4.10, preventing their annihilation, i.e. the complete transformation of
their time semisheaves of rings in their complementary $3D$-space semisheaves
of rings by the morphisms $(\gamma _{t\to r}\circ E)$~.\epr
\vskip 11pt

\begin{lm} The number of geometric points of all right (resp. left) time quanta  
$M^I_\mu(t)_{R,L}$ of rank $N$ is equal.\end{lm}
\vskip 11pt

\bpr The sections of the $1D$-time semisheaves of rings $\theta ^1_{R,L}(t)$
are generated by Eisenstein cohomology from symmetric splitting semifields
having the same number of simple roots according to definitions 1.1.2 and
1.1.4.  Consequently, all the sections $s_{\mu R,L}\in \theta ^1_{R,L}(t)$~,
generated from the specialization ideals $p_{\mu R,L}$ (see definition 1.1.3),
are composed  of functions on right (resp. left) time quanta having the same number of
geometric points \cite{L-N}.\epr
\vskip 11pt

This is true for semileptons.  But, if we take into account the following
proposition, it is also verified for semibaryons whose time structure
originates from symmetric splitting subsemifields $L^{\mp}_\mu $ whose number is greater than for
semileptons because the baryon masses are bigger than the lepton masses in a
given family of elementary particles.
\vskip 11pt

\begin{propo}  The right (resp. left) time quantum of a semiquark is a right
(resp. left) time  quantum of the baryonic core time quantum.\end{propo}
\vskip 11pt

\bpr Indeed, the smooth endomorphism $E_t$~, transforming the baryonic core
$1D$-time semisheaf of rings $\theta ^1_{R,L}(t_c)$ into the three
complementary $1D$-time semisheaves of rings $\theta ^1_{I_{R,L}}(t_i)_{q_i}$
of the three semiquarks, acts on $\theta ^1_{R,L}(t_c)$ prime ideal by prime
ideal through the action of the Galois
antiautomorphic group.

Consequently, the number of geometric points of a time quantum of the baryonic
core is equal to the number of the geometric points of a time
complementary  quantum of a semiquark.\epr
\vskip 11pt

\begin{propo} The number of geometric points of all right (resp. left) space
quanta $M^I_{\mu}(r)_{R,L}$ is equal.\end{propo}
\vskip 11pt

\bpr As by lemma 1.4.12, the number of geometric points of all time quanta $M^I_\mu(t)_{R,L}$ is equal and as the space quanta are
generated from the corresponding time quanta by the morphism $(\gamma _{t\to
r}\circ E)$~, we reach the thesis.\epr
\vskip 11pt

\begin{propo} The number of geometric points of a space quantum of a semiquark is
equal to the number of geometric points of a time quantum.\end{propo}
\vskip 11pt

\bpr By lemma 1.4.12 and proposition 1.4.13, we know that the number of
geometric points of a time quantum of the baryonic core is equal to the number of
the geometric points of a time  quantum of a semiquark. 
Considering on the one hand the three $(\gamma _{t_i\to r_i}\circ E_i)$
morphisms responsible for the generation of the space structure $\theta
^3_{I_{R,L}}(r_i)_{q_i}$ of the three semiquarks according to proposition
1.4.4 and on the other hand the $(\gamma _{t\to r}\circ E)$ morphism
responsible for the generation of the space semisheaf of rings $\theta
^3_{I_{R,L}}(r)_\ell $ of a semilepton,  for example, we get the thesis since
all time quanta have the same number of geometric\linebreak points.\epr
\vskip 11pt

\sub{} {\bf The quantification rules of the space-time structure} { of
semiparticles can then be envisaged by considering that:
\vskip 11pt 

\begin{enumerate}
\item the time structure $\theta ^1_{R,L}(t)_{\ell_1}$ of a semilepton
$\ell_1$ can lose time quanta by the action of the smooth endomorphism $E$
according to definition 1.2.4:
\[ E:\theta ^1_{R,L}(t)_{\ell_1}\to \theta
^{*1}_{I_{R,L}}(t)_{\ell_1}\txt\bigoplus\limits^m_{k=1}\widetilde M^I_k (t)_{R,L} \]
where $\widetilde M^I_k(t)_{R,L}$ are disconnected time quanta (functions) of rank $N$~.

These free right (resp. left) time quanta (functions) can then join the right (resp. left)
time semisheaf of rings $\theta ^1_{R,L}(t)_{\ell_2}$ of another semilepton,
labeled $\ell_2$~, and increase its time structure.
\vskip 11pt

\item similarly, space quanta $\widetilde M^I_k(r)_{R,L}$ can be disconnected from the
space structure $\theta ^3_{R,L}(r)_{\ell_1}$ of a semilepton $\ell_1$ by:
\[ E:\theta ^3_{R,L}(r)_{\ell_1}\to \theta
^{*3}_{I_{R,L}}(r)_{\ell_1}\txt\bigoplus\limits^{m'}_{k=1}\widetilde M^I_k(r)_{R,L}\] and increase the space structure $\theta ^3_{R,L}(r)_{\ell_2}$ of another
semilepton $\ell_2$~.
\vskip 11pt

\item the time semisheaves of rings $\theta ^1_{R,L}(t_i)_{q_i}$~, $1\le i\le
3$~, of the three semiquarks $q_i$ of a semibaryon $B_1$ can lose time quanta
$\widetilde M^I_{k_i}(t_i)_{R,L}$ by means of the smooth endomorphisms
\[ E_i :\theta ^1_{R,L}(t_i)_{q_i}\to \theta ^{*1}_{I_{R,L}}(t_i)_{q_i}
\txt\bigoplus\limits^{m_i}_{k_i=1}\widetilde M^I_{k_i}(t_i)_{R,L}\;.\]
These time quanta (functions) can then increase the time semisheaves of rings of the three
semiquarks of a semibaryon $B_2$~.

Similar conclusions are reached with space quanta.

\end{enumerate}}
\end{defi}
\vskip 11pt

Let us  note that quantification rules with right or left quanta are
not exact since only bisemiparticles have a real existence.  Consequently,
only quantification rules with biquanta can be considered as developed in
chapter 3, section 3. \vskip 11pt

It was demonstrated in \cite{Pie11} that the quantification rule  consisting in adding time or space quanta to a semisheaf of rings corresponds to a deformation of a modular Galois representation while the quantification rule referring to the removal of quanta from a semisheaf or rings corresponds to an inverse deformation of projective type associated to an endomorphism.

More concretely, let $s_{\mu_{R,L}}$ denote a section of a semisheaf of rings having a rank $n_\mu=(p+\mu)\cdot N$~.

Then, a deformation of $s_{\mu_{R,L}}$ corresponds to an equivalence class of lift:
\[\Ds^{[p+\mu]\to[p+\mu+\nu]}_{R,L}:\quad s_{\mu_{R,L}}\To s_{\mu+\nu_{R,L}}\]
sending $s_{\mu_{R,L}}$ to a section $s_{\mu+\nu_{R,L}}$ having a rank $n_{\mu+\nu}=(p+\mu+\nu)\cdot N$ and composed of $(p+\mu+\nu)$ quanta.  The deformation $\Ds^{[p+\mu]\to[p+\mu+\nu]}_{R,L}$ is associated to the exact sequence:
\[ 1\To \widetilde M^I_{\mu_{R,L}}\To s_{\mu+\nu_{R,L}}\To s_{\mu_{R,L}}\To 1\]
whose kernel is a quantum (function) $\widetilde M^I_{\mu_{R,L}}$~.

On the other hand, a section $s_{\mu+\nu_{R,L}}$ can be submitted to the inverse deformation
\[ \Ds_{R,L}^{[p+\mu +\nu ]\to[p+\mu ]}:\quad s_{\mu+\nu_{R,L}}\To s_{\mu_{R,L}}\]
which is a modular projective mapping sending a section $s_{\mu+\nu_{R,L}}$ of rank $n_{\mu +\nu }= (p+\mu +\nu )\cdot N$ to a section $s_{\mu_{R,L}}$ of rank $n_\mu =(p+\mu )\cdot N$ corresponding to an endomorphism of $s_{\mu+\nu_{R,L}}$ removing $\nu $ quanta which become ``free''.
\vskip 11pt

\setcounter{defi}{0}
\section{Deformations of the fundamental algebraic structure of semiparticles}

\subsection[Versal deformation and spreading-out isomorphism]{Versal deformation and spreading-out\protect\newline 
 isomorphism}

External perturbations can generate singularities on the sections of the
semisheaves of rings $\theta ^{1-3}_{R,L}(t,r)$~. This problem is analyzed in
this section by considering the versal deformation of a semisheaf of germs of
differentiable functions $\theta ^m_{R,L}(s_{R,L})$ of dimension $m$ having
isolated singularities.

The related question consisting in the algebraic extension of the quotient
algebra of the versal deformation is principally considered: it is essentially
the inverse problem of the versal deformation of a sheaf of germs of
differentiable functions.  This problem has some analogy with the resolution
\cite{Hir}, \cite{Thoma}, \cite{Tei} of the singularities of an algebraic
variety since it ``reduces'' the versal deformation.
\vskip 11pt

Under some external perturbation, singularities \cite{Tho1}, \cite{Lev} are
assumed to be generated on the sections $s_{\mu R,L}\in \theta ^m_{R,L}$~. We
then consider:

\begin{defis}[1. The division theorem]  { This theorem will be recalled for
germs of differentiable functions $s_{\mu R,L}$ having an isolated singularity
of corank 1~.  Remark that nonisolated singularities were investigated by
Siersma \cite{Sie} and Pellikaan \cite{Pel} who consider as starting point of
their developments the group of all local isomorphisms leaving the singular
locus invariant.

Let $(x_1,\cdots,x_{m-n},w_1,\cdots,w_n)$ denote the coordinates
in $(L^{\mp})^m$~.

A germ $s_{\mu R,L}(w_{R,L})\in \theta ^m_{R,L}(s_{R,L})$ has a singularity of corank 1
(then, $n=1$~) and order $p$ in $w_{R,L}$ if $s_{\mu R,L}(0,w_{R,L})=w^p_{\mu
R,L} e_\mu (w_{R,L})$ where $e_\mu (w_{R,L})$ is a differentiable unit, i.e.
verifying $e_\mu (0)\neq 0$~.
\vskip 11pt

Let $\theta ^{m}_{R,L}[w_{R,L}]$ be the algebra of polynomials in $w_{R,L}$
with coefficients $r_{i\mu }(x)_{R,L}$ being subfunctions of
$\theta ^{m-1}_{R,L}$ defined on a domain $D_{R,L}\subset B_{R,L}$ where
$B_{R,L}$ is a lower (resp. upper) half open ball of radius $b$ around $0\in
L^{m-1}$~.

If $s_{\mu R,L}\in \theta ^m_{R,L}(s_{R,L})$ has order $p$ in $w_{R,L}$~,
then, there exist a differentiable function $q_{\mu R,L}\in \theta
^m_{R,L}(q_{R,L})$ and a polynomial
\[ R'_{\mu R,L}=\txt\sum\limits^r_{i=1} r_{i\mu }(x)_{R,L}w^i_{\mu R,L}\in \theta
^m_{R,L}[w_{R,L}]\]
with degree $r<p$ such that
\[ f_{\mu R,L}=s_{\mu R,L}\cdot q_{\mu R,L}+R'_{\mu R,L}\]
is the versal unfolding of $s_{\mu R,L}$ and corresponds to the Malgrange division theorem for the right and left cases.  The Malgrange
division theorem \cite{Mal} is the differentiable version of the Weierstrass
division theorem \cite{G-R2}, \cite{G-K} valid for germs of analytic functions
\cite{Math1}.
\vskip 11pt

\noindent {\bf (2. The preparation theorem)}   Let
\[ w _{\mu R,L}=w^p_{R,L}+\txt\sum\limits^{p-1}_{i=0} b_{i\mu }(x)_{R,L}w^i_{R,L} \]
be the Weierstrass polynomial verifying $b_{1\mu }(0)=\cdots=b_{(p-1)\mu
}(0)=0$~.  If $s_{\mu R,L}\in \theta ^m_{R,L}(s_{R,L})$ has finite order $p$
in $w_{R,L}$~, then there exists a uniquely determined Weierstrass polynomial
$w _{\mu R,L}\in \theta ^{m}_{R,L}[w_{R,L}]$ and a unit $e_{\mu R,L}\in
\theta ^m_{R,L}(e)$ such that $s_{\mu R,L}=w _{\mu R,L}\cdot e_\mu (w_{R,L})$~.

If $s_{\mu R,L}\in \theta ^{m}_{R,L}[w_{R,L}]$~, then $e_{\mu R,L}\in \theta
^{m}_{R,L}[w_{R,L}]$ and we get the preparation theorem
\[ f_{\mu R,L}=s_{\mu R,L}\cdot q_{\mu R,L}+R_{\mu R,L}\]
where $q_{\mu R,L}\in \theta ^m_{R,L}(q)$ and
\[ R_{\mu R,L}=\txt\sum\limits^r_{i=1}a_{i\mu }(x)_{R,L}w^i_{\mu R,L}\in \theta
^{m}_{R,L}[w_{R,L}]\]
with $a_{i\mu }(x)_{R,L}\in \theta ^{m-1}_{R,L}(a_i)$ and $r<p$~.}
\end{defis}
\vskip 11pt

\begin{defis}[1.  Versal deformation]  { Let $\theta ^{m-1}_{R,L}(a_i)$ be
the semisheaf of  differentiable functions $a_{i\mu }(x)_{R,L}
\subset s_{\mu R,L}(x,w )$~, $x=(x_1,\cdots,x_{m-n})$~,
 and
$\theta ^1_{R,L}(w^i)$ be the $i$-th generator semisheaf of monomial functions 
$w^i_{\mu R,L}$~. Then, $\theta _{R,L}(s_{R,L})=\{\theta
^1_{R,L}(w^1_{R,L}),\cdots,$ $\theta ^1_{R,L}(w^i_{R,L}),\cdots,\theta
^1_{R,L}(w^r_{R,L})\}$ is the right (resp. left) family of semisheaves of the
right (resp. left) base $s_{R,L}$ of the versal deformation of the semisheaf
$\theta ^m_{R,L}(s_{R,L})$~.  Indeed, the versal deformation of the semisheaf
$\theta ^m_{R,L}(s_{R,L})$~, whose sections are the differentiable functions $s_{\mu R,L}$~, is
given by the product \cite{Trau1}, \cite{Trau2}:
\[\theta _{R,L}(f_{R,L})=\theta ^m_{R,L}(s_{R,L})\times \theta
^r_{R,L}(s_{R,L})\]
where $\theta ^r_{R,L}(s_{R,L})$ is the base semisheaf such that $\theta
_{R,L}(f_{R,L})$ is $s_{R,L}$-flat.
\vskip 11pt

Recall succinctly that a deformation \cite{Ill} is called versal if each
deformation of $\theta _{R,L}(s_{R,L})$ is isomorphic to another deformation
of $\theta _{R,L}(s_{R,L})$ induced by some transformation of the base
semisheaf $\theta _{R,L}(s_{R,L})$ \cite{Pala}.
\vskip 11pt

\noindent {\bf (2. Quotient algebra)\/}  The quotient algebra $\theta
_{R,L}[R_{w_{R,L}}]$ of the versal deformation of the semisheaf $\theta
^m_{R,L}(s_{R,L})$ is a finitely generated vector space of dimension $r$
whose elements are the polynomials $R_{\mu R,L}=\sum\limits^r_{i=1}a_{i\mu
}(x)_{R,L}w^i_{\mu R,L}$~.}\end{defis}
\vskip 11pt

\begin{defi}[Specialization semirings]  { Let $p(a_{i\mu })_{R,L}$ be the
specialization prime ideal of the subsemiring $A_{a_{i\mu R,L}}$ referring to
the generation of the function $a_{i\mu }(x)_{R,L}$~.  Then, $p(a_i)_{R,L}$ will
denote the set of specialization prime ideals $\{p(a_{i\mu })_{R,L}\}^q_{\mu
=1}$ of the semiring $A_{a_{i R,L}}$ referring to the generation of the
semisheaf $\theta ^{m-1}_{R,L}(a_i)$~.

Similarly, let $p(w^i_{\mu R,L})$ be the specialization prime ideal of the
subsemiring $A_{w^i_{\mu R,L}}$ referring to the generation of the $i$-th
base function of the polynomial $R_{\mu R,L}$~.  Then, $p(w^i)_{R,L}$ will
denote the set of specialization prime ideals $\{p(w^i_{\mu R,L})\}^q_{\mu
=1}$ of the semiring $A_{w^i_{R,L}}$ referring to the generation of the $i$-th
generator semisheaf $\theta  ^1_{R,L}(w^i_{R,L})$~.
\vskip 11pt

According to section 1.1.3, let $\beta (a_i)_{R,L}$ be the set of specialization ideals of the specialization
semiring $B_{a_{iR,L}}$ dividing the set of specialization ideals
$p(a_i)_{R,L}$~.  Similarly, let $\beta (w^i)_{R,L}$ be the set of specialization
ideals of the specialization semiring $B_{w^i_{R,L}}$ dividing the set of
specialization ideals $p(w^i)_{R,L}$~.

Then, $B_{a_{iR,L}}$ is the integral closure of $A_{a_{iR,L}}$ and $\theta ^{m-1}_{R,L}(a_i)$ is a semisheaf on the
free
$A_{a_{iR,L}}$-semimodule $B_{a_{i_{R,L}}}$~.

Similarly, $\theta
^1_{R,L}(w^i)$ is a semisheaf on the $A_{w^i_{R,L}}$-semimodule $B_{w^i_{R,L}}$~.}
\end{defi}
\vskip 11pt

\begin{lm} The semisheaves $\theta ^{m-1}_{R,L}(a_i)$~, $1\le
i\le r$~, and $\theta ^1_{R,L}(w^i)$ are
characterized by the same set of ranks.\end{lm}
\vskip 11pt

\bpr 
\Bena
\item Let $n_{w^i_{R,L}}=\{n_{w^i_1},\cdots,n_{w^i_\mu },\cdots,n_{w^i_q}\}$ be
the set of ranks corresponding to the set of subsemimodules $B
_{w^i_{R,L}}=\{B _{w^i_1},\cdots,B _{w^i_q}\}$ and let
$n_{a_{iR,L}}=\{n_{a_{i1}},\cdots,n_{a_{i\mu }},\cdots,$   $n_{a_{iq}}\}$ be the
set of ranks of the set of subsemimodules $B
_{a_{iR,L}}=\{B _{a_{i1}},\cdots,B _{a_{iq}}\}$~.

Each section $s_{\mu R,L}\in \theta ^m_{R,L}(s_{R,L})$~, $1\le \mu \le q$~,
having a singularity of order $p$ is $(p+1)$ determined \cite{Math2},
\cite{Tou}. Consequently, there exist $(r-1)$ embedded and sequential
subspaces of the quotient algebra $\theta _{R,L}[R_w]$ of the versal
deformation $\theta _{R,L}(f_{R,L})=\theta^m _{R,L}(s_{R,L})\times \theta^r
_{R,L}(s_{R,L})$ of the semisheaf $\theta^m _{R,L}(s_{R,L})$~: this also
reflects the finite determinacy \cite{Pie5} of the quotient algebra $\theta
_{R,L}[R_w]$~.

Considering that the quotient algebra develops according to:
\[\theta _{R,L}[R_w]=\txt\sum\limits^r_{i=1}\theta ^{m-1}_{R,L}(a_i)\times \theta
^1_{R,L}(w^i)\;,\]
each semisheaf direct product $\theta ^{m-1}_{R,L}(a_i)\times \theta
^1_{R,L}(w^i)$ must be finitely generated.

\item the section $a_{i\mu }(x)\RL\subset s_{\mu R,L}(x,w )$ being a subfunction of $s_{\mu R,L}(x,w )$ must be characterized by a rank
\[ n_{a_{i\mu }}=(h_\mu \cdot N)^{m-1}\]
where:
\Bi
\item the integer $h_\mu $ is a global residue degree verifying $h_\mu <\mu $ with $\mu $ being the global residue degree of the $\mu $-th conjugacy class of $T_m(\Aa_{L_v})$ or of $T^t_m(\Aa_{L_{\o v}})$ on which $s_{\mu R,L}(x,w )$ is defined.

Note that the rank $n_{s_\mu }$ of this $\mu $-th conjugacy class is $n_{s_\mu }=(\mu \cdot N)^m$ \cite{Pie9}.
\item $N$ is the rank of a real irreducible completion.
\Ei

\item As $\theta ^r\RL(s\RL)$ is projected onto the semisheaf $\theta ^{m-1}\RL(a_i)$ in 
such a way that the semisheaf $\theta ^1\RL(w^i\RL)$~, $1\le i\le r$~, be flat onto 
$\theta ^{m-1}\RL(a_i)$~, the monomial function 
$w^i_{\mu R,L}\in \theta ^1\RL(w^i\RL)$~, which is a normal crossings divisor, 
must have a rank $n_{w^i_\mu }$ proportional or equal to the rank $n_{a_{i\mu }}$ of $a_{i\mu }(x)\RL$~.

If $m-1\le 2$~, then we have that:
\Bi
\item \quad $n_{w^i_\mu }=(h_\mu \cdot N)^p$ where $p\ge m-1$~;
\item \quad $n_{w^i_\mu }\ge n_{a_{i\mu }}$~.\epr
\Ei
\Ee

\vskip 11pt

\begin{defi}[Singular ideal]  { The function $w^i_{\mu R,L}\in \theta
^1_{R,L}(w^i)$ can have an isolated singular point in the specialization ideal
$\beta _{j;w^i_\mu }$~.  Then, $\Delta\beta ^S_{j;w^i_\mu }=\beta _{j;w^i_\mu
}- \beta _{(j-1);w^i_\mu }$ will be called a singular ideal.  The
rank of $w^i_{\mu R,L}$ will be called the total rank,
noted $n^T_{w^i_\mu }$~, and will be equal to $n^T_{w^i_\mu }=(n_{w^i_\mu
}-1)+1\equiv n_{w^i_\mu }$ where the second term in the sum refers to the
singular ``rank'' of the singular ideal $\Delta \beta ^S_{j;w^i_\mu
}$~.}\end{defi}
\vskip 11pt

\begin{lm} Let $\theta ^1\RL(w^i\RL)$ be the $i$-th base semisheaf of the versal deformation of $\theta ^m\RL(s\RL)$~.

Let $f_i^{\max}$ be the maximal value of its global residue degree counting the irreducible subschemes of rank $N$~.

Then, the following smooth endomorphism 
\[ E_{w^i\RL}[\theta ^1\RL(w^i\RL)_{f_i^{\max}}]=\theta ^{*1}\RL(w^i\RL)_{f^*_{r_i}}\oplus \theta ^1_{I\RL}(w^i\RL)_{f^I_{r_i}}\;,\]
with $f^I_{r_i}=f_i^{\max}-f^*_{r_i}\in\NN$~, can be introduced on the semisheaf 
$\theta ^1\RL(w^i\RL)_{f_i^{\max}}$ in such a way that it decomposes into two non connected complementary semisheaves $\theta ^{*1}\RL(w^i\RL)_{f^*_{r_i}}$ and $\theta ^1_{I\RL}(w^i\RL)_{f^I_{r_i}}$~.
\end{lm}
\vskip 11pt

\bpr

\Bena
\item Referring to the rank $n_{w^i_\mu }=(h_\mu \cdot N)^p$ of the monomial function 
$w^i_{\mu R,L}\in \theta ^1\RL(w^i\RL)$ as given in lemma 2.1.4, we see that its unramified rank or global residue degree is given by:
\[ f_{w^i_\mu }=(h_\mu )^p=n_{w^i_\mu }\big/N^p\;.\]
The integer $f_{w^i_\mu }$ is the number of irreducible completions of rank $N$ on which $w^i_{\mu R,L}$ is defined.  So, $f_i^{\max}$ will be in the same manner the number of irreducible completions of rank $N$ on which $\theta ^1\RL(w^i\RL)$ is defined:
\[f_i^{\max}=\txt\bigoplus\limits_\mu \bigoplus\limits_{m_\mu}(h_{\mu ,m_\mu })^p\;.\]

\item The semisheaf $\theta ^{*1}\RL(w^i\RL)_{f^*_{r_i}}$ is a reduced semisheaf generated from $\theta  ^1\RL(w^i\RL)_{f_i^{\max}}$ under the action of the Galois antiautomorphic group according to the endomorphism $E_{w  ^i\RL}$ in such a way that:
\Bi
\item $\theta ^{*1}\RL(w^i\RL)_{f^*_{r_i}}$ is characterized by decreasing global residue degrees $f^*_{r_i}$~;
\item $\theta ^1_{I\RL}(w^i\RL)_{f^I_{r_i}}$ is characterized by increasing global residue $f^I_{r_i}$ verifying $f_i^{\max}=f^*_{r_i}+f^I_{r_i}$~.\epr
\Ei
\Ee
\vskip 11pt

\begin{propo} Every base semisheaf $\theta ^1\RL(w^i\RL)$ of the versal deformation of the semisheaf $\theta ^m\RL(s\RL)$ can generate under the smooth endomorphism $E_{w^i\RL}$ the elements of the category $c(\theta ^1_{w^i\RL})$ of the $(f_i-1)$ pairs of semisheaves of rings:
\begin{align*}
c(\theta ^1_{w^i\RL})
&= \{ (\theta ^{*1}\RL(w^i\RL)_{f_i^{\max}-1}\oplus \theta ^1_{I\RL}(w^i\RL)_1),\\
& \quad \cdots,
(\theta ^{*1}\RL(w^i\RL)_{f^*_{r_i}}\oplus \theta ^1_{I\RL}(w^i\RL)_{f^I_{r_i}}),\\
& \quad \cdots,
(\theta ^{*1}\RL(w^i\RL)_1\oplus \theta ^1_{I\RL}(w^i\RL)_{f_i^{\max}-1})\}\;,\quad
1\le f^*_{r_i}\le f_i^{\max}\;,\end{align*}
whose objects are two nonconnected semisheaves characterized by complementary global residue degrees verifying: $f_i^{\max}=f^*_{r_i}+f^I_{r_i}$~.
\end{propo}
\vskip 11pt

\bpr This is a generalization of lemma 2.1.6 where $(f_i^{\max}-1)$ smooth endomorphisms 
$E_{w^i\RL}$ are considered.\epr

\vskip 11pt

\begin{coro} Let $f^*_{r_{i}}$ denote the global residue degree set of the $i$-th 
reduced semisheaf of rings $ \theta
^{*1}_{R,L}(w ^i)$ and let $f^I_{r_{i}}$ denote the global residue degree set of the
$i$-th  complementary semisheaf of rings $ \theta ^1_{I_{R,L}}(w ^i)$~.

Then, the smooth endomorphism $E_{w^i\RL}$ applied on the 
semisheaf of rings $ \theta ^1_{R,L}(w^i)$  is maximal when $f^*_{r_{i}}=0$~.
\end{coro}
\vskip 11pt

\bpr Indeed, if $f^*_{r_{i}}=0$~, then $f^I_{r_{i}}=f_i^{\max}$~, which means that
the reduced semisheaf of rings $ \theta
^{*1}_{R,L}(w^i)$ has been completely transformed into the complementary semisheaf of
rings $ \theta ^1_{I_{R;L}}(w^i)$~.\epr
\vskip 11pt

\begin{propo} Let $\theta _{R,L}(f_{R,L})=\theta ^m_{R,L}(s_{R,L})\times
\theta ^r_{R,L}(s_{R,L})$ be the versal deformation of the semisheaf $\theta
^m_{R,L}(s_{R,L})$ having $\theta _{R,L}[R_w]=\sum\limits^r_{i=1}\theta
^{m-1}_{R,L}(a_i)\times \theta ^1_{R,L}(w^i)$ as quotient algebra.
\\
Then, there exists a family of isomorphisms $\Pi_s
(f^*_{r_{1_{R,L}}},\cdots,f^*_{r_{i_{R,L}}},\cdots,f^*_{r_{r_{R,L}}})$ given by:
\nobeqn
\lefteqn{\hspace{-1cm}\Pi_s
(f^*_{r_{1_{R,L}}},\cdots,f^*_{r_{i_{R,L}}},\cdots,f^*_{r_{r_{R,L}}}) : \theta
^m_{R,L}(s_{R,L})\times \theta ^r_{R,L}(s_{R,L})\to \theta
^m_{R,L}(s_{R,L})\times \theta'  _{R,L}(s_{R,L})}\\
\noalign{\vskip 6pt}
&\txt\bigcup & \{( \theta
^1_{I_{R,L}}(w^1))_{f^I_{r_{1_{R,L}}}},\cdots,(
\theta ^1_{I_{R,L}}(w^i))_{f^I_{r_{i_{R,L}}}},\cdots,( \theta
^1_{I_{R,L}}(w^r))_{f^I_{r_{r_{R,L}}}}\}\;, \quad 1\le i\le r\;,\noeeqn
where:
\begin{enumerate}
\item $( \theta
^1_{I_{R,L}}(w^i))_{f^I_{r_{i_{R,L}}}}$ is the $i$-th complementary semisheaf  having global residue degree set $f^I_{r_{i_{R,L}}}$ generated by the smooth endomorphism
$E_{w^i\RL}$ from the semisheaf $( \theta ^1_{{R,L}}(w^i))_{f _i}$ having global residue degree set 
$f_i$~.

\item $\theta ^m_{R,L}(s_{R,L})\times \theta' _{R,L}(s_{R,L})=(\theta
^m_{R,L}(s_{R,L})\times \theta ^r_{R,L}(s_{R,L}))- $\\
\hspace*{2cm} $\{( \theta
^1_{I_{R,L}}(w^1))_{f^I_{r_{1_{R,L}}}},\cdots, (
\theta ^1_{I_{R,L}}(w^i))_{f^I_{r_{i_{R,L}}}},\cdots,
 (
\theta ^1_{I_{R,L}}(w^r))_{f^I_{r_{r_{R,L}}}}\}$~, 
$1\le i\le r$~.
\end{enumerate}
\end{propo}
\vskip 11pt

\bpr 
\Bena
\item This proposition is a generalization of proposition 2.1.7 in such a way that the smooth endomorphism $E_{w^i\RL}$~, generating $(f_i^{\max}-1)$ pairs of semisheaves of the category $c(\theta ^1_{w^i\RL})$~, is extended to all the base semisheaves $\theta ^1\RL(w^i\RL)$~, $1\le i\le r$~, of the versal deformation.

\item The family of endomorphisms $\pi _s(f^*_{r_{1\RL}},\cdots,f^*_{r_{i\RL}},\cdots,
f^*_{r_{r\RL}})$ is such that:
\Bi
\item $f^I_{r_{1\RL}}$ irreducible subschemes of rank $N$ are disconnected from the base 
semisheaf $\theta ^1\RL(w^1\RL)_{f_1^{\max}}$ on $\theta ^m\RL(s\RL)$~;
\item $f^I_{r_{i\RL}}$ irreducible subschemes of rank $N$ are disconnected from the base semisheaf $\theta ^1\RL(w^i\RL)_{f_i^{\max}}$ on $\theta ^m\RL(s\RL)$~;
\item and so on, $1\le i\le r$~.
\Ei

\item The set of complementary residue degrees $(f^I_{r_{1\RL}},\cdots,f^I_{r_{i\RL}},\cdots,
f^I_{r_{r\RL}})$ varies in such  a way that:
\[ 1\le f^I_{r_{1\RL}}\le f_1^{\max}\;, \; \cdots\;,\; 1\le f^I_{r\RL}\le f^{\max}_r\]
implying for each set $(f^I_{r_{1\RL}},\cdots,f^I_{r_{r\RL}})$ a family of isomorphisms $\pi _s(f^*_{r_{1\RL}},\cdots,f^*_{r_{r\RL}})$~.

\item $\theta '\RL(s\RL)$ is the residue base semisheaf resulting from the disconnection of the set $\{\theta ^1_{I\RL}(w^1)_{f^I_{r_{1\RL}}},\linebreak \cdots,\theta ^1\RL(w^r)_{f^I_{r_{r\RL}}}\}$~.\epr
\Ee
\vskip 11pt

\begin{coro} The family of isomorphisms
\nobeqn
\lefteqn{\hspace{-1cm}\Pi_s^{\max (i)}(f^*_{r_{1_{R,L}}},\cdots,f^*_{r_{r_{R,L}}}) :
\theta ^m_{R,L}(s_{R,L})\times \theta ^r_{R,L}(s_{R,L})\to \theta
^m_{R,L}(s_{R,L})\times \theta ' _{R,L}(s_{R,L})}\\
\noalign{\vskip 6pt}
&\txt\bigcup & \{ (\theta
^1_{I_{R,L}}(w^1))_{f^I_{r_{1_{R,L}}}},\cdots,
(\theta ^1_{I_{R,L}}(w^i))_{f^I_{r_{i_{R,L}}}},\cdots,
 (\theta
^1_{I_{R,L}}(w^r))_{f^I_{r_{r_{R,L}}}}\}\noeeqn
is maximal in the $i$-th semisheaf  $( \theta ^1_{R,L}(w^i)$ if $f^*_{r_{i_{R,L}}}=0$~.\end{coro}
\vskip 11pt

\bpr If $f^*_{r_{i\RL}}=0$~, the $i$-th
semisheaf $ \theta
^1_{R,L}(w^i)$ has been completely transformed into its complementary
disconnected semisheaf  $ \theta
^1_{I_{R,L}}(w^i)$~.  Indeed, we have that: $f^I_{r_{I_{R,L}}}=f_i^{\max}$ if
$f^*_{r_{I_{R,L}}}=0$~.\epr
\vskip 11pt

\begin{coro} The family of isomorphisms $\Pi_s$ is maximal if:
\[
{\Pi_s^{\max}:\theta ^m_{R,L}(s_{R,L})\times \theta ^r_{R,L}(s_{R,L}))\to
\theta ^m_{R,L}(s_{R,L})}
\;\txt\bigcup\;  \{ ( \theta
^1_{I_{R,L}}w^1)_{f_1^{\max}},\cdots,( \theta
^1_{I_{R,L}}(w^r))_{f_r^{\max}}\}\;,\]
i.e. if $f^I_{r_{i_{R,L}}}=f_i$~, $\forall\ i$~, $1\le i\le r$~.\end{coro}
\vskip 11pt

\bpr Indeed, $f^I_{r_{i_{R,L}}}=f_i^{\max}$~, if $f^*_{r_{i_{R,L}}}=0$~, $\forall\ i$~.  In
this case, all the semisheaves 
$ \theta ^1_{R,L}(w^i)$~, $1\le i\le r$~, of the
quotient algebra have been disconnected from $\theta _{R,L}(f_{R,L})$~. 
Consequently, $\theta _{R,L}(f_{R,L})=\theta ^m_{R,L}(s_{R,L})\times \theta
^r_{R,L}(s_{R,L})$ reduces to the semisheaf $\theta ^m_{R,L}(s_{R,L})$~.\epr
\vskip 11pt

\begin{defi}[Category of vertical tangent vector bundles]  { Let
\[T_{V_{w}}= \{T_{V_{w_1}},\cdots,T_{V_{w_i}},\cdots,T_{V_{w_r}}\}\] denote
the family of tangent vector bundles obtained by considering the projection of
all complementary semisheaves  $( \theta ^1_{I_{R,L}}(w^i))_{f^I_{r_{i\RL}}}$~, $\forall\
i$~, $1\le i\le r$~, in the vertical tangent spaces $T_{V_{w_i}}$
characterized by the normal vector fields $\overrightarrow {w}_i$~.

Let $\tau_{V_{w_i}}$ be the proper projective map of the tangent vector
bundle $T_{V_{w_i}}$~:
\[ \tau_{V_{w_i}}:T_{V_{w_i}}( \theta
^1_{I_{R,L}}(w^i))_{f^I_{r_{i\RL}}}\to(\theta
^1_{I_{R,L}}(w^i))_{f^I_{r_{i\RL}}}\]
so that $\tau_{V_{w}}=\{\tau_{V_{w_i}}\}^r_{i=1}$~.
\\
To the category $c( \theta
^1_{I_{R,L}}(w^i))$ will then correspond the category $c(T_{V_{w_i}}( \theta ^1_{I_{R,L}}(w^i))$ of sections of tangent
vector bundles.}\end{defi}
\vskip 11pt

\begin{propo}
The extension of the quotient algebra of the versal deformation of the
semisheaf $\theta ^m_{R,L}(s_{R,L})$ having an isolated singularity of order
$p$ in each section $s_{\mu R,L}$ is realized by the spreading-out isomorphism
$SOT=(\tau_{V_{w}}\circ \Pi_s)$~.\end{propo}
\vskip 11pt

\bpr Let $I_{w_i}$ be the kernel of the normal vector bundle
$T_{V_{w_i}}$~.  

Then, the exact sequence
\[ 0\to I_{w_i}\to T_{V_{w_i}}( \theta
^1_{I_{R,L}}(w^i))_{f^I_{r_{i\RL}}}\stackrel{\tau_{V_{w_i}}}{\to} ( \theta ^1_{I_{R,L}}(w^i))_{f^I_{r_{i\RL}}}\to 0\]
represents an extension of the complementary semisheaf  $(\theta ^1_{I_{R,L}}(w^i)) _{f^I_{r_{i\RL}}}$ by the kernel 
$I_{w_i}$~.\epr
\vskip 11pt

\begin{propo} The spreading-out isomorphism $SOT$ is locally stable if the
complementary semisheaves $T_{V_{w_i}}( \theta ^1_{I_{R,L}}(w^i)) _{f^I_{r_{i\RL}}}$~, $\forall\
i$~,  $1\le i\le r$~, generated by $SOT$ from $\theta  _{R,L}(f_{R,L})$~, are locally
free semisheaves.\end{propo}
\vskip 11pt

\bpr If $T_{V_{w_i}}( \theta
^1_{I_{R,L}}(w^i)) _{f^I_{r_{i\RL}}}$ has a singular ideal in the sense of definition
2.1.5, then it is not stable, taking into account that a semisheaf is locally
stable if it is locally free.\epr
\vskip 11pt

\begin{propo} \qquad The maximal number of complementary semisheaves  
$T_{V_{w_i}}( \theta
^1_{I_{R,L}}(w^i)) _{f^I_{r_{i\RL}}}$ generated by the spreading-out isomorphism $SOT$
is equal to the codimension of the versal deformation of the semisheaf $\theta
^m_{R,L}(s_{R,L})$~.\end{propo}
\vskip 11pt

\begin{rem} { Let us recall that all the singularities of generic wave
fronts in spaces of dimension $\le 7$ are locally diffeomorphic to the
$A_{p-1}$ and $D_{p-1}$ singularities \cite{Arn1}, \cite{Arn2} whose simple
genotypes in $\rit^m\to \rit$ have the normal forms \cite{Mil1}:
\nobeqn
A_{p-1} &:& x^p+Q\;,\\
D_{p-1} &:& x^2y+y^{p-2}+Q' \noeeqn
where $Q$ and $Q'$ are nondegenerated quadratic forms respectively of $(m-1)$
and $(m-2)$ variables.

As we are concerned in this work essentially with spaces of dimension 3, the
only singularities to be considered are the corank one (i.e. with $n=1$~)
singularities $A_{p-1}$ and the corank two (i.e. with $n=2$~) singularities
$D_{p-1}$~.}
\end{rem}
\vskip 11pt

\begin{defis}[Corank 2 singularities]  { 1. The Malgrange preparation
theorem can be generalized to germs of differentiable functions $s_{\mu
R,L}\in \theta ^m_{R,L}(s_{R,L})$ having an isolated singularity of corank 2. 
Indeed, if a germ $s_{\mu R,L}$ has singularities of corank 2 and order $p$ in
the two indeterminates $w_1$ and $w_2$~, then:
\begin{enumerate}
\item $s_{\mu R,L}(0,w_1,w_2)=P_{\mu R,L}(w_1,w_2)\cdot e_{\mu R,L}(w_1,w_2)$
where $e_{\mu R,L}(w_1,w_2)$ is a differentiable unit and $P_{\mu
R,L}(w_1,w_2)$ is a Weierstrass polynomial of degree $p$~.
\item the quotient algebra of the Malgrange preparation theorem is a finitely
generated tensoriel space of type $(0,2)$ and dimension $r<p$~.
\end{enumerate}
\vskip 11pt

\noindent 2. The spreading-out isomorphism $SOT$ can clearly be applied to the
versal deformation of a semisheaf of germs of differentiable functions having
 singularities of corank 2 because the proposition 2.1.10 can be generalized
to this case.}\end{defis}
\vskip 11pt

\begin{defi}[Gluing-up of complementary semisheaves] { The category  of  
complementary semisheaf direct products $\{T_{V_{w_i}}( \theta ^1_{I_{R,L}}(w^i))\}^r_{i=1}$ does not
necessarily cover in a compact way the semisheaf $\theta
^m_{R,L}(s_{R,L})\times\theta' _{R,L}(s_{R,L})$ generated from the semisheaf
$\theta ^m_{R,L}(s_{R,L})$ by versal deformation and spreading-out isomorphism
$SOT$~.  However, we can define a gluing-up of semisheaves
$T_{V_{w_i}}( \theta
^1_{I_{R,L}}(w^i)) _{f^I_{r_{i\RL}}}$ which are above the semisheaf $\theta
^m_{R,L}(s_{R,L})\times\theta' _{R,L}(s_{R,L})$~.}\end{defi}
\vskip 11pt

As the sections of the semisheaves $T_{V_{w_i}}(\theta ^1_{I\RL}(w^i) _{f^I_{r_{i\RL}}}$ are constituted of normal crossings divisors having a rank $n_{w^i_\mu }=(h_\mu \cdot N)^p$~, with $p\ge m-1$~, according to lemma 2.1.4, we can say that the dimension of these semisheaves is approximatively equal to $m$~.

In this perspective,
let us denote the $i$-th and the $j$-th complementary semisheaves  
  by $(\theta ^m_i(D{(w_i)})_{R,L}$ and by $(\theta
^m_j(D{(w_j)})_{R,L}$ defined respectively on the domains $D(w_i)$ and
$D(w_j)$~.  Consider then the gluing-up of these two complementary
semisheaves in the following manner:

For each pair $(i,j)$~, let $\Pi_{ij}$ be an isomorphism from $\theta
^m_j(D(w_i)\cap D(w_j))_{R,L}$ to $\theta ^m_i(D(w_i)\cap
D(w_j))_{R,L}$~.  Then, there exists a semisheaf $\theta
^m(D(w_{i-j}))_{R,L}$ defined on the connected domain
$D(w_{i-j})=D(w_i)\cup D(w_j))$ and an isomorphism $n_i$ from $\theta
^m(D(w_i))_{R,L}$ to $\theta ^m_i(D(w_i))_{R,L}$ such that
$\Pi_{ij}=n_i\circ n^{-1}_j$ in each point of $(D(w_i)\cap D(w_j))$~,
$\forall\ i$~, $1\le i\le r$~: this is an adapted version of a proposition of
J.P. Serre \cite{Ser1}.

So, $\theta ^m(D(w_{i-j}))_{R,L}$ is the semisheaf corresponding to the
gluing-up of the semisheaves $\theta ^m_i(D(w_i))_{R,L}$ and $\theta
^m_j(D(w_j))_{R,L}$~.  It is then possible to envisage the gluing-up of some
complementary semisheaves  or of the complete category of these
complementary semisheaves  covering then by patches \cite{Tho2}
the semisheaf $\theta ^m_{R,L}(s_{R,L})\times \theta' _{R,L}(s_{R,L})$~.
\vskip 22pt

\begin{defi}[Sequence of spreading-out isomorphisms]  { Let  $\theta
^m_{SOT(1)_{R,L}}$  denote  the family of complementary semisheaves  
$\{\theta ^m_i(D(w_i))_{R,L}\}^p_{i=1}$~, $p\le r$~, covering $\theta
^m_{R,L}(s_{R,L})\times \theta' _{R,L}(s_{R,L})$ some of which can be glued
together.  $\theta ^m_{SOT(1)_{R,L}}$ is thus generated by the spreading-out
isomorphism $SOT(1)$~.

According to definition 2.1.5, the germs $w^i_{\mu R,L}\in \theta
^1_{R,L}(w^i)\subseteq \theta ^m_i(D(w_i))_{R,L}$ can be characterized by isolated degenerated
singular points.

Consequently, a versal deformation of $\theta ^m_{SOT(1)_{R,L}}$ can be
envisaged followed by a spreading-out isomorphism $SOT(2)$~.  The resulting
family of complementary semisheaves  will then be noted $\theta
^m_{SOT(2)_{R,L}}$~.

If we abbreviate $\theta ^m_{R,L}(s_{R,L})\times \theta ^r_{R,L} (s_{R,L})$ by $\theta
_{R,L}(s,s)_{R,L}$~, we have the exact sequence:
\nobeqn
\theta ^m_{R,L}(s)_{R,L} &\stackrel{Vd(1)}{\longrightarrow}& \theta _{R,L}(s,s_1)_{R,L}\\
&\stackrel{SOT(1)}{\longrightarrow }& (\theta _{R,L}(s,s_1)_{R,L})'\txt\bigcup \theta
^m_{SOT(1)_{R,L}}\\
&\stackrel{Vd(2)}{\longrightarrow}& (\theta _{R,L}(s,s_1)_{R,L})'\txt\bigcup \theta
_{R,L}(s_{SOT(1)},s_2)_{R,L}\\
&\stackrel{SOT(2)}{\longrightarrow}& (\theta _{R,L}(s,s_1)_{R,L})'\txt\bigcup (\theta
_{R,L}(s_{SOT(1)},s_2)_{R,L})'\txt\bigcup \theta ^m_{SOT(2)_{R,L}}\noeeqn
where
\begin{enumerate}
\item $Vd(1)$ and $Vd(2)$ denote the two successive versal deformations;
\item the versal deformation $Vd(2)$ of the semisheaf $\theta
^m_{SOT(1)_{R,L}}$ gives the semisheaf \\ $\theta _{R,L}(s_{SOT(1)},s_2)_{R,L}$ in
such a way that the dimension of its quotient algebra $q$ verifies $q<r$ where
$r$ is the dimension of the quotient algebra of $Vd(1)$~.
\end{enumerate}}\end{defi}
\vskip 11pt

\begin{propo}
 A sequence of maximum two successive spreading-out isomorphisms can be
envisaged from a given semisheaf of germs of differentiable functions $\theta
^m_{R,L}(s)_{R,L}$ where $m\le 3$~.\end{propo}
\vskip 11pt

\bpr As $m\le 3$~, the corank ``~$ck$~'' of the degenerated singularities on
$\theta ^m_{R,L}(s)_{R,L}$ is $ck\le 2$ according to the remarks 2.1.16 and
the codimension ``~$cd$~'' of the versal deformation of $\theta
^m_{R,l}(s)_{R,L}$ is $cd\le 3$~.  Consequently, the possible degenerated
singularities on the family of complementary semisheaves  $\theta
^m_{SOT(1)_{R,L}}$~, obtained from $\theta ^m_{R,L}(s)_{R,L}$ by the
$(SOT(1)\circ Vd(1))$ morphism, have a codimension $cd\le 2$~.  Thus, one and only
one supplementary $(SOT(2)\circ Vd(2))$ morphism can be envisaged from the semisheaf 
$\theta ^m_{SOT(1)_{R,L}}$~.\epr
\vskip 11pt

\subsection{The three embedded structures of semiparticles}

The aim of this section is to prove that the algebraic structure of
semiparticles is composed of three embedded semisheaves of rings whose most
internal is the space-time semisheaf of rings studied in chapter 1.
\vskip 11pt

If singularities are generated on this space-time semisheaf of rings, then a
sequence of maximum two successive spreading-out isomorphisms consecutive to
versal deformations can be considered leading to the generation of two
embedded semisheaves of rings covering the fundamental space-time semisheaf
of rings.

The developments will be made for semileptons because they are easier to
handle but they are also valid for semibaryons.

\vskip 11pt

\begin{defi}[Extension of the quotient algebra] { Consider the
$4D$-space-time   semisheaf of rings $(\theta ^{*1}_{R,L}(t)\oplus \theta
^3_{I_{R,L}}(r))_{ST}$~, noted ``~$ST$~'' for space-time, whose $q$ sections are differential functions isomorphic to
\[ (T^{*1}_\mu (t)_{R,L}\oplus T^1_{I_\mu }(r)_{R,L})\;,\qquad 1\le \mu \le
q\;.\]
Assume that under some external perturbations:
\begin{enumerate}
\item all the sections $s^{*1}_\mu (t)_{R,L}\in \theta ^{*1}_\mu (t)$ are
endowed with the same isolated singularities of corank 1 and codimension $r\le
3$~.
\item or/and that all the complementary $3D$-sections $s_{I_\mu
}(r)_{R,L}\in \theta ^3_{I_{R,L}}(r)$ have the same isolated singularities of
corank $ck\le 2$ and codimension $r\le 3$~.
\end{enumerate}

According to proposition 2.1.9, the versal deformation of the $q$ sections
$s^{*1}_\mu (t)_{R,L}\in \theta ^{*1}_{R,L}(t)$ defines the quotient algebra
\[ \theta _{R,L}[R_w]=\txt\sum\limits^r_{i=1}\theta ^0_{R,L}(a_i)\times \theta
^1_{R,L}(w^i)\]
where $\theta ^0_{R,L}(a_i)$ is a constant semisheaf and where $\theta
^1_{R,L}(w^i)$ is the $i$-th generator semisheaf of the versal unfolding, when the
versal deformation of the semisheaf of rings $\theta ^3_{I_{R,L}}(r)$ having
singularities of corank 1 gives rise to the quotient algebra
\[\theta ^2_{R,L}[R_w]=\txt\sum\limits^r_{i=1}\theta ^2_{R,L}(b_i)\times \theta
^1_{R,L}(w^i)\;.\]
But, if the singularities are of corank 2 on $\theta ^3_{I_{R,L}}(r)$~, the
quotient algebra of the versal deformation will be
\[ \theta ^1_{R,L}[R_{w_1,w_2}]=\txt\sum\limits^r_{i,j=1}\theta ^1_{R,L}(c_i)\times
\theta ^2_{R,L}(w_{1i},w_{2j})\;.\]
Then, the extension of the quotient algebra of the versal deformation realized
by the spreading-out isomorphism generates:
\begin{enumerate}
\item for the semisheaf of rings $\theta ^{*1}_{R,L}(t)_{ST}$ the category
$c(T_{w_i}( \theta ^1_{R,L}(w^i)))$ of sections of
vertical tangent bundles.

\item for the semisheaf of rings $\theta ^3_{I_{R,L}}(r)_{ST}$ having
singularities of corank 1 the category $c (T_{w_i}( \theta
^1_{R,L}(w^i))))$ of sections of vertical tangent bundles and for the semisheaf
of rings $\theta ^3_{I_{R,L}}(r)_{ST}$ having singularities of corank 2 the
category $c(T_{w_{ij}}( \theta
^2_{R,L}(w_{1i},w_{2j})))$ of sections of vertical tangent bundles.
\end{enumerate}}\end{defi}
\vskip 11pt

\begin{defi}[Covering the spreading-out isomorphism]  { Let \\$\{\theta
^1_i(D(w_i))_{R,L}\}^p_{i=1}$ (resp. $\{\theta
^3_i(D(w_i))_{R,L}\}^p_{i=1}$~) denote a family, i.e. with $p\le r$~, of
the category $c(T_{w_i}(\cdots ))$ (resp. 
$c(T_{w_i}(\cdots ))$  of vertical tangent bundle sections which are complementary semisheaves.

Assume that $\theta ^1_{R,L}(t_{res})_{ST}$ (resp. $\theta
^{3'}_{R,L}(r_{res})_{ST}$~), being the residue semisheaf after the versal
deformation and the spreading-out isomorphism $SOT(1)\circ Vd(1)$ (resp.
$SOT(1')\circ Vd(1')$~) (see definition 2.1.19) of $\theta ^{*1}_{R,L}(t)_{ST}$
(resp. $\theta ^{3'}_{I_{R,L}}(r)_{ST}$~), is partially covered \cite{Ful} by
the semisheaf $\theta ^1_{R,L}(t)_{MG}$ (resp. $\theta ^{3'}_{R,L}(r)_{MG}$~)
denoting one family $\{\theta ^1_i(D(w_i))_{R,L}\}^p_{i=1}$ (resp.
$\{\theta ^{3}_i(D(w_i))_{R,L}\}^p_{i=1}$~) or several families

\bt{ll}
& $\{\{\theta ^1_i(D(w_i))_{R,L}\}^p_{i=1},\cdots,\{\theta
^1_k(D(w_k))_{R,L}\}^s_{k=1}\}$\\
(resp. & $\{\{\theta ^3_i(D(w_i))_{R,L}\}^p_{i=1},\cdots,\{\theta
^3_k(D(w_k))_{R,L}\}^s_{k=1}\}$~)\te

\noindent where $p$ and $s$ are inferior or equal to the respective dimensions
of the versal deformations.

The covering by several families of semisheaves must be considered because
every section $s^{*1}_\mu (t)_{R,L}\in \theta ^{*1}_{R,L}(t)_{ST}$ (resp.
$s^{'}_{I_\mu} (r)_{R,L}\in \theta ^{3'}_{I_{R,L}}(r)_{ST}$~) can have several
isolated degenerated singularities. If there is a covering by one or several
families of semisheaves, some of these semisheaves can be glued together
according to definition 2.1.18.}
\end{defi}
\vskip 11pt

\begin{defi}[Embedded semisheaves of rings]  { Let $\theta ^1_{R,L}(t)_{MG}$
(resp. $\theta ^{3'}_{R,L}(r)_{MG}$~) be the semisheaf covering partially $\theta
^1_{R,L}(t_{res})_{ST}$ (resp. $\theta
^{3'}_{R,L}(r_{res})_{ST}$~) where ``~$MG$~'' is the abbreviated form for ``middle
ground''.  According to proposition 2.1.20, if the codimension of the
degenerated singularities on the semisheaves $\theta ^{*1}_{R,L}(t)_{ST}$
(resp. $\theta ^{3'}_{I_{R,L}}(r)_{ST}$~) is superior or equal to 3~, a versal
deformation and a spreading-out isomorphism $(SOT(2)\circ Vd(2))$ (resp.
$(SOT(2')\circ Vd(2'))$~) can be envisaged from $\theta ^1_{R,L}(t)_{MG}$
(resp. $\theta ^{3'}_{R,L}(r)_{MG}$~) leading to a semisheaf $\theta
^1_{R,L}(t)_M$ (resp. $\theta
^{3'}_{R,L}(r)_M$~) covering partially the residue semisheaf $\theta
^1_{R,L}(t_{res})_{MG}$ (resp. $\theta
^{3'}_{R,L}(r_{res})_{MG}$~), where ``~$M$~'' is the abbreviated form for ``mass''.}
\end{defi}
\vskip 11pt

If the semisheaves $\theta ^1_{R,L}(t_{res})_{MG}$ and $\theta ^1_{R,L}(t)_M$
(resp. $\theta ^{3'}_{R,L}(r_{res})_{MG}$ and $\theta ^{3'}_{R,L}(r)_M$~) are
generated by versal deformation and spreading-out isomorphism, then the
corresponding complementary semisheaves $\theta ^3_{I_{R,L}}(r)_{MG}$ and
$\theta ^3_{I_{R,L}}(r)_M$ (resp. $\theta ^{1'}_{I_{R,L}}(t)_{MG}$ and
$\theta ^{1'}_{I_{R,L}}(t)_M$~) can be generated respectively from\linebreak $\theta
^1_{R,L}(t_{res})_{MG}$ and $\theta ^1_{R,L}(t)_M$ (resp. $\theta
^{3'}_{R,L}(r_{res})_{MG}$ and $\theta ^{3'}_{R,L}(r)_M$~) by a $(\gamma _{t\to
r}\circ E)$ morphism (resp. a $(\gamma _{r\to t}\circ E')$ morphism according
to proposition 1.2.6 and corollary 1.2.8.
\vskip 11pt

The sequence of the two successive versal deformations and spreading-out
isomorphisms from the $1D$-semisheaf of rings $\theta ^1_{R,L}(t)_{ST}$ are
summarized in the two following diagrams A) and B):\\
A)
\[ \begin{array}{ccccc}
\hspace{-4mm}\theta ^1_{R,L}(t)_{ST} & \hspace{-4mm}\stackrel{SOT(1)\circ Vd(1)}{\longrightarrow }\hspace{-4mm} &\theta
^1_{R,L}(t_{res})_{ST}\cup \theta ^1_{R,L}(t)_{MG} &
\hspace{-4mm}\stackrel{SOT(2)\circ Vd(2)}{\longrightarrow}\hspace{-4mm} &\theta
^1_{R,L}(t_{res})_{ST}\cup \theta ^1_{R,L}(t_{res})_{MG}\cup \theta ^1_{R,L}(t)_M \\
\vfl{}{\gamma  ^{ST}_{t\to r}\circ E} && 
\vfl{}{\gamma  ^{MG}_{t\to r}\circ E} && 
\vfl{}{\gamma  ^{M}_{t\to r}\circ E} \\
\hspace{-4mm}\theta ^3_{I_{R,L}}(r)_{ST} & 
&\theta^3_{I_{R,L}}(r_{res})_{ST}\cup \theta ^3_{I_{R,L}}(r)_{MG} &
&\theta^3_{I_{R,L}}(r_{res})_{ST}\cup \theta ^3_{I_{R,L}}(r_{res})_{MG}\cup
\theta^3_{I_{R,L}}(r)_M 
\end{array}\]
B)
\[ \begin{array}{ccccc}
\hspace{-4mm}\theta ^1_{R,L}(t)_{ST} & 
&\theta^{1'}_{I_{R,L}}(t_{res})_{ST}\cup \theta ^{1'}_{I_{R,L}}(t)_{MG} &
&\theta^{1'}_{I_{R,L}}(t_{res})_{ST}\cup \theta ^{1'}_{I_{R,L}}(t_{res})_{MG}\cup
\theta ^{1'}_{I_{R,L}}(t)_M \\
\vfl{}{\gamma  ^{ST}_{t\to r}\circ E'} && 
\vupfl{}{\gamma  ^{MG}_{r\to t}\circ E'} && 
\vupfl{}{\gamma  ^{M}_{r\to t}\circ E'} \\
\hspace{-4mm}\theta ^{3'}_{I_{R,L}}(r)_{ST} & \hspace{-4mm} \stackrel{SOT(1')\circ
Vd(1')}{\longrightarrow}
\hspace{-4mm} &\theta
^{3'}_{R,L}(r_{res})_{ST}\cup \theta ^{3'}_{R,L}(r)_{MG} &
\hspace{-4mm}\stackrel{SOT(2')\circ Vd(2')}{\longrightarrow}\hspace{-4mm} &\theta
^{3'}_{R,L}(r_{res})_{ST}\cup \theta ^{3'}_{R,L}(r_{res})_{MG}\cup \theta
^{3'}_{R,L}(r)_M 
\end{array}\]
leading to the two sets of three embedded semisheaves of rings for the diagram A):
\nobeqn 
&\theta ^1_{R,L}(t_{res})_{ST}\subset \theta ^1_{R,L}(t_{res})_{MG}\subset
\theta ^1_{R,L}(t)_M&\\
&\theta ^3_{I_{R,L}}(r_{res})_{ST}\subset \theta
^3_{I_{R,L}}(r_{res})_{MG}\subset \theta ^3_{I_{R,L}}(r)_M&\noeeqn
in the sense that there is a topological embedding for all their $q$
sections, $1\le \mu \le q$~, i.e.:
\[\begin{array}{ccccc}
s_{\mu R,L}(t_{res})_{ST} &\subset & 
s_{\mu R,L}(t_{res})_{MG} &\subset & s_{\mu R,L}(t)_{M} \\
s_{\mu R,L}(r_{res})_{ST} &\subset & 
s_{\mu R,L}(r_{res})_{MG} &\subset & s_{\mu R,L}(r)_{M} \end{array}\]
where $s_{\mu R,L}(t_{res})_{ST}\in \theta ^1_{R,L}(t_{res})_{ST}$ and so on.

With respect to the diagram B), we have the corresponding embedding of semisheaves of
rings:
\nobeqn 
&\theta ^{1'}_{I_{R,L}}(t_{res})_{ST}\subset \theta
^{1'}_{I_{R,L}}(t_{res})_{MG}\subset
\theta ^{1'}_{I_{R,L}}(t)_M&\\
\noalign{\vskip 6pt}
&\theta ^{3'}_{R,L}(r_{res})_{ST}\subset \theta
^{3'}_{R,L}(r_{res})_{MG}\subset \theta ^{3'}_{R,L}(r)_M&\noeeqn
\vskip 11pt

\begin{propo} The semisheaves of rings $\theta ^3_{I_{R,L}}(r_{res})_{ST}\cup \theta
^3_{I_{R,L}}(r)_{MG}$~, generated by the morphism $\gamma ^{MG}_{t\to r}\circ E\circ
SOT(1)\circ Vd(1))$ from $\theta ^1_{R,L}(t)_{ST}$~, may be isomorphic to the
semisheaves of rings $\theta ^{3'}_{R,L}(r_{res})_{ST}\cup \theta
^{3'}_{R,L}(r)_{MG}$~, generated by the morphism $
(SOT(1')\circ Vd(1')\circ \gamma ^{ST}_{t\to r}\circ E')$ from $\theta
^1_{R,L}(t)_{ST}$ if and only if:
\Bean\item singularities of corank 1 and of the same codimension are at the origin of
the versal deformations $Vd(1)$ and $Vd(1')$~;
\item the singularities on the semisheaves $\theta ^1_{R,L}(t_{res})_{ST}\cup \theta
^1_{R,L}(t)_{MG}$ are conserved under the morphism $\gamma ^{MG}_{t\to r}\circ E$~.\Ee
\end{propo}
\vskip 11pt

\bpr By hypothesis, only singularities of corank 1 and of the same codimension are
taken into account in the versal deformations $Vd(1)$ and $Vd(1')$~: this is justified
physically by the fact that the same kind of perturbation must be envisaged on the
semisheaf $\theta ^1_{R,L}(t)_{ST}$ for the versal deformation $Vd(1)$ and on the
complementary semisheaf $\theta ^{3'}_{I_{R,L}}(t)_{ST}$ for the versal deformation
$Vd(1')$~.

The quotient algebra of the versal deformation $Vd(1)$ is
\[ \theta _{R,L}[R_w]=\txt\sum\limits^r_{i=1}\theta ^0_{R,L}(a_i)\times \theta
^1_{R,L}(w^i)\]
while it is
\[ \theta ^2_{R,L}[R_w]=\txt\sum\limits^r_{i=1}\theta ^2_{R,L}(b_i)\times \theta
^1_{R,L}(w^i)\]
for the versal deformation $Vd(1')$ according to definition 2.2.1.
\vskip 11pt

Now, under the hypothesis of the proposition, the generator semisheaves $\theta
^1_{R,L}(w^i)$  in $\theta _{R,L}[R_w]$ and in $\theta ^2_{R,L}[R_w]$ 
are composed of time quanta, i.e. time prime ideals of rank $N$~.  These time quanta are composed of the same number of geometric points as
the time quanta of the semisheaf
$\theta ^1_{R,L}(t)_{ST}$ because the semisheaves $\theta ^1_{R,L}(w^i)$ and $\theta
^1_{R,L}(t)_{ST}$ are supposed to be generated by Eisenstein cohomology from
$1D$-time symmetric splitting semifields (see definition 1.1.2).

Considering that the semisheaf $\theta ^2_{R,L}(b_i)\in \theta ^2_{R,L}[R_w]$ is
a stratum semisheaf of $\theta ^{3'}_{I_{R,L}}(r)_{ST}$~, we can admit that we reach
the thesis since the $\gamma _{t\to r}$ morphism is a morphism essentially
based on the inverse Kronecker's specialization \cite{Lan1} such that a ring
of irreducible polynomials in $n$ variables can be extended to a ring of
irreducible polynomials in $m$ variables, where $n<m$~.

Or, more directly, we have seen that the $3D$ semisheaves of rings $\theta ^{3'}\RL(r_{\rm res})_{ST}$ and
$\theta ^{3'}\RL(r_{\rm res})_{MG}$ degenerate into $1D$-semisheaves of rings according to proposition 1.2.6.
\epr
\vskip 11pt

\begin{coro} The semisheaves of rings $\theta ^3_{I_{R,L}}(r_{res})_{MG}\cup
\theta ^3_{I_{R,L}}(r)_{M}$ may be isomorphic to the semisheaves of rings
$\theta ^{3'}_{R,L}(r_{res})_{MG}\cup \theta ^{3'}_{R,L}(r)_{M}$ if and only if
\Bean\item singularities of corank 1 and of the same codimension are at the origin of
the versal deformations $Vd(2)$ and $Vd(2')$~;
\item the singularities on the semisheaves $\theta ^1_{R,L}(t_{res})_{MG}\cup
\theta ^1_{R,L}(t)_{M}$ are conserved under the mosphism $\gamma ^M_{t\to r}\circ
E$~.\Ee\end{coro}
\vskip 11pt

\begin{propo} \quad  Let $M^I_{R,L}(t)_{ST}\in \theta ^1_{R,L}(t_{res})_{ST}$~,
$M^I_{R,L}(t)_{MG}\in \theta ^1_{R,L}(t_{res})_{MG}$ and 
$M^I_{R,L}(t)_M\in \theta ^1_{R,L}(t)_{M}$ be the time quanta on which  the
corresponding semisheaves of rings $ST$~, $MG$ and $M$ are defined.  Then,
these time quanta have the same number of geometric points.
\end{propo}
\vskip 11pt

\bpr A time quantum is a  time submodule having a rank $N$~. As the
$MG$-time semisheaf $\theta ^1_{R,L}(t)_{MG}$ is generated from the $ST$-$1D$-time
semisheaf
$\theta ^1_{R,L}(t)_{ST}$ by the
$(SOT(1)\circ Vd(1))$ morphism, it results that the $MG$-time quanta
$M^I_{R,L}(t)_{MG}$ are submodules of the category $c(T_{aw_i}(\theta
^0_{R,L}(a_i)\times \theta ^1_{R,L}(w^i))$ according to definition 2.2.1.

Now, as the semisheaf $\theta ^1_{R,L}(w^i)$ is generated on and from irreducible completions $L_{v^1_\mu }$~, i.e. quanta
according
to section 1.1.4, we have that 
$ M^I_{R,L}(t)_{ST}$ has the same number of geometric points as $M^I_{R,L}(t)_{MG}$~.

Similarly, we can prove that $M^I_{R,L}(t)_{MG}$ has the same number of geometric
points as $M^I_{R,L}(t)_M$~.\epr
\vskip 11pt

\begin{coro}
If $M^I_{R,L}(r)_{ST}\in \theta ^3_{R,L}(r)_{ST}$~, 
$M^I_{R,L}(r)_{MG}\in \theta ^3_{R,L}(r)_{MG}$ and 
$M^I_{R,L}(r)_{M}\in \theta ^3_{R,L}(r)_{M}$ are space quanta, then
they have the same number of geometric points.
\end{coro}
\vskip 11pt

\bpr As the space quanta are generated from the time quanta by the $\gamma
_{t\to r}\circ E$ morphism and as the time quanta have the same number of geometric 
points, we have the thesis.\epr
\vskip 11pt

\begin{propo}
The semisheaves of rings $\theta ^1_{R,L}(t_{res})_{MG}$ and
$\theta ^1_{R,L}(t)_{M}$ as well as the semisheaves of rings $\theta
^3_{R,L}(r_{res})_{MG}$ and $\theta ^3_{R,L}(r)_{M}$ are not necessarily
compact and Zariski dense in such a way that their sections are open strings.\end{propo}
\vskip 11pt

\bpr Indeed, by construction (see definition 2.2.2), these semisheaves 
$\theta ^1\RL(t_{\rm res})_{MG}$ and $\theta ^1\RL(t)_{M}$  (idem for 
$\theta ^3\RL(t_{\rm res})_{MG}$ and $\theta ^3\RL(t)_{M}$~)
cover
partially by patches the semisheaf
$\theta ^1_{R,L}(t_{res})_{ST} $~.

As the sections of $\theta ^1\RL(t_{\rm res})_{MG}$ and of $\theta ^1\RL(t)_{M}$  (resp. of 
$\theta ^3\RL(r_{\rm res})_{MG}$ and $\theta ^3\RL(r)_{M}$~) are one-dimensional and cover partially the sections of $\theta ^1\RL(t_{\rm res})_{ST}$ (resp. $\theta ^3\RL(t_{\rm res})_{ST}$~), they are open strings.

More precisely, as the coefficients $a_{i_\mu }(x)$ of the quotient algebra of the versal 
deformation of germs $s_{\mu \RL}(w\RL)$ are (germs of) functions defined on domains 
$D_{\mu \RL}$ included into half open balls $B_{\mu \RL}$ whose radii increase in function 
of the global residue degrees of the sections $s_{\mu \RL}$~, the numbers of quanta $\widetilde M^I\RL (t)_{MG}$ and $\widetilde M^I\RL (t)_{M}$~, covering the sections $s_{\mu \RL}(t)_{ST}$ of the semisheaf 
$\theta ^1\RL(t_{\rm res})_{ST}$~, increase according to the global residue degrees $f_{\mu _{ST}}$ of
$s_{\mu \RL}(t)_{ST}$~. Thus, if $n_{M^I(t)_{MG}}(s_\mu )=f_{\mu _{MG}}$ and 
$n_{M^I(t)_{MG}}(s_{\mu +1})=f_{(\mu+1) _{MG}}$ denote the numbers of quanta, i.e. the corresponding global residue degrees, respectively of the $\mu $-th and $(\mu +1)$-th sections of the semisheaf $\theta ^1\RL(t_{\rm res})_{MG}$ covering the corresponding sections of the semisheaf $\theta ^1\RL(t_{\rm res})_{ST}$~, then 
$n_{M^I(t)_{MG}}(s_\mu )<
n_{M^I(t)_{MG}}(s_{\mu +1}) $~.

And, $n_{M^I(t)_{ST}}(s_\mu )\approx
n_{M^I(t)_{MG}}(s_{\mu }) $~, i.e. that the number of quanta
$n_{M^I(t)_{MG}}(s_\mu ) $ of the $\mu $-th section of $\theta ^1\RL(t_{\rm res})_{MG}$ is approximately equal to the number of quanta $n_{M^I(t)_{ST}}(s_\mu ) $ of the $\mu $-th section of $\theta ^1\RL(t_{\rm res})_{ST}$~.
\epr
\vskip 11pt

\begin{propo} 
Every semisheaf of rings  $\theta ^1_{R,L}(t_{res})_{ST}$~, 
$\theta ^1_{R,L}(t_{res})_{MG}$~, $\theta ^1_{R,L}(t)_{M}$~,  
$\theta ^3_{R,L}(r_{res})_{ST}$~, $\theta ^3_{R,L}(r_{res})_{MG}$ or
$\theta ^3_{R,L}(r)_{M}$ which is locally free corresponds to a Stein space.
\end{propo}
\vskip 11pt

\bpr A sheaf of rings $\theta _P$~, defined on a closed subset $P$ of a
topological space $X$~, is locally free if it has no degenerated singularity. 
Consequently, it cannot be submitted to a versal deformation and must satisfy
the condition $H^q(P,\theta _P)=0$~, $\forall\ q\ge 1$~, \cite{G-R3}.  If this
is the case, the sheaf of rings $\theta _P$ corresponds to a Stein space and
is locally free.\epr
\vskip 11pt

\begin{defi}[Semialgebras on $\aa_{R,L}$-semimodules]  { Let $\theta ^{1-3}_{R,L}(t,r)=\theta
^1_{R,L}(t)\oplus \theta ^3_{R,L}(r)$ denote the direct sum of the $1D$- and
$3D$-semisheaves of rings.  As  $\theta ^{1-3}_{R,L}(t,r)_{ST}$~, 
 $\theta ^{1-3}_{R,L}(t,r)_{MG}$ and  $\theta ^{1-3}_{R,L}(t,r)_M$ are semisheaves on
semimodules over $\aa_{R,L}$~,
they are semialgebras \cite{F-D}.}\end{defi}
\vskip 11pt

  This leads us to the
following proposition:
\vskip 11pt

\begin{propo} The  semialgebras $\theta ^{1-3}_{R,L}(t,r)_{ST}$~, 
 $\theta ^{1-3}_{R,L}(t,r)_{MG}$ and  $\theta ^{1-3}_{R,L}(t,r)_M$ are
commutative while the semialgebras

\bt{ll}
&$\theta ^{1-3}_{R,L}(t,r)_{ST}\oplus \theta ^{1-3}_{R,L}(t,r)_{MG}$~,\\
&$\theta ^{1-3}_{R,L}(t,r)_{ST}\oplus \theta ^{1-3}_{R,L}(t,r)_{MG}\oplus\theta
^{1-3}_{R,L}(t,r)_M$~,\\
and &$\theta ^{1-3}_{R,L}(t,r)_{MG}\oplus\theta ^{1-3}_{R,L}(t,r)_M$\te

\noindent extended from $\theta ^{1-3}_{R,L}(t,r)_{ST}$ by versal
deformation(s) and spreading-out isomorphism(s) are noncommutative.
\end{propo}
\vskip 11pt

\bpr 1. The semialgebra $\theta ^{1-3}_{R,L}(t,r)_{ST}$ is commutative by
construction (see chapter 1) since $\theta ^1_{R,L}(t)_{ST}$ is generated by
right (resp. left) Eisenstein cohomology from the $1D$-symmetric splitting
semifields $L^{\mp}_\mu $~.  The semialgebra $\theta ^3_{R,L}(r)_{ST}$ is also
commutative since it is generated by the $(\gamma _{t\to r}\circ E)$ morphism
from $\theta ^1_{R,L}(t)_{ST}$ according to proposition 1.2.6.

In fact, the semialgebras $\theta ^{1-3}_{R,L}(t,r)_{ST}$~,
$\theta ^{1-3}_{R,L}(t,r)_{MG}$ and $\theta ^{1-3}_{R,L}(t,r)_{M}$ are
commutative because it is possible for each one of these to define a unique
centralizator.
\vskip 11pt

\noindent 2. A  semialgebra extended by versal deformation and spreading-out
isomorphism, for example $\theta ^{1-3}_{R,L}(t,r)_{ST}\oplus
\theta ^{1-3}_{R,L}(t,r)_{MG}$~, is noncommutative because it is impossible
to define for it a unique centralizator.  Indeed, the generator semisheaves
$\theta _{R,L}(w^i)$ of the versal unfolding leading to the generation of the
$1D$-extended semisheaf $\theta ^1_{R,L}(t)_{MG}$ originate from the
specialization prime ideals $p(w^i_{\mu R,L})$ (see definition 2.1.3) while the
$1D$-semisheaf $\theta ^1_{R,L}(t)_{ST}$
 originates from specialization prime ideals $p_{\mu R,L}$ (see definition
1.1.3).  As these specialization ideals $p(w^i_{\mu R,L})$ and $p_{\mu R,L}$
are not equal, we reach the thesis.\epr
\vskip 11pt

\begin{defi}  {\bf The emission quantification\/} { of the space-time, middle
ground and mass structures of semiparticles can be envisaged by considering
that these three embedded structures are constituted by the three embedded time
semisheaves of rings $\theta ^1_{R,L}(t_{res})_{ST}\cup
\theta ^1_{R,L}(t_{res})_{MG}\cup\theta ^1_{R,L}(t)_{M}$~, noted in
abbreviated form $\theta ^1_{R,L}(t)_{ST-MG-M}$~, and by the three
embedded space semisheaves $\theta ^3_{R,L}(r)_{ST-MG-M}\equiv
\theta ^3_{R,L}(r_{res})_{ST}\cup \theta ^3_{R,L}(r_{res})_{MG}\cup
\theta ^3_{R,L}(r)_{M}$~.

Taking into account that the middle ground and mass semisheaves of rings are
above the space-time semisheaves of rings, the middle ground and mass quanta
will be above the space-time quanta.  Consequently, a smooth endomorphism
$E _{ST-MG-M}$~, acting simultaneously on the three embedded semisheaves
$ST$~, $MG$ and $M$~, can be defined by:
\[ E _{ST-MG-M} : \theta ^1_{R,L}(t)_{ST-MG-M}\to \theta
^{*1}_{I_{R,L}}(t)_{ST-MG-M}\txt\bigoplus\limits^m_{k=1} \widetilde M^I_{k_{R,L}}(t)_{ST-MG-M}\]
where $ \widetilde M^I_{k_{R,L}}(t)_{ST-MG-M}= \widetilde M^I_{k_{R,L}}(t)_{ST}\cup
 \widetilde M^I_{k_{R,L}}(t)_{MG}\cup M^I_{k_{R,L}}(t)_{M}$ are three ``disconnected''   functions on
time quanta from $\theta ^1_{R,L}(t)_{ST-MG-M}$ so that 
$ \widetilde M^I_{k_{R,L}}(t)_{MG}$ is above $ \widetilde M^I_{k_{R,L}}(t)_{ST}$ and
$ \widetilde M^I_{k_{R,L}}(t)_{M}$ is above $ \widetilde M^I_{k_{R,L}}(t)_{MG}$~.

This smooth endomorphism $E _{ST-MG-M}$ then represents a three stratum time
quantification of emission of semiparticles.

Similarly, a three stratum space quantification of emission would be introduced
by applying a smooth endomorphism $E _{ST-MG-M}$ on $\theta
^3_{R,L}(r)_{ST-MG-M}$ disconnecting space    quanta $ \widetilde M^I_{R,L}(r)_{ST-MG-M}$
from this space semisheaf.}\end{defi}
\vskip 11pt

\begin{propo} {\bf 1.\/} The standard quanta of quantum field
theory are the spatial left quanta $M^I_{L}(r)_M\in \theta
^3_{L}(r)_M$~.
\vskip 11pt

\noindent {\bf 2.\/} The Planck constant $h$ corresponds to the value of the integer $N$
in the mass unit system, where $N$ refers to the order of the global inertia subgroup. \end{propo}
\vskip 11pt

\bpr The  quanta of quantum theories are the spatial left
quanta $M^I_{L}(r)_M$ because only the mass structure of elementary
particles is presently observable and corresponds to the left semisheaf of rings
$\theta ^3_L(r)_M$ of left semiparticles.  However, right (resp. left) quanta are in
fact
spatial quanta
$M^I_{k_{R,L}}(r)_{ST-MG-M}\in
\theta ^3_{R,L}(r)_{ST-MG-M}$~.

On the other hand, according to axiom II 1.3.9, a ``real'' spatial quantum of an algebraic
quantum theory must be a biquantum given by the product of a right and a left spatial
quantum: $M^I_R(r)_{ST,MG,M}\times M^I_L(r)_{ST,MG,L}$~.
\vskip 11pt

\noindent 2) The Planck constant, introduced by Planck in Physics to take into
account the discontinued behavior of matter, must then correspond to the value of the integer $N$ in the system of units of the algebraic mass semisheaf of rings.\epr
\vskip 11pt

\begin{defi}[Vertical tangent semibundles]  { The $1D$-time semisheaves of
rings $\theta ^1_{R,L}(t_{res})_{MG}$ and $\theta ^1_{R,L}(t)_M$ as well as
the $3D$-space semisheaves of rings 
$\theta ^3_{R,L}(r_{res})_{MG}$ and $\theta ^3_{R,L}(r)_M$~, generated from
the semisheaf $\theta ^1_{R,L}(t)_{ST}$ by versal deformations, spreading-out
isomorphisms and $(\gamma _{t \raise -3pt\hbox{$\overrightarrow\leftarrow$} r}\circ E)$ morphisms according to definition
2.2.3, are total spaces respectively of the $1D$-middle ground (resp. mass)
vertical tangent bundle \\
\centerline{\bt{ll}
& $T^{(1)}_{MG_{R,L}}(\theta
^1_{R,L}(t_{res})_{MG},\theta ^1_{R,L}(t_{res})^B_{MG},\tau^{(1)}_{V_{MG}})$\\
(resp. &$T^{(1)}_{M_{R,L}}(\theta
^1_{R,L}(t)_{M},\theta ^1_{R,L}(t)^B_{M},\tau^{(1)}_{V_{M}})$~)\te}
\vskip 6pt

\noindent and of the $3D$-middle ground (resp. mass) vertical tangent bundle \\
\centerline{\bt{ll}
&$T^{(3)}_{MG_{R,L}}(\theta
^3_{R,L}(r_{res})_{MG},\theta ^3_{R,L}(r_{res})^B_{MG},\tau^{(3)}_{V_{MG}})$\\
(resp.& $T^{(3)}_{M_{R,L}}(\theta
^3_{R,L}(r)_{M},\theta ^3_{R,L}(r)^B_{M},\tau^{(3)}_{V_{M}})$~),\te}
\vskip 6pt

\noindent where
\begin{enumerate}
\item $\theta ^1_{R,L}(t_{res})^B_{MG}$ (resp. $\theta
^3_{R,L}(r_{res})^B_{MG}$~) is the basis of the vertical tangent bundle as
resulting globally from the isomorphism $\Pi_s$ (see proposition 2.1.10).
\vskip 11pt

\item $\theta ^1_{R,L}(t_{res})_{MG}$ (resp. $\theta
^3_{R,L}(r_{res})_{MG}$~) is the total space of the vertical tangent bundle \\
\centerline{\bt{ll}
& $T^{(1)}_{MG_{R,L}}(\theta
^1_{R,L}(t_{res})_{MG},\theta ^1_{R,L}(t_{res})^B_{MG},\tau^{(1)}_{V_{MG}})$\\
(resp. &$T^{(3)}_{MG_{R,L}}(\theta
^3_{R,L}(r_{res})_{MG},\theta ^3_{R,L}(r_{res})^B_{MG},\tau^{(3)}_{V_{MG}})$~)\te}
\vskip 6pt

\noindent obtained by considering the projection of the complementary semisheaf direct
products (i.e. $\theta ^1_{R,L}(t_{res})^B_{MG}$ (resp. $\theta
^3_{R,L}(r_{res})^B_{MG}$~) in the vertical tangent space, according to
definition 2.1.13.
\vskip 11pt

\item $\tau^{(1)}_{V_{MG}}$ (resp. $\tau^{(3)}_{V_{MG}})$ is the projective map.
\end{enumerate}}\end{defi}
\vskip 11pt

\begin{defi} {\bf The generators\/} { of the $1D$- and $3D$-translation groups of
the vertical tangent semibundles $T^{(1)}_{MG_{R,L}}$~, $T^{(3)}_{MG_{R,L}}$~,
$T^{(1)}_{M_{R,L}}$ and $T^{(3)}_{M_{R,L}}$ are respectively given by the following
elliptic differential operators:
\nobeqn
m_{0R,L;MG} &=& \pm i\hbar_{MG} \;\frac\partial{\partial t_0}\;, \\
\noalign{\vskip 11pt}
p_{R,L;MG} &=& \left\{ \pm i\frac{\hbar_{MG}}{c_{t\to
r;MG}}\;\frac\partial{\partial x},\pm  i\frac{\hbar_{MG}}{c_{t\to
r;MG}}\;\frac\partial{\partial y},\pm i\frac{\hbar_{MG}}{c_{t\to
r;MG}}\;\frac\partial{\partial z}\right\}\;, \\
\noalign{\newpage}
m_{0R,L;M} &=& \pm i\hbar_{M}\; \frac\partial{\partial t_0}\;, \\
\noalign{\vskip 11pt}
p_{R,L;M} &=& \left\{ \pm i\;\frac{\hbar_{M}}{c_{t\to
r;M}}\;\frac\partial{\partial x},\pm i\frac{\hbar_{M}}{c_{t\to
r;M}}\;\frac\partial{\partial y},\pm i\frac{\hbar_{M}}{c_{t\to
r;M}}\;\frac\partial{\partial z}\right\}\;, \noeeqn
where $\hbar_{MG}$ and $\hbar_{M}$ are constants corresponding to the integer $N$ in the ``~$MG$~'' and ``~$M$~'' unit systems.
$\hbar_M$ is the Planck constant $\hbar$~: it recalls that the total spaces $\theta
^1_{R,L}(t)_M$ and $\theta ^3_{R,L}(r)$ respectively of the vertical tangent
semibundles $T^{(1)}_{M_{R,L}}$ and $T^{(3)}_{M_{R,L}}$ are quantified since they are
composed of mass quanta.

The constant $c^{-1}_{t\to r;M}=\langle c^{-1}_{t\to r}(\rho )\rangle$ refers to an
average value of the quotient between algebraic Hecke characters according to
propositions 1.4.8 and 1.4.11 and gives a measure of the transformation of the
semisheaf $\theta ^3_{R,L}(r)_M$ from the semisheaf $\theta ^1_{R,L}(t)_M$~.  $c^{-1}_{t\to r;M}\approx c^{-1}$ where $c$ is the light velocity.

An equivalent interpretation can be given to the constants $\hbar_{MG}$ and
$c^{-1}_{t\to r;MG}$~.
}\end{defi}
\vskip 11pt

\begin{defi}  { The elliptic differential operators  $m_{0R,L}$ and $p_{R,L}$ can
be directional gradients, Lie derivatives or covariant derivatives \cite{B-G},
\cite{Kob}.  The covariant derivative $\Delta_{\vec V_M}$ of a semisheaf, for example
$\theta ^3_{R,L}(r)_M$~, along a vector field $\vec V_M$ is such that this semisheaf
if parallely transported along a family of geodesics orthogonal to it with tangent
vectors $\vec V_M$ \cite{Del1}, \cite{H-E}.}\end{defi}
\vskip 11pt

\subsection{Phase spaces associated to the vibrations of the three embedded structures and the vacuum of Quantum Field Theory}

\begin{propo} To each $1D$- and $3D$-space-time (~$ST$~), middle ground
(~$MG$~) and mass (~$M$~) semisheaf of rings corresponds a phase space which
is homeomorphic to $ \rit ^1\times \rit ^1$ or  $\rit ^3\times \rit ^3$ and which
has the structure of a    $F$-Steenrod bundle whose basis is given by the considered
semisheaf of rings.
\end{propo}
\vskip 11pt

\bpr Let, for example, $\theta ^3_{R,L}(r)_M$ be the  $3D$-space-mass semisheaf
of rings.  Then, its associated  $F$-Steenrod bundle is
given by $(\theta ^3_{R,L}(r,p)_M,\theta ^3_{R,L}(r)_M,pr^{(3)}_M)$
where $\theta ^3_{R,L}(r,p)_M$ is the total space and whose topological group
is $GL(3,\rit )$~.  $\theta ^3_{R,L}(r,p)_M=\theta ^3_{R,L}(r)_M\times\theta
^3_{R,L}(p)_M$ where the fiber $\theta ^3_{R,L}(p)_M$ has a $F$-structure where
$F=\rit ^3$~.  This  $\rit ^3$-structure is given by a set of homeomorphisms $\rit ^3\to
\theta ^3_{R,L}(p)_M$ so that each homeomorphism sends the action of the group 
$G=GL(3,\rit )$ from 
$F=\rit ^3$ to $\theta ^3_{R,L}(p)_M$~.\epr
\vskip 11pt

\begin{defi} \ {\bf (~$F$-equivalent fibers of a $F$-Steenrod bundle)\/} { 
\  Two fibers   $\theta ^3_{R,L}(p)^1_M$ and $\theta ^3_{R,L}(p)^2_M$ will be said to be
$F$-equivalent if they are homotopic, i.e. if there exists a continuous mapping from
the one to the other.
}\end{defi}
\vskip 11pt

\begin{propo} To each  $1D$- or $3D$- 
``~$ST$~'', ``~$MG$~'' or ``~$M$~'' $F$-Steenrod bundle corresponds a set of
 $F$-equivalent sections above a given basis related to a given frequency of
vibration of this basis.\end{propo}
\vskip 11pt

\bpr At a given basis of a $F$-Steenrod bundle corresponds a set of $F$-equivalent
fibers according to the definition 2.3.2 and thus a set of $F$-equivalent sections.

Each set of $F$-equivalent sections of a $F$-Steenrod bundle is then interpreted as
corresponding to all the possible vibrations of the basis at a given frequency.\epr
\vskip 11pt

\begin{propo} The frequencies of vibration of the $1D$ and $3D$ space-time
(~$ST$~), middle ground (~$MG$~) and mass (~$M$~) semisheaves of rings are
quantified.\end{propo}
\vskip 11pt

\bpr The semisheaves of rings ``~$ST$~'', ``~$MG$~'' and ``~$M$~'' are assumed to be
defined on quanta, i.e. submodules of rank $N$~.  Thus, the semisheaves of
rings $\theta ^1_{R,L}(t)_{ST,MG,M}$ and $\theta ^3_{R,L}(r)_{ST,MG,M}$ are
quantified.  As they are the basis of $F$-Steenrod bundles and as a given frequency
is associated to each basis of an $F$-Steenrod bundle according to the preceding
proposition, we reach the thesis.
\epr
\vskip 11pt

\begin{defi} {\bf The mass frequency\/} { of an elementary semiparticle is
an average measure of the vibration of all the points of the semisheaf of
rings $\theta ^3_{R,L}(r)_M$~.  From the preceding developments, it becomes clear that there exists a correspondence between the ranks of sections (i.e. classes of degrees of Galois extensions) and the integer numbers of the quantum mechanics referring to vibrations.  Indeed, these integer numbers $n_\mu$ refer to the numbers of quanta of $\mu$-th substates of a (semi)particle.}\end{defi}
\vskip 11pt

\begin{propo} The [semi]wave-[semi]particle duality of quantum theory results
from the quantification of the vibration frequency(ies) of the (~$ST$~, $MG$
and) $M$ [semi]sheaf(ves) of rings.\end{propo}
\vskip 11pt

\bpr Indeed, this duality is essentially traduced by the relations $E=h\nu $~,
$\vec p=\hbar\vec k$ between the dynamical variables related to the mass
structure of the semiparticles and the frequencies of the associated semiwaves
\cite{Mes}, \cite{deBro}.\epr
\vskip 11pt

\begin{rem} {\bf The vacuum\/} { in this algebraic quantum model is not
external to elementary semiparticles but is composed of their $4D$-``~$ST$~''
and ``~$MG$~'' semisheaves of rings which presently are unobservable and whose
spatial extension is of the order of the Planck length $\simeq 10^{-33}$~cm. 
The mass of a semiparticle is given by the  $4D$-``~$M$~'' semisheaf of rings
$\theta ^{1-3}_{R,L}(t,r)_M=\theta ^1_{R,L}(t)_M\cup \theta ^3_{R,L}(r)_M$
which is generated from the corresponding $4D$-``~$MG$~'' semisheaf of rings 
$\theta ^{1-3}_{R,L}(t,r)_{MG}=\theta ^1_{R,L}(t)_{MG}\cup \theta ^3_{R,L}(r)_{MG}$
by versal deformation $Vd(2)$ or $Vd(2')$~, spreading-out isomorphism $SOT(2)$
or $SOT(2')$ and $\gamma ^M_{t\to r}\circ E$ or $\gamma ^M_{r\to t}\circ E'$ morphism.

Consequently, the composition of morphisms:
\[ \gamma ^M_{t\to r}\circ E\circ SOT(2)\circ Vd(2):\theta ^{1-3}_{R,L}(t,r)_{MG}\to
\theta ^{1-3}_{R,L}(t_{res},r_{res})_{MG}\cup \theta ^{1-3}_{R,L}(t,r)_{M}\]
corresponds to the creation operator of Quantum Field Theory.}\end{rem}
\vskip 11pt

\begin{propo} The $4D$-``~$M$~'' semisheaf of rings of a semiparticle is
observable while the $4D$-``~$ST$~'' and ``~$MG$~'' semisheaves of rings are
unobservable because the vibration frequencies of the ``~$M$~'' semisheaf is
inferior   to the vibration frequencies of the ``~$ST$~'' and ``~$MG$~''
semisheaves.\end{propo}
\vskip 11pt

\bpr The ``~$M$~'' semisheaf of rings, being generated by versal deformation
and spread\-ing-out isomorphism from the ``~$MG$~'' semisheaf of rings, is
characterized by a set of ranks $n_{\theta _M}$ inferior or equal to the set
of ranks $n_{\theta _{MG}}$ of the ``~$MG$~'' semisheaf of rings
since the codimension of the singularities on the ``~$MG$~'' semisheaf of rings
is inferior to the codimension of the singularities on the ``~$ST$~'' semisheaf
of rings.

According to proposition 2.3.4, the vibration frequency of the ``~$M$~''
semisheaf of rings must thus be inferior to the vibration frequency of the
``~$MG$~'' semisheaf of rings since:
\Bi
\item the frequency vibrations of the ``~$MG$~'' and
``~$M$~'' semisheaves are quantified;
\item the sections of the ``~$M$~'' semisheaf are open strings covering partially from outside the open strings of the ``~$MG$~'' semisheaf.\epr
\Ei
\vskip 11pt

\begin{rem}[Dark energy] { If the semiparticles are  composed of the 
$4D$-``~$ST$~'' semisheaves of rings $\theta
^{1-3}_{R,L}(t_{res},r_{res})_{ST}$ or of the $4D$-``~$ST$~'' and ``~$MG$~''
semisheaves of rings  $\theta
^{1-3}_{R,L}(t_{res},r_{res})_{ST}$ $\cup \theta
^{1-3}_{R,L}(t,r)_{MG}$~, noted $\theta
^{1-3}_{R,L}(t,r)_{ST-MG}$~, they are massless and
unobservable and could contribute to the dark energy of the Universe.}\end{rem}
\vskip 11pt

\subsection{The electric charge and the existence of three families of
semiparticles}

Let $\theta ^{1-3}_{R,L}(t,r)_{ST-MG-M}$ denote the three embedded
$4D$-semishaves of rings. \\ Consider that an external perturbation generates on
each section of the semisheaves of rings\linebreak 
$\theta ^{1-3}_{R,L}(t_{res},r_{res})_{ST}$~, $
\theta ^{1-3}_{R,L}(t_{res},r_{res})_{MG}$ and
$\theta ^{1-3}_{R,L}(t,r)_{M}$ an isolated degenerated singularity of corank 1.

Then, it will be seen that singularities of codimension 1 on $\theta
^1_{R,L}(t)_{ST-MG-M}$ may be interpreted as being at the origin of the time structure
of the electric charge and that singularities of codimension 2 and 3 on $\theta
^3_{R,L}(r)_{ST-MG-M}$ or on $\theta
^1_{R,L}(t)_{ST-MG-M}$ are at the origin of the generation of the second and of
the third family of elementary semiparticles.

Note that the time structure of the electrical charge is supposed to be generated by
versal deformation and spreading-out isomorphism because it must have a permanent
structure on the contrary of the magnetic moment of a semiparticle which is
generated only on the basis of the smooth endomorphism ``~$E$~'' as it will be
seen in the following.
\vskip 11pt

\begin{defi}[The time structure of the electric charge]  { Let $\theta
^1_{R,L}(t)_{ST-MG-M}$ $\equiv\theta ^1_{R,L}(t_{res})_{ST}\linebreak \cup
\theta ^1_{R,L}(t_{res})_{MG}\cup \theta ^1_{R,L}(t)_{M}$ denote the $1D$-time
``~$ST$~'', ``~$MG$~'' and ``~$M$~'' semisheaves of rings of a  semilepton or of
a semiquark.  Consider that each section of these  semisheaves is endowed with
an isolated degenerated singularity of codimension one due to an external
perturbation.
\\
Then, the versal deformation and spreading-out isomorphism, applied to $\theta
^1_{R,L}(t)_{ST-MG-M}$ gives:
\[SOT(e)\circ Vd(e):\theta ^1_{R,L}(t)_{ST-MG-M}\to 
\theta ^1_{R,L}(t_{res})_{ST-MG-M}\cup\theta ^1_{R,L}(t)^{(e)}_{ST-MG-M}\]
where $\theta ^1_{R,L}(t)^{(e)}_{ST-MG-M}$ is interpreted as the time
structure of the electric charge of a semilepton or of a semiquark; its
$3D$-spatial structure is given by a $3D$-semisheaf of rings
$\theta ^3_{L,R}(r)^{(e)}_{ST-MG-M}$ composed of $3D$-left (resp. right) quanta
generated by the smooth endomorphism $E_{ST-MG-M}$ (see definition 2.2.12)
acting simultaneously on the $3D$-semisheaves of rings $\theta
^3_{L,R}(r)_{ST-MG-M}$ of its associated semiparticle.

If $\theta ^1_{R,L}(t)^{(e)}_{ST-MG-M}$ represents the time structure of the
electric charge of a semiquark, then the ranks of the ``~$ST$~'', 
``~$MG$~'' and ``~$M$~'' semisheaves $\theta ^1_{R,L}(t)^{(e)}_{ST-MG-M}$ are
equal to $\frac 13$ or $\frac 23$ \cite{L-P-F} of the ranks of the
corresponding $1D$-electric semisheaves of a semilepton because the electric
charge must be conserved.}\end{defi}
\vskip 11pt

\begin{propo} Only three families of elementary semiparticles can exist in the
above-mentioned mathematical frame.\end{propo}
\vskip 11pt

\bpr Let $\theta ^3_{R,L}(r)^{(A)}_{ST-MG-M}=
\theta ^3_{R,L}(r_{res})_{ST}\cup \theta ^3_{R,L}(r_{res})_{MG}\cup
\theta ^3_{R,L}(r)_{M}$ be the three embedded $3D$-semisheaves of rings
``~$ST$~'', 
``~$MG$~'' and ``~$M$~'' of a semilepton or of a semiquark of the first family
$A$~, i.e. a semielectron or a semiquark ``up''.

Under some strong external perturbation, each section of the $3D$-semisheaves
or rings ``~$ST$~'', 
``~$MG$~'' and ``~$M$~'' is assumed to have one or a set of degenerated
singularities of corank 1 and of the same codimension $cd=2$~.

Then, the versal deformation and the spreading-out isomorphism of the three
embedded semisheaves of rings of a semiparticle of the first family ``~$A$~'',
$\theta^3_{R,L}(t,r)^{(A)}_{St-MG-M}$ generate the three embedded semisheaves
of rings of a semiparticle of the second family ``~$B$~'' according to:
\[ SOT(A)\circ Vd(A):\theta ^3_{R,L}(r)^{(A)}_{ST-MG-M}\to
\theta ^3_{R,L}(r_{res})^{(A)}_{ST-MG-M}\cup
\theta ^3_{R,L}(r)^{(B)}_{ST-MG-M}\]
where $\theta ^3_{R,L}(r_{res})^{(A)}_{ST-MG-M}\cup
\theta ^3_{R,L}(r)^{(B)}_{ST-MG-M}$ represents the three embedded structures
of this semiparticle ``~$B$~''.

But, if the singularities on the sections of $\theta
^3_{R,L}(r)^{(A)}_{ST-MG-M}$ are of corank 1 and codimension 3, then the three
embedded semisheaves of rings of a semiparticle of the third family ``~$C$~''
can be generated by versal deformation and spreading-out isomorphism from the
remaining degenerated singularities of corank 1 and codimension 1 on the
sections of $\theta ^3_{R,L}(r)^{(B)}_{ST-MG-M}$~. 
We then have:
\[ SOT(B)\circ Vd(B):\theta ^3_{R,L}(r)^{(B)}_{ST-MG-M}\to
\theta ^3_{R,L}(r_{res})^{(B)}_{ST-MG-M}\cup
\theta ^3_{R,L}(r)^{(C)}_{ST-MG-M}\]
where $\theta ^3_{R,L}(r_{res})^{(A)}_{ST-MG-M}\cup
\theta ^3_{R,L}(r_{res})^{(B)}_{ST-MG-M}\cup
\theta ^3_{R,L}(r)^{(C)}_{ST-MG-M}$ represents the three embedded spatial
structures of a semiparticle of the third family $C$~.

Finally, the corresponding $1D$-time semisheaves of rings are obtained from
the $3D$-semisheaves of rings by the morphisms $(\gamma _{r\to t}\circ E)$
(see definition 2.2.3).\epr
\vskip 11pt

\begin{rem} { As the three embedded semisheaves of rings of semiparticles
of the second and of the third family are supercompact by construction, they
are highly distorted.  Consequently, the three embedded structures of
semiparticles of these families $B$ and $C$ are highly unstable which explains
their rapid decays.}\end{rem}
\vskip 11pt

\begin{propo} The heavy semiquark of a given family can be obtained from
the lighter semiquark of the same family by versal deformation and
spreading-out isomorphism of the singularities of corank 1 and codimension 1
on the sections of the three embedded semisheaves of rings of this lighter
semiquark.\end{propo}
\vskip 11pt

\bpr Let $\theta ^3_{R,L}(r)^{(Li)}_{ST-MG-M}$ be the three embedded
$3D$-spatial semisheaves of rings of a light semiquark of a given family
``~$A$~'', ``~$B$~'' or ``~$C$~''.  Assume that the sections of 
$\theta ^3_{R,L}(r)^{(Li)}_{ST-MG-M}$ are endowed with singularities of corank
1 and codimension 1 under some external perturbation.  Then, under versal
deformation and spreading-out isomorphism, $\theta ^3_{R,L}(r)^{(Li)}_{ST-MG-M}$
is transformed according to:
\[ SOT(Li)\circ Vd(Li):\theta ^3_{R,L}(r)^{(Li)}_{ST-MG-M}\to
\theta ^3_{R,L}(r_{res})^{(Li)}_{ST-MG-M}\cup
\theta ^3_{R,L}(r)^{(He)}_{ST-MG-M}\]
where $\theta ^3_{R,L}(r_{res})^{(Li)}_{ST-MG-M}\cup
\theta ^3_{R,L}(r)^{(He)}_{ST-MG-M}$ represents the three embedded
$3D$-semi\-sheaves of rings of the heavier semiquark.  The corresponding
$1D$-semisheaves of rings are obtained by the $(\gamma _{r\to t}\circ E)$
morphism.\epr
\vskip 11pt

\begin{rem} { As the middle ground (~$MG$~) and mass (~$M$~) structures of
semiparticles are generated from the space-time (~$ST$~) structure by versal
deformation and spreading-out isomorphism, we shall not consider that the
creation of these ``~$MG$~'' and ``~$M$~'' structures correspond to an axiom
which, otherwise, would have been an homotopy axiom according to M. Atiyah
\cite{Ati3}.}\end{rem}
\vskip 11pt

\setcounter{defi}{0}\section{Bialgebras of von Neumann, probability calculus and quantification
rules}

The main purpose of this chapter is to introduce the bialgebras of von
Neumann and to restore in this manner the classical probability calculus in
quantum theories dealing thus with the sixth problem of Hilbert which consists
in the ontological meaning of the theory of probabilities.

In this context, the spectral representation of a (bi)operator is explicitly given as:
\Bi
\item corresponding to the representation of the general bilinear semigroup $GL_{2(n)}(\aa_R\times \aa_L)$  in the $G_R(\aa_R)\times G_L(\aa_L)$ bisemimodule $(M_R\otimes M_L)$ where $M_{R,L}$ is  $3(n)$-dimensional;
\item resulting from the representation of the Lie algebra $\gl_{2(n)}(\aa_R\times \aa_L)$  of the general bilinear semigroup
$GL_{2(n)}(\aa_R\times \aa_L)$   in the shifted $G_R(\aa_R)
\times G_L(\aa_L)$-bisemimodule $(M^a_R\otimes M^a_L)$ which is a perverse bisemisheaf.
\Ei

As our objective is the study of the space-time structure of elementary
particles which become bisemiparticles in this mathematical frame and as a
massive bisemiparticle is composed of a left and a right semiparticle whose
structure is given by the three embedded structures ``~$ST$~'', ``~$MG$~'' and 
``~$M$~'', we shall have to consider a bialgebra of von Neumann on each of
these three structures.
\vskip 11pt

\subsection{Hilbert, magnetic and electric bilinear spaces}

We thus begin this section by introducing the structure of a massive
bisemiparticle and the space on which it is defined.

\vskip 11pt

\begin{defi}[Structure of a massive right and left semiparticle] 
{ The three  embedded $4D$-\linebreak structures of a right and a left semiparticle, i.e.
essentially of a semilepton or of a semiquark, is given respectively by the
three embedded right $4D$-semisheaves of rings
\[ \theta ^{1-3}_R(t,r)_{ST-MG-M}=\theta ^{1-3}_R(t,r)_{ST}\cup
\theta ^{1-3}_R(t,r)_{MG}\cup \theta ^{1-3}_R(t,r)_{M}\] and by the three
embedded left $4D$-semisheaves of rings
\[ \theta ^{1-3}_L(t,r)_{ST-MG-M}=\theta ^{1-3}_L(t,r)_{ST}\cup
\theta ^{1-3}_L(t,r)_{MG}\cup \theta ^{1-3}_L(t,r)_{M}\]
as developed in chapter 2, section 2.

As $ \theta ^{1-3}_{R,L}(t,r)_{ST}\cap
\theta ^{1-3}_{R,L}(t,r)_{MG}=\emptyset$~, $ \theta ^{1-3}_{R,L}(t,r)_{ST}\cap
\theta ^{1-3}_{R,L}(t,r)_{M}=\emptyset$ and
$ \theta ^{1-3}_{R,L}(t,r)_{MG}\cap
\theta ^{1-3}_{R,L}(t,r)_{M}=\emptyset$~,  we shall envisage  the
direct sum of the three embedded semisheaves  ``~$ST$~'', ``~$MG$~'' and 
``~$M$~'':
\[ \Theta_{R,L}:\theta ^{1-3}_{R,L}(t,r)_{ST-MG-M}\to
\theta ^{1-3}_{R,L}(t,r)_{ST}\oplus \theta ^{1-3}_{R,L}(t,r)_{MG}\oplus
\theta ^{1-3}_{R,L}(t,r)_{M}\]
noted $\theta ^{1-3}_{R,L}(t,r)_{ST\oplus MG\oplus M}$~.

$\theta ^{1-3}_{R,L}(t,r)_{ST\oplus MG\oplus M}$ is then defined on  3 embedded
topological spaces of dimension 4.}\end{defi}
\vskip 11pt

\begin{defi} {\bf (Structure of a massive bisemiparticle)} { Massive
elementary  stable objects of Nature are in fact biobjects, i.e. bisemiparticles
according to axiom II 1.3.9.  Their space-time structure is crudely given by the
tensor product between the three embedded right and left $4D$-semisheaves of
rings:
\[ \theta ^{1-3}_{R}(t,r)_{ST\oplus MG\oplus M}\otimes
\theta ^{1-3}_{L}(t,r)_{ST\oplus MG\oplus M}\]
which allows to generate interactions between the right and left structures,
i.e. between the right and left ``~$ST$~'', ``~$MG$~'' and 
``~$M$~'' semisheaves of rings.

Consider the condensed notation $\theta ^4_{R;ST}$ for $\theta
^{1-3}_R(t,r)_{ST}$~.

This tensor product then develops according to:
\nobeqn 
\lefteqn{\theta ^{1-3}_{R}(t,r)_{ST\oplus MG\oplus M}\otimes
\theta ^{1-3}_{L}(t,r)_{ST\oplus MG\oplus M} \equiv 
 \theta ^{4}_{R;ST\oplus MG\oplus M}\otimes
\theta ^{4}_{L;ST\oplus MG\oplus M}} \\ &= &
 (\theta ^{4}_{R;ST}\oplus \theta ^4_{R;MG}\oplus \theta ^4_{R;M})\otimes
(\theta ^{4}_{L;ST}\oplus \theta ^4_{L;MG}\oplus \theta ^4_{L;M})\\
&= &
 (\theta ^{4}_{R;ST}\otimes \theta ^{4}_{L;ST})\oplus (\theta ^4_{R;MG}\otimes
\theta ^4_{L;MG})\oplus (\theta ^4_{R;M}\otimes \theta ^4_{L;M})
 \\
&&\qquad \oplus (\theta ^4_{R;ST}\otimes \theta ^4_{L;MG})
\oplus (\theta ^4_{R;MG}\otimes \theta ^4_{L;ST})
\oplus (\theta ^4_{R;ST}\otimes \theta ^4_{L;M})
\\
&&\qquad \oplus (\theta ^4_{R;M}\otimes \theta ^4_{L;ST})
\oplus (\theta ^4_{R;MG}\otimes \theta
^4_{L;M})
\oplus (\theta ^4_{R;M}\otimes \theta ^4_{L;MG})\noeeqn
where the three first tensor products refer to the ``~$ST$~'', ``~$MG$~'' and 
``~$M$~'' structures of the considered bisemiparticle while the six other
tensor products refer to the interactions between the right and left
``~$ST$~'', ``~$MG$~'' and 
``~$M$~'' structures.}\end{defi}
\vskip 11pt

\begin{defi}[Duality of semisheaves] { Let $ \widetilde M_R$ and $\widetilde M_L$ denote
 a  right semisheaf $\theta ^4_{R;ST}$~,
$\theta ^4_{R;MG}$ or $\theta ^4_{R;M}$ and
a  left semisheaf $\theta ^4_{L;ST}$~,
$\theta ^4_{L;MG}$ or $\theta ^4_{L;M}$~.

Their tensor product is given by the bisemimodule $(\widetilde M_R\otimes \widetilde M_L)$ which
decomposes under the blowing-up isomorphism $S_L$ (see proposition 1.3.2) into
the direct sum of
\begin{enumerate}
\item the diagonal bisemisheaf $(\widetilde M_R\otimes_D \widetilde M_L)$~,
\item the magnetic bisemisheaf $(\widetilde M^S_R\otimes_{\rm magn} \widetilde M^S_L)$~,
\item the electric bisemisheaf $(\widetilde M^{T-(S)}_R\otimes_{\rm elec}
\widetilde M^{S-(T)}_L)$~,
\end{enumerate}
where $\widetilde M^S_{R,L}$ is a $3D$-spatial subsemisheaf and where $\widetilde M^T_{R,L}$ 
is a $1D$-time subsemisheaf.

For the facility of notations, \bt[t]{l}
$(\widetilde M^S_R \otimes_{\rm magn} \widetilde M^S_L)$ will be
written $(\widetilde M_R\otimes_{ m} \widetilde M_L)$ and \\
$(\widetilde M^{T-(S)}_R\otimes_{\rm elec}
\widetilde M^{S-(T)}_L)$ will be written $(\widetilde M_R\otimes_{ e}
\widetilde M_L)$~.\te
\vskip 6pt

If we consider the projective linear map:
\[ p_L:\widetilde M_R\otimes_{D,m,e} \widetilde M_L\to \widetilde M_{R(P)/_{D,m,e}L}\]
of the right semisheaf $\widetilde M_R$ onto the left semisheaf $\widetilde M_L$ with respect to
the diagonal, magnetic or electric metric, then $\widetilde M_{R(P)}$ is the dual semisheaf
of $\widetilde M_L$ and is called a coleft semisheaf whose elements are coleft
differential functions.

But, if we take into account the projective linear map
\[ p_R:\widetilde M_R\otimes_{D,m,e} \widetilde M_L\to \widetilde M_{L(P)/_{D,m,e}R}\]
projecting the left semisheaf $\widetilde M_L$ onto the right semisheaf $\widetilde M_R$~, then
$\widetilde M_{L(P)}$ is the dual semisheaf of $\widetilde M_R$ and will be called a coright semisheaf
whose elements are coright functions.}\end{defi}
\vskip 11pt

\begin{rms} {{\bf 1.}
The ``~$ST$~'', ``~$MG$~'' or  
``~$M$~'' diagonal (bi)structure of a bisemiparticle is thus given by the
diagonal bisemisheaf $\widetilde M_{R(P)/_{D}L}$ (resp. $\widetilde M_{L(P)/_DR}$~) constituted by the
diagonal tensor product between the left (resp. right) semisheaf $\widetilde M_L$ (resp.
$\widetilde M_R$~) of the left (resp. right) semiparticle and the projected right (resp.
left) semisheaf $\widetilde M_{R(P)}$ (resp. $\widetilde M_{L(P)}$~) of the projected right (resp.
left) semiparticle.

The projected right (resp. left) semisheaf $\widetilde M_{R(P)}$ (resp. $\widetilde M_{L(P)}$~) is
thus called a coleft (resp. coright) semisheaf and the projected right
(resp. left) semiparticle is then called a coleft (resp. coright)
semiparticle.
\vskip 11pt

{\bf 2.} The following developments about bilinear Hilbert spaces concern the bisemisheaves $(\widetilde M_R\otimes \widetilde M_L)$ as well as the $G_{R\times L}(\Aa_R\times \Aa_L)$-bisemimodules $(M_R\otimes M_L)$ on which they are defined.
}\end{rms} 
\vskip 11pt

\begin{defi} {\bf (Algebraic external Hilbert, magnetic and electric bilinear  
spaces) \quad 1.}  { By the projective linear map $p_L$ (resp. $p_R$~), the
diagonal bisemisheaf $(\widetilde M_R\otimes _D\widetilde M_L)$ is transformed into $\widetilde M_{R(P)/_DL}$ (resp.
$\widetilde M_{L(P)/_D R}$~).  If we endow $\widetilde M_{R(P)/_D L}$ (resp. $\widetilde M_{L(P)/_DR}$~) with an
external scalar product characterized by an euclidian metric $\delta ^\alpha
_\beta $ of type $(1,1)$~, $0\le \alpha ,\beta \le 3$~, then we get a left
(resp. right) external bilinear Hilbert space noted $\Hs_L^a$ (resp. $\Hs_R^a$~)
\cite{Pie4}, which is of algebraic nature.
\vskip 11pt

\noindent {\bf 2.}  Similarly, the projective linear map $p_L$ (resp. $p_R$~) transforms the
magnetic bisemisheaf $(\widetilde M_R\otimes_m\widetilde M_L)$ into a left (resp. right) external
bilinear magnetic bisemisheaf $\widetilde M_{R(P)/_mL}$ (resp. $\widetilde M_{L(P)/_m R}$~) which
becomes a left (resp. right) external bilinear magnetic space, noted $V^{m;a}_L$
(resp. $V^{m;a}_R$~), if it is endowed with an external magnetic product $\langle
\phi_{R(P)},\phi_L\rangle_m$ (resp. $\langle
\phi_{L(P)},\phi_R\rangle_m$~) defined from $(\widetilde M_{R(P)}\times_m\widetilde M_L)$ to $\cit $
(resp. from $(\widetilde M_{L(P)}\times_m\widetilde M_R)$ to $\cit $~) and  characterized by a
noneuclidian magnetic metric $g^\alpha _\beta $~, $\forall\ \alpha \neq \beta $~,
$1\le \alpha ,\beta \le 3$~, of type $(1,1)$~.
\vskip 11pt

\noindent {\bf 3.}  The electric bisemisheaf $(\widetilde M_R\otimes_e\widetilde M_L)$ is
transformed by the projective linear map $p_L$ (resp. $p_R$~) into the left
(resp. right) external bilinear electric bisemisheaf $\widetilde M_{R(P)/_e L}$ (resp.
 $\widetilde M_{L(P)/_e R}$~) which becomes a left (resp. right) external bilinear
electric space, noted $V^{e;a}_L$ (resp. $V^{e;a}_R$~), if it is endowed with an
external electric product $\langle \phi_{R(P)},\phi_L\rangle_e$ (resp.
$\langle \phi_{L(P)},\phi_R\rangle_e$~) defined from $(\widetilde M_{R(P)}\times_e\widetilde M_L)$
to $\cit $ (resp. from $(\widetilde M_{L(P)}\times_e\widetilde M_R)$
to $\cit $~) and characterized by a noneuclidian electric metric $g^\alpha _\beta $ of
type $(1,1)$ with $\alpha =0$ and $1\le \beta \le 3$ or with $1\le \alpha \le
3$ and $\beta =0$~.}\end{defi}
\vskip 11pt

\begin{propo} The left and right external bilinear Hilbert spaces $\Hs_L^a$ and
$\Hs_R^a$ are characterized by bilinear orthogonal $4D$-basis while the left and right
external bilinear electric and magnetic spaces are characterized by
$3D$-basis.\end{propo}
\vskip 11pt

\bpr 1.  The bilinear Hilbert spaces $\Hs_L^a$ and $\Hs_R^a$ are characterized by
$4D$-orthogonal bilinear basis since they result from the diagonal bisemisheaf
$(\widetilde M_R\otimes_D\widetilde M_L)$~.
\vskip 11pt

\noindent 2.  The electric basis is three-dimensional and not
six-dimensional because the set of electric basis bivectors $\{e^0\otimes
f_\beta \}^3_{\beta =1}$ are orthogonal to the electric basis bivectors
$\{e^\beta \otimes
f_0 \}^3_{\beta =1}$~; indeed, we have that $
\langle (e^0)^*\otimes f_\beta ,(e^\beta )^*\otimes f_0)\rangle=0$ implying 
$\langle (e^0)^*,f_0\rangle\langle (e^\beta)^*,f_\beta\rangle=0$ since
$\langle (e^0)^*, f_0 \rangle =\langle(e^\beta )^*, f_\beta
)\rangle=0$~, with $1\le \beta \le 3$~, by hypothesis on the electric metric.
\vskip 11pt

\noindent 3.  Similar conclusions are obtained for the external bilinear
magnetic spaces $V^{m;a}_L$ and $V^{m;a}_R$~.\epr
\vskip 11pt

\begin{defi} {\bf (Algebraic internal Hilbert, magnetic and electric bilinear  
spaces)}\quad 
 { Let
$B_L:M_{R(P)}\to M_L$ (resp. $B_R:M_{L(P)}\to M_R$~) be the bijective linear
isometric map from $M_{R(P)}$ (resp. $M_{L(P)}$~) to $M_L$ (resp. $M_R$~)
mapping each covariant element of $M_{R(P)}$ (resp. $M_{L(P)}$~) into a
contravariant element of $M_L$~, noted $M_{L_R}$ (resp. of $M_R$~, noted
$M_{R_L}$~) as introduced in proposition  1.3.6.

Then, $B_L$ (resp. $B_R$~) transforms:
\begin{description}
\item{1.} the left (resp. right) external bilinear Hilbert space $\Hs_L^a$
(resp. $\Hs_R^a$~) into the left (resp. right) internal bilinear Hilbert space
$\Hs^+_a$ (resp. $\Hs^-_a$~) in such a way that:
\begin{enumerate} \item the bielements of $\Hs^+_a$ (resp. $\Hs^-_a$~) are bivectors, i.e. two
confounded vectors;
\item each external scalar product of $\Hs_L^a$ (resp. $\Hs_R^a$~) is transformed
into an internal scalar product defined from $M_{L_R}\times_D M_L$ (resp.
$M_{R_L}\times_D M_R$~) to $\cit$~.
\end{enumerate}
\vskip 11pt

\item{2.} the left (resp. right) external bilinear magnetic space $V^{m;a}_L$
(resp. $V^{m;a}_R$~) into the left (resp. right) internal bilinear magnetic space
$V^+_{m;a}$ (resp. $V^-_{m;a}$~) in such a way that the external magnetic product of
$V^{m;a}_L$ (resp. $V^{m;a}_R$~) be transformed into an internal magnetic product
defined from $M_{L_R}\times_mM_L$ (resp. $M_{R_L}\times_mM_R$~) to
$\cit$~.  This internal magnetic space $V^+_{m;a}$ (resp. $V^-_{m;a}$~) is
characterized by a noneuclidian metric $g_{\alpha \beta }$  of type $(0,2)$~,
$\forall\ \alpha \neq
\beta $~, $1\le \alpha ,\beta \le 3$~.
\vskip 11pt

\item{3.} the left (resp. right) external bilinear electric space $V^{e;a}_L$
(resp. $V^{e;a}_R$~) into the left (resp. right) internal bilinear electric space
$V^+_{e;a}$ (resp. $V^-_{e;a}$~) such that the external electric product of
$V^{e;a}_L$ (resp. $V^{e;a}_R$~) be transformed into an internal electric product
defined from
$M_{L_R}\times_eM_L$ (resp. $M_{R_L}\times_eM_R$~) to $\cit$~.
\end{description}}\end{defi}
\vskip 11pt

\begin{defis} {\bf (1. Algebraic extended external bilinear Hilbert spaces $H^a_L$
and
$H^a_R$~)\/}
 { Let $\widetilde M_R$ and $\widetilde M_L$ denote respectively the $4D$-right semisheaf
and the $4D$-left semisheaf.  Then, we consider on the noneuclidian
  bisemisheaf $\widetilde M_R\otimes \widetilde M_L$ the projective linear
map:
\nobeqn	&&p_L:\widetilde M_R\otimes \widetilde M_L\to \widetilde M_{R(P)/_cL} \qquad (\ ``~c~'':\mbox{\ for
complete)}\\
&\mbox{or}& p_R:\widetilde M_R\otimes \widetilde M_L\to \widetilde M_{L(P)/_cR}\noeeqn
of the right (resp. left) semisheaf $\widetilde M_R$ (resp. $\widetilde M_L$~) on the left (resp. right)
semisheaf $\widetilde M_L$ (resp. $\widetilde M_R$~).

If we endow the bisemisheaf $\widetilde M_{R(P)/_cL}$ (resp. $\widetilde M_{L(P)/_cR}$~) with a complete
external bilinear form defined from $\widetilde M_{R(P)}\times \widetilde M_L$ (resp. $\widetilde M_{L(P)}\times
\widetilde M_R$~) to $\cit$~, we get a left (resp. right) extended external bilinear Hilbert
space $H^a_L$ (resp. $H^a_R$~) characterized by a nonorthogonal basis.
\vskip 11pt

\noindent {\bf (2. Algebraic extended internal bilinear Hilbert spaces $H^+_a$ and
$H^-_a$~)\/}  The left (resp. right) extended external bilinear Hilbert space
$H^a_L$ (resp. $H^a_R$~) is transformed into the left (resp. right) extended
internal bilinear Hilbert space $H^+_a$ (resp. $H^-_a$~) by means of a bijective
bilinear isometric map $B _L$ (resp. $B _R$~) from $\widetilde M_{R(P)}$ (resp.
$\widetilde M_{L(P)}$~) to $\widetilde M_L$ (resp. $\widetilde M_R$~).

The complete external bilinear form of $H^a_L$ (resp. $H^a_R$~) is then transformed
into a complete internal bilinear form of $H^+_a$ (resp. $H^-_a$~).}
\end{defis}
\vskip 11pt

\begin{defi}\  {\bf (Analytic Hilbert, magnetic and electric bilinear
spaces)} { Let 
$X^s_{R,L}$ be the analytic semivariety associated to the semispace $\partial\ \o{\!
S}_{K_{R,L}}$ and let $\widetilde M^s_{R,L}$ be an analytic semisheaf on $X^s_{R,L}$~.

From the complete, diagonal, magnetic or electric tensor product between the right
and left semisheaves $\widetilde M^s_R$ and $\widetilde M^s_L$~, we can construct by application of the
composition of maps $B_L\circ p_L$ (resp. $B_R\circ p_R$~) on the bisemisheaves
$\widetilde M^s_R\otimes \widetilde M^s_L$~, $\widetilde M^s_R\otimes_D \widetilde M^s_L$~, $\widetilde M^s_R\otimes_m \widetilde M^s_L$ or
$\widetilde M^s_R\otimes_e \widetilde M^s_L$ an analytic left (resp. right) internal bilinear extended
Hilbert space $H^+_h$ (resp. $H^-_h$~):

{\setlength{\unitlength}{1mm}
\begin{picture}(120,15)


\put(0,0){$\begin{array}[b]{ccccc}
\{ \widetilde M^s_R,\widetilde M^s_L\} &\hfl{}{} & \widetilde M^s_R\otimes \widetilde M^s_L 
& \hfl{B_L\circ p_L}{} & \widetilde M^s_{L_R}\otimes \widetilde M^s_L\subset H^+_h\\
&&&{}^{B_R\circ p_R}\hspace{1cm} & \widetilde M^s_{R_L}\otimes \widetilde M^s_R\subset H^-_h\\
\end{array}$}
\put(65,7){\vector(2,-1){10}}

\end{picture}}

\noindent an analytic left (resp. right) internal bilinear (diagonal) Hilbert space
$\Hs^+_h$ (resp. $\Hs^-_h$~):

{\setlength{\unitlength}{1mm}
\begin{picture}(120,15)


\put(0,0){$\begin{array}[b]{ccccc}
\{ \widetilde M^s_R,\widetilde M^s_L\} &\hfl{}{} & \widetilde M^s_R\otimes_D \widetilde M^s_L 
& \hfl{B_L\circ p_L}{} & \widetilde M^s_{L_R}\otimes_D \widetilde M^s_L\subset \Hs^+_h\\
&&&{}^{B_R\circ p_R}\hspace{1cm} & \widetilde M^s_{R_L}\otimes_D \widetilde M^s_R\subset \Hs^-_h\\
\end{array}$}
\put(68,7){\vector(2,-1){10}}

\end{picture}}

\noindent an analytic left (resp. right) internal bilinear magnetic space
$V^+_{m;h}$ (resp.
$V^-_{m;h}$~):

{\setlength{\unitlength}{1mm}
\begin{picture}(120,15)


\put(0,0){$\begin{array}[b]{ccccc}
\{ \widetilde M^s_R,\widetilde M^s_L\} &\hfl{}{} & \widetilde M^s_R\otimes_m \widetilde M^s_L 
& \hfl{B_L\circ p_L}{} & \widetilde M^s_{L_R}\otimes_m \widetilde M^s_L\subset V^+_{m;h}\\
&&&{}^{B_R\circ p_R}\hspace{1cm} & \widetilde M^s_{R_L}\otimes_m \widetilde M^s_R\subset V^-_{m;h}\\
\end{array}$}
\put(68,7){\vector(2,-1){10}}

\end{picture}}

\noindent an analytic left (resp. right) internal bilinear electric space
$V^+_{e;h}$ (resp.
$V^-_{e;h}$~):

{\setlength{\unitlength}{1mm}
\begin{picture}(120,15)


\put(0,0){$\begin{array}[b]{ccccc}
\{ \widetilde M^s_R,\widetilde M^s_L\} &\hfl{}{} & \widetilde M^s_R\otimes_e \widetilde M^s_L 
& \hfl{B_L\circ p_L}{} & \widetilde M^s_{L_R}\otimes_e \widetilde M^s_L\subset V^+_{e;h}\\
&&&{}^{B_R\circ p_R}\hspace{1cm} & \widetilde M^s_{R_L}\otimes_e \widetilde M^s_R\subset V^-_{e;h}\\
\end{array}$}
\put(68,7){\vector(2,-1){10}}

\end{picture}}

All these internal bilinear spaces are endowed with the corresponding internal
bilinear forms in complete analogy with which was developed in definition 3.1.7.}
\end{defi}
\vskip 11pt

\begin{defi} {\bf (Diagonal, complete, magnetic and electric products of right  
and left Eisenstein cohomologies)\/} \quad { In chapter 1, section 1, right and
left Eisenstein cohomologies $H^*_{R,L}(\partial\ \o S_{K_{ R,L}},\widetilde M_{R,L})$
defined on the right and left semispaces $\partial\ \o S_{K_{ R,L}}$ and
associated to the generation of the right and left semisheaves $\widetilde M_{R,L}$ were
studied.

This allows to generate a diagonal, complete, magnetic or electric bisemisheaf\\
$\widetilde M_R\otimes_{(D),m,e}\widetilde M_L$ on $\partial \ \o S_{K_{R}}\times_{(D),m,e}\partial
\ \o S_{K_{L}}$ by the diagonal, complete, magnetic or electric product of right
and left Eisenstein cohomology groups:
\nobeqn
H^E_{R\times_{(D),m,e}L}&:&H^*_R(\partial\ \o{\!
S}_{K_R},\widetilde M_R)\times_{(D),m,e} H^*_L(\partial \ \o S_{K_L},\widetilde M_L)\\
&&\qquad \to  H^*_{R\times_{(D),m,e}L}(\partial \ \o S_{K_R}
\times_{(D),m,e}\partial\ \o S_{K_L},\widetilde M_R\otimes_{(D),m,e}\widetilde M_L)\;.\noeeqn}
\end{defi}
\vskip 11pt

\begin{propo}  The bilinear Eisenstein cohomology
\[ H^*_{R\times_{(D),m,e}L}\ (\partial\ \o S_{K_R}\times_{(D),m,e}\partial\ \o
{\! S}_{K_L},\widetilde M_R\otimes_{(D),m,e}\widetilde M_L)\]
associated to the coefficient system $\widetilde M_R\otimes_{(D),m,e}\widetilde M_L$ decomposes into sum
of products of one-dimensional eigenspaces according to:

\begin{align*} 
&
H^*_{R\times_{(D),m,e}L}\ (\partial\ \o S_{K_R}\times_{(D),m,e}\partial\ \o {\! S}_{K_L}
,\widetilde M_R\otimes_{(D),m,e}\widetilde M_L)\\[11pt]
&\quad=\textstyle \bigoplus\limits_\mu \bigoplus\limits_{m_\mu} \bigoplus\limits_{1_\ell}
\Ind^{G_{R\times_{(D),m,e}L}(\aa_R\times \aa_L)}_{K^D_{R\times_{(D),m,e}L} (\o\ZZ^2_{pq})}\\
&\quad \qquad \times
H^{1_\ell,1_\ell}_{R\times_{(D),m,e}L} (S^{M_{R\times_{(D),m,e}L}} ,
H^{1_\ell,1_\ell}(\tilde u_{K^D_{R\times_{(D),m,e}L}} ,\widetilde M_R^{1_\ell}(\mu ,m_\mu )\otimes_{(D),m,e} \widetilde M^{1_\ell}(\mu ,m_\mu )))
\end{align*}
where the sum over $1_\ell$~, $1\le\ell\le n$~, refers to the decomposition of the $n^{(2)}$-dimensional bisemisheaf $\widetilde M_R\otimes \widetilde M_L$ into products of $1$-dimensional subsemisheaves
$\widetilde M^{1_\ell}_R\otimes \widetilde M_L^{1_\ell}$ on the representation of $GL_{2_\ell}(\aa_R\times \aa_L)$ \cite{Pie9}.
\end{propo}
\vskip 11pt

\bpr  This immediately results from 
proposition 1.1.22 and from the reducible  Langlands program developed in \cite{Pie9} in such a 
way that \begin{equation}
\Rep ( GL_{2n=2_1+\cdots+2_\ell+\cdots+2_n}(\aa_R\times \aa_L))=\mathop{\boxplus}\limits^n_{\ell=1}\Rep GL_{2_\ell}(\aa_R\times \aa_L)\;.\tag*{\eop}\end{equation}
\vskip 11pt

\begin{defi}[The analytic de Rham cohomology] { As in the algebraic case,  
the analytic cohomology $H^*(X^s_{R,L},\widetilde M^s_{R,L})$ can be computed through the
analytic de Rham complex.

We can also define a diagonal, complete, magnetic or electric product of right and
left analytic cohomology groups:
\nobeqn 
H^h_{R\times_{(D),m,e}L}& :&
H^*(X^s_R,\widetilde M^s_R)\times_{(D),m,e}H^*(X^s_L,\widetilde M^s_L)\\
&&\qquad\to 
H^*_{R\times_{(D),m,e}L}(X^s_R\times_{(D),m,e}X^s_L,\widetilde M^s_R\otimes_{(D),m,e}\widetilde M^s_L)\noeeqn
with coefficients in the respective product $\widetilde M^s_R\otimes_{(D),m,e}\widetilde M^s_L$ of the
analytic semisheaves $\widetilde M^s_R$ and $\widetilde M^s_L$~.}
\end{defi}
\vskip 11pt

\begin{propo}  There is an isomorphism:
\nobeqn
\lefteqn{i_{H^*(\partial\ \o{\!
S}_{K_{R\times_{(D),m,e}L}})-H^*(X^s_{R\times_{(D),m,e}L}}) \ :\ }\\
&&H^*_{R\times_{(D),m,e}L}(X^s_R\times_{(D),m,e}X^s_L,\widetilde M^s_R\otimes_{(D),m,e}\widetilde M^s_L)\\
&&\qquad \to H^*_{R\times_{(D),m,e}L}(\partial\ \o S_{K_R}\times_{(D),m,e}\
\partial\
\o S_{K_L},\widetilde M_R\otimes_{(D),m,e}\widetilde M_L)\noeeqn
between products of Eisenstein cohomologies and analytic de Rham cohomologies.
\end{propo}
\vskip 11pt

\bpr According to Grothendick \cite{Gro3}, there is an isomorphism between the de
Rham cohomologies of $\Omega^*$-smooth differential forms with respect to $\partial\
\o S_{K_{R,L}}$ and $X^s_{R,L}$~:
\[ H^*(\Omega^*_{\partial\  \o S_{K_{R,L}}})\simeq H^*(\Omega^*_{X^s_{R,L}})\]
leading to the following isomorphism
\[ H^*(\partial\  \o S_{K_{R,L}},\widetilde M_{R,L})\simeq H^*(X^s_{R,L},\widetilde M^s_{R,L})\]
and thus to the thesis.\epr
\vskip 11pt

\subsection{Bialgebras of Von Neumann}

\begin{defi} {\bf (Diagonal, complete, magnetic and electric products of operators) 
\quad 1.\/} { Let $(\widetilde M_{L_R}\otimes_{(D),m,e}\widetilde M_L)$ (resp.
$(\widetilde M_{R_L}\otimes_{(D),m,e}\widetilde M_R)$~) be the algebraic diagonal, complete, magnetic or
electric bisemisheaf respectively of the left (resp. right) algebraic internal
diagonal Hilbert, extended Hilbert, magnetic or electric bilinear space
$\Hs^{\pm}_a$~, $H^{\pm}_a$~, $V^{\pm}_{m;a}$ or $V^{\pm}_{e;a}$~.

Similarly, let $(\widetilde M^s_{L_R}\otimes_{(D),m,e}\widetilde M^s_L)$ (resp.
$(\widetilde M^s_{R_L}\otimes_{(D),m,e}\widetilde M^s_R)$ be the corresponding analytic bisemisheaves of
the analytic bilinear spaces $\Hs^{\pm}_h$~, $H^{\pm}_h$~, $V^{\pm}_{m;h}$ or
$V^{\pm}_{e;h}$~.
\vskip 11pt

\noindent {\bf 2.\/} Consider the diagonal, complete, magnetic or electric
tensor product between a right and a left elliptic (linear differential) operator
$D_R$ and $D_L$ acting respectively on a right and a left algebraic or analytic
semisheaf
$\widetilde  M^{(s)}_{L_R}$ or $\widetilde M^{(s)}_R$ and  $\widetilde M^{(s)}_{L}$
or $\widetilde M^{(s)}_{R_L}$ of $\Hs^{\pm}_{h,a}$~, $H^{\pm}_{h,a}$~, $V^{\pm}_{m;h,a}$ or
$V^{\pm}_{e;h,a}$~:
\nobeqn
\{ D_R,D_L\}\{ \widetilde M^{(s)}_{L_R},\widetilde M^{(s)}_L\}&\to &
(D_R\otimes_{(D),m,e}D_L)(\widetilde M^{(s)}_{L_R}\otimes_{(D),m,e}\widetilde M^{(s)}_L) \\
\{ D_R,D_L\}\{ \widetilde M^{(s)}_{R_L},\widetilde M^{(s)}_R\}&\to &
(D_R\otimes_{(D),m,e}D_L)(\widetilde M^{(s)}_{R_L}\otimes_{(D),m,e}\widetilde M^{(s)}_R)\;.\noeeqn
The index \cite{Ati3} of a diagonal, complete, magnetic or electric product of a
right and a left elliptic operators is given by:
\[\gamma _{R\times L}(D_R\otimes_{(D),m,e}D_L)=\gamma _R(D_R)\times \gamma _L(D_L)\]
taking into account that $\gamma _{R,L}(D_{R,L})$ is the index of a right (resp.
left) operator.

Furthermore, we have that:
\[\gamma _{R,L}(D_R\otimes_{(D),m,e}D_L)=\dim \Ker (D_R\otimes_{(D),m,e}D_L) -
\dim{\rm co}\Ker (D_R\otimes_{(D),m,e}D_L)\;.\]
\vskip 11pt

\noindent {\bf 3.\/}  If the complete, diagonal, magnetic or electric tensor product
between a right and a left operator is bounded and has a finite-dimensional kernel
and cokernel, then it is a complete, diagonal, magnetic or electric Fredholm
bioperator, noted $(T_{F_R}\otimes_{(D),m,e}T_{F_L})$~.
\vskip 11pt

\noindent {\bf 4.\/}  Let $\Ls^B_{R,L}(\widetilde M^{(s)}_{R,L})$ denote the algebra of right
(resp. left) bounded operators $T_{R,L}$ acting respectively on the right or left
semisheaf
$\widetilde M^{(s)}_R$ or $\widetilde M^{(s)}_L$~.

Then, the algebra of right (resp. left) self-adjoint bounded operators $T_{R,L}$
acting on $\Hs^{\pm}_{a,h}$~, $H^{\pm}_{a,h}$~, $V^{\pm}_{m;a,h}$ or
$V^{\pm}_{e;a,h}$ will be noted $\Ls^B_{R,L}(\Hs^{\pm}_{a,h})$~,
$\Ls^B_{R,L}(H^{\pm}_{a,h})$~,
$\Ls^B_{R,L}(V^{\pm}_{m;a,h})$ or 
$\Ls^B_{R,L}(V^{\pm}_{e;a,h})$ while the bialgebra of diagonal, complete, magnetic
or electric tensor product of a right and a left bounded operators
$(T_R\otimes_{(D),m,e}T_L)$ acting on the corresponding bilinear spaces will be
noted $\Ls^B_{R\times_D L}(\Hs^{\pm}_{a,h})\equiv\Ls^B_R\otimes_D
\Ls^B_L(\Hs^{\pm}_{a,h})$~, $\Ls^B_{R\times L}(H^{\pm}_{a,h})$~,
$\Ls^B_{R\times_mL}(V^{\pm}_{m;a,h})$ or
$\Ls^B_{R\times_eL}(V^{\pm}_{e;a,h})$~.
}\end{defi}
\vskip 11pt

\begin{lm} The bialgebra $\Ls^B_{R\times_DL}(\Hs^{\pm}_a)$ is abelian.\end{lm}
\vskip 11pt

\bpr 
Considering that $\Hs^{\pm}_a$ is characterized by a diagonal metric, the
bialgebra  $\Ls^B_{R\times_DL}(\Hs^{\pm}_a)$ must then be abelian.\epr
\vskip 11pt

\begin{defis}[1.  Self-adjointness]  { Consider that the right and left bounded
operators $T_R$ and $T_L$ are self-adjoint, i.e. that we have
$T_R\equiv T^{\dagger}_L=T_L$~.

A left and a right involutions are then defined by:
\nobeqn i_L &:& T_L\to T^{\dagger}_L\equiv T_R\;, \\ i_R &:& T_R\to
T^{\dagger}_R\equiv T_L\;.\noeeqn
The physical interpretation of the self-adjointness consists in the fact that the
action of the self-adjoint operator $T_R$ on the co-left semisheaf $\widetilde  M_{L_R}$ is
equal to its antiunitary involutary action on the left semisheaf $\widetilde M_L$~.

The mathematical origin of the self-adjointness results from the fact that the
centralizer of the co-left semimodule $M_{L_R}$ is $Z_0(L^-)$ while the
centralizer of the left semimodule $M_L$ is $Z_0(L^+)$ according to \cite{Pie4}.
\vskip 11pt

\noindent {\bf (2.  Complete, diagonal, magnetic or electric norm topologies)}  The
complete, diagonal, magnetic or electric norm topology on
$(T_R\otimes_{(D),m,e}T_L)$ will be given by:
\[\|T_R\otimes_{(D),m,e}T_L\|=\sup\ \frac{(T_R\psi _{L_R},T_L\psi
_L)_{(D),m,e}}{(\psi _{L_R},\psi _L)_{(D),m,e}}\quad \forall\ \psi _{L_R}\in
\widetilde M_{L_R}\ , \; \forall \ \psi _L\in\widetilde  M_L\]
where $(\cdot ,\cdot )_{(D),m,e}$ is respectively a complete, diagonal, magnetic or
electric internal bilinear form as introduced in definition 3.1.7 and characterized
by a complete, diagonal, magnetic or electric metric.
\vskip 11pt

\noindent {\bf (3.)}  A {\bf weight\/} on the algebra $\Ls^B_{R,L}(H^+_a)$ is given
by the positive bilinear form $(T_R\psi _{L_R},\psi _L)$ or $(\psi _{L_R},T_L\psi
_L)$ which is a map from $\Ls^B_{R,L}(M_{L_R}\times M_L)$ into $ \cit$ for every
section $\psi _{L_R}\in \widetilde M_{L_R}$ and $\psi _L\in \widetilde M_L$~.

Similarly, a weight on the bialgebra $(\Ls^B_R\otimes \Ls^B_L)(H^+_a)$ is given by
the positive bilinear form $(T_R\psi _{L_R},T_L\psi _L)$ which is a map from
$(\Ls^B_R(M_{L_R})\otimes \Ls^B_L(M_L))$ into $ \cit$ for all $T_{R,L}\in
\Ls^B_{R,L}$~.}
\end{defis}
\vskip 11pt

\begin{propo} The extended bilinear Hilbert spaces $H^{\pm}_{a,h}$ are the natural
representation spaces for the algebras and the bialgebras of bounded operators.
\end{propo}
\vskip 11pt

\bpr The representation of a group $G$ in a linear Hilbert space $h$ is an
application such that to each element $g$ of $G$ corresponds a linear operator
$T(g)$~.  In the finite-dimensional case, $T(g)$ is defined by a matrix of
$M_n(K)$~.

On the other hand, the enveloping algebra $M^e$ of the semimodule $M^{(s)}_{R,L}$
is given by
\[ M^e_{(s)}=M^{(s)}_R\otimes_{\aa_R\times \aa_L}M^{(s)}_L\]
where $M^{(s)}_R$ (resp. $M^{(s)}_L$~) must be considered as the opposite 	algebra
of $M^{(s)}_L$ (resp. $M^{(s)}_R$~).

If $M^{(s)}_{R,L}$ is a projective right (resp. left) semimodule of dimension $n$~,
then $M^{(s)}_{R,L}\simeq \aa_{R,L}^n$ and we have:
\[ M^e_{(s)}\simeq \End_{\aa_R\times \aa_L}(M^{(s)}_{R,L})\simeq \End_{\aa_R\times \aa_L}(
((\aa_R\times \aa_L)^n)=M_n( \aa_R\times \aa_L)\]
where $M_n(\aa_R\times \aa_L)$ is the ring of matrices of order $n$ over $\aa_R\times \aa_L$~.

The homomorphism $E_{M^e}:M^{(s)}_{R,L}\to M_n(\aa_{R,L})$ is the $n$-dimensional
representation of $M^{(s)}_{R,L}$~.

As the extended bilinear Hilbert space $H^{\pm}_{a,h}$ is composed of a bisemisheaf defined on a 
bisemimodule $(M^{(s)}_{L_R}\otimes M^{(s)}_L)$ or $(M^{(s)}_{R_L}\otimes
M^{(s)}_R)$ which is an enveloping algebra isomorphic to  $M_n(\aa_R\times \aa_L)$~,we have that
$H^{\pm}_{a,h}$ is the natural representation space for the algebras and the bialgebras of
bounded operators  acting on the above
defined semisheaves or bisemisheaves.\epr
\vskip 11pt

\begin{defi} {\bf (Algebras and bialgebras of von Neumann on extended bilinear
Hilbert spaces) \quad 1)\/} { A right (resp. left) algebra of von
Neumann
$\mit^{a,h}_{R,L}(H^{\pm}_{a,h})$ in the representation algebraic or analytic
extended bilinear Hilbert space $H^{\pm}_{a,h}$ is an involutive subalgebra of
$\Ls^B_{R,L}(H^{\pm}_{a,h})$ having a closed norm topology.
\vskip 11pt

\noindent {\bf 2)\/} A bialgebra of von Neumann $\mit^{a,h}_{R\times L}(H^+_{a,h})$
in the representation space $H^{\pm}_{a,h}$ is an involutive subalgebra of
$\Ls^B_{R\times L}(H^{\pm}_{a,h})$ having a closed norm topology.
\vskip 11pt

\noindent {\bf 3)\/}  A bialgebra of von Neumann
$\mit^{a,h}_{R\times_DL}(\Hs^{\pm}_{a,h})$~,
$\mit^{a,h}_{R\times_mL}(V^{\pm}_{m;a,h})$ or
$\mit^{a,h}_{R\times_eL}(V^{\pm}_{e;a,h})$ is an involutive subalgebra of
respectively $\Ls^B_{R\times_DL}(\Hs^{\pm}_{a,h})$~,
$\Ls^B_{R\times_mL}(V^{\pm}_{m;a,h})$ or $\Ls^B_{R\times_eL}(V^{\pm}_{e;a,h})$
having a closed norm topology.}
\end{defi}
\vskip 11pt

\begin{propo} Between the algebraic and analytic von Neumann algebras and
bialgebras, we have the following isomorphisms:
\nobeqn
i_{\ptmit^a_{R,L}-\ptmit^h_{R,L}}&:& \mit^a_{R,L}(H^{\pm}_a)\to
\mit^h_{R,L}(H^{\pm}_h)\;,
\\
i_{\ptmit^a_{R\times L}-\ptmit^h_{R\times L}}&:& \mit^a_{R\times L}(H^{\pm}_a)\to
\mit^h_{R\times L}(H^{\pm}_h)\;,
\\
i_{\ptmit^a_{R\times_DL}-\ptmit^h_{R\times_DL}}&:&
\mit^a_{R\times_DL}(\Hs^{\pm}_a)\to
\mit^h_{R\times_DL}(\Hs^{\pm}_h)\;,
\\
i_{\ptmit^a_{R\times_mL}-\ptmit^h_{R\times_mL}}&:&
\mit^a_{R\times_mL}(V^{\pm}_{m;a})\to
\mit^h_{R\times_mL}(V^{\pm}_{m;h})\;,
\\
i_{\ptmit^a_{R\times_eL}-\ptmit^h_{R\times_eL}}&:&
\mit^a_{R\times_eL}(V^{\pm}_{e;a})\to
\mit^h_{R\times_eL}(V^{\pm}_{e;h})\;.\noeeqn
\end{propo}
\vskip 11pt

\bpr This results immediately from the isomorphisms between the Eisenstein and the
analytic de Rham cohomologies according to proposition 3.1.13.\epr
\vskip 11pt

\begin{defis} {\bf (Shifted actions of (bi)operators on the functional representations of (bi)linear 
semigroups)}\quad {\bf 1)\/}  Let $T_{R,L} \in\mit^h_{R,L}(H^{\pm}_h)$ be a right
(resp. left) bounded linear operator of the algebra of von Neumann
$\mit^h_{R,L}(H^{\pm}_h)$~.
It can be assumed that this operator $T_{R,L}$ is a differential operator of the form $T_{R,L}=\sum\limits_n\sum\limits_r c_{mr}U_r^m$ where $U$ is the unitary translation operator.  This operator is supposed to be a regular
representation of the discrete compact triangular semigroup $T^t_m(\cit)$ (resp. $T_m(\cit)$~) in the extended
bilinear Hilbert space $H^{\pm}_h$ such that $T^t_m(\cit)$ (resp. $T_m(\cit)$~) acts 
 on the right (resp. left) $n$-dimensional semisheaf $\widetilde M^{(s)}_{R,L}$ of $H^{\pm}_h$ with $m\le n$~.

Similarly, let $(T_R\otimes T_L)$ be the tensor product of a right and a left
bounded linear operators acting on the bisemisheaf of the extended bilinear
Hilbert space $H^{\pm}_h$~.  So, $(T_R\otimes T_L)$ belongs to the bialgebra of von
Neumann $\mit^h_{R\times L}(H^{\pm}_h)$~.  This bioperator $(T_R\otimes T_L)$ is
supposed to be the regular representation of the product $ GL_m(\cit\times \cit)=T^t_M(\cit)\times T_m(\cit)$ 
of the compact semigroups $T^t_m(\cit)$ and $T_m(\cit)$~.
\vskip 11pt

\noindent {\bf 2)\/} More concretely, a differential bioperator $(T_R\otimes T_L)$~, being the regular representation of $GL_m(\cit\times\cit)$ in a bisemisheaf $\widetilde M_R\otimes \widetilde M_L$ on a $GL_n(\aa_R\times \aa_L)$-bisemimodule $M_R\otimes M_L$~, has a representation in the bilinear Lie algebra $\gl_m(\cit\times\cit)$ of the bilinear Lie semigroup $GL_m(\cit\times\cit)$~.  Then, the action of the differential bioperator $(T_R\otimes T_L)$ on the bisemisheaf $\widetilde M_R\otimes \widetilde M_L$ is equivalent to:
\Bean
\item consider a shift in $(m\times m)$-dimensions of the bisemisheaf $\widetilde M_R\otimes \widetilde M_L$ constituting a functional
representation of the bilinear Lie semigroup $GL_n(\aa_R\times \aa_L)$ leading to the homomorphism of the functional representation of the bilinear semigroup:
\begin{align*}
T_R\otimes T_L:\quad &\widetilde M_R\otimes \widetilde F_L=\FRep(GL_n(\aa_R\times \aa_L))\\
&\qquad \To \widetilde M^a_{R_n[m]}\otimes \widetilde M^a_{L_n[m]}
=\FRep(GL_{n[m]}((\aa_R\otimes\cit)\times 
(\aa_L\otimes \cit)))\end{align*}
where $\FRep(GL_{n[m]}((\aa_R\otimes \cit)\times (\aa_L\otimes \cit))$ denotes the functional representation of the bilinear semigroup $GL_n(\aa_R\times \aa_L)$ shifted in $(m\times m)$ dimensions.

\item map $\widetilde M_R\otimes \widetilde M_L$ in the bisemisheaf 
$\widetilde M^a_{Rn[m]} \otimes \widetilde M^a_{Ln[m]} $ shifted in $(m\times m)$ dimensions such that 
$\widetilde M^a_{Rn[m]} \otimes \widetilde M^a_{Ln[m]} $ be a perverse bisemisheaf, i.e. an object of the derived category $D(\widetilde M_R\otimes \widetilde M_L,\cit)$ \cite{Pie12}.

$\widetilde M^a_{Rn[m]} \otimes \widetilde M^a_{Ln[m]} $ will be written in condensed form $\widetilde M^a_{R} \otimes \widetilde M^a_{L} $~.
\Ee
\vskip 11pt

\noindent {\bf 3)\/} Similarly, we have on the bilinear subsemigroup $K^D_{R\times L;n}(\o\ZZ^2_{pq})$ the following shifted action resulting from the action of the differential bioperator:
\begin{align*}
&T_R\otimes T_L:\quad 
K^D_{R\times L;n}(\o\ZZ^2_{pq}) = 
D_n(\o\ZZ^2_{pq})\times [UT^t_n(\o\ZZ_{pq})\times UT_n(\ZZ_{pq})]\\
&\qquad \To 
K^D_{R\times L;n[m]}(\o\ZZ^2_{pq}\otimes\cit^2)\\
&\hspace{3cm}  = 
D_{n[m]}(\o\ZZ^2_{pq}\otimes \cit^2)\times [UT^t_{n[m]}\o\ZZ_{pq}\otimes\cit)\times UT_{n[m]}(\o\ZZ_{pq}\otimes\cit)]\end{align*}
where
\Bi
\item \quad $D_{n[m]}(\o\ZZ^2_{pq}\otimes \cit)=D_n(\o\ZZ^2_{pq})\times D_{n[m]}(\cit^2)$

such that $D_{n[m]}(\cit^2)$ is the subgroup of diagonal matrices of order $n$ shifted in $m$ dimensions, i.e. whose elements $d_{n[m]}(\cit^2)$ are
\[\BM \times & & & & \raisebox{-4mm}{\Huge 0} & \\
         & \times &&&& \\
& & 1 & &&\\
&\raisebox{-4mm}{\Huge 0}&&&1 & \\
&&&&&1\EM \quad \begin{array}{l}
\left] \begin{array}{l} \mbox{}\\m \\ \mbox{}\\
\end{array} \right.\\
\left] \begin{array}{l} \mbox{}\\n-m \\ \mbox{}\\
\end{array} \right.\end{array}\]
The $m$ shifts of $d_{n[m]}(\cit^2)$ are the squares of the infinitesimal generators of the Lie algebra of the diagonal subgroup $D_m(\cit)$ of order $m$~.

\item \quad $UT_{n[m]}(\o\ZZ_{pq}\otimes \cit)=UT_n(\o\ZZ_{pq})\times UT_{n[m]}(\cit)$

such that the shifts in $m$ dimensions of  $UT_{n[m]}(\cit)$ correspond to the generators of the nilpotent Lie algebra.
\Ei

\noindent {\bf 4)\/} Under the action of $(T_R\otimes T_L)$~, the functional representation of the bilinear parabolic subgroup $P_n(\Aa_{L^T_{\o v^1}}\times \Aa_{L^T_{v^1}})$ is shifted in $(m\times m)$ dimensions according to:
\begin{align*}
&T_R\otimes T_L : \;
\FRep(P_n( \Aa_{L^T_{\o v_1}} \times \Aa_{L^T_{v^1}} )) = \FRep(D_n(\Aa_{L^T_{\o v^1}}\times \Aa_{L^T_{v^1}})\times
[UT^t_n(\Aa_{L^T_{\o v^1}})\times UT_n(\Aa_{L^T_{v^1}})])\\
&\quad \To 
\FRep(P_{n[m]}((\Aa_{L^T_{\o v^1}}\otimes\cit)\times (\Aa_{L^T_{v^1}} \otimes\cit))) \\
& \qquad \qquad  = \FRep(D_{n[m]}
((\Aa_{L^T_{\o v^1}}\otimes\cit)\times (\Aa_{L^T_{v^1}}\otimes\cit))
[UT^t_{n[m]}(\Aa_{L^T_{\o v^1}}\otimes\cit)\times UT_{n[m]}(\Aa_{L^T_{v^1}}\otimes\cit)])\end{align*}
where: \[D_{n[m]}
(\Aa_{L^T_{v^1}}\otimes\cit)= D_{n}
(\Aa_{L^T_{\o v^1}})\times D_{n[m]}(\cit)\;.\]
\end{defis}\vskip 11pt

\begin{sub}{\bf Shifted Shimura bisemivariety}\end{sub}
Under the action of the differential bioperator $(T_R\otimes T_L)$~, the functional representation of the Shimura bisemivariety
$\partial\o S_{K^D_{R\times L,n}} $ given by the bisemisheaf $ \widetilde  M_R\otimes  \widetilde  M_L
=\FRep( \partial\o S_{K^D_{R\times L,n}} )=\FRep(P_n(\Aa_{L^T_{\o v^1}}\times \Aa_{L^T_{v^1}})\setminus GL_n(\aa_R\times \aa_L)\Big/ K^D_{R\times L;n}(\o\ZZ^2_{pq}))$
is shifted in $(m\times m)$ dimensions according to:
\[T_R\otimes T_L : \quad \widetilde  M_R\otimes \widetilde  M_L=
\FRep(\partial\o{S}_{K^D_{R\times L;n}})\To 
 \widetilde  M^a_R\otimes \widetilde  M^a_L=\FRep(
\partial\o S_{K^D_{R\times L;n[m]}})\]
where the shifted Shimura bisemivariety $\partial\o S_{K^D_{R\times L;n[m]}}$ is given by:
\begin{align*}
 \partial\o S_{K^D_{R\times L;n[m]}}
&= P_{n[m]}(\Aa_{L^T_{\o v^1}}\otimes\cit)\times (\Aa_{L^T_{v^1}}\otimes\cit))
\setminus\\
&\qquad GL_n((\aa_R\otimes\cit)\times (\aa_L\otimes \cit))
\Big/ K^D_{R\times L;n[m]}(\o\ZZ^2_{pq}\otimes\cit^2)\;.\end{align*} \vskip 11pt

\begin{propo} The semimodules $M_{L_{R,L}}$~, $M^a_{L_{R,L}}$ and $M^s_{L_{R,L}}$
have a basis of dimension $i=t$ corresponding to the upper degree of the Galois
extensions.
\end{propo}
\vskip 11pt

\bpr Under the automorphisms $\sigma _{R,L}$ of the algebraic semigroup
$T_n^t(\aa_R)$ (resp. $T_n(\aa_L)$~), the semimodule $M_{L_{R,L}}$ decomposes into:
\[ M_{L_{R,L}}={\textstyle\bigoplus\limits^t_{\sigma _{R,L}=1}}M_{L_{R,L}}(\sigma
_{R,L})\] where the number $t$ of automorphisms is the degree of the Galois
extension.

Now, under the cross action of 
$T_{n[m]}^t(\aa_R\otimes\cit)$ (resp. $T_{n[m]}(\aa_L\otimes \cit)$~), the semimodule $M^a_{L_{R,L}}$ decomposes into:
\[ M^a_{L_{R,L}}={\textstyle\bigoplus\limits_{\sigma _{R,L} }}M^a_{L_{R,L}}(\sigma _{R,L})\]
where the number of cross automorphisms is also $t$~, corresponding to the same
upper degree of Galois extension as for the semimodule $M_{L_{R,L}}$~.

So, the semimodules $M_{L_{R,L}}$ and $M^a_{L_{R,L}}$ have a basis with the same
dimension $i=t$~.

Referring to the isomorphism between the Eisenstein cohomology and the analytic de
Rham cohomology, it appears that the semisheaf $ \widetilde  M^s_{L_{R,L}}$ must have a basis  $\{ e^i_{R,L(s)}\}^t_{i=1}$ with the same dimension $i=t$ as the algebraic basis
$\{ e^i_{R,L(a)}\}^t_{i=1}$ of the semisheaf $  \widetilde  M^a_{L_{R,L}}$~.\epr
\pagebreak

\begin{sub}{\bf Shift of the Eisenstein bicohomology}\end{sub}
Let $(T_R\otimes T_L)$ be the tensor product of bounded differential operators of the von Neumann bialgebra $\MM^a_{R\times L}(H_a^\pm)$~.  Its shifted action on the bilinear Eisenstein cohomology will be:
\[T_R\otimes T_L : \quad H^*_{R\times L}(\partial\o S_{K^D_{R\times L;n}},  \widetilde  M_R\otimes  \widetilde  M_L)
\To H^*_{R\times L}(\partial\o S_{K^D_{R\times L;[n[m]}}, \widetilde  M^a_R\otimes  \widetilde  M^a_L)\]
such that $H^*_{R\times L}(\cdot,\cdot)$ decomposes into the double sum $\bigoplus\limits_\mu
\bigoplus\limits_{m_\mu}$~, associated to the places $\mu$ with multiplicities $m_\mu$ of the semifield $L^T_v$ (or $L^T_{\o v}$~), according to:
\begin{align*}
&H^*_{R\times L}(\partial\o S_{K^D_{R\times L;[n[m]}},  \widetilde  M^a_R\otimes  \widetilde  M^a_L)\\
&\quad =
\txt\bigoplus\limits_\mu \bigoplus\limits_{m_\mu} 
\Ind^{GL_{n[m]}((\aa_R\otimes \ptcit)\times (\aa_L\otimes \cit))}
_{K^D_{R\times L;n[m]}(\o\ZZ^2_{pq}\otimes\cit^2)}
H^*_{R\times L}(S^{M(\aa_R\otimes\cit)\times M(\aa_L\otimes\cit)},
H^*(\tilde u_{K^D_{R\times L;n[m]}},  \widetilde  M^a_R\otimes  \widetilde  M^a_L))\end{align*}
where 
\begin{align*}
& S^{M(\aa_R\otimes\ptcit)\times M(\aa_L\otimes\cit)}\\
&\quad =D_{n[m]}((\Aa_{L^T_{\o v^1}}\otimes\cit)\times (\Aa_{L^T_{v^1}}\otimes\cit))\setminus D_{n[m]}((\aa_R\otimes\cit)\times (\aa_L\otimes\cit))\Big/ D_{n[m]}(\o\ZZ^2_{pq}\otimes\cit^2)\end{align*}
following the notations of definition 1.1.11.

The coefficient system given by the Lie algebra cohomology $H^*(\tilde u_{K^D_{R\times L;n[m]}}, \widetilde  M^a_R\otimes  \widetilde  M^a_L)$ decomposes according to the cosets of $GL_{n[m]}((\aa_R\otimes\cit)\times(\aa_L\otimes\cit))\big/K^D_{R\times L;n[m]}(\o\ZZ^2_{pq}\otimes\cit^2)$ generating the set of subrepresentatives
$\{ \widetilde  M^a_{\o v_{\mu,m_\mu}}\otimes  \widetilde  M^a_{v_{\mu,m_\mu}}\}^q_{\mu=1}$ on $GL_{n[m]}((\aa_R\otimes\cit)\times (\aa_L\otimes\cit))$~.

Note that it was proved in \cite{Pie12} that the shifted bilinear Eisenstein cohomology 
$H^*_{R\times L}(\partial\o S_{K^D_{R\times L;[n[m]}},\linebreak \widetilde  M^a_R\otimes  \widetilde  M^a_L)$ is isomorphic to the 
adjoint functional representation  ${\rm Ad\,}\FRep(GL_n(\aa_R\times\aa_L))$ which corresponds to 
$\FRep(GL_{n[m]}((\aa_R\otimes\cit)\times(\aa_L\otimes\cit))$ where $\FRep(\cdot)$ denotes the functional representation of the considered bilinear semigroup. \vskip 11pt

\begin{propo}  Applying the Kostant's theorem, we can decompose the bilinear Eisenstein cohomology
$H^*_{R\times L}(\partial\o S_{K^D_{R\times L;[n[m]}},\widetilde M^a_R\otimes \widetilde M^a_L)$ into sums of products
of pairs of one-dimensional eigenspaces following:
\begin{align*}
&H^*_{R\times L}(\partial\o S_{K^D_{R\times L;[n[m]}}, \widetilde  M^a_R\otimes  \widetilde  M^a_L)\\
&\quad =
\txt\bigoplus\limits_\mu \bigoplus\limits_{m_\mu}\bigoplus\limits_{1_\ell}
 \Ind^{GL_{n[m]}
((\aa_R\otimes \cit)\times (\aa_L\otimes \cit))}
_{K^D_{R\times L;n[m]}}\\
&\hspace{2cm}\times 
H^{1_\ell,1_\ell}_{R\times L}
(S^{M(\aa_R\otimes\cit)\times M(\aa_L\otimes\cit)},
H^{1_\ell,1_\ell}(\tilde u_{K^D_{R\times L;n[m]}}
, \widetilde  M^{a_{1_\ell}}_R(\mu ,m_\mu )
\otimes  \widetilde  M^{a_{1_\ell}}_L (\mu ,m_\mu ) ))\;.\end{align*}
Then, the decomposition of the Lie algebra cohomology 
$H^{1_\ell,1_\ell}(\tilde u_{K^D_{R\times L;n[m]}}
, \widetilde  M^{a_{1_\ell}}_R(\mu ,m_\mu ) \otimes  \widetilde  M^{a_{1_\ell}}_L(\mu ,m_\mu ) )$ into sums of products of pairs of one-dimensional 
eigenspaces involves a decomposition of the bilinear Hilbert space $H^{\pm}_ h$ into a tower of 
embedded bilinear Hilbert subspaces $H^{\pm}_h\{\mu\}^q_1$ decomposing into pairs of one-dimensional subspaces.
\end{propo} \vskip 11pt

\bpr 1.  The decomposition of the shifted bilinear Eisenstein cohomology into sums of products of pairs of one-dimensional eigenspaces results from proposition 3.1.11. \vskip 11pt

\noindent 2.  The embedded representation subspaces $H^\pm_a\{\mu\}$ of $H^{\pm}_a\simeq  \widetilde  M_{R(P)/_cL}$ forms a Jordan-H\"older serie for the homomorphism
\[
\Pi_{H-\gl}:\quad H^*_{R\times L}(\tilde u_{K^D_{R\times L;n[m]}}, \widetilde  M^{a_{1_\ell}}_R\otimes  \widetilde  M^{a_{1_\ell}}_L)\To \gl(H^\pm_a)\]
of the Lie algebra $H^*_{R\times L}(\cdot,\cdot)$ into the Lie algebra $\gl(H^\pm_ a)$ of the automorphisms of $H^\pm_a$ isomorphic to $H^\pm_h$~.  We thus have a sequence of embedded bilinear Hilbert subspaces:
$H^\pm_h\{1\}\subset \cdots \subset H^\pm_h\{\mu\}\subset \cdots\subset H^\pm_h\{q\}$ where
\[ H^\pm_h\{\mu \}=\txt\bigoplus\limits^\mu _{\nu =1}H^\pm_h(\nu )\]
with $H^+_h(\nu )$ the extended bilinear Hilbert subspace constituted by the $\nu$-th subbisemisheaf $ \widetilde  M^s_{L_R}(\nu)\otimes  \widetilde  M^s_L(\nu)$ corresponding to the $\nu$-th biplace of $L^T_{\o v}\times L^T_v$~.
\epr \vskip 11pt

\begin{rms}  In order to include the  above-mentioned cases in a uniform presentation, we shall admit until the end of Section 3.2 that the integer ``~$i$~'' refers to:
\Bd \item[a)] a Galois extension degree related to the dimension of the basis of the semimodule $M^a_{L_R,L}$~;
\item[or b)] a class of degrees of Galois extensions which corresponds to the global class residue degree $f_{v_i}$ (see 1.1.4) labelling the $i$-th coset of $GL_{n[m]}((\aa_R\otimes\cit)\times(\aa_L\otimes\cit))  \big/K^D_{R\times L;,[m]}(\o\ZZ^2_{pq}\otimes \cit^2)$ (in this case, $i=\mu$~).
\Ed
\end{rms}
\vskip 11pt

\begin{defi} {\bf (Random bioperators on analytic bilinear Hilbert spaces)}  
{ Let
$T_R\otimes T_L$ be the tensor product of a right and a left bounded linear
operators being the regular representation of $GL_m(\cit\times\cit)$ in $(\widetilde M_R\otimes \widetilde M_L)$~. 

 $GL_m(\cit\times\cit)$ has for bilinear semigroup of inner automorphisms \cite{Kac} $\Int \Gamma ^h_R\times \Int\Gamma ^h_L$ (see definition 3.2.7) and has the inner conjugacy biclasses noted $g^h_R\times g^h_L$ if the fixed bielement is of dimension 1 with respect to the basis of $( \widetilde  M^a_R\otimes  \widetilde  M^a_L)$ in the case a) of 3.2.12.

$GL_m(\cit\times\cit)$ has $ \Gamma ^h_R\times \Gamma ^h_L$ for bilinear semigroup of modular 
automorphisms and has the modular conjugacy biclasses  $\gamma ^h_R\times \gamma ^h_L$ if the fixed bielement, which is a normal bilinear subsemigroup, is of dimension $N^2$ with respect to the (algebraic) basis of $(M^a_R\otimes M^a_L)$ in the case b) of 3.2.12.

The right (resp. left) bounded linear operator $T_{R,L}(\Gamma^h_{R,L})$ is a
random operator if it decomposes into a set of right (resp. left) bounded linear
operators $\{ T_{R,L}(g^h_{R,L}(i))\}$~, $\forall\ g^h_{R,L}(i)\in \Int(\Gamma^h_{R,L})$ or $\{T_{R,L}(\gamma^h_{R,L}(i))\}$~, $\forall\ \gamma^h_{R,L}(i)\in \Gamma^h_{R,L}$~.

So, the tensor product $(T_R(\Gamma^h_R)\otimes T_L(\Gamma^h_L))$ of a right and
a left bounded linear operators is a random bioperator if it decomposes into a
tensor product of a set of right and left bounded linear operators:
\begin{align*}
 T_R(\Gamma ^h_R)\otimes T_L(\Gamma ^h_L)&=\{ T_R(g^h_R(i))\otimes
T_L(g^h_L(i'))\}^t_{i,i'=1}\\
\mbox{(resp.}\qquad T_R(\Gamma^h_R)\otimes T_L(\Gamma^h_L) &= \{T_R(\gamma^h_R(i))\otimes T_L(\gamma^h_L(i'))\}^q_{i,i'=1}\;), \qquad t\ge q\;.\end{align*}
Let
\begin{alignat*}{3}
g^h_{R,L}\{i\} &= \txt\bigoplus\limits^i_{j=1} g^h_{R,L}(j)\;, \qquad &&1\le i\le t
\\
\mbox{(resp.}\qquad 
\gamma ^h_{R,L}\{i\} &= \txt\bigoplus\limits^i_{\nu =1} g^h_{R,L}(\nu )\;, \qquad &&1\le \nu \le q 
\;), \end{alignat*}
denote the sum of inner (resp. modular) conjugacy classes of $\Gamma  ^h_{R,L}$~.  This leads to define a sum of inner (resp. modular) random operators by:
\begin{align*}
T^D_{R,L}(g^h_{R,L}\{i\}) &= \txt\bigoplus\limits^i_{j=1} T^D_{R,L}(g^h_{R,L}(j))
\\[11pt]
\mbox{(resp.}\qquad 
T^D_{R,L}(\gamma ^h_{R,L}\{i\}) &= \txt\bigoplus\limits^i_{\nu =1} T^D_{R,L}(g^h_{R,L}(\nu ))
\;), \end{align*}
such that
\begin{align*}
T^D_{R,L}(g^h_{R,L}\{i\}) &\in \MM^{h({\rm in})}_{R,L}(H^{\mp}_h\{i\})
\\[11pt]
\mbox{(resp.}\qquad 
T^D_{R,L}(\gamma ^h_{R,L}\{i\}) &\in \MM^{h({\rm mod})}_{R,L}(H^{\mp}_h\{i\})
\;), \end{align*}
where $\MM^{h({\rm in})}_{R,L}(H^{\mp}_h\{i\})$ (resp. $\MM^{h({\rm mod})}_{R,L}(H^{\mp}_h\{i\})
$~) is an inner (resp. modular) von Neumann subalgebra referring to the $i$-th sum of inner 
(resp. modular) random operators.

So, a tower of inner (resp. modular) von Neumann subalgebras can be intoduced by:
\begin{align*}
 \MM^{h({\rm in})}_{R,L}(H^{\mp}_h\{1\}) \subset \cdots \subset
&\MM^{h({\rm in})}_{R,L}(H^{\mp}_h\{i\}) \subset \cdots \subset
\MM^{h({\rm in})}_{R,L}(H^{\mp}_h\{t\})\\[11pt]
\text{(resp.} \quad
 \MM^{h({\rm mod})}_{R,L}(H^{\mp}_h\{1\}) \subset \cdots \subset
&\MM^{h({\rm mod})}_{R,L}(H^{\mp}_h\{i\}) \subset \cdots \subset
\MM^{h({\rm mod})}_{R,L}(H^{\mp}_h\{q\})\; ),\end{align*}
such that:
\begin{align*}
\MM^{h({\rm in})}_{R,L}(H^{\mp}_h\{i\})
&= \txt\bigoplus\limits^i_{j=1} \MM^{h({\rm in})}_{R,L}(H^{\mp}_h(j))\\
\text{(resp.} \quad 
\MM^{h({\rm mod})}_{R,L}(H^{\mp}_h\{i\})
&= \txt\bigoplus\limits^i_{\nu =1} \MM^{h({\rm mod})}_{R,L}(H^{\mp}_h(\nu ))\;).\end{align*}
}
\end{defi}
\vskip 11pt

\begin{propo} Let $ T_{R, L}(g^h_{R,L}(t))$ and $
T_{R,L}(g^h_{R,L}(r))\}$ be two right or left inner random operators such that
$t<r$~.

Then, the random bioperator $ T_R(g^h_R(r))\otimes
T_L(g^h_L(r))\}$ is an extension of the random bioperator $ T_R(g^h_R(t))\otimes
T_L(g^h_L(t))$ corresponding to a Galois extension of degree $(r-t)$~.

Let $T_{R,L}(\gamma^h_{R,L}(q))$ and $T_{R,L}(\gamma^h_{R,L}(s))$ bet two right or left modular random operators such that $q<s$~.

Then, the random bioperator $T_R(\gamma^h_R(s))\otimes T_L(\gamma^h_L(s))$ is an extension of the random bioperator $T_R(\gamma^h_R(q))\otimes T_L(\gamma^h_L(q))$ corresponding to a Galois extension of class of degree $(s-q)$~.
\end{propo}
\vskip 11pt

\bpr Indeed, $g^h_{R,L}(i)$ (resp. $\gamma^h_{R,L}(i)$~) is a inner (resp. modular) conjugacy class of the discrete semigroup $T^{(t)}_m(\cit)$ whose representation semispace $M^s_{R,L}$ has a basis of dimension $t$
whose entire number $t$ (resp. $q$~) corresponds to a Galois extension of degree $t$ or a class of Galois extension degrees $q$~.  
\epr
\vskip 11pt 

\begin{propo} Let $\mit^h_{R\times L}(H^{\pm}_h)$ be the von Neumann bialgebra of
bounded self-adjoint bioperators on the analytic extended bilinear Hilbert space
$H^{\pm}_h$~.

Let $\mit^h_{R\times L}(H^{\pm}_h\{ i\})$ be the von Neumann bialgebra of random
bioperators on the analytic extended bilinear subspace $H^{\pm}_h\{i\}$ and let
$\mit^h_{R\times L}(\Hs^{\pm}_h\{ i\})$ be the corresponding von Neumann bialgebra
on the analytic internal bilinear subspace $\Hs^{\pm}_h\{i\}$~.

Then, the discrete (diagonal) spectrum $\sigma _D(T_R\otimes T_L)$ of a bioperator
$T_R\otimes T_L\!\in\! \mit^h_{R\times L}(H^{\pm}_h)$ is obtained by the
isomorphism:
\nobeqn
i^h_{\{ i\}^D_{R\times L}}\circ i^h_{\{ i\}_{R\times L}}\ : \
\mit^h_{R\times L}(H^{\pm}_h)&\to & \{\mit^h_{R\times L}(\Hs^{\pm}_h\{ i\})\}_i\;, \\
T_R\otimes T_L&\to & \sigma _D(T_R\otimes T_L)\;, \noeeqn
where the isomorphisms $i^h_{\{ i\}_{R\times L}}$ and $i^h_{\{ i\}^D_{R\times L}}$
are defined by
\nobeqn i^h_{\{ i\}_{R\times L}} &:& \mit^h_{R\times L}(H^{\pm}_h)\ \to \
\{\mit^h_{R\times L}(H^{\pm}_h\{ i\})\}_i\;, \\
i^h_{\{ i\}^D_{R\times L}} &:& \{\mit^h_{R\times L}(H^{\pm}_h\{ i\})\}_i\ \to \
\{\mit^h_{R\times L}(\Hs^{\pm}_h\{ i\})\}_i\;.\noeeqn
\end{propo}
\vskip 11pt

\bpr The isomorphism $i^h_{\{ i\}_{R\times L}}$ is an isomorphism transforming the
bounded bioperator $(T_R(\Gamma ^h_R)\times T_L(\Gamma ^h_L))$ into the set of
bounded bioperators $\{ T_R(g^h_R\{i\})\otimes T_L(g^h_L\{i'\})\}$ (resp. $\{ T_R(\gamma^h_R\{i\})\otimes T_L(\gamma^h_L\{i'\})\}$~).

On the other hand, the isomorphism $i^h_{\{ i\}^D_{R\times L}}$ is an isomorphism
transforming the nonabelian von Neumann subbialgebras $\{\mit^h_{R\times L}(H^{\pm}_h\{
i\})\}_i$ into the abelian or diagonal von Neumann subbialgebras\linebreak $\{\mit^{h}_{R\times
L}(\Hs^{\pm}_h\{ i\})\}_i$ of random bioperators acting on the ``diagonal'' enveloping
algebra $(\Hs^{\pm}_h\{ i\})$~.
\\
$\{\mit^h_{R\times L}(\Hs^{\pm}_h\{ i\})\}_i$ is thus the spectral algebra of the bounded
bioperator $(T_R\otimes T_L)$~.\epr
\vskip 11pt

\begin{coro} Let $\mit^h_{R\times_DL}(H^{\pm}_h)$ be the diagonal bialgebra of von
Neumann on the analytic extended bilinear Hilbert space
$H^{\pm}_h$~.

Then, the discrete  spectrum $\sigma
_{D}(T_R\otimes_{D}T_L)$ of the bioperator $T_R\otimes _{D}T_L\in
\mit^h_{R\times_{D}L}(H^{\pm}_h)$ is obtained  by the isomorphism:
\[
i^h_{\{ i\}^D_{R\times_D L}}\circ i^h_{\{ i\}_{R\times_D L}}\ : \
\mit^h_{R\times_D L}(H^{\pm}_h) \to  \{ \mit^h_{R\times_D L}(\Hs^{\pm}_h\{ i\})\}_i\;.\]
\end{coro}
\vskip 11pt

\bpr This proposition is a generalization of the preceding one to the von Neumann
bialgebra\linebreak $\mit^h_{R\times_{D}L}(H^{\pm}_h)$~.

 The corresponding spectrum is then defined on the von Neumann bialgebra 
$\{\mit^h_{R\times_DL}(\Hs^{\pm}_h\{ i\})\}_i$ with a spectrum characterized by a diagonal
metric.\epr
\vskip 11pt

\begin{propo}  There exists a set of spectral bimeasures $\{ \mu
_R(i)\times_{D}\mu _L(i)\}$ on the spectrum $\sigma
_{D}(T_R\otimes_{D}T_L)$ such that every 
bivector of the space $\Hs^{\pm}_h\{ i\}$  of the von Neumann bialgebra
$\mit^h_{R\times_DL}(\Hs^{\pm}_h\{ i\})$ be an eigenbivector of the
bioperator $(T_R\times_{D}T_L)$ where $  i   $ is a degree of Galois
extension or a class of degrees of Galois extensions.\end{propo}
\vskip 11pt

\bpr    The existence of spectral bimeasures $\{ \mu _R(i)\times_{D}\mu
_L(i)\}$ on the spectrum  $\sigma _{D}(T_R\otimes_{D}T_L)$ is a consequence
of the isomorphisms $i^h_{\{ i\}^{D}_{R\times_{D}L}}\circ
i^h_{\{ i\}_{R\times_{D}L}}$ introduced in proposition 3.2.15.\epr
\vskip 11pt

\stepcounter{defi}
\paragraph{Results concerning the von Neumann (bi)algebras  \thedefi}\quad  {\bf 1)\/}\quad 
 If the integer ``~$i$~'' refers to a class of Galois extension degrees related to a coset of 
$GL_{n[m]}((\aa_R\otimes\cit)\times (\aa_L\otimes \cit))\big/K^D_{R\times L;n[m]}(\o\ZZ^2_{pq}\otimes\cit^2)
$~, then   the   algebra of von Neumann $\mit^h_{R\times L}(H^{\pm}_h)$ decomposes into $\mit^h_{R\times L}(h^{\pm}_ h)=\bigoplus\limits_i \mit^h_{R\times L}(\Hs^{\pm}_h\{i\})$~.

 The spectrum $\sigma_D(T_R\otimes T_L)$ is degenerated if there is an action of the decomposition group in the sense of section 3.2.10.\vskip 11pt

\noindent {\bf 2)\/} If the integer ``~$i$~'' refers to a class of Galois extension degrees related to  
one-dimensional cosets of $GL_{n[m]}((\aa_R\otimes\cit)\times (\aa_L\otimes \cit))\big/K^D_{R\times L;n[m]}(\o\ZZ^2_{pq}\otimes\cit^2)
$~, then  the algebra of von Neumann $\mit^h_{R\times L}(H^{\pm}_h)$ decomposes into a direct sum of factors such that the set of integers $\{1,\cdots,i,\cdots,q\}$ are the entire dimensions of a von Neumann (bi)algebra of type $I_{q}$~.  The multiplicity of the spectrum of $(T_R\otimes T_L)$ results from the action of the decomposition group as introduced in proposition 3.2.11.
\vskip 11pt

\noindent {\bf 3)\/} The spectrum of the operator $T_R\otimes T_L\in \mit^h_{R\times L}(H^{\pm}_h)$ is obtained through the isomorphism:
\[ i ^h_{\{i\}^D_{R\times L}} \circ i ^h_{\{i\}_{R\times L}} :\mit^h_{R\times L}(H^{\pm}_h) \to \{ \mit^h_{R\times L}(\Hs^{\pm}_h\{i\})\}_i\]
such that we have the embedding of the $\Hs^{\pm}_h\{i\} $~:
\[ \Hs^{\pm}_h\{1\} \subset \cdots \subset \Hs^{\pm}_h\{i\} \subset \cdots \subset \Hs^{\pm}_h\{q\} \;,\]
and the development of the $i$-th eigenbifunction $\psi_{L_R}(i)\otimes_D \psi_L(i)\in \Hs^{\pm}_h\{i\}$  following:
\[\psi_{L_R}(i)\otimes_D \psi_L(i)= \txt\sum\limits_i \txt\sum\limits_{m_i} c_{i_R}\phi_{L_R}(i)\otimes_D c_{i_L}\phi_L(i)\]
where 
\Bi
\item $\psi_{L_R}(1)\otimes_D\psi_L(1)\equiv \phi_{L_R}(1)\otimes_D\phi_L(1)$ is the first eigenfunction in $\Hs^{\pm}_h\{1\}$~;
\item $\psi_{L_R}(i)\otimes_D\psi_{L}(i)$ is (isomorphic to) a $n$-dimensional truncated global elliptic bisemimodule;
\item $\phi _{L_R}(i)\otimes_D\phi _L(i)$ is a section of $ \widetilde M^s_{L_R}\otimes_D\widetilde M^s_L\in\Hs^+_h$ (see section 3.1.9).
\Ei

Indeed, the bioperator $(T_R\otimes T_L)$ maps the bisemisheaf 
$(\widetilde M_R\otimes\widetilde M_L)$ over 
the $GL_n(\aa_R\otimes \aa_L)$-bisemimodule $(M_R\otimes M_L)$ into the 
perverse bisemisheaf $(\widetilde M^a_R\otimes \widetilde M^a_L)$ over the 
shifted $GL_{n[m]}((\aa_R\otimes\cit)\times(\aa_L\otimes\cit))$-bisemimodule $(M^a_R\times M^a_L)$ decomposed into sums over the conjugacy classes $i$ with multiplicities $m_i$ according to section 3.2.7.  Now, the Langlands program \cite{Pie9}, \cite{Pie12}, succinctly introduced in 1.1.23, sets up bijections between:
\Bi
\item $(M_R\otimes M_L)$ and the $n$-dimensional global elliptic bisemimodule $\phi _R(s_R)\otimes_D \phi _L(s_L)$ (see proposition 1.1.19);
\item $(M^a_R\otimes M^a_L)$ and the $n$-dimensional shifted global elliptic bisemimodule $\phi ^a_R(s_R)\otimes_D \phi^a _L(s_L)$~.
\Ei \vskip 11pt

This leads to the following proposition: \vskip 11pt

\begin{propo} Let $\phi _R(s_R)\otimes_D \phi _L(s_L)$ be a $n$-dimensional global elliptic bisemimodule constituting an analytic representation of the $GL_n(\aa_R\times \aa_L)$-bisemimodule $M_R\otimes M_L$~.

Let $\phi ^a_R(s_R)\otimes_D \phi^a _L(s_L)$ denote the corresponding $n$-dimensional shifted global elliptic bisemimodule constituting the analytic representation of the perverse bisemimodule $M^a_R\otimes M^a_L$~.

Then, the action of the bioperator $(T_R\otimes T_L)$ is such that:
\[(T_R\otimes T_L) : \quad 
\phi _R(s_R)\otimes_D \phi _L(s_L) \To \phi ^a_R(s_R)\otimes_D \phi^a _L(s_L)\;.\]
The shifted global elliptic bisemimodule $\phi ^a_R(s_R)\otimes_D \phi^a _L(s_L)$ gives rise to the eigenbivalue equation:
\[
\phi ^a_R(s_R)\otimes_D \phi^a _L(s_L)=\lambda _R(n,i)\cdot \lambda _L(n,i)(\phi _R(s_R)\otimes_D \phi _L(s_L))\]
rewritten following:
\[(T_R\otimes_D T_L)  
(\phi _R(s_R)\otimes_D \phi _L(s_L) ) =\lambda _R(n,i)\cdot \lambda _L(n,i)(\phi _R(s_R)\otimes_D \phi _L(s_L))\]
where the \rl eigenvalue $\lambda_{R,L}(n,i)$ was interpreted in \cite{Pie12} as a set of shifts in $m$ dimensions of Hecke characters, i.e. infinitesimal generators of the considered Lie algebra.
\end{propo} \vskip 11pt

\bpr  As the bialgebra of von Neumann $\MM_{R\times L}(H^\pm_h)$ can be considered as a solvable bialgebra, i.e. implying a sequence of embedded subalgebras:
\[ \MM_{R\times L}(H^\pm_h\{1\})\subset \cdots \subset
\MM_{R\times L}(H^\pm_h\{i\})\subset \cdots \subset
\MM_{R\times L}(H^\pm_h\{q\})\;,\]
the set of eigenvalues of $(T_R\otimes T_L)$ forms an embedded sequence:
\[ \lambda _R(n,1)\cdot\lambda _L(n,1) \subset \cdots \subset
\lambda _R(n,i)\cdot\lambda _L(n,i)  \subset \cdots \subset
\lambda _R(n,q)\cdot\lambda _L(n,q)  \]
in one-to-one correspondence with the set of embedded eigenbifunctions:
\[ \psi _{L_R}(1)\otimes_D\psi_L(1) \subset \cdots \subset
\psi _{L_R}(i)\otimes_D\psi_L(i) \subset \cdots \subset
\psi _{L_R}(q)\otimes_D\psi_L(q) \;,\]
where $\psi _{L_R}(q)\otimes_D\psi _L(q)$ is isomorphic to a $n$-dimensional truncated global elliptic bisemimodule given by
\begin{equation}
\phi _{R}(s_R)\otimes_D\phi_L(s_L)= \txt\sum\limits^q_{i,m_i=1}\phi (s_R)_{i,m_i}\ e^{-i\pi i(p+i)z}
\otimes_D \txt\sum\limits^q_{i,m_i=1}\phi (s_L)_{i,m_i}\ e^{i\pi i(p+i)z}\;, \quad z\in\rit^n\;.\tag*{\eop}
\end{equation}
\vskip 11pt

\begin{propo}  Let $\mit^a_{R\times_{(D)}L}(H^{\pm}_a)$ be the
complete (resp. diagonal) von Neumann bialgebra of bounded
self-adjoint bioperators on the algebraic extended bilinear Hilbert space
$H^{\pm}_a$~.

Let $\mit^a_{R\times L}(\Hs^{\pm}_a\{ i \})$ and  $\mit^a_{R\times_D
L}(\Hs^{\pm}_a\{ i \})$ be the complete and diagonal subbialgebras of von Neumann
 on the closed algebraic internal bilinear subspaces $\Hs^{\pm}_a\{
\sigma \}$~.
 Then, the discrete spectrum of the bioperator $T_R\otimes_{(D)}T_L\in
\mit^a_{R\times_{(D)}L}(H^{\pm}_a)$ is obtained through the isomorphism(s):
\nobeqn
i^a_{\{ \sigma \}^D_{R\times L}}\circ i^a_{\{ \sigma \}_{R\times L}}\ : \
\mit^a_{R\times L}(H^{\pm}_a)&\to & \{ \mit^a_{R\times L}(\Hs^{\pm}_a\{ i \})\}_i\;,
\\ i^a_{\{ \sigma \}^D_{R\times_D L}}\circ i^a_{\{ \sigma \}_{R\times_D L}}\ : \
\mit^a_{R\times_D L}(H^{\pm}_a)&\to & \{ \mit^a_{R\times_D L}(\Hs^{\pm}_a\{ i
\})\}_i\;.\noeeqn
\end{propo}
\vskip 11pt

\bpr This proposition is the algebraic correspondent of proposition 3.2.15 and
corollary 3.2.16 and results from the isomorphisms between analytic and algebraic
von Neumann bialgebras as developed in proposition  3.2.6.\epr
\vskip 11pt

\subsection[Quantification rules, probability calculus, spin, $PC T$ map
and relativity invariants]{{\boldmath Quantification rules, probability calculus, spin, $PC T$ map
and relativity invariants}}

As 
the entire dimensions of the von Neumann bialgebras can correspond to classes of degrees of Galois extensions, biquanta $\widetilde M^I_{k_R}\otimes_{D,m,e}\widetilde M^I _{k_L}$~, i.e. $1D$-irreducible closed subschemes of rank $N^2$~, can be emitted from (or absorbed by) the algebraic bisemisheaf $(\theta
^{1-3}_{R;ST,MG,M}\otimes_{D,m,e}\theta ^{1-3}_{L;ST,MG,M})$~.
\vskip 11pt

\begin{defi}[``~$ST$~'', ``~$MG$~'' and ``~$M$~'' bistructures]  { Referring to the
structure of a massive bisemiparticle as described in definitions 3.1.2 and 3.1.3,
we recall that:

Each $ST$~, $MG$ and $M$ structure $(\theta ^4_{R;ST}\otimes \theta ^4_{L;ST})$~, 
$(\theta ^4_{R;MG}\otimes \theta ^4_{L;MG})$ and $(\theta ^4_{R;M}\otimes \theta
^4_{L;M})$ of a bisemiparticle decomposes under the blowing-up isomorphism into:
\Bean\item a diagonal bisemisheaf $(\theta ^4_{R;ST,MG,M}\otimes_D \theta
^4_{L;ST,MG,M})$~, 
\item a magnetic bisemisheaf $(\theta ^3_{R;ST,MG,M}\otimes_m \theta
^3_{L;ST,MG,M})$~, 
\item an electric bisemisheaf $(\theta ^{1-(3)}_{R;ST,MG,M}\otimes_e \theta
^{3-(1)}_{L;ST,MG,M})$~, \Ee
where $ST$~, $MG$~, $M$ means ``~$ST$~'', ``~$MG$~'' or ``~$M$~''.}\end{defi}
\vskip 11pt

\begin{propo} The quantification rules of emission of biquanta on the
$ST\oplus MG\oplus M$ bistructure of a bisemiparticle are obtained by
considering the diagonal, magnetic or electric products of the right and left
smooth endomorphisms \ $(E_{R;ST\oplus MG\oplus M}\ \times_{D,M,E}\ 
E_{L;ST\oplus MG\oplus M})$\ \ applied  on  $(\theta ^{1-3}_{R;ST\oplus MG\oplus
M} \otimes_{D,m,e}  \theta ^{1-3}_{L;ST\oplus MG\oplus M})$  until  the 
fundamental\ rank sets  $n ^0_{D,m,e;ST,MG, M}$ are reached.
\end{propo}
\vskip 11pt

\bpr This proposition is an adaptation of the emission quantification rules
introduced in 1.4.16 and in definition 2.2.12 to the ``~$ST\oplus MG\oplus M$~'' bistructure
of a bisemiparticle.  We then have
\nobeqn 
\lefteqn{E_{R;ST\oplus MG\oplus M}\times_{D,m,e} E_{L;ST\oplus MG\oplus M} :}\\
&&\theta ^{1-3}_{R;ST\oplus MG\oplus M}
\otimes_{D,m,e} \theta ^{1-3}_{L;ST\oplus MG\oplus M}\\
&&\qquad \to (\theta ^{*1-(3)}_{R;ST\oplus MG\oplus M}
\otimes_{D,m,e} \theta ^{*1-(3)}_{L;ST\oplus MG\oplus M}){\textstyle
\bigoplus\limits_{k}} (\widetilde M^I _{k_{R;ST\oplus MG\oplus M}}
\otimes_{D,m,e} \widetilde M^I _{k_{L;ST\oplus MG\oplus M}})\noeeqn
where $( \widetilde M^I _{k_{R;ST\oplus MG\oplus M}}
\otimes_{D,m,e} \widetilde M^I _{k_{L;ST\oplus MG\oplus M}})$ are $1D$-time or
$1D$-space diagonal, magnetic or electric biquanta on the three bistructures
$ST\oplus MG\oplus M$~.\epr
\vskip 11pt

\begin{rms} { 1. The standard quantification rules of quantum (field)
theory would be obtained by considering the smooth endomorphism
\[ E_{L;M} : \theta ^3_{L;M} \to \theta ^{*3}_{L;M}{\textstyle\bigoplus\limits^m_{k=1}}
\widetilde M^I_{k_{L;M}}\]
applied on the mass (``~$M$~'') left-$3D$-semisheaf of rings $\theta ^3_{L;M}$~.
\vskip 11pt

\noindent 2.  The quantum theories work essentially with analytic functions.  Due
to the hypothesis considered in this work, namely that the quantum nature is
algebraic, algebraic (semi-) sheaves of rings have been essentially taken into account.}
\end{rms}
\vskip 11pt

According to the preceding section, algebraic semisheaves of rings were considered
as isomorphic to analytic semisheaves: this is among others a consequence of the
(iso)mor\-phism of J.P. Serre \cite{Ser7}.  Thus, if we want to reach the
mathematical objects of quantum theories, we have to consider bijections between algebraic semisheaves of rings and analytic global elliptic semimodules following the Langlands program.\vskip 11pt

\begin{defi} {\bf (Bispectrum of Fredholm diagonal bioperators)\/} 
{ Let  
$(T_{R}\otimes_D T_{L})$ be a diagonal Fredholm bioperator acting from
$\Hs^+_{h;M}$ to $\Hs^+_{h;M}$~.

  Let $\{\psi_{L_R(i)}\otimes_D
\psi_{L(i)}\}^q_{i=1}$~, $q\le \infty $~, be the set of eigenbivectors of
$(T_{R}\otimes_D T_{L})$ and let $\{\lambda_{R(i)}\times\lambda
_{L(i)}\}^q_{i=1}=\{\lambda ^2_i\}^q_{i=1}$ be the corresponding set of
eigenbivalues occurring with probability measures $\{P_{\lambda _i}=\mu
_{R(i)}\times_D \mu _{L(i)}\}$~.

Now, the probability measure $P_{\lambda _i}$ can be written in Dirac
terminology \cite{Dir4} following: \[ P_{\lambda _i}
=\langle \psi_{L_R}(i)\mid \Psi \rangle \langle \Psi\mid \psi _L(i)\rangle\]
where $\langle\cdot\mid\cdot\rangle$ is an internal scalar product between a
bra eigenvector $\langle \psi_{L_R}(i)\mid$ and the total left wave function.

As we are working in the frame of an orthogonal geometry with a diagonal
metric, abbreviated by ``~$\times_D$~'', the eigenbivectors are orthogonal
between themselves \cite{Pie4}.}
\end{defi}
\vskip 11pt

\begin{propo} The semisheaf $\widetilde M_L$ (resp. $\widetilde M_R$~) on the $G_L(\aa _L)$-left semimodule $M_L$ (resp. the $G_R(\aa _R)$-right semimodule $M_{L_R}$~)  constitutes an algebraic  representation of a left (resp. right) wave function which is a ket (resp. bra) vector in the terminology of Dirac.  The wave function has then an algebraic structural interpretation in terms of algebraic eigenvectors and a statistical interpretation as given classically in the quantum theories.
\end{propo}
\vskip 11pt

\bpr In the terminology introduced in definition 3.1.3, a ket vector is a left vector and a bra vector is a coleft vector.  The 
 bisemisheaf $\widetilde M_{L_R}\otimes_D \widetilde M_L$ ``at the mass level''  has an automorphic irreducible representation in terms of global elliptic bisemimodule as developed in 1.1.14 to 1.1.20 and in 3.2.10 to 3.2.20.

Let $\{\psi_{L_R}(i)\otimes \psi_L(i)\}_i$ be the set of eigenbivectors of an operator $T_{R}\otimes T_{L}$ as defined in results 3.2.18 and in definition 3.3.4.  Then, the  semisheaf $\widetilde M_L$ (resp. $\widetilde M_{L_R}$~) has for spectral representation the left (resp. right) wave function $\mid \Psi\rangle$ (resp. $\langle   \Psi\mid$~) developed following:
\[ \mid \Psi\rangle =\txt\sum\limits_i d_i\mid \psi_L(i)\rangle\qquad \mbox{(resp.\ } \langle \Psi\mid  =\txt\sum\limits_i d^*_i\langle \psi_{L_R}(i)\mid\; )\]
if we refer to definition 3.3.4 where $d_i$ (resp. $d^*_i$~) is given by
\begin{equation}d_i =\langle \psi_{L_R}(i)\mid \Psi\rangle \qquad \mbox{(resp.\ } d^*_i =\langle \Psi\mid \psi_L(i)\rangle\;).\tag*{\eop}\end{equation}
\vskip 11pt

\begin{rem} { Referring to proposition  3.3.5, we notice that
 the coefficients $d_i$ and $d^*_i$ are probability measures.  We
shall then see that the traditional calculus of probability amplitudes of
quantum theories \cite{B-vonN} is replaced in this context by the probability
calculus with intensities.}\end{rem}
\vskip 11pt

\begin{propo}  The traditional calculus with the amplitudes of probability
{ \cite{Fey1},  \cite{Dir5}\/} of quantum (field) theory is replaced by a calculus
with intensities of probability in the context of this algebraic quantum
theory.
\end{propo}
\vskip 11pt

\bpr If we realize on a bisystem, an elementary bisemiparticle for example, an
observation ``~$A$~'', corresponding to the Fredholm bioperator
$T_{R}\otimes_D T_{L}$ defined from $\Hs^+_{h;M}$ to $\Hs^+_{h;M}$~, we
shall obtain the eigenbivalue $\lambda ^2_ a$ (or more exactly $+\sqrt{\lambda
^2_a}$ since the associated coleft particle is unobservable) with probability
$P_{\lambda _a}=\langle \psi_{L_R}(a) \mid \Psi\rangle \langle \Psi \mid \psi_{L}(a)\rangle$~.  An
observation ``~$B$~'' on the same bisystem will give the eigenbivalue $\lambda
^2_b$ with probability $P_{\lambda _b}=\langle \psi_{L_R}(b) \mid \Psi\rangle \langle \Psi \mid
\psi_{L}(b)\rangle$~.

Thus, $P_{\lambda _b}\cdot P_{\lambda _a}=P_{\lambda _b\cdot\lambda _a}
=\langle \psi_{L_R}(b) \mid \Psi\rangle \langle \Psi \mid \psi_{L}(b)\rangle
\langle \psi_{L_R}(a) \mid \Psi\rangle \langle \Psi \mid \psi_{L}(a)\rangle$ will correspond to the
probability of two successive measurements ``~$A$~'' and ``~$B$~'' on a bisystem.

This differs from the ordinary calculus with amplitudes of probability
\cite{Fey1} $\psi_{\lambda _b\cdot\lambda _a}=\langle \psi_{L_R}(b) \mid \Psi\rangle \langle \Psi \mid
\psi_{L}(a)\rangle$ of quantum theory dealing with elementary particles and
not with elementary bisemiparticles as considered here.

As we have $P_{\lambda _b\cdot\lambda _a}=\psi^*_{\lambda _b\cdot\lambda
_a}\cdot \psi_{\lambda _b\lambda _a}$~, we see that the classical probability
calculus with intensities is restored in quantum theory if bisystems
throughout bisemiparticles are taken into account.\epr
\vskip 11pt

\begin{defi} {\boldmath \bf The $PC T$ map\/} { of quantum field theory
(\cite{B-D}, \cite{Lud}, \cite{Wig1},  \cite{W-W-W}) transforms the fields of
particles into the fields of antiparticles and vice-versa.  Its equivalent in
this AQT model is the following set of maps:
\begin{enumerate}
\item $B_R\circ p_R : \widetilde M_L\to \widetilde M_{R_L}$~,\\
transforming the left semisheaf $\widetilde M_L$ of the left semiparticle into the
(involuted) coright-semisheaf $\widetilde M_{R_L})$~;
\vskip 11pt

\item $p^{-1}_L\circ B^{-1}_L : \widetilde M_{L_R}\to \widetilde M_R$~,\\
transforming the coleft semisheaf $\widetilde M_{L_R}$ of the right semiparticle into the
(involuted) right semisheaf $\widetilde M_R$~.
\end{enumerate}
\vskip 11pt

The maps $p_{R,L}$ and $B_{R,L}$ are described in definitions 3.1.5 and
3.1.7.}\end{defi}
\vskip 11pt

Then, the left bisemisheaf $\widetilde M_{L_R}\otimes_D \widetilde M_L$ of a left bisemiparticle,
associated to the left internal bilinear Hilbert space $\Hs^+_a$~, is
transformed into the right bisemisheaf $\widetilde M_{R_L}\otimes_D\widetilde M_R$ of a right
bisemiparticle, associated to the right internal bilinear Hilbert space
$\Hs^-_a$~, according to:
\[ (p^{-1}_L\circ B^{-1}_L)\otimes_D (B_R\circ p_R) : \widetilde M_{L_R}\otimes_D\widetilde M_L\to
\widetilde M_{R_L}\otimes_D\widetilde M_R\;.\]
$(p^{-1}_L\circ B^{-1}_L)\otimes_D (B_R\circ p_R)$ is thus a parity time bimap whose physical 
meaning is given in 1.3.10.

$\widetilde M_{L_R}\otimes_D\widetilde M_L$ is the ``physical field'' of the left (bisemi)particle and 
$\widetilde M_{R_L}\otimes_D\widetilde M_R$ is the `` physical field'' of the right (bisemi)particle according to section 1.1.6.

If the bisemiparticle is electrically charged, a supplementary set of maps
$(p^{-1}_L\circ B^{-1}_L)\otimes_{m,e} (B_R\circ p_R)$ must be applied on the
magnetic and electric bisemisheaves reversing then the electric charge and the
magnetic moment; this will thus correspond to a charge conjugation.

Note that the intrinsic parity-time of a $4D$-semisheaf corresponds to its
orientation: this results from its generation by Eisenstein cohomology from
the symmetric splitting semifield $L^{\mp}$ (see definition 1.1.2). \vskip 11pt

\begin{defis}[1. Right and left $4D$-elliptic operators]\mbox{}  { As the
cohomology  $H^*(\Gamma  )$ of an arithmetic subgroup $\Gamma  $ may be
identified with the cohomology of $\Gamma  $-invariant smooth differential
forms of de Rham \cite{Bor1}, \cite{Gro3}, we shall assume that the
$4D$-  differential operator
\nobeqn \lefteqn{T_{R,L;ST,MG,M}}\\
\noalign{\vskip 6pt}
&=&\left\{ \pm i\hbar_{ST,MG,M}\ dt_0,\pm
i\frac{\hbar_{ST,MG,M}}{c_{t\to r;ST,MG,M}}\ dx,\pm
i\frac{\hbar_{ST,MG,M}}{c_{t\to r;ST,MG,M}}\ dy, \pm
i\frac{\hbar_{ST,MG,M}}{c_{t\to r;ST,MG,M}}\ dz\right\}\noeeqn
can apply on the $4D$-semisheaf of rings $\theta ^{1-3}_{R,L}(t,r)_{ST}$~,
$\theta ^{1-3}_{R,L}(t,r)_{MG}$ or $\theta ^{1-3}_{R,L}(t,r)_{M}$~, \\
where
\begin{enumerate}
\item the ``~$+$~'' or ``~$-$~'' sign is a convention depending on the sense of
rotation of the considered semisheaf of rings;
\item $c^{-1}_{t\to r;ST,MG,M}$ is an average parameter equal to the ratio of
algebraic Hecke characters (see definition 1.4.10).
\end{enumerate}
\vskip 11pt

On the other hand, as the semisheaves of rings   $\theta
^{1-3}_{R,L}(t,r)_{MG}$ and  $\theta
^{1-3}_{R,L}(t,r)_{M}$ are the basis of the vertical tangent semibundles
$T^{(1-3)}_{MG_{R,L}}$ and $T^{(1-3)}_{M_{R,L}}$ according to definition 2.2.14,
their projective maps are given by the elliptic operators:
\nobeqn \lefteqn{DT_{R,L;MG,M}}\\
\noalign{\vskip 6pt}
&=&\left\{ \pm i\hbar_{MG,M}\ \frac
\partial{\partial t_0},\pm i\frac{\hbar_{MG,M}}{c_{t\to r;MG,M}}\ \frac
\partial{\partial x},\pm i\frac{\hbar_{MG,M}}{c_{t\to r;MG,M}}\ \frac
\partial{\partial y}, \pm i\frac{\hbar_{MG,M}}{c_{t\to r;MG,M}}\
\frac \partial {\partial z}\right\}\noeeqn
where $\hbar_{MG,M}$ corresponds to the order of the global inertia subgroup respectively in
the  ``~$MG$~'' or ``~$M$~'' system of units.  In particular,
$\hbar_M\equiv \hbar$~, i.e. the Planck's constant.
\vskip 11pt

\noindent {\bf (2. Tensor products of right and left elliptic operators)\/}
 Let, for example, the bioperators $(T_{R;ST}\otimes T_{L;ST})$~, 
$(DT_{R;MG}\otimes DT_{L;MG})$ and $(DT_{R;M}\otimes DT_{L;M})$ act
respectively on the bisemisheaves $(\theta ^{1-3}_R(t,r)_{ST}\otimes 
\theta ^{1-3}_L(t,r)_{ST})$~, $(\theta ^{1-3}_R(t,r)_{MG}\otimes 
\theta ^{1-3}_L(t,r)_{MG})$ and $(\theta ^{1-3}_R(t,r)_M\otimes 
\theta ^{1-3}_L(t,r)_M)$~.

Then, the bioperator $(T_{R;ST}\oplus DT_{R;MG}\oplus DT_{R;M})\otimes
(T_{L;ST}\oplus DT_{L;MG}\oplus DT_{L;M})$ will act on the bisemisheaf:
\nobeqn
\lefteqn{\theta ^{1-3}_R(t,r)_{ST-MG-M}\otimes \theta
^{1-3}_L(t,r)_{ST-MG-M}}\\
&=& (\theta ^{1-3}_R(t,r)_{ST}\oplus \theta ^{1-3}_R(t,r)_{MG}\oplus 
\theta ^{1-3}_R(t,r)_{M})\\
&&\qquad \otimes (\theta ^{1-3}_L(t,r)_{ST}\oplus \theta
^{1-3}_L(t,r)_{MG}\oplus  \theta ^{1-3}_L(t,r)_{M})\noeeqn
representing the complete massive structure of a bisemiparticle.

According to the development of the bisemisheaf\ 
$(\theta ^{1-3}_R(t,r)_{ST-MG-M}\otimes$\\ $ \theta ^{1-3}_L(t,r)_{ST-MG-M})$ in
direct sums of bisemisheaves as given in definition 3.1.2, the bioperator
$(T_{R;ST}\oplus DT_{R;MG}\oplus DT_{R;M})\otimes (T_{L;ST}\oplus
DT_{L;MG}\oplus DT_{L;M})$ will decompose into:
\nobeqn
\lefteqn{(T_{R;ST}\oplus DT_{R;MG}\oplus DT_{R;M}) \otimes (T_{L;ST}\oplus DT_{L;MG}\oplus DT_{L;M})}\\
&=& (T_{R;ST}\otimes T_{L;ST})+(DT_{R;MG}\otimes DT_{L;MG})+ 
(DT_{R;M}\otimes DT_{L;M})\\
&&\qquad + (T_{R;ST}\otimes DT_{L;MG}) + (DT_{R;MG}\otimes T_{L;ST})
+ (T_{R;ST}\otimes DT_{L;M})\\ 
&&\qquad + (DT_{R;M}\otimes T_{L;ST}) + (DT_{R;MG}\otimes DT_{L;M})
+ (DT_{R;M}\otimes DT_{L;MG})\;.\noeeqn
$(T_{R;ST}\otimes DT_{L;MG})$ has for $3D$-spatial off-diagonal components
\[ L_L(k) = -i (dr_ip^j_{MG}+dr_jp^i_{MG})\;,\qquad 1\le i,j\le 3\;,\]
and $(DT_{R;MG}\otimes T_{L;ST})$ has for $3D$-spatial off-diagonal components
\[ L_R(k) = +i (p_{i;MG}dr^j+p_{j;MG}dr^i)\;,\]
where
\begin{enumerate}
\item $\displaystyle{dr_i = \frac{\hbar_{ST}}{c_{t\to r;ST}}\ dx_i }$ so that,
if $\begin{array}[t]{ll} i=1\;, & x_i\equiv x\;,\\
i=2\;, & x_i\equiv y\;,\\
i=3\;, & x_i\equiv z\;;\end{array}$
\vskip 11pt

\item $\displaystyle{p_{i;MG}=DT_{(i)_{R;MG}}=+i\frac{\hbar_{MG}}{c_{t\to r;MG}}\ 
\frac \partial{\partial x_i}\;;}$
\vskip 11pt

\item $\displaystyle{p^i_{MG}=DT_{(i)_{L;MG}}=-i\frac{\hbar_{MG}}{c_{t\to
r;MG}}\ 
\frac \partial{\partial x^i}\;.}$\end{enumerate}}
\end{defis}
\vskip 11pt

\begin{defi}[Right and left internal angular momenta] \quad  { $L_{L;MG(ST)}(k)$  
is  interpreted as the components of the angular momentum vector $\vec
L_{L;MG(ST)}$ of the left middle-ground structure of the left semiparticle
and $L_{R;MG(ST)}(k)$ is interpreted as the components of 
the angular momentum vector $\vec L_{R;MG(ST)}$ of the right middle-ground structure of
the right
semiparticle. Thus, $\vec L_{R,L;MG(ST)}$ represents the ``angular momentum''
of all the sections of the right (resp. left) semisheaf of rings $\theta
^3_{R,L}(r)_{MG}$ with respect to the left (resp. right) semisheaf of rings
$\theta ^3_{L,R}(t)_{ST}$~.

Similarly, the $3D$-spatial off-diagonal components of $(T_{R;ST}\otimes
DT_{L;M})$ (resp. of $(DT_{R;M}\otimes
T_{L;ST})$~) will be $L_{L;M(ST)}(k)$ (resp. $L_{R;M(ST)}(k)$~).
$L_{L;M(ST)}(k)$ (resp. $L_{R;M(ST)}(k)$~) are thus the components of the
angular momentum vector $\vec L_{L;M(ST)}$ (resp. $\vec L_{R;M(ST)}$~)
interpreted as the angular momentum of the ``mass'' structure $\theta
^3_{L;R}(r)_M$ of the left (resp. right) semiparticle with respect to the
right (resp. left) space-time structure $\theta ^3_{R,L}(r)_{ST}$ of the right
(resp. left) semiparticle.

To each right (resp. left) ``~$ST$~'', ``~$MG$~'' or ``~$M$~'' semisheaf or rings
corresponds a right (resp. left) internal angular momentum vector $\vec L_{R;ST,MG,M}$ (resp. $\vec
L_{L;ST,MG,M}$~) which indicates its angular velocity and its sense of
rotation with respect to its associated corresponding left (resp. right)
semisheaf of rings.  The right (resp. left) internal momentum vector 
$\vec
L_{R,L;ST,MG,M}$ then corresponds to the spin concept \cite{Pau}, \cite{Dir3}
of quantum (field) theory.}\end{defi}
\vskip 11pt

\begin{propo} A right and a left semiparticle rotate in opposite senses and
have only two possible spin states.
\end{propo}
\vskip 11pt

\bpr In definitions 3.3.9 and 3.3.10, we have defined the right and the left
internal angular momentum components of the right and the left ``~$MG$~''
semisheaf of rings $\theta ^3_{R,L}(r)_{MG}$ by
\nobeqn
L_R(k) &=& +i (p_{i;MG} dr^j+p_{j;MG}dr^i)\\
\noalign{\noindent and by}
L_L(k) &=& -i (dr_ip^j_{MG} +dr_jp^i_{MG})\;.\noeeqn
It is then evident that $L_R(k)=-L_L(k)$ which proves that:
\begin{enumerate}
\item a right and a left associated semiparticle have opposite rotation senses;
\item two senses of rotation can only exist for a right and a left
semiparticle and also for a bisemiparticle since only the left semiparticle is
observable in a bisemiparticle and thus only its own left internal angular
momentum.
\end{enumerate}
\vskip 11pt

Indeed, it can be remarked that the sign of $L_{R,L}(k)$ depends on the sign
of $dr^i$~, $1\le i\le 3$~, which is reflected by the mapping:
\[ \phi_{\pm} : \begin{array}[t]{ccc}
\rit^{\pm} &\to & \rit^{\pm}(\pm i)\;, \\
{\pm}dr^i &\leadsto & {\pm}dr^i(\pm i)\;.\end{array}\]
of the positive (resp. negative) reals in the positive or negative pure
imaginary reals.
\vskip 11pt

\bt[t]{c}
A left handled rotation corresponds\\
\noalign{\vskip -6pt}
 to the mapping:\\
\noalign{\vskip 6pt}
$\phi_+ : (\rit^+,\rit^-) \to (\rit^+(+i),\rit^-(+i))\;,$\\
\noalign{\vskip 11pt}
{\setlength{\unitlength}{1mm}
\begin{picture}(40,30)
\put(0,15){\vector(1,0){40}}
\put(20,0){\vector(0,1){30}}
\put(0,10){$\rit^-$}
\put(35,10){$\rit^+$}
\put(14,10){$\searrow$}
\put(22,18){$\nwarrow \phi^+$}
\put(22,0){$+i\rit^-$}
\put(22,28){$+i\rit^+$}

\end{picture}
}\te\hfill
\bt[t]{c}
and a  right handled rotation corresponds \\
\noalign{\vskip -6pt}
to the mapping:\\
\noalign{\vskip 6pt}
$\phi_- : (\rit^+,\rit^-) \to (\rit^+(-i),\rit^-(-i))\;.$\\
\noalign{\vskip 11pt}
{\setlength{\unitlength}{1mm}
\begin{picture}(40,30)
\put(0,15){\vector(1,0){40}}
\put(20,0){\vector(0,1){30}}
\put(0,10){$\rit^-$}
\put(35,10){$\rit^+$}
\put(22,10){$\swarrow$}
\put(7,18){$\phi ^- \nearrow$}
\put(22,0){$-i\rit^+$}
\put(22,28){$-i\rit^-$}

\end{picture}}
\te\qquad\qquad\\
\mbox{}\epr
\vskip 11pt

\begin{sub}{\bf Spin-statistics and supersymmetry}
\end{sub}

According to proposition 3.3.11, each elementary (semi)particle has two senses of rotation: this is the case for:
\Bi
\item the elementary leptons: $e^-$~, $\mu ^-$~, $\tau ^-$ and their neutrinos $\nu _{e^-}$~, $\nu _{\mu ^-}$~, $\nu _{\tau ^-}$~;
\item the quarks: $u^+$~, $d^-$~, $s^-$~, $c^+$~, $b^-$~, $t^+$~;
\item the photons.
\Ei
So, this algebraic quantum (field) theory, which does not refer to a (non)abelian gauge theory, takes up the spin concept differently from quantum field theories.  However, it seems evident that elementary (semi)fermions must always obey the Fermi-Dirac statistics while the photons behave in accordance with the Bose-Einstein statistics since they can increase their quanta number as developed in section 1.4.16.

Consequently, the supersymmetry, whose aim is the transformation of half integer spin particles into interger spin particles, does not seem essential in the present context and will not be taken into account.
\vskip 11pt

\begin{rem}[Interpretation of special relativity invariants] \ { Let
$T_{R,L;ST,MG,M}$\\ be the right or left $4D$-differential operator acting on
the ``~$ST$~'', ``~$MG$~'' or ``~$M$~'' $4D$-semisheaf of rings $\theta
^{1-3}_{R,L}(t,r)_{ST,MG,M}$ and let
\[ dt^2 = (T_{R;ST,MG,M},T_{L;ST,MG,M}) =
\hbar^2_{ST,MG,M}(dt_0^2+c^{-2}_{t\to r}\ dx^2
+c^{-2}_{t\to r}\ dy^2+c^{-2}_{t\to r}\ dz^2)\]
be their internal scalar product which is an additional structure of the
corresponding ``~$ST$~'', ``~$MG$~'' or ``~$M$~'' internal bilinear Hilbert space
$\Hs^{\mp}_{ST,MG,M}$~.

The corresponding Minkowsky space-time differential form of special relativity
\cite{Ein2} is
\[ dt^2_0=dt^2-c^{-2}\ dx^2-c^{-2}\ dy^2-c^{-2}\ dz^2\;.\]
It is an invariant whose meaning in view of the developments of this paper can
be interpreted as follows: if we remember that the Eisenstein cohomology
classes are represented by differential forms in bijection with Eisenstein series, we
can deduce from it that every increasing or decreasing of $dt^2$~, i.e.
finally of $(\theta ^{1-3}_R(t,r)_{ST,MG,M}\otimes_D \theta
^{1-3}_L(t,r)_{ST,MG,M})$~, happens by external capture or loosing of 
biquanta $(\widetilde M^I_R(r)_{ST,MG,M}\otimes_D \widetilde M ^I_L(r)_{ST,MG,M})$ throughout
the smooth biendomorphism $(E_{R;ST,MG,M}\times_D E_{L;ST,MG,M})$ according to proposition 3.3.2: indeed, this corresponds to the increasing or to the
decreasing of $c^{-2}(dx^2+dy^2+dz^2)$~.

On the other hand, the euclidian invariant
\[ dt^2=\hbar^2_{ST,MG,M}(dt^2_0+c^{-2}_{t\to r}(dx^2+dy^2+dz^2))\;,\] valid for
a closed system and essentially envisaged in this work, can be interpreted in
function of the internal morphism $((\gamma _{t\raise -3pt\hbox{$\overrightarrow\leftarrow$}r}\circ
E_{R;ST,MG,M})\times_D (\gamma _{t\raise -3pt\hbox{$\overrightarrow\leftarrow$}r}\circ
E_{L;ST,MG,M}))$ applied on $(\theta ^{1-3}_R(t,r)_{ST,MG,M}\otimes_D
\theta ^{1-3}_L(t,r)_{ST,MG,M})$~.  Indeed, if $dt^2$ is invariant, then time
biquanta can be transformed into $3D$-spatial biquanta and
vice-versa.}\end{rem}
\vskip 11pt

\setcounter{defi}{0}
\section{Second order differential bilinear equations}

This chapter is devoted to the study of the differential equations relative to the
bisemiparticles.  It is thus necessary to classify the bisemiparticles with respect to the
presently observed elementary particles and in function of their general structure as developed in
the preceding chapters: this is the object of this first section.  We shall take for reference the
traditional statistical classification of fermions and bosons.
\vskip 11pt

\subsection{Classification of bisemiparticles}

\begin{defi}[Bisemifermions and bisemibosons] { Let $(\theta ^{1-3}_R(t,r)_{ST-MG-M}\otimes$ \\ $
\theta ^{1-3}_L(t,r)_{ST-MG-M})$ denote the three embedded structures of 
a massive bisemiparticle as
developed in definition 3.1.2.  According to definition 3.1.1, this bistructure corresponds
essentially to a bisemilepton or to a bisemiquark, i.e. to an elementary massive bisemifermion. 
In definition 3.1.2, this tensor product has been decomposed into the direct sum of three tensor
products referring to the ``~$ST$~'', ``~$MG$~'' and ``~$M$~'' bistructures and of six other tensor
products referring to the interactions between the right and the left ``~$ST$~'', ``~$MG$~'' and
``~$M$~'' structures.  Taking into account the general structure of a bisemiparticle as given
above, we can classify the bisemiparticles in the following four categories:

\Bena
\item  {\bf An elementary massive bisemifermion} has a bistructure given by $(\theta
^{1-3}_R(t,r)_{ST-MG-M}\otimes\linebreak
\theta ^{1-3}_L(t,r)_{ST-MG-M})$ whose each of the nine constitutive tensor products:
$(\theta ^{1-3}_R(t,r)_{ST}\otimes
\theta ^{1-3}_L(t,r)_{ST}),\linebreak \cdots, \; (\theta ^{1-3}_R(t,r)_{M}\otimes
\theta ^{1-3}_L(t,r)_{MG})$~, (noted in a general abbreviated form
$(\theta ^{1-3}_R(t,r)\otimes
\theta ^{1-3}_L(t,r))$~), (see definition 3.1.2) decomposes under the 
blowing-up isomorphism $S_L$
into:\\[6pt]
\Be
\item a diagonal bisemisheaf $(\theta ^{1-3}_R(t,r)\otimes_D
\theta ^{1-3}_L(t,r))$ giving in the case of 
$(\theta ^{1-3}_R(t,r)_{ST,MG,M}\otimes_D\linebreak
\theta ^{1-3}_L(t,r)_{ST,MG,M})$ the diagonal central bistructure of the three embedded
bisemisheaves of rings ``~$ST$~'', ``~$MG$~'' or ``~$M$~'';
\vskip 11pt

\item a magnetic bisemisheaf $(\theta ^{3}_R(r)^{(m)}\otimes_m
\theta ^{3}_L(r)^{(m)})$ which is composed in the case of\linebreak
$(\theta ^{3}_R(r)^{(m)}_{ST,MG,M}\otimes_m
\theta ^{3}_L(r)^{(m)}_{ST,MG,M})$ of nonorthogonal ``~$ST$~'', ``~$MG$~'' or ``~$M$~'' magnetic
biquanta $( \widetilde  M^{I(3)} _{k_{R;ST,MG,M}}\otimes_m \widetilde  M^{I(3)} _{k_{L;ST,MG,M}})$~.\\
In fact, these magnetic biquanta are generated by the magnetic smooth biendomorphism according to
proposition 3.3.2:
\nobeqn
\lefteqn{\hspace{-1cm}E_{R;ST,MG,M}\times_m E_{L;ST,MG,M} : 
\theta ^{1-3}_R(t,r)_{ST,MG,M}\otimes
\theta ^{1-3}_L(t,r)_{ST,MG,M}}\\
\noalign{\vskip 6pt}
&\to & (\theta ^{1-(3)^*}_R(t,r)_{ST,MG,M}\otimes
\theta ^{1-(3)^*}_L(t,r)_{ST,MG,M}) \\
\noalign{\vskip 6pt}
 &&\qquad \textstyle\bigoplus\limits^m_{k=1} (\widetilde  M
^{I(3)}_{k_{R;ST,MG,M}}\otimes_m \widetilde  M ^{I(3)}_{k_{L;ST,MG,M}})\noeeqn
and constitute the magnetic moment of the considered bisemifermion;
\vskip 11pt

\item an electric bisemisheaf $(\theta ^{1-(3)}_R(t,(r))^{(e)}\otimes_e
\theta ^{3-(1)}_L((t),r)^{(e)})$ which is composed in the case of 
$(\theta ^{1-(3)}_R(t,(r))^{(e)}_{ST,MG,M}\otimes_e
\theta ^{3-(1)}_L((t),r)^{(e)}_{ST,MG,M})$ of  ``~$ST$~'', ``~$MG$~'' or ``~$M$~''
electric time-space biquanta $(\widetilde  M^{I(1)} _{k_{R;ST,MG,M}}\otimes_e \widetilde  M^{I(3)} _{k_{L;ST,MG,M}})$ or
space-time biquanta $( \widetilde  M^{I(3)} _{k_{R;ST,MG,M}}\otimes_e \widetilde  M^{I(1)} _{k_{L;ST,MG,M}})$ such that the
right (resp. left) time quanta are generated by versal deformation and spreading-out isomorphism
$SO(e)\circ Vd(e)$  according to definition 2.4.1 while the left (resp. right) space quanta are generated
by smooth endomorphism $E_{L,R;ST,MG,M}$~.  The electric bisemisheaf  $(\theta
^{1-(3)}_R(t,(r))^{(e)}_{ST-MG-M}\otimes_e
\theta ^{3-(1)}_L((t),r)^{(e)}_{ST-MG-M})$ constitutes the electric charge of the considered
massive bisemifermion.
\Ee
\vskip 11pt

\item {\bf A bisemiphoton} has a spatial structure given by the tensor product 
$(T ^{1}_R(r_k)_{ST-MG-M} \otimes\linebreak
T ^{1} _L(r_k)_{ST-MG-M})$ of a right semiphoton by a left semiphoton which can split under the blowing-up isomorphism $S_L$ into:\vskip 6pt
${S_L : (T ^{1}_R(r_k)_{ST-MG-M}\otimes
T ^{1}_L(r_k)_{ST-MG-M})}$\\[6pt]
$\;\to\;  (T ^{1}_R(r_k)_{ST-MG-M}\otimes_D
T ^{1}_L(r_k)_{ST-MG-M}) \oplus (T ^{1}_R(r_k)^{(m)}_{ST-MG-M}\otimes_m
T ^{1}_L(r_k)^{(m)}_{ST-MG-M})$\\[6pt]
where $(T ^{1}_R(r_k)_{ST-MG-M}\otimes_D
T ^{1}_L(r_k)_{ST-MG-M})$ refers to the three embedded diagonal bisections
``~$ST$~'', ``~$MG$~'' and ``~$M$~'' representing its central space bistructure and where
$(T ^{1}_R(r_k)^{(m)}_{ST-MG-M}\otimes_m
T ^{1}_L(r_k)^{(m)}_{ST-MG-M})$ refers to the three embedded magnetic bisections of rings
``~$ST$~'', ``~$MG$~'' and ``~$M$~'' representing its magnetic structure composed of
 magnetic space biquanta.
\vskip 11pt

\item {\bf A bisemiboson of magnetic structure} is an electrically neutral meson which will be
proved in chapter 5 to be generated by a magnetic biendomorphism from a bisemiquark.  In this
category, we may also include the magnetic biquanta whose structure is given by the magnetic
bisemisheaf $(\theta ^{3}_R(r)^{(m)}_{ST-MG-M}\otimes_m
\theta ^{3}_L(r)^{(m)}_{ST-MG-M})$~.
\vskip 11pt

\item {\bf An electrically charged bisemiboson} is an electrically charged meson generated from a
bisemiquark as it will be seen in chapter 5.  In this category, we may also include the electric
charge whose structure is given by the electric bisemisheaf: 
$(\theta
^{1-(3)}_R(t,(r))^{(e)}_{ST-MG-M}\otimes_e
\theta ^{3-(1)}_L((t),r)^{(e)}_{ST-MG-M})$ where $\theta ^{1-(3)}$ means a $1D$-time or $3D$-space
semisheaf of rings.
\Ee}\end{defi}
\vskip 11pt

\begin{defi}[Annihilation of a semilepton pair] \qquad  { Let \quad $(\theta
^{1-3}_R(t,r)_{ST-MG-M}\otimes_D$\\
$\theta ^{1-3}_L(t,r)_{ST-MG-M})\oplus (\theta
^{3}_R(r)^{(m)}_{ST-MG-M}\otimes_m
\theta ^{3}_L(r)^{(m)}_{ST-MG-M})\oplus (\theta
^{1-(3)}_R(t,(r))^{(e)}_{ST-MG-M}\otimes_e$\linebreak  $
\theta ^{3-(1)}_L(t,r)^{(e)}_{ST-MG-M})$ be the ``~$ST\oplus MG\oplus M$~''-semisheaves of rings
constituting the massive structure of a bisemilepton.
Under some external perturbation, a breaking of the diagonal bisemisheaves 
$(\theta
^{1-3}_R(t,r)_{ST-MG-M}\otimes_D
\theta ^{1-3}_L(t,r)_{ST-MG-M})$ can occur such that the right and left semisheaves are no more
localized in some open ball of radius $R$ where $R$ is the radius of the topological domain on
which the constitutive bisemisheaf $(\theta
^{1-3}_R(t,r)_{ST-MG-M}\otimes
\theta ^{1-3}_L(t,r)_{ST-MG-M})$ is defined.}\end{defi}
\vskip 11pt

If the right and left semisheaves $\theta
^{1-3}_R(t,r)_{ST-MG-M}$ and $
\theta ^{1-3}_L(t,r)_{ST-MG-M}$ are no more localized in the same open ball, a new electric and
magnetic ``~$ST- MG-M$~'' bisemisheaf can be generated.  Indeed, a right and a left electric and
magnetic bisemisheaves, corresponding to a positive and a negative electric charges and magnetic
moments, can exist simultaneously because they are no more orthogonal according to proposition
3.1.6.  The result is that a pair of semileptons is generated such that each semilepton is endowed
with an electric charge and a magnetic moment of opposite signs.

But this pair of semileptons can annihilate.  Indeed, by electromagnetic attraction, this pair
will be again concentrated in a same open ball which involves that the two electric bisemisheaves,
representing their electric charges, become orthogonal.  Consequently, they cannot conserve the
same structure as remarked above: they must then transform themselves into magnetic bisemisheaves
as follows:

\nobeqn
\gamma _{t\to r}\circ E : 
(\theta
^{1-(3)}_R(t,(r))^{(e)}_{ST-MG-M}\otimes_e
\theta ^{3-(1)}_L((t),r)^{(e)}_{ST-MG-M})&& \\
\noalign{\vskip 6pt}
\to \; (\theta
^{3-(3)}_{I_R}(r,(r))^{(m)}_{ST-MG-M}\otimes_m
\theta ^{3-(3)}_L((r),r)^{(m)}_{ST-MG-M})\;.&&\noeeqn
The resulting magnetic bisemisheaves can be transformed later in diagonal bisemisheaves.

On the other hand, as the time structures of the electric charge are generated by the morphisms
$SOT(e)\circ Vd(e)$ from the right and the left semisheaves 
$\theta ^{1-3}_{R,L}(t,r)_{ST-MG-M}$  of the considered semileptons according to
definition 2.4.1 and as the $1D$-time semisheaves of the electric charge must be transformed into
their complementary $3D$-space semisheaves by the fact of the collision, it is reasonable to admit
that the external perturbation provoking a $(\gamma _{t\to r}\circ E)$ morphism
on the $1D$-time structures of the electric charge will also provoke a $(\gamma
_{t\to r}\circ E)$ morphism on the two semisheaves $\theta
^{1-3}_{R,L}(t,r)_{ST-MG-M}$ constituting the central massive structure of the
two semileptons.  We then will have:
\[ (\gamma _{t\to r}\circ E_{R,L}) : \theta ^{1-3}_{R,L}(t,r)_{ST-MG-M} \to \theta
^{3}_{R,L}(r)_{ST-MG-M}\simeq T^1_{R,L}(r_k)_{ST-MG-M}\]
transforming the central $4D$-structures of the pair of semileptons into $3D$-structures of
semiphoton(s).

Indeed, we have finally that the pair of semileptons annihilate into a pair of semiphotons
according to:
\nobeqn
\lefteqn{[(\theta ^{1-3}_{R}(t,r)_{ST-MG-M})\oplus (\theta ^{1}_{R}(t)^{(e)}_{ST-MG-M}\otimes_e
\theta ^{3}_{L}(r)^{(e)}_{ST-MG-M}) }\\
\noalign{\vskip 6pt}
\lefteqn{\oplus \theta ^{3}_{R}(r)^{(m)}_{ST-MG-M}\otimes_m
\theta ^{3}_{L}(r)^{(m)}_{ST-MG-M})]}\\
\lefteqn{\cup [(\theta ^{1-3}_{L}(t,r)_{ST-MG-M})\oplus (\theta
^{3}_{R}(r)^{(e)}_{ST-MG-M}\otimes_e
\theta ^{1}_{L}(t)^{(e)}_{ST-MG-M})}\\
\noalign{\vskip 6pt}
\lefteqn{ \oplus (\theta ^{3}_{R}(r)^{(m)}_{ST-MG-M}\otimes_m
\theta ^{3}_{L}(r)^{(m)}_{ST-MG-M})]}\\
\noalign{\vskip 6pt}
&\to & [(T ^{1}_{I_R}(r_k)_{ST-MG-M})\oplus (\theta ^{3}_{R}(r)^{(m)}_{ST-MG-M}\otimes_m
\theta ^{3}_{L}(r)^{(m)}_{ST-MG-M}) ]\\
\noalign{\vskip 6pt}
&&\qquad \cup [(T ^{1}_{I_L}(r_k)_{ST-MG-M})\oplus (\theta
^{3}_{R}(r)^{(m)}_{ST-MG-M}\otimes_m
\theta ^{3}_{L}(r)^{(m)}_{ST-MG-M})]\;.\noeeqn
\vskip 11pt

\begin{rem}[Hypothesis concerning the structure of semineutrinos]  { Consider that a pair of semileptons, endowed each
one with its electric charge and magnetic moment, comes into collision in such a way that almost all the ``mass'' quanta of the semisheaves of rings $\theta ^{1-3}_{R,L}(t,r)_{ST-MG-M}$ blow up by an endomorphism as described in section 1.2 such that $\theta ^{1-3}_{R,L}(t,r)_{ST-MG-M}$ be reduced to $\theta ^{1-3}_{R,L}(t,r)_{ST-MG-(M\to 0)}$~.  Then, the (semi)lepton central structure $\theta ^{1-3}_{R,L}(t,r)_{ST-MG-M}$ has been transformed into a (semi)neutrino central structure $\theta ^{1-3}_{R,L}(t,r)_{ST-MG-(M\to 0)}$~.
}\end{rem}
\vskip 11pt

\subsection{Second order elliptic bilinear equations on extended bilinear Hilbert spaces}

 For the facility of manipulations and notations, the elliptic differential
bilinear equations will be considered for the mass (``~$M$~'') structure of the lightest massive
bisemilepton, i.e. the bisemielectron or classical electron, considering that the elliptic
differential bilinear equations relative to the other bisemiparticles and to the other structures
``~$ST$~'' and ``~$MG$~'' are exactly of the same type.
\vskip 11pt

\begin{defi}[Bisections of bisemisheaves]  { Let $(\theta
^{1-3}_{L_R}(t,r)_{M}\otimes \theta ^{1-3}_{L}(t,r)_{M})$ be the left-bisemisheaf
defined on the left extended internal bilinear Hilbert space $H^+_a$ and representing
the mass structure of a bisemielectron.  This left-bisemisheaf then results from the
bisemisheaf 
$(\theta
^{1-3}_{R}(t,r)_{M}\otimes \theta ^{1-3}_{L}(t,r)_{M})$ by application of the composition of maps
$B_L\circ p_L$ according to:
\[ B_L\circ p_L : (\theta
^{1-3}_{R}(t,r)_{M}\otimes \theta ^{1-3}_{L}(t,r)_{M})\to
(\theta
^{1-3}_{L_R}(t,r)_{M}\otimes \theta ^{1-3}_{L}(t,r)_{M})\;.\]
The right (resp. left) semisheaf 
$\theta
^{1-3}_{L_R}(t,r)_{M}$ (resp. $\theta
^{1-3}_{L}(t,r)_{M}$~) is composed of $q$ sections which are (isomorphic to) differentiable
right (resp. left) functions $\phi _{L_{R\mu }}(t,r)$ (resp. $\phi _{L_{\mu }}(t,r)$~), $1\le \mu
\le q$~, defined on a compact domain homeomorphic to a compact domain $D_{e^+}$ of
$(\rit^+)^1\times (\rit^+)^3$ centered in the upper half plane with respect to the
emergence point.

The right (resp. left) function $\phi _{L_{R\mu }}(t,r)$ (resp. $\phi _{L_{\mu }}(t,r)$~) can be
decomposed following:
\nobeqn
\phi _{L_{R\mu }}(t,r) &=& \phi _{L_{R\mu }}(t)\oplus \phi _{L_{R\mu }}(r)\\
\mbox{(resp.\ } \phi _{L_{\mu }}(t,r) &=& \phi _{L_{\mu }}(t)\oplus \phi _{L_{\mu }}(r)\;)\noeeqn
where ``~$r$~'' denotes the triple of spatial variables $\{ x_1,x_2,x_3\}$~.

This is a consequence of the generation of the three-dimensional semisheaf of rings $\theta^3_{L_R,L}(r)$ from the one-dimensional semisheaf of rings $\theta^1_{L_R,L}(t)$ according to Section 1.2.

The left bisemisheaf $(\theta
^{1-3}_{L_R}(t,r)_{M}\otimes \theta ^{1-3}_{L}(t,r)_{M})$ is composed of
algebraic bifunctions isomorphic to
$(\phi _{L_{R\mu }}(t,r)\otimes\phi _{L_{\mu }}(t,r))$ defined on the left extended internal bilinear
Hilbert space $H^+_h$~.

These bifunctions are defined on a curved space-time domain homeomorphic to a domain
$D_{e^+}\times D_{e^+}$ of $(\rit^+)^4\times (\rit^+)^4\simeq (\rit^+)^{10}$ into
$\rit$~.}\end{defi}
\vskip 11pt

\begin{defi}[$4D$-elliptic differential operator]  { As we are dealing with the 
vertical tangent semibundles $T^{(1)}_{M_{R,L}}$ and $T^{(3)}_{M_{R,L}}$ according to
definition 2.2.15, the elliptic differential right (resp. left)
operator to be considered is the following:
\[ \MM_{R,L} =\{ m_{0_{R,L;M}},\vec p_{R,L;M}\}  \]
whose explicit development is given by:
\[ \MM_{R,L}=\left\{ \pm i\hbar_M\ {\frac\partial{\partial t_0}},\pm i \  \frac{\hbar_M}{c_{t\to r;M}}
\ {\frac\partial{\partial x_1}},\cdots,\pm i\ {\frac{\hbar_M}{c_{t\to r;M}}}
\ {\frac\partial{\partial x_3}}\right\}\;.\]
But we must take into account the spin of the bisemielectron, i.e. the rotation of the sections of
$\theta
^{1-3}_{L_R}(t,r)_{M}$ and $ \theta ^{1-3}_{L}(t,r)_{M}$ (see definition 3.3.11). 
This can be achieved by considering that the elliptic differential operators
$m_{0_{R,L;M}}$ and
$\vec p_{R,L;M}$ are respectively $1D$- and $3D$-directional gradients
$s_{0_{R,L}}\nabla_{0_{R,L}}$ and $\vec s_{R,L}\overrightarrow{\nabla}_{R,L}$ where
$s_{0_{R,L}}$ and $\vec s_{R,L}$ are $1D$- and $3D$-unit vectors referring to the spin with
direction cosines $\{ s_{0_{R,L}}\}$ and $\{ s_{1_{R,L}},s_{2_{R,L}},s_{3_{R,L}}\}$~.

Indeed, the spin cannot be introduced judiciously by $\gamma $ or $\sigma $ matrices because bisemi\-fermions are
characterized by complete bilinear forms of $H^+_a$~.

The $4D$-elliptic self-adjoint differential mass operator will thus be written as follows:
\[ \MM_{R,L}=\left\{ \pm i\hbar s_0\ {\frac\partial{\partial t_0}},\pm i \ {\frac{\hbar}{c}} s_1
\ {\frac\partial{\partial x_1}},\pm i \ \frac{\hbar}{c} s_2
\ \frac\partial{\partial x_2},\pm i \ \frac{\hbar}{c} s_3
\ \frac\partial{\partial x_3}\right\}\]
where $\hbar$ and $c$ are abbreviated notations respectively for $\hbar_M$ and $c_{t\to
r;M}$~.}\end{defi}
\vskip 11pt

\begin{defi}[Second order differential bilinear equation]  { The differential  mass
bioperator will be $(\MM_{R\mu }\otimes \MM_{L\mu })$ acting on the differentiable
bifunction
$(\phi _{L_{R\mu }}(t,r)\otimes \phi _{L\mu }(t,r))$ such that the $\mu $-th mass second order
elliptic differential bilinear equation to be considered is:
\[ (\MM^2_\mu -\EE^2_\mu )(\phi _{L_{R\mu }}\cdot \phi _{L\mu })=0\]
if we take into account the self-adjointness of the right (resp. left) elliptic operator $\MM_{R\mu
}$ (resp. $\MM_{L\mu }$~).

This differential bilinear equation can be explicitly written according to:
\nobeqn
{\txt\sum\limits^3_{i,j=1} A^{ij}_\mu \ \frac{\partial^2(\phi_{L_{R\mu }}(r)\cdot
\phi_{L\mu }(r)}{\partial x_i\ \partial x_j}
\ +\  \txt\sum\limits^3_{i=1} A^{0i}_\mu \ \frac{\partial\phi_{L_{R\mu }}(t)}{\partial t}
\cdot \ \frac{\partial\phi_{L\mu }(r)}{\partial x_i}}&&\\
\noalign{\vskip 11pt}
\ + \  \txt\sum\limits^3_{i=1} A^{0i}_\mu \ \frac{\partial\phi_{L_{R\mu }}(r)}{\partial x_i}
\cdot \ \frac{\partial\phi_{L\mu }(t)}{\partial t}\ + \
 A^{00}_\mu \ \frac{\partial^2(\phi_{L_{R\mu }}(t)\cdot \phi_{L\mu }(t
))}{\partial t^2}&&\\
\noalign{\vskip 11pt}
\ - \  \EE^2_\mu (\phi_{L_{R\mu }}(t,r)\cdot \phi_{L\mu }(t,r))& = & 0\noeeqn
where 
\[
A^{ij}_\mu  = -\ \frac{\hbar^2}{c^2} s^i_\mu s^j_\mu \;,\qquad
A^{00}_\mu  = -\hbar^2s^0_\mu s^0_\mu \;,\qquad
A^{0i}_\mu  =-\ \frac{\hbar^2}{c} s^0_\mu s^i_\mu \;\cdotp\]
It is a second order elliptic differential bilinear equation which is 
degenerated because the bilinear form $(p_{R\mu }\phi_{L_{R\mu }}(r), 
p_{L\mu }\phi_{L_\mu }(r))_c$ is degenerated.

$p_{R\mu ,L_\mu }$ stands for the $3D$-linear momentum operator given by
\[ p_{R\mu,L\mu }=\left\{ -i\ \frac\hbar c s_{1\mu }\ \frac{\partial}{\partial
x_1},-i\ \frac\hbar c s_{2\mu }\ \frac{\partial}{\partial
x_2},-i\ \frac\hbar c s_{3\mu }\ \frac{\partial}{\partial
x_3}\right\} \]
(for the literature on the second order elliptic differential linear
equation, see \cite{K-N1}, \cite{K-N2}, \cite{G-I-L}).}\end{defi}
\vskip 11pt

\begin{propo} A right and a left isomorphisms transform the second order
degenerated elliptic differential bilinear equation:
\nobeqn
{\txt\sum\limits^3_{i,j=1} A^{ij}_\mu \ \frac{\partial^2(\phi_{L_{R\mu }}(r)\cdot
\phi_{L\mu }(r)}{\partial x_i\ \partial x_j}
\ +\  \txt\sum\limits^3_{i=1} A^{0i}_\mu \ \frac{\partial\phi_{L_{R\mu }}(t)}{\partial t}
\cdot \ \frac{\partial\phi_{L\mu }(r)}{\partial x_i}}&&\\
\noalign{\vskip 11pt}
\ + \ \txt\sum\limits^3_{i=1} A^{0i}_\mu \ \frac{\partial\phi_{L_{R\mu }}(r)}{\partial x_i}
\cdot \ \frac{\partial\phi_{L\mu }(t)}{\partial t}\ + \
  A^{00}_\mu \ \frac{\partial^2(\phi_{L_{R\mu }}(t)\cdot \phi_{L\mu }(t
))}{\partial t^2}&&\\
\noalign{\vskip 11pt}
\ - \ \EE^2_\mu (\phi_{L_{R\mu }}(t,r)\cdot \phi_{L\mu }(t,r))& = & 0\noeeqn
into the canonical second order elliptic-parabolic differential bilinear
equation:
\nobeqn
{\bar A^{33}_\mu \ \frac{\partial^2(\phi_{L_{R\mu }}(z)\cdot
\phi_{L\mu }(z)}{\partial z^2}
\ +\  \bar A^{03}_\mu \ \frac{\partial\phi_{L_{R\mu }}(t)}{\partial t}
\cdot \ \frac{\partial\phi_{L\mu }(z)}{\partial z}}&&\\
\noalign{\vskip 11pt}
\ + \ \bar A^{03}_\mu \ \frac{\partial\phi_{L_{R\mu }}(z)}{\partial z}
\cdot \ \frac{\partial\phi_{L\mu }(t)}{\partial t}\ + \
 \bar A^{00}_\mu \ \frac{\partial^2(\phi_{L_{R\mu }}(t)\cdot \phi_{L\mu }(t
))}{\partial t^2}&&\\
\noalign{\vskip 11pt}
\ -\ \EE^2_\mu (\phi_{L_{R\mu }}(t,z)\cdot \phi_{L\mu }(t,z))& = & 0\noeeqn
where
\[
\bar A^{33}_\mu  = \ \frac{-\hbar^2}{c^2} (s^2_1+s^2_2+s_2^3)\ \equiv\
-\ \frac{-\hbar^2}{c^2} s^2_\mu \;,\qquad
\bar A^{03}_\mu  = \ \frac{\hbar^2}{c} s_\mu \;.\]
\end{propo}
\vskip 11pt

\bpr The canonical form of the second order degenerated elliptic differential
bilinear equation is obtained for a fixed ``~$\mu $~'' throughout the following
change of variables \cite{K-S-G}:
\begin{enumerate}
\item $\displaystyle{\xi_k =\txt\sum\limits^3_{i=1} x_i\beta ^i_k}$ such that
\[ \xi_k\xi_\ell = \txt\sum\limits^3_{i,j=1} \beta ^i_kx_ix_j\beta ^j_\ell=\beta _Lx
x^T\beta ^{\dagger}_R\;,\]
with $\beta \in SO(3)$~, $x$ a $3D$-vector, $x^T$ its transposed and
such that
$B^{ij}x_ix_j$ be transformed into $\bar B^{k\ell}\xi_k\xi_\ell$~, where
\[ B^{ij} = \left( \begin{array}{ccc}
1&1&1\\ 1&1&1 \\ 1&1&1\end{array}\right) \qquad \tim{and}\qquad
\bar B^{k\ell} = \beta _LB^{ij}\beta ^{\dagger}_R\;;\]
\vskip 11pt

\item $\displaystyle{\ \frac\partial{\partial x_i} =\txt\sum\limits^3_{k=1} \ \frac\partial
{\partial \xi_k}\cdot \ \frac{\partial \xi_k}{\partial x_i} =\txt\sum\limits^3_{k=1}
\ \frac\partial{\partial \xi_k}\alpha ^k_i}$ such that
$\displaystyle{ A^{ij}\ \frac{\partial^2}{\partial x_i\ \partial x_j}}$ be
transformed into\\  $\displaystyle{\bar A^{k\ell} \ \frac\partial{\partial \xi_k}\cdot
\ \frac \partial{\partial \xi_\ell}}$
where
\nobeqn
\bar A^{k\ell} &=&\txt\sum\limits^3_{i,j=1} \alpha ^k_i A^{ij}\alpha ^\ell_j =\alpha
_L\ A \alpha ^{\dagger}_R\;,\\
\noalign{\vskip 11pt}
\bar A^{0k} &=&\txt\sum\limits^3_{i=1} A^{0i}\alpha ^k_i\;.\noeeqn
\end{enumerate}
\vskip 11pt

The transformed equation then becomes:
\nobeqn
{\txt\sum\limits^3_{k,\ell=1}\bar A^{k\ell} \ \frac{\partial^2(\phi_{L_{R
}}(\rho )\cdot \phi_{L }(\rho ))}{\partial \xi_k\ \partial \xi_\ell}
\ +\ \txt\sum\limits^3_{k=1} \bar A^{0k} \ \frac{\partial\phi_{L_{R }}(t)}{\partial t}
\cdot \ \frac{\partial\phi_{L }(\rho )}{\partial \xi_k}}&&\\
\noalign{\vskip 11pt}
\ +\ \txt\sum\limits^3_{k=1}\bar A^{k0} \ \frac{\partial\phi_{L_{R }}(\rho )}{\partial \xi_k}
\cdot \ \frac{\partial\phi_{L }(t)}{\partial t}\ + \
 \bar A^{00} \ \frac{\partial^2(\phi_{L_{R}}(t)\cdot \phi_{L }(t
))}{\partial t^2}&&\\
\noalign{\vskip 11pt}
\ -\ \EE^2 (\phi_{L_{R }}(t,\rho )\cdot \phi_{L }(t,\rho ))& = &0\noeeqn
where $\rho =\{\xi_1,\xi_2,\xi_3\}$ stands for the triple of $\xi$-spatial
variables.  The canonical second order elliptic parabolic differential
bilinear equation reduces if we remark that:
\Bena
\item $\bar B^{k\ell}=\beta _LB^{ij}\beta ^\dagger_R=E$ corresponds to a left
and a right unitary transformations, such that the eigenvalue diagonal matrix $\lambda $ has
for unique eigenvalue $e=\sqrt 3$ (this point was outlined to me by G. Raseev). 
Consequently,
$\phi_{L_R}(\rho )\cdot \phi_L(\rho )$ becomes $\phi_{L_R}(\xi_3)\cdot
\phi_L(\xi_3)$ rewritten $\psi(z)=\phi_{L_R}(z)\cdot \phi_L(z)$ if $z\equiv
\xi_3$~.
\vskip 11pt

\item $\bar A^{k\ell}=\alpha _L\ A \alpha ^\dagger_R=D$ also
corresponds to a left and a right unitary transformations such that $D$ is the
diagonal matrix whose unique eigenvalue different from zero is: 
\vskip 11pt

\hfill $a=-\ \frac{\hbar^2}{c^2}(s_1^2+s_2^2+s_3^2)=s^2\cdot \frac{\hbar^2}{c^2}\;.$\epr

\end{enumerate}
\vskip 11pt

Notice that the differential bilinear equation over the set of four variables $\{t,x_1,x_2,x_3\}$ reduces to a canonical differential bilinear equation over the set of two variables $\{t,z\}$~.  Indeed, the three-dimensional spatial section $\phi_{L_{R_\mu}}(r)$ (resp. $\phi_{L_{\mu}}(r)$~)  degenerates into the one-dimensional  function $\phi_{L_R}(z)$ (resp. $\phi_L(z)$~) justifying the decomposition of the shifted Eisenstein bicohomology into pairs of one-dimensional eigenspaces following proposition 3.2.11.
\vskip 11pt

\begin{defi}[Differential bilinear equation of the bisemielectron] \mbox{}\\{ In a
first step, we shall suppose that the time variable is constant. 
Consequently, the canonical second order elliptic-parabolic differential
bilinear equation, relative to a ``mass'' irreducible section of the bisemielectron, will
become \cite{Pie2}:
\[ -\ \frac{\hbar^2}{c^2}S^2\ \frac{\partial^2\psi(z)}{\partial z^2}
-2i\ \frac\hbar c m_0S\ \frac{\partial \psi(z)}{\partial
z}+(m_0^2-\EE^2)\psi(z)=0\;.\] 
This equation is $1$-dimensional, and is thus defined on
the left internal bilinear Hilbert space $\Hs^+_h$~.}\end{defi}
\vskip 11pt

\begin{propo}  The $1D$-mass equation for one irreducible bisection of the bisemielectron
\[ -\ \frac{\hbar^2}{c^2}S^2\ \frac{\partial^2\psi}{\partial z^2}
-2i\ \frac\hbar c m_0S\ \frac{\partial \psi}{\partial
z}+(m_0^2-\EE^2)\psi=0\] is the equation of a damped harmonic oscillator whose general
solution consists in the superposition of two damped waves in phase opposition with
frequencies given by
\[ \EE=+\ \frac{\hbar}{c} \nu S\]
and whose general motion corresponds to a damped sinusoidal motion whose
dephasage is proportional to the linear momentum $\vec p_{\mu _{R,L}}$ of the
considered section of the right or left semielectron.

The energy $\EE$ of a section $\phi _{L_\mu }(r)$ at $(p+\mu)$ quanta is equal to $(p+\mu)$-times the quantum energy $\EE^I_\mu$ which can be found from the corresponding nontrivial zero of the Riemann zeta function.\end{propo}
\vskip 11pt

\bpr This equation was already worked out elsewhere in \cite{Pie2}.  However,
we shall briefly give the following elements of the proof:
\Bena
\item Rewrite the $1D$-mass biwave equation of the bisemielectron in the form:
\[ \ \frac{d^2\psi}{dz^2} +k\ \frac{d\psi}{dz} +\omega ^2\psi=0\]
where
\[
k =  2i\ \frac{c}{\hbar} S^{-1} m_0\;,\qquad
\omega ^2 =  \ \frac{c^2}{\hbar^2}S^{-2}(\EE^2-m_0^2)\;.\]
This equation is the one of a damped harmonic oscillator.  The nature of the
solution depends on the characteristic roots \cite{Stru}:
\[ \lambda =-\ \frac k2 \pm \left[\left(\ \frac k2\right)^2-\omega
^2\right]^{1/2}\;.\]
If $\omega ^2>\left(\ \frac k2\right)^2$~, i.e. if $\EE^2>0$~, then $\omega
_1=\left[ \omega ^2-\left( \ \frac k2\right)^2\right]^{1/2}$ has to be
considered such that $\lambda =-\ \frac k2\pm i\omega _1$~.

Using 
\[ \omega ^2=\ \frac{c^2}{\hbar^2} S^{-2}(\EE^2-m_0^2)\;,\]
we find that $\omega _1=\ \frac c\hbar S^{-1}\EE$ or that $\EE=
\ \frac h c \nu
S$ if $\omega _1=2\pi\nu $~.

We then recover the famous relation of the Broglie except for the spin factor
$S$~.  Note that this formula concerns the total energy ``~$\EE$~'' of a
section of a left or right semielectron which can then be interpreted as follows:

If this section is the $\mu$-th section $s_{\mu_{R,L}}$ having $\mu_p=p+\mu$ 
quanta (see definition 1.2.17), then the energy $\EE_\mu$ of this section will be 
$\EE_\mu=\mu_p \EE^I_\mu$ where $\EE^I_\mu$ is the energy of one quantum in $s_{\mu_{R,L}}$~.  
Now, $\EE^I_\mu=h\nu_{\tau_\mu}$ where $h$ is the Planck constant associated with the degree of 
Galois extension $N$ and where $\nu_{\tau_\mu}$ is assumed to be the frequency of a prime ideal 
corresponding to one Galois automorphism. One then has $\EE_\mu=\mu_p \EE^I_\mu=\mu_p h
\nu_{\tau_\mu}=h\nu_\mu$ where $\nu_\mu=\mu_p\nu_{\tau_\mu}$ is the frequency of $\mu_p$ 
prime ideals.

Remark that it was proved in \cite{Pie10} that the energy $\EE^I_\mu$ of a quantum 
$M^I_\mu$ in a section $s_{\mu_{R,L}}$ can be obtained from the corresponding nontrivial 
zero $\lambda (4\nu ^2,i^2,\EE^I_{4\nu^2})$ 
of the Riemann zeta function $\zeta (s)=\sum\limits_n n^{-s}$~.  Indeed, the trivial zeros of $\zeta (s)$ are the negative integers $-2$~, $-4$~, \ldots, $-2\nu $~, \ldots, $-2\eta $ such that the even integer $\mu_p =2\nu $ be the global class residue degree of the section $s_{\mu_{R,L}}$ (see 1.1.4).  Now, the nontrivial zeros can be obtained from the corresponding trivial ones by the action of the Lie algebra of the decomposition group, whose coset representative is:
\[ D_{4\nu ^2,i^2}=\left[\BM 1 & i\\ 0&1\EM\ \BM 1&0\\ i & 1\EM\right]\;, \]
such that the eigenvalues of:
\[ D_{4\nu ^2,i^2}\cdot \varepsilon _{4\nu ^2}\cdot \alpha _{4\nu ^2}=\left[\BM 1 & i\\ 0&1\EM\ 
\BM 1&0\\ i & 1\EM\right] \BM E^I_{4\nu ^2} & 0\\ 0&1\EM \ \BM 4\nu ^2 & 0\\ 0&1\EM\]
be the nontrivial zeros:
\[\lambda _\pm (4\nu ^2,i^2,E^I_{4\nu ^2})=\F{1\pm i\sqrt{16\nu ^2\cdot E^I_{4\nu ^2}-1}}2=\half \pm i\gamma _{\mu _p}\;,\]
where $\gamma _{\mu _p}=(16\nu ^2(E^I_{2\nu })^2-1)^{\half}\big/2$ with $E^I_{4\nu ^2}\equiv (E^I_{2\nu })^2$~.

We thus have that:
\[ \lambda _+(4\nu ^2,i^2,E^I_{4\nu ^2})\cdot \lambda _-(4\nu ^2,i^2,E^I_{4\nu ^2})
= 4\nu ^2(E^I_{2\nu })^2=\F14+\gamma ^2_{\mu _p}\]
leading to:
\[2\nu E^I_{2\nu }\equiv\mu _pE^I_{2\nu }=\L(\F14+\gamma ^2_{\mu _p}\R)^{\half}
\simeq \gamma \quad \text{for} \quad \gamma ^2_{\mu _p}\gg\F14\;.\]
So, $\mu _pE^I_{\nu }$~, which is the energy of the $\mu $-th section $s_\mu $ or of a photon at $\mu _p$ quanta, is approximatively equal to the imaginary part $\gamma _{\mu _p}$ of the nontrivial zero $\lambda _\pm(4\nu ^2ni^2,E^I_{4\nu ^2})$~.

\item The solution of the $1D$-mass biwave equation of the bisemielectron is:
\nobeqn
\psi(z) &=&\ \frac{\psi_0}2 \exp \left(-i\ \frac c\hbar
S^{-1}m_0z\right)\\
\noalign{\vskip 11pt}
&&\left[\left(1+\ \frac{m_0}\EE\right)\exp
\left(i\ \frac c\hbar S^{-1}\EE z\right)+\left(1-\ \frac{m_0}\EE\right)\exp
\left( -i\ \frac c\hbar S^{-1}\EE z\right)\right]\noeeqn
where
\begin{itemize}
\item the Cauchy initial conditions are at $z=0$~, $\psi=\psi_0$ and
$\ \frac{d\psi}{dz}=0$~;
\item $\exp\left(-i\ \frac c\hbar S^{-1}m_0z\right)$ is a damping factor
depending on the rest mass $m_0$~.
\end{itemize}

The solution $\psi(z)$ is thus the sum of two damped waves in phase
opposition: the positive frequency wave refers to the left semielectron and
the negative frequency wave refers to the right semielectron.

This solution $\psi(z)$ can be written in the following form:
\[ \psi(z) = \psi_0\left( 1-\ \frac{m_0^2}{\EE^2}\right) \exp \left(-i\ \frac c\hbar
S^{-1}m_0z\right) \sin \left(\ \frac c\hbar S^{-1}\EE z+\tan^{-1}\ \frac
\EE{im_0}\right)\]
which corresponds to a damped sinusoidal motion of period
\[ T=\ \frac{2\pi\hbar S^{+1}}{cE}\]
and dephasage
\[ \Delta = \tan^{-1}\ \frac \EE{im_0}\simeq \tan^{-1}p\;.\]
The mass structure of a one-dimensional bisection of the bisemielectron behaves thus globally
like a damped  harmonic oscillator having a dephasage $\Delta$ proportional to the
linear momentum $\vec p$ of the $\mu $-th section of the right or left
semielectron.\epr
\end{enumerate}
\vskip 11pt

\begin{propo}  The total energy ``~$\EE_{eR,L}$~'' of a right (resp. left)
semielectron is given by
\[ \EE_{eR,L}=\txt\sum\limits^q_{\mu =1}\txt\sum\limits_{m_\mu} \EE_{\mu,{m_{\mu eR,L}}}\]
where $\EE_{\mu,m_{\mu eR,L}}$~, noted $\EE$ above, is the energy of a one-dimensional irreducible subsection of the $\mu $-th
section.  The total linear momentum $p_{eR,L}$ is similarly given by
\[ p_{eR,L}=\txt\sum\limits^q_{\mu =1} \txt\sum\limits_{m_\mu}  p_{\mu,m_{\mu eR,L}}\;.\]\end{propo}
\vskip 11pt

\begin{defi}[Generic biconnexion]  { By a generic connexion of a fiber
bundle  $T_{M_{R;L}}$~, we mean a distribution which admits in each point $P$ of the
total space of the fibration of $T_{M_{R;L}}$ an
horizontal direction tangent to this total space, and which is transversal to the fiber in
$P$~.}\end{defi}
\vskip 11pt

We shall consider a generic connexion associated to the mass vertical tangent
semibundle
$T^{(1-3)}_{M_{R;L}}$ and generated by the
$4D$-multiplicative operator:
\[ \partial A(t,r)_{R,L}=\left\{ +\hbar Ze\ \frac 1 {t_0},+\ \frac\hbar c Ze\ \frac
1{x_1},+\ \frac\hbar c Ze\ \frac
1{x_2},+\ \frac\hbar c Ze\ \frac
1{x_3}\right\}\]
corresponding to the action of a strong external (super) heavy nuclear system
of charge $Ze$~.

If the time variable is supposed to be constant, the mass bioperator of the
bisemielectron $(\MM_R,\MM_L)_C$~, endowed with the generic biconnexion $(\partial
A(r)_R,\partial A(r)_L)_C$~, will be:
\nobeqn
\lefteqn{(\MM_R+\partial A_R,\MM_L+\partial A_L)_C}\\
\noalign{\vskip 11pt}
&=& \left(\left\{m_0,-i\ \frac \hbar c s_1\ \frac\partial{\partial x_1}+\ \frac
\hbar cZe\ \frac 1{x_1},-i\ \frac \hbar c s_2\ \frac\partial{\partial x_2}+\ \frac
\hbar cZe\ \frac 1{x_2},-i\ \frac \hbar c s_3\ \frac\partial{\partial x_3}+\ \frac
\hbar cZe\ \frac 1{x_3}\right\},\right.\\
\noalign{\vskip 11pt}
&& \left.\left\{m_0,-i\ \frac \hbar c s_1\ \frac\partial{\partial x_1}+\ \frac
\hbar cZe\ \frac 1{x_1},-i\ \frac \hbar c s_2\ \frac\partial{\partial x_2}+\ \frac
\hbar cZe\ \frac 1{x_2},-i\ \frac \hbar c s_3\ \frac\partial{\partial x_3}+\ \frac
\hbar cZe\ \frac 1{x_3}\right\}\right)_C\noeeqn
where $(\cdot,\cdot)_C$ is a complete bilinear form of the left extended
internal bilinear Hilbert space $H^+_h$~.
\vskip 11pt

\begin{coro}  The mass equation of the bisemielectron endowed
with the ``strong'' biconnexion\linebreak $(\partial A_R,\partial A_L)_C$ is  a second
order degenerated elliptic differential bi(linear) equation whose canonical
form has a set of particular solutions obtained with the condition
\[ m_0^2-\ \frac 1{2^n} (m_0^2-\EE^2)=0\;,\qquad n\in \nit\;,\]
or $\EE=\sqrt{2^n-1}m_0$ which allows to find the energy levels of the right or
left semielectron in the strongly perturbated confining phase.
\end{coro}
\vskip 11pt

\bpr The treatment of the mass equation of the bisemielectron endowed with the
biconnexion\linebreak $(\partial A_R,\partial A_L)_C$ was developed in \cite{Pie2}.

Let us note that the first calculated energy levels of the (semi)electron given
by the formula
\[ \EE_{eR,L}=\sqrt{2^n-1}\ m_{0_{eR,L}}\]
corresponds quite well to the observed values.\epr
\vskip 11pt

\begin{defi}[{\boldmath $S_L$-isomorphism}]  { Instead of considering that the
bisemielectron  mass structure is defined on the left extended internal Hilbert
space $H^+_a$ as done since definition 4.2.1, we could consider that it is
described by the sum of the three bisemisheaves, according to definition 4.1.1:
\[ \left( \theta ^{1-3}_R(t,r)_M\otimes_D \theta ^{1-3}_L(t,r)_M\right)\oplus
\left( \theta ^{3}_R(r)^{(m)}_M\otimes_m \theta ^{3}_L(r)^{(m)}_M\right)\oplus
\left( \theta ^{1-(3)}_R(t,(r))^{(e)}_M\otimes_e \theta
^{3-(1)}_L((t),r)^{(e)}_M\right)\]
obtained by application of the $S_L$-isomorphism on the complete tensor
product  $
\theta ^{1-3}_R(t,r)_M\otimes \theta ^{1-3}_L(t,r)_M$ between the right
semielectron  semisheaf $\theta ^{1-3}_R(t,r)_M$ and the left
semielectron  semisheaf $ \theta ^{1-3}_L(t,r)_M$~.}\end{defi}
\vskip 11pt

According to definitions 3.1.5 and 3.1.7, this sum of three bisemisheaves can
be transformed by means of the $B_L\circ p_L$ map into:
\[ \left( \theta ^{1-3}_{L_R}(t,r)_M\otimes_D \theta
^{1-3}_L(t,r)_M\right)\oplus \left( \theta ^{3}_{L_R}(r)^{(m)}_M\otimes_m \theta
^{3}_L(r)^{(m)}_M\right)\oplus \left( \theta
^{1-(3)}_{L_R}(t,(r))^{(e)}_M\otimes_e \theta ^{3-(1)}_L((t),r)^{(e)}_M\right)\]
such that $\left( \theta ^{1-3}_{L_R}(t,r)_M\otimes_D \theta
^{1-3}_L(t,r)_M\right)$ be defined on the left internal bilinear Hilbert space\linebreak
$\Hs^+_a$~, $\left( \theta ^{3}_{L_R}(r)^{(m)}_M\otimes_m \theta
^{3}_L(r)^{(m)}_M\right)$ on the left internal bilinear magnetic space $V^+_{m;a}$
and $\left( \theta
^{1-(3)}_{L_R}(t,(r))^{(e)}_M\otimes_e\R.$\linebreak $\L.\theta
^{3-(1)}_L((t),r)^{(e)}_M\right)$ on the left internal bilinear electric space
$V^+_{e;a}$~.
\vskip 11pt

\begin{defi}[Bilinear diagonal, magnetic and electric wave equations] \mbox{}\\{ 
We shall then obtain a set of three second order elliptic differential
bilinear equations:
\begin{enumerate}
\item a central mass biwave equation:
\[ \txt\sum\limits^3_{i=0} A^{ii} \ \frac{\partial^2\psi(t,r)}{\partial
x_i^2}-\EE_D^2\psi(t,r)=0\]
where
\begin{itemize}
\item $\displaystyle{A^{ii}=-\ \frac{\hbar^2}{c^2}s^is^i\;,\; 1\le i\le 3\;;
\quad A^{00}=-\hbar^2s^0s^0}$~;
\vskip 11pt

\item $\displaystyle{\psi(t,r)=\phi_{L_R}(t,r)\otimes_D\phi_L(t,r)\in \theta
^{1-3}_{L_R}(t,r)_M\otimes_D\theta ^{1-3}_L(t,r)_M}$ is a diagonal bisection
such that $\psi(t,r)$ be defined on a compact euclidian domain of
$\rit^4$ of the left internal Hilbert space
$\Hs^+_h$~.  
\end{itemize}

This equation corresponds to the Klein-Gordon equation except that the metric
$\delta ^i_i$ is euclidian and not pseudo-euclidian or of Minkowsky type.
\vskip 11pt

\item a bilinear magnetic mass biwave equation:
\[ \txt\sum\limits^3_{{i,j=1}\atop{i\neq j,i>j}} A^{ij} \ \frac{\partial
\phi_{L_R}(r)^{(m)}}{\partial x_i}\cdot  \ \frac{\partial
\phi_{L}(r)^{(m)}}{\partial x_j}-\mu (\phi_{L_R}(r)^{(m)}\cdot
\phi_L(r)^{(m)})=0\]
where
\begin{itemize}
\item $\phi_L(r)^{(m)}$ (resp. $\phi_{L_R}(r)^{(m)}$~) refers to a left (resp.
right) magnetic section of $\theta ^3_L(r)^{(m)}_M$ (resp. of
$\theta ^3_{L_R}(r)^{(m)}_M$~);
\vskip 11pt

\item $\displaystyle{A^{ij}=-\ \frac{\hbar^2}{c^2} s^is^j}$~, $i\neq j$~;
\vskip 11pt

\item $\mu $ refers to the magnetic moment of the bisemielectron.
\end{itemize}
\vskip 11pt

\item a bilinear electric mass biwave equation:
\[ \txt\sum\limits^3_{i=1} A^{i0} \ \frac{\partial
\phi_{L_R}(r)^{(e)}}{\partial x_i}\cdot  \ \frac{\partial
\phi_{L}(t)^{(e)}}{\partial t}-e (\phi_{L_R}(r)^{(e)}\cdot
\phi_L(t)^{(e)})=0\]
where
\begin{itemize}
\item $\phi_L(t)^{(e)}$ (resp. $\phi_{L_R}(r)^{(e)}$~) refers to a left (resp.
right) electric time (resp. space) section of $\theta ^1_L(t)^{(e)}_M$ (resp. of $\theta
^3_{L_R}(r)^{(e)}_M$~);
\vskip 11pt

\item $\displaystyle{A^{i0}=-\ \frac{\hbar^2} c s^i s^0}$~;
\vskip 11pt

\item $e$ corresponds to the electric charge of the bisemielectron.
\end{itemize}
\end{enumerate}}\end{defi}
\vskip 11pt

\begin{propo} If the elliptic differential mass biwave equation of the
bisemielectron as given in definition 4.2.3 splits into the set of three
elliptic differential equations:
\begin{enumerate}
\item a central mass biwave equation;
\item a bilinear magnetic mass biwave equation;
\item a bilinear electric mass biwave equation,
\end{enumerate}
the general bilinear solution of the elliptic differential mass biwave
equation is given by the sum of the solutions of the three split elliptic
differential equations, i.e. by:
\[ \phi_{L_R}(t,r)\cdot \phi_L(t,r)=\psi(t,r)+\phi_{L_R}(r)^{(m)}\cdot
\phi_L(r)^{(m)}+\phi_{L_R}(r)^{(e)}\cdot
\phi_L(t)^{(e)}\]
where
\begin{itemize}
\item $\psi(t,r)$ is defined on the left internal bilinear Hilbert space
$\Hs^+_h$~;
\item $\phi_{L_R}(r)^{(m)}\cdot
\phi_L(r)^{(m)}$ is defined on the left internal bilinear 
magnetic space $V^+_{m;a}$~;
\item $\phi_{L_R}(r)^{(e)}\cdot
\phi_L(t)^{(e)}$ is defined on the left internal bilinear  
electric space $V^+_{e;a}$~.
\end{itemize}
\end{propo}
\vskip 11pt

\paragraph{Outline of the proof} : The fact that the general bilinear solution
of the elliptic differential mass biwave equation of the bisemielectron (see
definition 4.2.3)
\[ (\MM^2-E^2)\phi_{L_R}(t,r)\cdot \phi_L(t,r)=0\]
can be developed as the sum of the solutions of the three split bilinear
elliptic differential equations results from the $S_L$-isomorphism according
to definition 4.1.1.\epr
\vskip 11pt

\begin{defi}[Bisemiphoton wave equation] { As the bisemiphoton mass
structure is given on the left extended internal bilinear Hilbert space
$H^+_a$ by a bisection isomorphic to $T ^1_{L_R}(r)_M\otimes T
^1_L(r)_M$ according to definition 4.1.1, the elliptic differential mass
biwave equation of the mass structure of a bisemiphoton at $(p+\mu)$ biquanta $\widetilde  M^I_\mu(r)_R\otimes \widetilde  M^I_\mu(r)_L$
will be:
\[ \txt\sum\limits^3_{i,j=1} A^{ij}\ \frac{\partial^2(\phi_{L_R}(r)\cdot
\phi_L(r))}{\partial x_i\ \partial x_j} -E^2_\mu(\phi_{L_R}(r)\cdot \phi_L(r))=0\]
where
\Bi
\item \quad $\phi_{L_R}(r)\cdot
\phi_L(r)\simeq T^1_{L_R,\mu}(r)_M\times T^1_{L,\mu}(r)_M$~,
\item \quad $\ds A^{ij}=-\ \frac{\hbar^2}{c^2}s^is^j$~,
\Ei
so that $s^i$ is the $i$-th component of a $3D$ unit vector of polarization of
the semiphoton referring to the two possible different rotations of its sections.

This equation is a second order differential elliptic-parabolic bilinear
equation which is degenerated.}\end{defi}
\vskip 11pt

\begin{defi}[Canonical wave equation of the bisemiphoton]  { A right and a
left unitary inner automorphisms transform the degenerated second order
differential elliptic bilinear equation of the bisemiphoton into the
$1D$-canonical second order elliptic differential equation:
\[ \bar A^{33} \ \frac{\partial^2(\phi_{L_R}(z)\cdot \phi_L(z))}{\partial z^2}
-E_\mu^2 (\phi_{L_R}(z)\cdot \phi_L(z))=0\]
where
\[ A^{33} =-\ \frac{\hbar^2}{c^2} (s_1^2+s_2^2+s_3^2)
=-\ \frac{\hbar^2}{c^2}S^2\;,\qquad z\equiv x_3\;.\]
This is the equation of an harmonic oscillator:
\[ \ \frac{\partial^2\psi(z)}{\partial z^2} +\omega_\mu ^2\psi(z)=0\]
where
\[ \omega_\mu ^2=\ \frac{c^2}{\hbar^2}S^{-2}E_\mu^2\quad \tim{and} \quad \psi(z) =
\phi_{L_R}(z)\cdot \phi_L(z)\;.\]

The general solution of the harmonic oscillator equation of the bisemiphoton
consists in the superposition of two waves in phase opposition having
frequencies given by
\[ \omega_\mu =\ \frac c\hbar S^{-1} E_\mu\]
leading to the well-known relation of Einstein:
\[ E_\mu=\ \frac hc \nu_\mu S\;, \]
excepting the factor $S$~, where $\nu _\mu $ is the frequency of $\mu_p$ prime ideals as introduced in proposition 4.2.6.}\end{defi}
\vskip 11pt

This general solution $\psi(z)$ is explicitly given by:
\nobeqn
\psi(z) &=& c_1 \exp \left( i\ \frac c\hbar S^{-1}E_\mu z\right)+
c_2\exp \left(-i\ \frac c\hbar S^{-1}E_\mu z\right)\\
\noalign{\vskip 11pt}
&=& A\sin\left(\ \frac c\hbar S^{-1}E_\mu z+\delta \right)\ = \ A\sin (\omega_\mu 
z+\delta )\noeeqn
corresponding to a sinusoidal motion.

\begin{defi} The solution $\psi(z)$ is a linear combination of two one-dimensional waves corresponding to the one-dimensional irreducible components of the bisemiphoton.  The coefficients $c_1$ and $c_2$ allow to define the radii of the tori $\exp \left( \pm i\ \frac c\hbar S^{-1}E_\mu z\right)$ according to proposition 1.1.18.

Assume that  $\psi(z)$ has $p_\mu=(p+\mu)$ biquanta.  Then, the limit condition gives in $z\equiv a=$ radius:
\[\psi(z) \simeq A\sin (\omega_\mu z)=0\]
whose solution can be obtained only if $\omega_\mu a=\mu \pi$~, i.e. if $\omega_\mu =\frac{\mu \pi}a$~.\end{defi}
\vskip 11pt

\begin{defi}[Kinetic energy of a bisemilepton] { Consider now the
central  mass diagonal bioperator $(\MM_{L_R},\MM_L)=\MM^2_D$ of the central mass biwave
equation of a bisemilepton defined on the internal Hilbert space $\Hs^+_h$ (see
definition 4.2.11). It is:
\[ \MM^2_D=m_0^2+\txt\sum\limits^3_{i=1}p_i^2\]
if we do not take the spin vector into account.

The norm of the central mass $M_D$ of a bisemilepton is then:
\[ \|\MM_D\| =\|E^2\|=m_0\left( 1+\ \frac 1{m_0^2}\txt\sum\limits^3_{i=1}p_i^2\right)^{1/2} \simeq m_0+\omega \]
if $\omega \ll m_0$~.  We thus have that:
\[ E^2 =m_0^2+\txt\sum\limits^3_{i=1}p_i^2\simeq m_0^2+2m_0\omega +\omega ^2\]
which implies that:
\[ \omega =\ \frac 1{2m_0}\txt\sum\limits^3_{i=1}p_i^2\;.\]
This term is the kinetic energy of the semilepton and corresponds to the
harmiltonian of the Schr\"odinger equation in a zero potential.}\end{defi}
\vskip 11pt

The following proposition results from this.
\vskip 11pt

\begin{propo}  The kinetic energy of a right or left semilepton is equal to
the normed (by the factor $\ \frac 1{2m_0}$~) inner product between the
linear momenta of the right and left semileptons.\end{propo}
\vskip 11pt

\bpr Indeed, the inner product $(p_R,p_L)$ between the linear momenta of the
right and left semileptons is $\sum\limits^3_{i=1}p_i^2$~.  Thus, this inner
product $(p_R,p_L)$ normed by the factor $\ \frac 1{2m_0}$ corresponds to the
kinetic energy of a semilepton $\omega =\ \frac 1{2m_0}(p_R,p_L)$~.\epr
\vskip 11pt

\begin{defi}[The concept of field in algebraic quantum theory] {
The nature of the field in quantum field theory proceeds from the treatment of the harmonic oscillator in classical mechanisms.  It is considered that an infinite number of harmonic oscillators brings us to a field theory with the field at each point of space considered as independent generalized coordinate and that the field quantization results from the quantization of an infinite assemblage of harmonic oscillators \cite{B-D}.

In this model of algebraic quantum theory, each pair of right and left one-dimensional sections, isomorphic to one-dimensional tori, can be considered as a (damped) harmonic oscillator according to propositions 4.2.6, 4.2.12 and 4.2.14.  Let us recall that each elementary bisemiparticle (bisemifermion or bisemiboson) has a central (i.e. diagonal) ``mass'' spatial structure composed of pairs of right and left one-dimensional sections (which are in fact one-dimensional waves or strings) of the mass bisemisheaf $(\theta ^3_{L_R}(r)_M\otimes_D\theta ^3_L(r)_M)$~, the electric charge and the magnetic moment being also composed of pairs of one-dimensional sections characterized by an electric and a magnetic metric as developed in proposition 3.1.6.

Thus, every elementary bisemiparticle has a central ``mass'' spatial structure given by a corresponding field 
$\theta ^3_{L_R}(r)_M\otimes_D \theta ^3_L(r)_M$ 
behaving like a sum of independent harmonic oscillators if we refer for example to the definition of the wave (bi)function of an elementary particle having a spectral decomposition of algebraic type as introduced in proposition 3.3.5.  By this way, we recover the classical concept of field of the quantum theories \cite{Wein2}.

However, let us note that the electron field is given in quantum field theory by
\[ \psi(x)=\txt\sum\limits_ku_k(\vec x)e^{-i\omega_kt}a_k\]
where $\{u_k(\vec x)e^{-i\omega_kt}\}$ is a set of orthonormal plane-wave solutions of Dirac equation and where the $a_k$ are annihilation operators.  The annihilation and creation operators were defined in this algebraic quantum model as morphisms of type $\gamma^M_{t\to r}\circ E\circ SOT(2)\circ Vd(2)$ (see remark 2.3.7) generating the ``mass'' semisheaves of rings from the vacuum composed of the internal semisheaves of rings ``~$ST$~" and ``~$MG$~''.

In quantum field theory, one is dealing with a system of an infinite number of degrees of freedom, leading to a nonseparable Hilbert space \cite{Wigh1}.  This refers to the old problem of quantum mechanics consisting in the difference between its discontinued space ``~$Z$~'' (~$=1,2,\cdots$~) of discrete values of the index $\mu$ and its configuration space ``~$\Omega$~'' which is continued with $k$ dimensions where $k$ is the number of degrees of freedom of the system \cite{V.Neu1}.

If we consider the algebraic spectral decomposition of the wave bifunction in terms of pairs of one-dimensional sections which correspond to the degrees of freedom of the envisaged system, it appears that the spaces ``~$Z$~'' and ``~$\Omega$~'' are in one-to-one correspondence since:
\Bi
\item the global class residue degree $f_\mu$ of an irreducible one-dimensional section at $\mu_p$ quanta is a degree of freedom of the system;
\item the index $\mu$ of the configuration space ``~$\Omega $~'' is the integer $\mu$ labelling the $\mu$-th eigenbifunction $\psi _{L_R}(\mu)\otimes \psi _L(\mu)\in\MM^h_{R\times L}(\Hs^+_h\{\mu\})$ (see 3.2.18 and 3.2.19).
\Ei 
This brings a new light on the Hilbert spaces of infinite dimensions of quantum field theory since the spectral decomposition of a wave (bi)function can have an infinite dimension.}
\end{defi}
\vskip 11pt

\begin{defi}[Invariance of elements of bisemimodules]  { We shall now
envisage the invariance of the space-time structure of bisemifermions and more
particularly of bisemileptons.  Referring to definition 4.1.1, the
space-time left bisemisheaf of a bisemilepton defined on the left extended
internal bilinear Hilbert space $H^+_a$ splits under the $S_L$ isomorphism into
the set of the three disconnected left bisemisheaves:
\vskip 6pt

\Bena
\item the $4D$-space-time diagonal left bisemisheaf $\theta
^{1-3}_{L_R}(t,r)\otimes_D \theta ^{1-3}_L(t,r)$ whose elements are the
diagonal bielements $\phi_{L_R}\otimes_D\phi_L\in \Hs^+_a$ characterized by a
$4D$(-euclidian) metric of type $\delta ^\alpha _\alpha $~, $0\le \alpha \le 3$~.

These bielements are invariant under a (bi)representation of $SO(4,\rit)\times
SO(4,\rit)$~.

Note that $SO(n,\rit)$ is the orthogonal  unimodular group of order $n$
acting linearly on a left or right $n$-dimensional semimodule.
\vskip 11pt

\item the $3D$-space magnetic left bisemisheaf $\theta
^3_{L_R}(r)^{(m)}\otimes_m \theta ^3_L(r)^{(m)}$ whose bielements are
characterized by a $3D$ metric of type $g_{\alpha \beta }$~,
$1\le \alpha $~, $\beta \le 3$~, $\alpha \ge \beta $~, $\alpha \neq \beta $~.

The magnetic bielements are then invariant under a representation of
$SL(3,\rit)\times SL(3,\rit)$ where $SL(3,\rit)$ is the unimodular special
linear group of order 3~.

The group of left or right magnetic invariance is $SL(3,\rit)$ because, if
$g_m\in SL(3,\rit)$~, then the magnetic invariance condition is
$(g^T_m)_R(g_m)_L=h_m$ where
\[ h_m=\left( \begin{array}{ccc} 0&1&0 \\ 0&0&1\\ 1&0&0\end{array}\right)
\quad\tim{or}\quad 
h_m=\left( \begin{array}{ccc} 0&0&1 \\ 1&0&0\\ 0&1&0\end{array}\right)\]
has determinant equal to one.
\vskip 11pt

\item the $3D$-space-time electric left bisemisheaf
\[ \theta ^{1-(3)}_{L_R}(t,(r))^{(e)}\otimes_e \theta ^{(1)-3}_L((t),r)^{(e)}\]
whose bielements are characterized by a $3D$  metric of type
$g_{0\alpha }$ or $g_{\alpha 0}$~, $1\le \alpha \le 3$~.

The electric bielements are then invariant under a representation of
$SL(1,\rit)\times SL(3,\rit)$ corresponding to a $1D$-right (resp. $3D$-left)
translation on the right (resp. left) semisheaf and to a $3D$-left (resp.
$1D$-right) translation on the left (resp. right) semisheaf.
\end{enumerate}}\end{defi}
\vskip 11pt

Note that the space-structure of a bisemiphoton defined on $H^+_a$ is given by
the complete bisection $T ^1_{L_R}(r)\otimes T ^1_L(r)$ which
splits under the $S_L$-isomorphism into the sum of the two disconnected
bisections:
\vskip 6pt

\Bena
\item the $3D$-space diagonal bisection $T ^1_{L_R}(r)\otimes_D T
^1_L(r)$ whose diagonal bielements are invariant under a representation of
$SO(3,\rit)\times SO(3,\rit)$~;
\vskip 11pt

\item the $3D$-space magnetic bisection $T
^1_{L_R}(r)^{(m)}\otimes_m T ^1_L(r)^{(m)}$ whose bielements are invariant
under a representation of $SL(3,\rit)\times SL(3,\rit)$~.
\end{enumerate}
\vskip 11pt

\begin{propo}  The $4D$ (resp. $3D$~) diagonal bielements of a $4D$ (resp.
$3D$~) diagonal left bisemisheaf defined on a bilinear internal Hilbert space
$\Hs^+_a$ are invariant under a right and a left action of $SO(4,\rit)\times
SO(4,\rit)$ (resp.  of $SO(3,\rit)\times
SO(3,\rit)$~) which correspond to a right and a left inner automorphism.

The $3D$ magnetic bielements of a $3D$  magnetic bisemisheaf are
invariant under a right and a left action of $SL(3,\rit)\times SL(3,\rit)$~.

The $3D$ electric bielements of a $3D$  electric bisemisheaf
are invariant under a right (resp. left) and a left (resp. right) action of 
 $SL(1,\rit)\times
SL(3,\rit)$ (resp. of  $SL(3,\rit)\times
SL(1,\rit)$~).
\end{propo}
\vskip 11pt

\subsection{The bidynamics of bisemiparticles}

\begin{defi}[Bidynamics and bisemiflow] { Let $\theta
^{1-3}_R(t,r)\otimes_{D,m,e} \theta ^{1-3}_L(t,r)$ be  the diagonal, magnetic
or electric tensor product between the  right semisheaf 
$\theta ^{1-3}_R(t,r)$~, referring to a right semiparticle, and the
 left semisheaf $\theta ^{1-3}_L(t,r)$~, referring to a left
semiparticle.

The dynamics of this bisemisheaf is  a bidynamics corresponding to right and
left diffeomorphisms at the one parameter time ``~$t$~'' applied respectively
to the envisaged  right and  left semisheaves.  We are then
led to define a local bisemisheaf as follows:
\vskip 11pt

Let $\Gamma(\theta ^1_{R,L}(t))$ and $\Gamma(\theta ^3_{R,L}(r))$ be the
right or left $1D$ and $3D$ sections of $\theta ^1_{R,L}(t)$ and $\theta
^3_{R,L}(r)$ above respectively $1D$-time and $3D$-space domains.

By a local bisemiflow for a densely defined self-adjoint bioperator, for
example,
\[ p_R\otimes_Dp_L =\left\{ +i\ \frac \hbar c s_1\ \frac\partial{\partial x_1},
\cdots,  +i\ \frac \hbar c s_3\ \frac\partial{\partial x_3}\right\}
\otimes_D
\left\{ -i\ \frac \hbar c s_1\ \frac\partial{\partial x_1},
\cdots,  -i\ \frac \hbar c s_3\ \frac\partial{\partial x_3}\right\}\]
acting on $\Gamma(\theta ^3_R(r))\otimes_D\Gamma(\theta ^3_L(r))$~, we mean a
bijective bilinear map:
\[ F(p_R)_{-t}\otimes_D F(p_L)_{+t} :\Gamma(\theta
^3_R(r))\otimes_D\Gamma(\theta ^3_L(r))\to \Gamma(\theta
^3_R(r))\otimes_D\Gamma(\theta ^3_L(r))\]
such that
\[ F(p_R)_{-(t_1+t_2)}\otimes_D F(p_L)_{+(t_1+t_2)} =
F(p_R)_{-t_1}\cdot F(p_R)_{-t_2}\otimes_D
F(p_L)_{+t_1}\cdot F(p_L)_{+t_2}\]
be a geodesic bisemiflow corresponding to a ``time'' translation on the right
and the left $1D$-sections of
$\theta ^3_R(r)$ and $\theta ^3_L(r)$ \cite{Lan2}, \cite{Sma1}.

A bivector field with domain of $\Gamma(\theta ^3_R(r))\otimes_D \Gamma(\theta
^3_L(r))$ is a bimap:
\[ p_R\otimes_D p_L :\Gamma(\theta ^3_R(r))\otimes_D\Gamma(\theta ^3_L(r))
\to \Gamma(T^{(3)}_R(\theta ^3_R(r)))\otimes_D\Gamma(T^{(3)}_L(\theta ^3_L(r)))\]
into the diagonal tangent bibundles $(T^{(3)}_R\otimes_D T^{(3)}_L)$~.

We can thus generalize the Stone theorem \cite{C-M} in the case of a local
bisemiflow.}\end{defi}
\vskip 11pt

\begin{propo} The local bisemiflow $F(O_R)_{-t}\otimes_{D,m,e} F(O_L)_{+t}$ 
for a self-adjoint bioperator $O_R\otimes_{D,m,e}O_L$ is
given by the right and left actions $U_R(-t)\otimes_{D,m,e} U_L(+t)$ of the
continuous one-parameter unitary Lie group such that:
\[ U_R(-t)\otimes_{D,m,e} U_L(+t)=e^{-itO_R}\otimes_{D,m,e}
e^{+itO_L}\quad \tim{for} t\in\rit^+\;.\]
\end{propo}
\vskip 11pt

\bpr This proposition generalizes the Stone theorem in the case of 	 a
bisemiflow.   Remark that $O_R$ is a right operator semibounded from above
and $O_L$ is a left operator semibounded from below.

The bigenerator $O_R\otimes_{D,m,e} O_L$ of the local bisemiflow
$F(O_R)_{-t}\otimes_{D,m,e} F(O_L)_{+t}$ is given by:
\[ O_R\otimes_{D,m,e}O_L = \left.\ \frac d{dt} F(O_R)_{-t}\right|_{t=0}
\otimes_{D,m,e} \left.\ \frac d{dt} F(O_L)_{+t}\right|_{t=0}\;.\]
We then have for the bisemimodule $\theta _R(-t)\otimes_{D,m,e}\theta
_L(+t)$~:
\[ \theta _R(-t)\otimes_{D,m,e}\theta
_L(+t) =F(O_R)_{-t}\theta _R(0) \otimes_{D,m,e}F(O_L)_{+t}\theta _L(0)\;.\]\epr
\vskip 11pt

\begin{rms} {\bf Evolution of bisemiparticles and the classical symplectic 
structure}  { 
\Bena
\item This bidynamics is a bidiffeomorphism with respect to the time variable
describing the (bi)evolution of bisemiparticles from an initial event
localized at the time $T=0$~. This initial event may be the big-bang of
physics.

Consequently, the bisemiflow should take into account this delayed
(bi)evolution by a parameter (or a constant following the traditional
terminology) depending on time, called $B(\pm t)$ and related likely to the
Hubble constant $H(\pm t)$~, such that:
\[ U(-t)\otimes_{D,m,e} U(+t)=e^{-itB(-t)O_R} \otimes_{D,m,e}
e^{+itB(+t)O_L}\;.\]

\item This bidynamics must be envisaged simultaneously on the three embedded
structures ``space-time'', ``middle-ground'' and ``mass'' of the bisemiparticles
giving their irreversible evolution in time.

It can happen that, under some external perturbation during a small interval
of time $dt$~, there are fixed points on the bisemiparticle structures ``~$ST$~,
``~$MG$~'' or ``~$M$~'' with respect to the bidiffeomorphism at the one
parameter time group.  These fixed points then correspond to degenerated
singularities of genotype attractors, problem which was developed elsewhere
\cite{Pie8}.
\vskip 11pt

\item It is natural to regard the complex Hilbert space as the analog of the
cotangent bundle in a classical system \cite{Sim}.  Endowed with a symplectic 
form which is the imaginary part of the inner product, we get a symplectic 
structure \cite{God}, \cite{Sha}, \cite{Lich}, from which the classical (and
quantum) Lagrangian and hamiltonian dynamics proceed.

Let us note that the (bi)dynamics, developed in this paper, does not result
from a Lagrangian or an hamiltonian method which presents the following
difficulties:
\begin{enumerate}
\item The Lagrangian action is not well understood in mathematics \cite{C-M}
likely because the ``~$q$~'' and ``~$p$~'' variables are inextricably mixed
\cite{Tho3}.  This could be explained by the fact that classical and quantum
theories ``work'' on a single structure or level while, in the present $AQT$~,
three embedded structures have been taken into account leading to a
noncommutative algebra \cite{Gui}, \cite{Con1}.
\vskip 11pt

\item The automorphisms of the symplectic  structure generate a space of
infinite dimension \cite{Kib} while the automorphisms of the euclidian
structure leads to a space of finite dimension.
\vskip 11pt

\item The second order elliptic linear Laplace equation
$\ \frac{\partial^2u}{\partial t^2}+\Delta u=0$ leads to no really decent Banach
space on which this equation generates a flow and an energy of definite sign
\cite{C-M}, which is not the case here.
\end{enumerate}
\vskip 11pt

\item It is commonly admitted that the classical mechanics is the limit case
of the quantum mechanics providing that the Planck constant satisfies $\hbar
\to 0$~.  Now, quantum theories deal with the discontinued behavior of the
matter while the classical theories are only concerned with continuous
objects.  As the discontinued behavior of matter is described here
algebraically, it becomes evident that the structure of the quantum theories
must be described by coherent algebraic sheaves of rings while the structure
of classical theories could refer to coherent analytic sheaves, i.e. ring of
(germs of) holomorphic functions.  It then results that the isomorphism from
quantum theories to classical theories given by the condition $\hbar\to 0$
corresponds likely to the isomorphism of J.P. Serre between coherent algebraic
sheaves and coherent analytic sheaves \cite{Ser7}.

Indeed, according to definition 1.1.24, a quantum $\widetilde M^I_{\mu \RL}$ is  a continuous subfunction of the algebraic semisheaf of rings $\theta ^1\RL$ over a big point centered on $M^I_{\mu \RL}$~.  So, when $\hbar \to 0$~, the ``big point'' on which is centered $M^I_{\mu \RL}$ and to which the Zariski topology corresponds tends to an ``ordinary point'' associated with the ordinary finer topology.  Consequently, the algebraic semisheaf of rings $\theta ^1\RL$ is transformed into an analytic sheaf.
\end{enumerate}}\end{rms}
\vskip 11pt

\begin{defi}[Physical internal machinery of a bisemiparticle] { The
internal  machinery of a bisemiparticle allows to justify the absorption and
the emission of right and left quanta from the space-time, middle-ground or
mass bisemisheaves of rings, noted in abbreviated form $\theta ^{1-3}_R\otimes
\theta ^{1-3}_L$~.  According to definition 2.2.12, the emission of right or
left quanta occurs by means of the smooth endomorphism $E_{R,L}$~.

The considered bisemiparticle will be a bisemilepton for simplicity and the
following developments will be envisaged for the spatial bistructure $\theta
^3_R\otimes \theta ^3_L$~.

It was proved in proposition 3.3.11 that the right and left semiparticles
rotate in opposite senses.  This means that each spatial bisection of the 
``~$ST$~,
``~$MG$~'' or ``~$M$~'' bisemisheaf of rings $\theta
^3_R\otimes_D \theta ^3_L$~, and thus that the ``~$ST$~,
``~$MG$~'' or ``~$M$~'' spatial bistructure, behaves globally like two adjacent
gyroscopes having opposite torques $\tau_R$ and $\tau_L$ so that the right and
left torques are defined at the points $P_{R,L}\in\theta ^3_{R,L}$ by
$\tau_{R,L}=\ \frac{dL_{R,L}}{dt}$ where $L_{R,L}$ is the right (resp. left)
angular momentum.

Now, it is well-known that the centripetal force 
\[F_{p_{R,L}}=-\ \frac{mv^2_J}r\]
acting on a point, having a linear momentum
\[ p_{R,L}=(mv)_{R,L}=(m\omega  r)_{R,L}\;,\]
can decompose into the sum of the three forces \cite{F-L-S}:
\[ F_{p_{R,L}}=-\ \frac{mv_J^2}r =-\ \frac{mv_M^2}r-2mv_M\omega -m\omega ^2r\]
where
\begin{enumerate}
\item the velocity $v_J$ of the rotating point $P_{R,L}$ is the sum of the
rotational velocity $v_M$ and of an additional angular velocity $\omega
r$~:
\[ v_J=v_M+\omega  r\;;\]
\item $r$ is the distance from the point $P_{R,L}$ to the emergence point of
the bisemiparticle;
\item $F_D=-\ \frac{mv_M^2}r$ is the ``diagonal'' centripetal force which is in
fact independent of the rotation.  $F_M=-2mv_M\omega $ is the Coriolis force
responsible for the torque $\tau=\ \frac{dL}{dt}=F_M\times r$ in action in
the gyroscope.

$F_E=-m\omega ^2r$ is the centripetal force acting on points $P_{R,L}$ even
still in $\theta ^3_{R,L}$~.
\end{enumerate}}\end{defi}
\vskip 11pt

\begin{propo}  \Bena \item The space of diagonal biquanta
$ \widetilde M^I_R\otimes_D  \widetilde M^I_L$ are generated from the bisemisheaf 
$\theta ^3_R\otimes_D\theta ^3_L$ under the action of the diagonal centripetal
biforce $F_{D_R}\otimes_D F_{D_L}$~, where $F_D=-\ \frac{mv_M^2}r$~, responsible
for the smooth biendomorphism $E_R\otimes_D E_L$ acting on
$\theta ^3_R\otimes_D \theta ^3_L$~.
\vskip 11pt

\item The magnetic space biquanta $ \widetilde M  ^I_R\otimes_m  \widetilde M  ^I_L$ are
generated from the bisemisheaf $\theta ^3_R\otimes_D \theta ^3_L$ under the
action of the Coriolis biforce $F_{M_R}\otimes F_{M_L}$~, where
$F_M=-2mv_M\omega $~, responsible for the smooth biendomorphism $E_R\otimes_m
E_L$ acting on
$\theta ^3_R\otimes_D \theta ^3_L$~.
\vskip 11pt

\item The emission (and the subsequent reabsorption) of left and right
magnetic quanta by the left and right semisheaves $\theta ^3_L$ and $\theta
^3_R$
 of a left and right semiparticles having different magnitudes of rotational
velocities results from the differences between right and left torques and
generates by reaction a global movement of translation of the bisemiparticle.
\end{enumerate}\end{propo}
\vskip 11pt

\bpr\Bena \item The emission of diagonal biquanta 
$ \widetilde M^I_R\otimes_D  \widetilde M^I_L$ must result from the diagonal centripetal force
$F_{D_R}\otimes_D F_{D_L}$~.  Indeed, a diagonal centripetal force $F_D$ must
provoke Galois antiautomorphisms on an algebraic semigroup leading to the
smooth endomorphism $E_{R,L} $~.
\vskip 11pt

\item Knowing that the Coriolis force is a force acting sidewise, it seems
natural to attribute to this force the cause of the emission of magnetic
quanta.  Indeed, magnetic quanta would be emitted in order to balance any
variation of the rotational kinetic energy between a left and a right
semisheaves because the work done against the centrifugal force ought to agree
with the difference in rotational energy. If it was not the case, the
centrifugal force would not be equilibrated and would run out.

Note also in this context that the magnetic moment of the electron is equal to $\mu=\F e{2mc}\cdot \ell$ and is thus proportional to its angular momentum $\ell$~.
\vskip 11pt

\item Consider that a left and a right $3D$-space semisheaves do not have the
same magnitude of rotational velocity.  Then we can assign to every point
$P_R\in \theta ^3_R$ a torque $\tau_R$ whose length and direction are different from
those of a torque $\tau_L$ corresponding to a point $P_L\in \theta ^3_L$~.

We thus have a resulting torque $\delta \tau=\tau_R-\tau_L$ at the ``bipoint''
$P_R\times P_L$~.  It is reasonable to admit that the action of the resulting
torque $\delta \tau$ will be at the origin of a smooth endomorphism
$E_{R,L}$~.  

As a consequence, a left or a right magnetic quantum will be emitted which
will provoke by reaction a movement of translation of the bisemiparticle.

Thus, the Coriolis force which is in fact apparent becomes here effective by
the emission of magnetic quanta.  The magnetic quanta, emitted towards the
emergence point of the bisemiparticle, are reabsorbed later on.  Thus, a
process of emission-reabsorption of magnetic left and right quanta generates a
global movement of translation of the bisemiparticle.\epr
\end{enumerate}
\vskip 11pt

\begin{rms} { \Bena \item It is likely that the centripetal force
$F_e=-m\omega ^2r$ is responsible for the emission of $3D$-space electric
quanta.

\item The assertions of proposition 4.3.5 are valid for bisemiphotons, which
explains why a set of bisemiphotons generates a magnetic field.
\end{enumerate}}\end{rms}
\vskip 11pt

\setcounter{defi}{0}
\section{The gravito-electro-magnetic interactions
between bisemiparticles}

It was seen in chapters 3 and 4 that the structure of bisemiparticles is
given by bisemisheaves so that an action-reaction process is generated in
a bisemiparticle by the interactions between the right-semisheaves of the
right semiparticle and the left-semisheaves of the left semiparticle.

Generalizing this concept to a set of bisemiparticles, one easily
demonstrates that the interactions between a set of bisemiparticles
result from the interactions between the right and the left semisheaves
belonging to different bisemiparticles leading to a set of mixed
action-reaction processes of bilinear nature.

Mathematically, if we have a set of $N$ (distinguishable) bisemiparticles, their $3D$ spatial structure is given by the completely reducible representation 
$\Rep (GL_{2N}(\aa_R\times \aa_L))$ of the bilinear general semigroup 
$GL_{2N}(\aa_R\times \aa_L)$~. Indeed, given a partition $2N=2_1+2_2+\cdots+2_N$ of $2N$~, 
the tensor product $\Rep(GL_{2_1}(\aa_R\times \aa_L)) \otimes \cdots \otimes
\Rep(GL_{2_N}(\aa_R\times \aa_L))$ has an irreducible quotient given by the formal sum 
\begin{align*}
&\Rep(GL_{2N=2_1+\cdots+2_i\cdots+2_N}(\aa_R\times \aa_L))\\
&\quad = \Rep(GL_{2_1}(\aa_R\times \aa_L))\boxplus\cdots\\
&\qquad \qquad \boxplus
\Rep(GL_{2_i}(\aa_R\times \aa_L)) \boxplus\cdots  \boxplus
\Rep(GL_{2_N}(\aa_R\times \aa_L))\;, \tag*{$1\le i\le N$~,}\end{align*}
which constitutes a completely reducible orthogonal representation of 
$GL_{2_N}(\aa_R\times \aa_L)$~.

The nonorthogonal completely reducible representation of $GL_{2_N}(\aa_R\times \aa_L)$ is reached if we add to $\Rep(GL_{2N=2_1+\cdots+2_i\cdots+2_N}(\aa_R\times \aa_L))$ the direct sum of off-diagonal irreducible bilinear representations $\Rep(T^t_{2_{i_R}} (\aa_R)\times T_{2_{j_L}}(\aa_L))$ for all pairs of semiparticle indices, $i\neq j$~.

These mathematical considerations allow to develop Langlands global bilinear correspondences for reducible representations of $GL(2N)$ \cite{Pie9} and to introduce the general mathematical frame of the interactions between bisemiparticles as studied in section 5.1.

\vskip 11pt

\subsection{Interactions between bisemiparticles}

\begin{defi}\  {\bf (Interacting bisemiparticles)\/} \  { Let $(\theta _{R_i}\otimes
\theta _{L_i})$ be a bisemisheaf\\ ``~$ST$~'', ``~$MG$~'' or ``~$M$~'' of 	a
bisemiparticle ``~$i$~''.  The total right (resp. left) semisheaf $\Theta
_{R_N}$ (resp. $\Theta
_{L_N}$~) of a set of $N$ right (resp. left) semiparticles 
on $GL_{2N}(\Aa_R)$ (resp. $GL_{2N}(\Aa_L)$~)
is the union
of the disconnected right (resp.  left) semisheaves:
\[ \Theta _{R,L_N} = \theta _{R,L_1}\cup \cdots \cup \theta _{R,L_i}\cup
\cdots \cup \theta _{R,L_N}\]
given by the direct sum \cite{Art} of all right (resp. left) semisheaves
$\Theta _{R,L_N} = \bigoplus\limits^N_{i=1} \theta _{ R,L_i}$ if and only
if $\theta _{R,L_i}\cap \theta _{R,L_{i+1}}=\emptyset$~.

Then, the bisemisheaf $(\Theta _{R_N}\otimes \Theta _{L_N})$ of a set of
$N$ interacting bisemiparticles on $GL_{2N}(\Aa_R\times \Aa_L)$ will be:
\[ \Theta _{R_N}\otimes \Theta _{L_N}
=\left(\textstyle\bigoplus\limits^N_{i=1} \theta _{ R_i}\right)\otimes
\left(\textstyle\bigoplus\limits^N_{j=1} \theta _{ L_j}\right)=
\textstyle\bigoplus\limits^N_{i,j=1}
\left(\theta _{ R_i}\otimes \theta _{ L_j}\right)\]
which can be decomposed following:
\[ \Theta _{R_N} \otimes \Theta
_{L_N}=\textstyle\bigoplus\limits^N_{i=1}\left(\theta _{ R_i}\otimes
\theta _{ L_i}\right) 
\textstyle\bigoplus\limits^N_{{i,j=1}\atop{i\neq j}}\left(\theta _{
R_i}\otimes
\theta _{ L_j}\right)\]
where
\Be
\item the direct sum $\bigoplus\limits^N_{i=1}\left(\theta _{ R_i}\otimes
\theta _{ L_i}\right)$ refers to the total bisemisheaf of
$N$-noninteracting (i.e. free) bisemiparticles verifying the condition
of nonconnectivity between the bisemisheaves ``~$i$~'' and ``~$j$~'':
$(\theta _{R_i}\otimes \theta _{L_i})\cap (\theta
_{R_j}\otimes \theta _{ L_j})=\emptyset$~.

\item the ``mixed'' direct sum $\bigoplus\limits^N_{i,j=1} (\theta _{
R_i}\otimes
\theta _{L_j})$ refers to the bilinear interactions between the right
semisheaves $\theta _{ R_i}$ of the $N$ right semiparticles and the left
semisheaves $\theta _{L_j}$ of the $N$ left semiparticles.\\
The bisemisheaf of a ``mixed'' direct sum is thus an interference
bisemisheaf between the $N$ interacting bisemiparticles.
\Ee}
\end{defi}
\vskip 11pt

\begin{defi}\ {\bf (Interaction bisemisheaves of interacting bisemileptons)\/}
{ The  $(i-j)$-th interaction bisemisheaf $(\theta _{R_i}\otimes
\theta _{L_j})$ of the ``mixed'' direct sum of the total bisemisheaf
$(\Theta _{ R_N}\otimes\Theta _{L_N})$ of a set of $N$ interacting
bisemileptons represents the interactions between the right semisheaf
$\theta _{R_i}$ of the $i$-th bisemilepton and the left semisheaf
$\theta _{L_j}$ of the $j$-th bisemilepton and  can be developed
following:
\nobeqn
\lefteqn{\theta _{R_i}\otimes \theta _{L_j} \equiv 
\theta _{R_i}^{1-3}(t_i,r_i)\otimes \theta _{L_j}^{1-3}(t_j,r_j)}\\
\noalign{\vskip 6pt}
&\stackrel{S_L}{\to}& (\theta _{R_i}^{1-3}(t_i,r_i)\otimes_D \theta
_{L_j}^{1-3}(t_j,r_j))\\
\noalign{\vskip 6pt}
&&\txt\bigoplus\limits (\theta _{R_i}^{3}(r_i)^{(m)}\otimes_m \theta
_{L_j}^{3}(r_j)^{(m)}) \txt\bigoplus\limits
(\theta _{R_i}^{1-(3)}(t_i,(r_i))^{(e)}\otimes_e \theta
_{L_j}^{(1)-3}((t_j),r_j)^{(e)}\noeeqn
if we take into account the $S_L$  isomorphism (see definition 4.1.1).
\vskip 11pt

Note that
\Bena
\item $\theta _{R_i}^{1-3}(t_i,r_i)\otimes_D \theta
_{L_j}^{1-3}(t_j,r_j)$ is a ``mixed'' diagonal $4D$-space-time
bisemisheaf composed of:

\Be
\item ``mixed''  orthogonal space biquanta $ \widetilde M
^I_{R_i}(r_i)\otimes_D \widetilde M ^I_{L_j}(r_j)$ generated by a spatial smooth
biendomorphism $E_R\otimes_D E_L$ (see proposition 4.3.5);

\item ``mixed'' time biquanta $\widetilde M ^I_{R_i}(t_i)\otimes \widetilde M
^I_{L_j}(t_j)$ generated by a time smooth biendomorphism $E_R\otimes
E_L$~.
\Ee
\vskip 11pt

\item $\theta ^3_{R_i}(r_i)^{(m)}\otimes_m \theta ^3_{L_j}(r_j)^{(m)}$ is
a ``mixed'' $3D$ magnetic bisemisheaf composed of ``mixed''
 magnetic biquanta $\widetilde M ^I_{R_i}(r_i)^{(m)}\otimes_m \widetilde M
^I_{L_j}(r_j)^{(m)}$ of a magnetic field.
\vskip 11pt

\item $\theta _{R_i}^{1-(3)}(t_i(,r_i))^{(e)}\otimes_e \theta
_{L_j}^{(1)-3}((t_j),r_j)^{(e)}$ is a ``mixed'' electric
bisemisheaf composed of electric biquanta $(\widetilde M
^I_{R_i}(t_i)^{(e)}\otimes_e \widetilde M ^I_{L_j}(r_j)^{(e)})$ or
$(\widetilde M
^I_{R_i}(r_i)^{(e)}\otimes_e \widetilde M ^I_{L_j}(t_j)^{(e)})$ 
of an electric field.
\Ee
\vskip 11pt

Remark that all these ``mixed'' biquanta of interaction are localized
between the bisemi\-leptons labelled ``~$i$~'' and ``~$j$~''.}
\end{defi}
\vskip 11pt

\begin{defi}\ {\bf (Interaction bisemisheaves of interacting bisemiphotons)\/}\   
{ Similarly, the $(i-j)$-th interaction bisection $(T
_{R_i}\otimes
T _{L_j})$ of the ``mixed'' direct sum of the total bisemisheaf
$(T _{ R_M}\otimes T _{L_M})$ of a set of $M$ interacting
bisemiphotons is given by:
\nobeqn
\lefteqn{T _{R_i}\otimes T _{L_j} \equiv 
T _{R_i}^{1}(r_i)\otimes T _{L_j}^{1}(r_j)}\\
\noalign{\vskip 6pt}
&\stackrel{S_L}{\to} (T _{R_i}^{1}(r_i)\otimes_D T
_{L_j}^{1}(r_j))
\txt\bigoplus\limits (T _{R_i}^{1}(r_i)^{(m)}\otimes_m T
_{L_j}^{1}(r_j)^{(m)})\noeeqn
where
\Bena
\item $T _{R_i}^{1}(r_i)\otimes_D T
_{L_j}^{1}(r_j)$ is a ``mixed'' diagonal space
bisection composed of ``mixed''  orthogonal space biquanta $\widetilde M
^I_{R_i}(r_i)\otimes_D \widetilde M ^I_{L_j}(r_j)$~.
\vskip 11pt

\item $T ^1_{R_i}(r_i)^{(m)}\otimes_m T ^1_{L_j}(r_j)^{(m)}$ is
a ``mixed'' magnetic bisection composed of 
 magnetic biquanta of the magnetic field of the
interacting bisemiphotons.
\Ee}\end{defi}
\vskip 11pt

\begin{defi}\ {\bf (Bisemileptons interacting with bisemiphotons)\/}\ { The
total}  { bisemisheaf $\Theta _{R_{N-M}}\otimes \Theta _{L_{N-M}}$ of a set
of $N$ interacting bisemileptons interacting with a set of
$M$ interacting bisemiphotons is given by
\nobeqn
\lefteqn{\Theta _{R_{N-M}} \txt\bigotimes\limits \Theta _{L_{N-M}} = \left(
\textstyle\bigoplus\limits^N_{i=1} \theta _{R_i}
\textstyle\bigoplus\limits^M_{k=1} T _{R_k}\right) \txt\bigotimes\limits
\left(
\textstyle\bigoplus\limits^N_{j=1} \theta _{L_j}
\textstyle\bigoplus\limits^M_{h=1} T _{L_h}\right)}\\
\noalign{\vskip 11pt}
&\equiv&
\left(
\textstyle\bigoplus\limits^N_{i=1} \theta _{R_i}^{1-3}(t_i,r_i)
\textstyle\bigoplus\limits^M_{k=1} T _{R_k}^1(r_k)\right)\txt\bigotimes\limits
\left(
\textstyle\bigoplus\limits^N_{j=1} \theta _{L_j}^{1-3}(t_j,r_j)
\textstyle\bigoplus\limits^M_{h=1} T _{L_h}^1(r_j)\right)\\
\noalign{\vskip 11pt}
&&=\ \textstyle\bigoplus\limits^N_{i,j=1}\left( \theta
_{R_i}^{1-3}(t_i,r_i)\otimes  \theta_{L_j}^{1-3}(t_j,r_j)\right)
\textstyle\bigoplus\limits^M_{k,h=1}\left( T
_{R_k}^{1}(r_k)\otimes  T_{L_h}^{1}(r_h)\right)\\
\noalign{\vskip 11pt}
&&\qquad \textstyle\bigoplus\limits^{N,M}_{i,h=1}\left( \theta
_{R_i}^{1-3}(t_i,r_i)\otimes  T_{L_h}^{1}(r_h)\right)
\textstyle\bigoplus\limits^{N,M}_{k,j=1}\left( T _{R_k}^{1}(r_k)\otimes 
\theta_{L_j}^{1-3}(t_j,r_j)\right)\noeeqn where
\Be
\item the direct sum $\bigoplus\limits^N_{i,j=1}$ refers to the total
bisemisheaf of $N$ interacting bisemileptons;
\vskip 11pt

\item $\bigoplus\limits^M_{k,h=1}$ refers to the total bisemisheaf of
$M$ interacting bisemiphotons;
\vskip 11pt

\item the mixed direct sum $\bigoplus\limits^{N,M}_{i,h=1}$ refers to the
bilinear interactions between the right semisheaves $\theta 
^{1-3}_{R_i}(t_i,r_i)$ of the $N$ right semileptons and the left
semisheaves $T ^1_{L_h}(r_h)$ of the $M$ left semiphotons.\\
The $(i-h)$-th interaction bisemisheaf of this direct sum
$\bigoplus\limits^{N,M}_{i,h=1}$ decomposes under the $S_L$-isomorphism
into the bisemisheaves:
\nobeqn
\lefteqn{S_L : \theta
_{R_i}^{1-3}(t_i,r_i)\otimes  T_{L_h}^{1}(r_h)}\\
\noalign{\vskip 6pt}
&\to&\left( \theta
_{R_i}^{3}(r_i)\otimes_D  T_{L_h}^{1}(r_h)\right) \txt\bigoplus\limits
\left(\theta 
_{R_i}^{3}(r_i)^{(m)}\otimes_m  T_{L_h}^{1}(r_h)^{(m)}\right)\txt\bigoplus\limits \left(
\theta _{R_i}^{1}(t_i)^{(e)}\otimes_e  T_{L_h}^{1}(r_h)^{(e)}\right)\noeeqn
where
\Be
\item $\theta 
_{R_i}^{3}(r_i)\otimes_D T_{L_h}^{1}(r_h)$ is a mixed diagonal
 space bisemisheaf composed of mixed  orthogonal space biquanta
$\widetilde M  ^I_{R_i}(r)\otimes_D\widetilde M ^I_{L_h}(r)$~;
\item $\theta 
_{R_i}^{3}(r_i)^{(m)}\otimes_m  T_{L_h}^{1}(r_h)^{(m)}$ is a mixed
$3D$ magnetic bisemisheaf generating a magnetic  field;
\item $\theta 
_{R_i}^{1}(t_i)^{(e)}\otimes_e  T_{L_h}^{1}(r_h)^{(e)}$ is a mixed
$3D$   electric bisemisheaf generating an electric field.
\Ee
\vskip 11pt

\item the mixed sum $\bigoplus\limits^{M,N}_{k,j=1}$ refers to the
bilinear interactions between the right sections $T
^1_{R_k}(r_k)$ of the $M$ right semiphotons and the left semisheaves
$\theta ^{1-3}_{L_j}(t_j,r_j)$ of the $N$ left semileptons.\\
The $(k-j)$-th interaction bisemimodule of this direct sum
$\bigoplus\limits^{M,N}_{k,j=1}$ can be handled under the
$S_L$-isomophism similarly as in c).\Ee}
\end{defi}
\vskip 11pt

\begin{lm} Let $\theta ^3_{R_i}(r_i)\otimes_D \theta ^3_{L_j}(r_j)$ be
the mixed $3D$ diagonal space bisemisheaf of interaction between the
$3D$ space right semisheaf $\theta ^3_{R_i}(r_i)$ of a right
semiparticle (semifermion or semiphoton) and the $3D$ space left
semisheaf $\theta ^3_{L_j}(r_j)$ of a left semiparticle (semifermion or
semiphoton).

Then, the mixed  orthogonal space biquanta:
\[ \widetilde M ^I_{R_i}(r_i)\otimes_D \widetilde M ^I_{L_j}(r_j) \in \theta 
^3_{R_i}(r_i)\otimes_D \theta ^3_{L_j}(r_j)\]
are gravitational biquanta of the
gravitational field between the $i$-th right semiparticle and the $j$-th
left semiparticle.

Respectively, the mixed $1D$ time biquanta $\widetilde M ^I_{R_i}(t_i)\otimes
\widetilde M ^I_{L_j}(t_j)$ will be assumed to generate a scalar gravitational
field between the $i$-th right semifermion and the $j$-th left
semifermion.
\end{lm}
\vskip 11pt

\bpr The gravitational biquanta $\widetilde M ^I_{R_i}(r_i)\otimes_D \widetilde M ^I_{L_j}(r_j)$
are elements of the gravitational field because they belong to the
bisections of the tangent bibundles $T^{(3)}_{R_i;ST,MG,M}\otimes_D
T^{(3)}_{L_j;ST,MG,M}$ whose inverse projective bimaps
$(p_{R_i}\otimes_D p_{L_j})$ are the $3D$ diagonal momentum bioperators
(see definition 4.2.2).

Now, these bioperators $(p_{R_i}\otimes_D p_{L_j})$ can be considered as
operators of ``mixed'' acceleration if the tangent bibundles 
$T^{(3)}_{L_{R_i;ST,MG,M}}\otimes_D
T^{(3)}_{L_j;ST,MG,M}$ are defined on the internal bilinear Hilbert
spaces $\Hs^+_a$ (see definition 3.1.7).

Knowing that the intensity of the gravitational field is proportional to
an acceleration, we have the thesis.\epr
\vskip 11pt

We are thus led to the following proposition:
\vskip 11pt

\begin{propo}  A set of bisemileptons interact by means of a
gravito-electro-  magnetic field.

A set of bisemiphotons interact by means of a gravito-magnetic field.

A set of bisemileptons and of bisemiphotons interact by means of a
gravito-electro-magnetic field.\end{propo}
\vskip 11pt

\bpr This proposition results from the definitions 5.1.2 and 5.1.4 and
from the lemma 5.1.5.  However, let us remak that:
\Be
\item the mixed diagonal interactions generate the gravitational field
while the mixed off-diagonal interactions generate the electro-magnetic
field.

\item if we work in the context of a bilinear quantum theory, then the
bisemiphotons interact between themselves while in the standard linear
quantum theory, the interactions are generated by gauge theories (see for
example \cite{A-L}, \cite{Langa}) as in the $U(1)$ abelian gauge theory,
excluding any interaction between photons.\epr\Ee
\vskip 11pt

\begin{propo} Let us adopt the convention that the structure of a
negative electric charge is given by an electric bisemisheaf of type
$\theta ^1_R(t)^{(e)}\otimes_e\theta  ^3_L(r)^{(e)}$ and that the
structure of a positive electric charge is characterized by an electric
bisemisheaf of type $\theta ^3_R(r)^{(e)}\otimes_e \theta 
^1_L(t)^{(e)}$~.

Then, we have that:
\Be
\item a set of $N$ electric charges of the same sign of $N$ interacting
bisemifermions interact by means of an electric field;

\item a set of $N$ electric charges of opposite sign  of $N$ interacting
bisemifermions interact by means of a magnetic field and by means of a
time scalar gravitational field;

\item a set of $N$ magnetic moments of $N$ interacting bisemifermions
interact by means of a magnetic field.\Ee\end{propo}
\vskip 11pt

\bpr\Be
\item Let $\theta ^1_{R_i}(t)^{(e)}\otimes_e\theta  ^3_{L_i}(r)^{(e)}$
and $\theta ^1_{R_j}(t)^{(e)}\otimes_e\theta  ^3_{L_j}(r)^{(e)}$ be two
bisemisheaves characterizing two electric charges of the same sign (here
negative) of two interacting bisemifermions labelled ``~$i$~'' and ``~$j$~.
\\
These two negative electric bisemisheaves interact between themselves by
means of the two interaction bisemisheaves: 
$\theta ^1_{R_i}(t)^{(e)}\otimes_e\theta  ^3_{L_j}(r)^{(e)}$
and $\theta ^1_{R_j}(t)^{(e)}\otimes_e\theta  ^3_{L_i}(r)^{(e)}$
according to definition 5.1.1.
\\
Clearly, these two interaction bisemisheaves are of electric nature
composed of electric biquanta which are elements of an electric
field.  As these two interaction bisemisheaves are of the same nature,
they will be of
repulsive nature.
\vskip 11pt

\item The two electric bisemisheaves $\theta ^1_{R_i}(t)^{(e)}\otimes_e
\theta  ^3_{L_i}(r)^{(e)}$
and $\theta ^3_{R_j}(r)^{(e)}\otimes_e\theta  ^1_{L_j}(t)^{(e)}$
characterizing two electric charges of opposite sign  of two interacting
bisemifermions ``~$i$~'' and ``~$j$~'' interact by means of the two
interaction bisemisheaves:
\Be\item $\theta ^3_{R_j}(r)^{(e)}\otimes_e\theta  ^3_{L_i}(r)^{(e)}$
which is a $3D$ space bisemisheaf, i.e. a magnetic
field; we thus have the identity:
\[ \theta ^3_{R_j}(r)^{(e)}\otimes_e\theta  ^3_{L_i}(r)^{(e)}\equiv
\theta
^3_{R_j}(r)^{(m)}\otimes_m\theta  ^3_{L_i}(r)^{(m)}\]
so that the magnetic bisemisheaf $\theta
^3_{R_j}(r)^{(m)}\otimes_m\theta  ^3_{L_i}(r)^{(m)}$ is composed of
magnetic biquanta of a magnetic field;

\item $\theta ^1_{R_i}(t)^{(e)}\otimes_e\theta  ^1_{L_i}(t)^{(e)}$ which
is a mixed time bisemisheaf composed of scalar gravitational time
biquanta that are of attractive nature.
\Ee
\vskip 11pt

\item From the definitions 1.1.5 and 5.1.2, it appears that a set of $N$
magnetic moments of $N$ interacting bisemifermions interact by means of a
magnetic field.\epr\Ee
\vskip 11pt

\begin{defi}[Internal bilinear mixed Hilbert space] { According to 
definition  5.1.1, the total bisemisheaf $(\Theta _{R_N}\otimes \Theta
_{L_N})$ of a set of $N$ interacting bisemifermions or bisemiphotons is
given by:
\[ \Theta _{R_N}\otimes \Theta _{L_N} =\textstyle\bigoplus\limits^N_{i=1}
(T _{R_i}\otimes T  _{L_i})
\textstyle\bigoplus\limits^N_{{i,j=1}\atop{i\neq j}}
(T _{R_i}\otimes T  _{L_j})\;.\]

Let $p_L$ be a projective linear map, mapping the right semisheaf $\Theta
_{R_N}$ onto the left semisheaf $\Theta _{L_N}$ and let $B_L$ be a
bijective linear isometric map from the projected right semisheaf
$\Theta _{R_N(P)}$ to $\Theta _{L_N}$ so that the bisemisheaf $\Theta
_{R_N}\otimes \Theta _{L_N}$ be transformed as follows \cite{Pie4}:
\[ (B_L\circ p_L) : \Theta _{R_N}\otimes \Theta _{L_N}\to 
\Theta _{L_{R_N}}\otimes \Theta _{L_N}\]
leading to
\[ \Theta _{L_{R_N}}\otimes \Theta _{L_N} =
\textstyle\bigoplus\limits^N_{i=1}
(T _{L_{R_i}}\otimes T _{L_i})
\textstyle\bigoplus\limits^N_{{i,j=1}\atop{i\neq j}} 
(T _{L_{R_i}}\otimes T _{L_j})\;.\]

Then, the diagonal bisections $\{T _{L_{R_i}}\otimes_D T _{L_i}\}^N_{i=1}$ and 
$\{T _{L_{R_i}}\otimes_D T _{L_j}\}^N_{i,j=1}$~, obtained under
the $S_L$-isomorphism, are defined respectively on the internal bilinear
left Hilbert space $\Hs^+_i$ and on the internal bilinear mixed left
Hilbert space $\Hs^+_{(i-j)}$~.}
\end{defi}
\vskip 11pt

\begin{defi}\ {\bf (What differentiate bisemifermions from bisemiphotons)\/}\ The   { 
bisemisheaves $\{\theta _{L_{R_i}}\otimes \theta _{L_i}\}^N_{i=1}$ of
the bisemifermions differ from the corresponding bisections of the
bisemiphotons by the existence of the electric bisemisheaves at the
origin of the electric charges of the bisemifermions.  Similarly, the
bisemifermion interaction bisemisheaves 
$\{\theta _{L_{R_i}}\otimes \theta _{L_j}\}^N_{{i,j=1}\atop{i\neq j}}$
differ from the corresponding bisemiphoton interaction bisections by
the existence of electric bisemisheaves at the origin of an electric
field.

This way of handling interactions between bisemiparticles differs from
the standard description, in linear quantum theory, of a set of $N$
fermions given by an antisymmetric Fock space over the Hilbert space
while a set of $N$ photons is described by a symmetric Fock space over the
Hilbert space \cite{Foc}, \cite{VN-M}.}\end{defi}
\vskip 11pt

Let us be more explicit by working out the standard Hilbert space at two
particles and the present bilinear Hilbert space at two bisemiparticles. 
For the purpose of simplicity, we shall exclude the magnetic and electric
bisemisheaves of the bisemifermions and the magnetic bisemisheaves of the
bisemiphotons.  We then have:
\vskip 11pt

\begin{propo}[Bilinear Hilbert space] The standard Hilbert space at two
particles is given by $\Hs_2=\Hs_{1}\otimes\Hs_{2}$ while the
internal bilinear left Hilbert space at two bisemiparticles is:
\[ \Hs^+_2=\Hs^+_{(1)}\oplus \Hs^+_{(2)}\oplus [\Hs^+_{(1-2)}\oplus
\Hs^+_{(2-1)}]_{\rm int}\;.\]
$\Hs_2$ is thus a Hilbert space at two free particles while $\Hs^+_2$
leads naturally to a bilinear Hilbert space at two interacting
bisemiparticles.\end{propo}
\vskip 11pt

\bpr The structure of the Hilbert space $\Hs^+_2$ at two bisemiparticles
results directly from definitions 5.1.1 and 5.1.8.

Let us point out the difference between $\Hs_2$ and $\Hs^+_2$ by
developing the additional structures of these two Hilbert spaces.

\Be 
\item If $\{ f_\gamma \}^{n_1}_{\gamma =1}$ and $\{ f_\delta 
\}^{n_2}_{\delta  =1}$ are the orthonormal basis of $\Hs_{1}$ and
$\Hs_{2}$~, then a vector $
\phi $ of $\Hs_2=\Hs_1\otimes \Hs_2$ will write:
\[ \phi =\txt\sum\limits^{n_1,n_2}_{\gamma ,\delta =1}c_{\gamma \delta } (f_\gamma
\otimes f_\delta )\]
and the scalar product $(\phi ,\phi )$ defined on $\Hs_2\times \Hs_2$
will be given by \cite{R-S}:
\nobeqn
(\phi ,\phi ) &=& (\txt\sum\limits c_{\gamma \delta }(f_\gamma \otimes f_\delta ),
\txt\sum\limits c_{\mu \nu }(f_\mu \otimes f_\nu))\\
\noalign{\vskip 6pt}
&=& \txt\sum\limits \o c_{\gamma \delta } c_{\mu \nu }(f_\gamma ,f_\mu )\cdot
(f_\delta ,f_\nu )\;;\noeeqn

\item on the other hand, a bivector $\phi _{L_{R_2}}\otimes_D\psi _{L_2}$
of
\[ \Hs^+_2 \simeq \theta _{L_{R_2}}\otimes_D\theta
_{L_2}=(\theta_{L_{R(1)}}\oplus \theta _{L_{R(2)}})\otimes_D
(\theta_{L(1)}\oplus \theta _{L(2)})\]
(see definition 5.1.1) will write:
\nobeqn 
\phi _{L_{R_2}}\otimes_D \psi _{L_2} &=& (\phi_{L_{R(1)}}\oplus
\phi_{L_{R(2)}})\otimes_D (\psi_{L(1)}\oplus
\psi_{L(2)})\\
\noalign{\vskip 6pt}
&=& \left( \txt\sum\limits_\alpha \o b_\alpha (e^\alpha )^*\oplus 
\txt\sum\limits_\beta  \o c_\beta  (f^\beta  )^*\right) \otimes_D
 \left( \txt\sum\limits_\gamma   b_\gamma  (e_\gamma )\oplus 
\txt\sum\limits_\delta     c_\delta  (f_\delta  )\right)\noeeqn
if $\{ (e^\alpha )^*\}^{n_1}_{\alpha =1}$ and
$\{ (f^\beta  )^*\}^{n_2}_{\beta  =1}$ are orthonormal basis
respectively of the right semimodules $\theta _{L_{R(1)}}$ and $\theta
_{L_{R(2)}}$ while $\{ e_\gamma  \}^{n_1}_{\gamma  =1}$ and
$\{ f_\delta \}^{n_2}_{\delta  =1}$ are orthonormal basis respectively
of the left semimodules $\theta _{L(1)}$ and $\theta _{L(2)}$~.

Then, the internal scalar product on $\theta _{L_{R_2}}\otimes_D \theta
_{L_2}$ will be given by:
\nobeqn
(\phi _{L_{R_2}},\psi _{L_2}) &=& \txt\sum\limits_{\alpha =\gamma } \o b_\alpha 
b_\gamma (e^\alpha ,e_\gamma )+ \txt\sum\limits_{\beta  =\delta  } \o c_\beta 
c_\delta  (f^\beta  ,f_\delta  )\\
\noalign{\vskip 11pt}
&&+ \txt\sum\limits_{\alpha =\delta } \o b_\alpha 
c_\delta  (e^\alpha ,f_\delta  )+\txt\sum\limits_{\beta  =\gamma } \o c_\beta 
b_\gamma (f_\beta  ,e_\gamma )\;.\noeeqn

Remark that
\nobeqn
\phi _{L_{R_2}}\otimes_D \psi _{L_2} &=& 
(\phi _{L_{R(1)}}\otimes_D \psi _{L(1)})\oplus (\phi _{L_{R(2)}}\otimes_D
\psi _{L(2)})\\
\noalign{\vskip 6pt}
&&\txt\bigoplus\limits (\phi _{L_{R(1)}}\otimes_D \psi _{L(2)})\oplus
(\phi _{L_{R(2)}}\otimes_D \psi _{L(1)})\noeeqn
where $(\phi _{L_{R(1)}}\otimes_D \psi _{L(2)})\in \Hs^+_{(1-2)}$ and
$(\phi _{L_{R(2)}}\otimes_D \psi _{L(1)})\in \Hs^+_{(2-1)}$~.

$[\Hs^+_{(1-2)}\oplus \Hs^+ _{(2-1)}]_{\rm int}$ is the direct sum of the
interaction bilinear Hilbert spaces between the two bisemiparticles $(1)$
and $(2)$~.

Then we have that
\Be\item if $[\Hs^+ _{(1-2)}\oplus \Hs^+ _{(2-1)}]_{\rm int}=0$~, the two
bisemiparticles do not interact;
\item If $[\Hs^+ _{(1-2)}\oplus \Hs^+ _{(2-1)}]_{\rm int}\neq 0$~, the two
bisemiparticles interact by means of a gravitational field (see lemma
5.1.5) generated by:
\begin{itemize}
\item gravitational biquanta from the $3D$  mixed bilinear Hilbert spaces
\[ [\Hs^{\rm space}_{(1-2)}\oplus \Hs^{\rm space}_{(2-1)}]\subset
\Hs^+ _{(1-2)}\oplus \Hs^+ _{(2-1)}\;;\]
\item time mixed biquanta of a scalar gravitational field from   $1D$ 
mixed bilinear Hilbert spaces
\begin{equation} [\Hs^{\rm time}_{(1-2)}\oplus \Hs^{\rm time}_{(2-1)}]\subset
\Hs^+ _{(1-2)}\oplus\Hs^+ _{(2-1)} \;.\tag*{\eop} \end{equation}
\end{itemize}
\Ee\Ee
\vskip 11pt

\begin{defi}[Wave equations of interacting bisemiparticles]   { We shall
now}  { introduce the ``wave'' equation of $N$ interacting bisemiparticles.

Referring to definition 5.1.1, the total bisemisheaf $(\Theta
_{R_N}\otimes \Theta _{L_N})$ of a set of $N$ interacting bisemiparticles
is given by:
\[ \Theta _{R_N}\otimes \Theta
_{L_N}=\textstyle\bigoplus\limits^N_{i,j=1} (\theta _{R_i}\otimes \theta
_{L_j})\;.\]
According to definition 3.1.8, this bisemisheaf 
$(\Theta
_{R_N}\otimes \Theta _{L_N})$ will generate the extended internal
bilinear Hilbert space $H^+ _N$ if we apply to it the $(B_L\circ p_L)$ map
transforming it into the bisemisheaf $\Theta
_{L_{R_N}}\otimes \Theta _{L_N}$~.}\end{defi}
\vskip 11pt

If we take into account the section 2 of chapter 4 where the ``mass''
second order elliptic differential bilinear equation was introduced, we
can state that the ``mass'' biwave equation of $N$ interacting
bisemiparticles can be developed following:
\[ \txt\sum\limits^N_{{i,j=1}\atop{\mu, \nu }} [( \MM _{R_{i\mu }}\otimes \MM _{L_{j\nu }}) -
(E_{R_{i\mu }}\otimes E_{L_{j\nu }})][\phi _{L_{R_{i\mu }}}(t,r)\otimes
\phi _{L_{j\nu }}(t,r)]=0\]
where
\Be
\item $\MM _{R_{i\mu }}$ (resp. $\MM _{L_{j\nu  }}$~) is the mass differential
right (resp. left) operator, given explicitly in definition 4.2.2, acting
on the $\mu $-th right (resp. $\nu $-th left) section 
$\phi _{L_{R_{i\mu }}}(t,r)$ (resp. $
\phi _{L_{j\nu }}(t,r)$~) of the right (resp. left) semisheaf $\theta 
_{L_{R_i}}$ (resp. $\theta  _{L_j}$~) defined on the bilinear Hilbert
space $H^+ _N$~;
\vskip 11pt

\item $(E_{R_{i\mu }}\otimes E_{L_{j\nu }})$ is the corresponding
eigenbivalue.\Ee
\vskip 11pt

We then have the following proposition:
\vskip 11pt

\begin{propo}  The biwave equation of $N$ interacting bisemiparticles
separates automatically into $N_q$ biwave equations of the $N_q$
bisections of $N$ bisemiparticles and into $((N_q)^2-N_q)$ biwave
equations referring to the interactions between the right and left
sections of these $N$ bisemiparticles.
\end{propo}
\vskip 11pt

\bpr The mass biwave equation of $N$ interacting bisemiparticles on the
extended internal bilinear Hilbert space $H^+ _N$~:
\[ \txt\sum\limits^N_{{i,j=1}\atop{\mu ,\nu }} [(\MM _{R_{i\mu }}\phi _{L_{R_{i\mu
}}}(t,r)\otimes \MM _{L_{j\nu }}\phi _{L_{j\nu }}(t,r)) - (E_{R_{i\mu
}}\otimes E_{L_{j\nu }})(\phi _{L_{R_{i\mu }}}(t,r)\otimes
\phi _{L_{j\nu }}(t,r))]=0\]
decomposes into:\\[6pt]
$\ds
 \txt\sum\limits_{i,\mu } [(\MM _{R_{i\mu }}\phi _{L_{R_{i\mu
}}}(t,r)\otimes \MM _{L_{i\mu  }}\phi _{L_{i\mu  }}(t,r)) - (E_{R_{i\mu
}}\otimes E_{L_{i\mu  }})(\phi _{L_{R_{i\mu }}}(t,r)\otimes
\phi _{L_{i\mu  }}(t,r))]$\\[11pt]
$\; \ds+
\txt\sum\limits_{{i,j,\mu ,\nu }\atop{i\neq j,\mu\neq \nu }} [(\MM _{R_{i\mu }}\phi
_{L_{R_{i\mu }}}(t,r)\otimes \MM _{L_{j\nu }}\phi _{L_{j\nu }}(t,r)) -
(E_{R_{i\mu }}\otimes E_{L_{j\nu }})(\phi _{L_{R_{i\mu }}}(t,r)\otimes
\phi _{L_{j\nu }}(t,r))] =0$\\[6pt]
where\Be
\item $\sum\limits_{i,\mu }$ refers to the sum of $N_q$ biwave
equations relative to the $N_q$ bisections of the $N$ free
bisemiparticles;
\item $\sum\limits_{i,j,\mu ,\nu }$ refers to the sum of $[(N_q)^2-(N_q)]$
biwave equations relative to the interactions between the right and left
sections of the $N$ bisemiparticles.\epr
\Ee
\vskip 11pt

\begin{defi}[Biwave equation between two different sections] { The}  
{ 
biwave equation relative to a bisection of a free bisemiparticle was
already handled in  section 2 of chapter 4.  We shall now develop the
biwave equation between two different right and left sections, i.e. when
$\mu \neq \nu $~: this corresponds to a term b) of the proof of
proposition 5.1.12:
\[  [(\MM_{R_{i\mu }}\phi _{L_{R_{i\mu
}}}\otimes \MM_{L_{j\nu }}\phi _{L_{j\nu }}) - (E_{R_{i\mu
}}\otimes E_{L_{j\nu }})(\phi _{L_{R_{i\mu }}}\otimes
\phi _{L_{j\nu }})]=0\;.\]
Proceeding as in definition 4.2.3, this biwave equation becomes:
\nobeqn
\left[ \txt\sum\limits^3_{p,q=1} A^{pq}_{\mu \nu } \frac{\partial^2(\phi
_{L_{R_{i\mu }}}(r_{i\mu })\phi
_{L_{j\nu }}(r_{j\nu  }))}{\partial x_{i\mu _p}\ \partial x_{j\nu _q}}+
\txt\sum\limits^3_{p=1} A^{p0}_{\mu \nu } \frac{\partial\phi
_{L_{R_{i\mu }}}(r_{i\mu })}{\partial x_{i\mu _p}}\cdot \frac{\partial\phi
_{L_{j\nu }}(t_{j\nu  })}{\partial t_{j\nu }}\right.&&\\
\noalign{\vskip 11pt}
+\left.\txt\sum\limits^3_{q=1} A^{0q}_{\mu \nu } \frac{\partial\phi
_{L_{R_{i\mu }}}(t_{i\mu })}{\partial t_{i\mu }}\cdot \frac{\partial\phi
_{L_{j\nu }}(r_{j\nu  })}{\partial x_{j\nu_q }} +
A^{00}_{\mu \nu } \frac{\partial^2(\phi
_{L_{R_{i\mu }}}(t_{i\mu })\phi
_{L_{j\nu }}(t_{j\nu  }))}{\partial t_{i\mu }\ \partial t_{j\nu }}\right]
&&\\
\noalign{\vskip 11pt}
- [(E_{R_{i\mu }}\cdot E_{L_{j\nu }})(\phi _{L_{R_{i\mu }}}(t_{i\mu
},r_{i\mu })\cdot
\phi _{L_{j\nu }}(t_{j\nu 
},r_{j\nu }))]&=&
0\noeeqn
where
\[ A^{pq}_{\mu \nu } =-\frac{\hbar^2}{c^2}\ s^p_\mu \ s^q_\nu \;,
\qquad  A^{00}_{\mu \nu } =-\hbar^2\ s^0_\mu \ s^0_\nu \;,\]

\[r_{i\mu } =\{ x_{i\mu _1},x_{i\mu _2},x_{i\mu _3}\}\;.\]}\end{defi}
\vskip 11pt

It is a second order degenerated elliptic differential bilinear equation
which can be solved by separation of variables.  This equation allows to find the
interaction energy, which is of gravitational, electric or magnetic nature according
to proposition 5.1.6.
\vskip 11pt

\begin{rms} { \Bena
\item In \cite{Pie1}, Green's propagators for bisemiparticles, i.e.
Green's bipropagators, were evaluated which allowed to develop the
$S$-matrix for bisemiparticles.

It was then demonstrated that the traditional Feynman graphs \cite{B-D},
like the electron self-mass and the vacuum polarisation, ``open'' and
split giving rise to new Feynman ``bigraphs'' in the context of
bisemiparticles: the result is that the Feynman graphs, which are
divergent in the context of quantum field theory \cite{Schwe}, are
transformed into corresponding bisemiparticle bigraphs which were
proved not to be more divergent.
\vskip 11pt

\item This way of handling the interactions between $N$ bisemiparticles,
and more particularly between $N$ bisemielectrons, clears up the problem
of the electronic correlation between $N$ electrons (see for example
\cite{Low}).
\Ee}\end{rms}
\vskip 11pt

\subsection{The gravito-electro-magnetism}

Instead of considering the interactions between a set of $N$ well
defined and localized bisemiparticles as done until now, it is possible
to envisage the interaction of a given bisemiparticle with an external
field representing the global influence of the set of $(N-1)$
remaining bisemiparticles.
\vskip 11pt

\begin{defi}[The tensor of the gravito-bifield]  { 
This}  { external field will be given by the generic biconnexion
$(A_R(t,r)\otimes A_L(t,r))$ such that $A_{R,L}(t,r)$ be a right (resp.
left) connexion, i.e. a right (resp. left) distribution at the
considered point $P_{R,L}(t,r)$ (see definition 4.2.8).

\[A_{R,L}(t,r)=\{A^t_{R,L},A^x_{R,L},A^y_{R,L},A^z_{R,L}\}\]
is a four-vectorial distribution whose components $A^m_{R,L}$~,
$m= x,y,z $~, are given for example by \cite{B-D}:
\[ A^m_{R,L} (t,r) = \int d^3k_{{R,L}}A_{R,L}(\vec k,\vec S)e^{\pm i\vec k\vec r}\varepsilon (k_{R,L},\lambda )\]
where $\varepsilon (k_{R,L},\lambda )$ is the polarization unit vector depending on the integer $\lambda =1,2$ referring to the two transverse polarization modes of the semiphotons.

The mass bisemisheaf of a bisemiparticle in an external field on the
left extended internal bilinear Hilbert space $H^+ _a$ will then be written:
\[ [(\MM _R+eA_R(t,r))\otimes (\MM _L+eA_L(t,r))](\theta_{L_R}\otimes
\theta_L)\]
where ``~$e$~'' is the classical charge parameter modulating the
connexion $A_{R,L}(t,r)$ in order to have an interaction between two
proportional charges throughout an infinitesimal right or left connexion.

The mass bioperator $(\MM _R\otimes \MM _L)$ of a bisemiparticle  endowed
with the infinitesimal biconnexion $(eA_R(t,r)\otimes eA_L(t,r))$~,
noted $(A_R\otimes A_L)$~, will develop according to:
\[ (\MM _R+A_R)\otimes (\MM _L+A_L) =(\MM _R\otimes \MM _L)+(A_R\otimes
A_L)+((\MM _R\otimes A_L)+(A_R\otimes \MM _L))\]
where $((\MM _R\otimes A_L)+(A_R\otimes \MM _L))$ represents the interaction
bioperator between the mass bioperator of a bisemiparticle and the
infinitesimal biconnexion giving the global influence of an external
bifield.  This interaction bioperator is the sum of two tensors of the
same type and will be noted $\MM A_{mn}$~.

The interaction tensor $\MM A_{mn}$ is a tensor whose components are:
\[ \MM A_{mn}=\MM _mA_n+A_m\MM _n\;,\]
with $\MM _m=\{m_0,p_x,p_y,p_z\}$ and $A_m=\{A_t,A_x,A_y,A_z\}$~.

This interaction tensor $\MM A_{mn}$~, called the gravito-electro-magnetic
tensor or GEM tensor, can be explicitly written as follows:
\[ \MM A_{mn} =\left[ \begin{array}{cccc}
G_t & E^-_x & E^-_y & E^-_z \\
E^+_x & G_x & B^-_z & B^+_y \\
E^+_y & B^+_z & G_y & B^-_x\\
E^+_z & B^-_y & B^+_x & G_z\end{array}\right] \]
where
\Be
\item $\vec E^{\pm}=\{E^{\pm}_x,E^{\pm}_y,E^{\pm}_z\}$ is a
$3D$-positively (resp. negatively) charged electric field vector;

\item $\vec B^{\pm}=\{B^{\pm}_x,B^{\pm}_y,B^{\pm}_z\}$ is a
$3D$-positive (resp. negative) magnetic field vector;

\item $  G=\{G_x,G_y,G_z\}$ is a
$3D$ gravitational field diagonal tensor and $G_t$ is a scalar
gravitational field.
\Ee}\end{defi}
\vskip 11pt

\begin{propo} The interaction tensor $\MM A_{mn}$ is transformed into the
antisymmetric tensor $F_{mn}$ of electromagnetism if $\MM A_{mn}$ is
submitted to the bijective antisymmetric map $C:\MM A_{mn}\to F_{mn}$
transforming the right components of $A_m$ into their corresponding left
components and the left components of $\MM _m$ into their corresponding
right components, which corresponds to a map transforming a symplectic metric into an orthogonal metric.\end{propo}
\vskip 11pt

\bpr \Be \item The off-diagonal left electric components of the
interaction tensor $\MM A_{mn}$ are:
\[ E^-_i \equiv E_{L_i}=m_0A_i+A_tp_i\simeq +i\hbar \
\frac\partial{\partial t}\ A_i -A_t\cdot i\ \frac \hbar c\
\frac\partial{\partial i}\;,\quad i=x,y,z\;.\]
The ``~$C$~'' map defined as
\[ m_0A_i+A_tp_i\to m_0A_i+p_iA_t\]
transforms $E_{L_i}$ into $-E_i$~:
\[ C : E^-_i \equiv E_{L_i}\to -E_i\]
where 
\[-E_i=+i\left( \frac{\partial A_i}{\partial t}-\frac{\partial
A_t}{\partial i}\right)\]
in the $c=\hbar=1$ systems of units.

It then appears that $\{-E_i\}_{i=x,y,z}$ are the components similar at
a sign of the $3D$ negatively charged left electric field vector:
\[ -\vec E =+\vec\nabla A_t+\frac{\partial \vec A}{\partial t}\]
of classical electromagnetism.

Similarly, 
\[ E^+_i \equiv E_{R_i}=p_iA_t+A_im_0\simeq +i\frac\hbar c \
\frac\partial{\partial i}\ A_t -i\hbar A_i\ 
\frac\partial{\partial t}\]
can be transformed by the ``~$C$~'' map into:
\[ C : E^+_i\equiv E_{R_i}\to +E_i\]
where
\[E_i=+i\left( \frac{\partial A_t}{\partial i}-\frac{\partial
A_i}{\partial t}\right)\;.\]

Clearly, $\{+E_i\}_{i=x,y,z}$ are very closed at a sign to the
components of the $3D$ positively charged right electric field vector
\[ \vec E = -\vec \nabla A_t-\frac{\partial \vec A}{\partial t}\]
of electromagnetism.
\vskip 11pt

\item In a similar way, the off-diagonal left magnetic components of the
GEM interaction tensor $\MM A_{mn}$ are:
\[ B^-_k \equiv B_{L_k}=p_iA_j+A_ip_j\simeq +i\ \frac\hbar c \
\frac\partial{\partial i}\ A_j - i\ \frac \hbar c\ A_i\
\frac\partial{\partial j}\;,\quad \{i,j,k\}\Leftrightarrow\{x,y,z\}\;.\]

The ``~$C$~'' map transforms $B^-_k\equiv B_{L_k}$ into
\[ -B_k = +i\left( \frac{\partial A_j}{\partial i}-\frac{\partial
A_i}{\partial j}\right)\]
in the $c=\hbar=1$ system of units.

$\{-B_k\}_{k=x,y,z}$ are the components of the $3D$ left negative
magnetic field vector $-\vec B=-\vec\nabla \times \vec A$ of
electromagnetism.

Similarly, 
\[ B^+_k\equiv B_{R_k}=p_jA_i+A_jp_i\simeq +i\ \frac\hbar c\ \frac
\partial{\partial j}\ A_i -i\ \frac \hbar c\ A_j\ \frac \partial
{\partial i}\;.\]

The ``~$C$~'' map transforms $B^+_k\equiv B_{R_k}$ into
\[ +B_k = +i\left( \frac{\partial A_i}{\partial j}-\frac{\partial
A_j}{\partial i}\right)\]
where $\{B_k\}_{k=x,y,z}$ are the components of the $3D$ right positive
magnetic field vector $+\vec B=\vec \nabla \times \vec A$~.
\vskip 11pt

\item Finally, the diagonal components $G_t=m_0A_t+A_tm_0$ and
$G_i=p_iA_i+A_ip_i$~, $i=x,y,z$~, of the interaction tensor $\MM A_{mn}$
are the components respectively of a scalar gravitational field $G_t$
and of a $3D$ gravitational field diagonal tensor $G$ because $G_t$ and
$G$ are ``mixed'' $1D$ and $3D$ diagonal bioperators acting respectively
on ``mixed'' $1D$ and $3D$ space orthogonal bisemisheaves which are
gravitational bisemisheaves according to lemma 5.1.5.

The ``~$C$~'' map transforms $G_t=m_0A_t+A_tm_0$ and
$G_i=p_iA_i+A_ip_i$ respectively into 
\[\underset{\sim}{G}_t = +i\hbar\left( \frac{\partial A_t}{\partial
t}-\frac{\partial A_t}{\partial t}\right)\]
and into
\[\underset{\sim}{G}_i = +i\ \frac\hbar c\ \left( \frac{\partial
A_i}{\partial i}-\frac{\partial A_i}{\partial i}\right)\;;\]
so, if the right components of 
$A_m$ anticommute with the left components of $\MM _m$~, then
\[ \underset{\sim}{G}_t = \underset{\sim}{G}_i =0\;.\]

This explains why the gravitational field is so hardly observable and why
it does not appear in the tensor $F_{mn}$ of electromagnetism as we
should expect it.

Summarizing, we have:
\[ C : \quad \MM A_{mn}\to F_{mn}\]
where
\nobeqn
\MM A_{mn} &=& \left[ \begin{array}{cccc}
G_t & E^-_x & E^-_y & E^-_z \\
E^+_x & G_x & B^-_z & B^+_y \\
E^+_y & B^+_z & G_y & B^-_x \\
E^+_z & B^-_y & B^+_x & G_z \end{array}\right]\\
\noalign{and}
F_{mn} &=& \left[ \begin{array}{cccc}
0 & -E_x & -E_y &-E_z \\
+E_x & 0 & -B_z & +B_y \\
+E_y & +B_z & 0 & -B_x \\
+E_z & -B_y & +B_x & 0 \end{array}\right]\;.\noeeqn
\epr\Ee
\vskip 11pt

\begin{propo} The ``GEM'' gravito-electro-magnetic tensor $\MM A_{mn}$ is reduced
to the ``GM'' gravito-magnetic tensor $\MM A^p_{ij}$~, $i,j= x,y,z $~, in
the case of bisemiphotons, i.e. when a bisemiphoton interacts with an
external field.\end{propo}
\vskip 11pt

\bpr According to proposition 5.1.6, bisemiphotons interact by means of
a gravito-magnetic field.  Consequently, the tensor $\MM A_{mn}$ reduces to
the tensor

\[ \MM A^p_{ij} =\left[ \begin{array}{ccc}
G_x & B^-_z & B^+_y \\
B^+_z & G_y & B^-_x \\
B^-_y & B^+_x & G_z\end{array}\right] \]
which is transformed into the tensor
\[ F^p_{ij} =\left[ \begin{array}{ccc}
0 & -B_z & +B_y \\
+B_z & 0 & -B_x \\
-B_y & +B_x & 0\end{array}\right] \]
under the action of the ``~$C$~'' map.\epr
\vskip 11pt

\begin{rem}  { The bisemifermions interact by means of gravitational,
electric and magnetic biquanta which could be interpreted as virtual photons
\cite{Fey2}, \cite{Fey3} like in quantum electrodynamics.

Indeed, the external field is given by the generic biconnexion
$A_R(t,r)\otimes A_L(t,r)$ which could be interpreted as a ``bi''semiphoton
gauge field as in quantum electrodynamics.

Thus, the existence of a local bilinear gauge transformation on a physical field is equivalent to consider a deformation of this field following 1.4.16 and \cite{Pie11}.}\end{rem}
\vskip 11pt

\begin{propo} The condition $4D$-nul divergence: $\partial^n\MM A_{mn}=0$~,
i.e. 
\[ (1\otimes \delta _L)[(\MM _R\otimes A_L)+(A_R\otimes \MM _L)]=0\;,\]
applied to the GEM tensor $\MM A_{mn}$ leads to a set of formal
differential equations:
\[ \left\{ \begin{array}{lll}
\vec \nabla \cdot \vec E &=& \displaystyle{\frac{\partial G_t}{\partial
t}}\;,\\ \noalign{\vskip 11pt}
\vec \nabla \times \vec B &=& \displaystyle{\vec\nabla \cdot
G+\frac{d\vec E}{dt}}\;\cdotp \end{array}\right.\]
analog to the second set of Maxwell equations:
$\partial^nF_{mn}=4\Pi j_m$~, or
\[ \left\{ \begin{array}{lll}
\vec \nabla \cdot \vec E &=&\rho\;,\\ \noalign{\vskip 6pt}
\vec \nabla \times \vec B &=& \displaystyle{\vec j +\frac{d\vec
E}{dt}}\;, \end{array}\right.\]
where $j_m=\{\rho,j_x,j_y,j_z\}$~.
\end{propo}
\vskip 11pt

\bpr As $A_{R,L}$ is a right (resp. left) connexion in contrast with
the vector potential $A_m(r,t)=\{\phi,A_x,A_y,A_z\}$ of
electro-magnetism given classically by \cite{F-L-S}:
\nobeqn
\phi(1,t) &=&\int\ \frac{\rho(2,t-r_{12/c})}{4\Pi\varepsilon
_0r_{12}}\ dv_2\;,\\
\vec A(1,t) &=& \int\ \frac{\vec j(2,t-r_{12/c})}{4\Pi\varepsilon
_0c^2r_{12}}\ dv_2\;,\noeeqn
i.e. defined respectively from the charge density $\rho(2,\cdots)$ and
from the current density $j(2,\cdots)$~, the $1D$-divergence
$\frac{\partial G_t}{dt}$ and the $3D$-``divergence''
\[ \vec \nabla\cdot G = \frac{\partial G_x}{\partial x}+
\frac{\partial G_y}{\partial y}+\frac{\partial G_z}{\partial z}\]
of the gravitational field $G_m$ appears formally in the set of
equations:
\[ \left\{ \begin{array}{lll}
\vec\nabla \cdot \vec E &=& \displaystyle{\frac{\partial G_t}{\partial
t}}\;,\\
\noalign{\vskip 11pt}
\vec\nabla \times \vec B &=& \vec\nabla \cdot
G+\displaystyle{\frac{d\vec E}{dt}}\;\cdotp\end{array}\right.\]

The conditions of $4D$-nul divergence $\partial^n\MM A_{mn}=0$ of the
tensor $\MM A_{mn}$ leads to the conditions $(\delta _L,A_L)=(\delta
_L,\MM _L)=0$~, where $\delta _L$ is a $4D$-left divergence and
$(\cdot,\cdot)$ is a scalar product.  Now, $(\delta _L,A_L)=0$
corresponds to the radiation gauge or to the Lorentz condition
\cite{B-D} of electromagnetism while $(\delta _L,\MM _L)=0$ is a condition
of conservation of the left mass of the reference left semiparticle or
is a condition of nonaccelerated (i.e. uniform) motion.

This set of differential equations gives the possibility of generating a
$1D$ and a $3D$ gravitational field respectively from an electric field
and from an electromagnetic field.

The transformation
\[ \vec\nabla\cdot\vec E =\frac{\partial G_t}{\partial t}\]
is realized through a $(\gamma _{r\to t}\circ E)$ morphism applied to
$\vec\nabla\cdot\vec E$~.\epr
\vskip 11pt

\begin{defi}[$4D$-external current]  { If the tensor $\MM A_{mn}$ is no
more conserved, i.e. if $(\delta _L,A_L)\neq 0$ and if $(\delta
_L,M_L)\neq 0$~, then we have:
\[ (1\otimes \delta _L)[(\MM _R\otimes A_L)+(A_R\otimes \MM _L)]=J_R\]
or
\[\partial^n\MM A_{mn}=J_m\]
where $J_m=\{J_t,J_x,J_y,J_z\}$ is a $4D$-external perturbating right
current.

This condition $\partial^n\MM A_{mn}=J_m$ leads to the set of differential
equations:
\[ \left\{ \begin{array}{lll}
J_t+\vec \nabla\cdot \vec E &=& \displaystyle{\frac{\partial
G_t}{\partial t}}\;,\\
\noalign{\vskip 11pt}
\vec J+\vec\nabla \times \vec B &=& \vec\nabla\cdot
G+\displaystyle{\frac {\partial \vec E}{\partial
t}}\;,\end{array}\right. \]
with $\vec J = \{J_x,J_y,J_z\}$~.}\end{defi}
\vskip 11pt

\begin{defi}[$3D$ external current]  { The condition of $3D$ nul
divergence:
\[ \nabla^j \MM A^p_{ij}=0\;,\qquad i,j= x,y,z \;,\]
applied to the GM tensor $\MM A^p_{ij}$ of bisemiphotons leads to the set
of differential equations:
\[ \vec\nabla\times \vec B =\vec\nabla\cdot G\]
which gives the possibility of generating a gravitational field $G$ from
a magnetic field $\vec B$~.

If the tensor $\MM A^p_{ij}$ is no more conserved, i.e. if
$(\nabla_L,A_L)\neq 0$ and if $(\nabla_L,p_L)\neq 0$ where $\nabla_L$ is
a $3D$ divergence, then we have:
\[\nabla^j\MM A^p_{ij}=J^p_i\]
where $\vec J^p=\{J^p_x,J^p_y,J^p_z\}$ is a $3D$ external right current.

The condition $\nabla^j\MM A^p_{ij}=J^p_i$ leads to the set of differential
equations
\[ \vec J^p+\vec\nabla\times \vec B=\vec\nabla G\;.\]}
\end{defi}
\vskip 11pt

\begin{rem}  { It is commonly assumed that light waves are
electromagnetic waves.  However, considering the preceding
developments, it appears that isolated light waves, i.e. bisemiphotons,
generate only a magnetic field.  It is only when light waves interact
with bisemifermions that an electromagnetic field of interaction is
produced according to proposition 5.1.6.

On the other hand, bisemiphotons could not have a proper mass (i.e. components depending on their proper time) strictly equal to zero following proposition 1.4.8, because, otherwise, the velocity of light would be infinite.  But, the proper mass of the bisemiphotons is too tiny to generate an internal electric field.}\end{rem}
\vskip 11pt

\begin{propo}  The gravitational field is of attractive nature while the
electromagnetic field is of repulsive and/or of attractive nature.
\end{propo}
\vskip 11pt

\bpr This results from the fact that the gravitational field consists of
diagonal biquanta while the electromagnetic field is composed of pairs
of off-diagonal magnetic and electric biquanta, generating a positive or
a negative field following the sense of rotation of the sections of
the magnetic and electric bisemisheaves of the corresponding magnetic
and electric fields.\epr
\vskip 11pt

\begin{rem}[The gravitational field and the theory of general
relativity]  { Let us finally\linebreak make two remarks concerning the
gravitational field.
\vskip 6pt

\Bena
\item In this algebraic quantum model (AQT), the gravitation results
directly from the diagonal interactions between bisemiparticles.  The
question is now to find some connexion between the way by which
gravitation has been introduced in AQT and the way by which it was
described by A. Einstein in general relativity \cite{Ein3}, \cite{Ein4}.

The solution is not immediate.  Indeed, it appears that there are two
fundamental tools in general relativity:
\Be
\item the metric tensor $g_{\mu \nu }$~, interpreted as a gravitational
potential leads to a description of gravitation in terms of curvature of space-time throughout the Ricci tensor $R_{\mu\nu}$ without really attending to the cause of gravitation;
\item the equation of Poisson: $\Delta\phi=4\Pi \kappa\rho$ having been
used as a guiding principle for deriving the equations of general
relativity:
\[ R_{\mu \nu } -\half\ g_{\mu \nu }R=-\kappa T_{\mu \nu }\;\cdotp\]
\Ee

Now it appears that the concept of interaction between (bi)objects at
the basis of the generation of the gravitational field could be related
to the basic metric tensor $g_{\mu \nu }$~.  On the other hand, 
Poisson's equation, describing the ``dynamics'' of the production of the
density of matter $\rho$ from the gravitational field, likely refers in
AQT to the transformation of $3D$ graviphotons into $3D$ bisemiphotons.
\vskip 11pt

\item It was demonstrated that every massive elementary left or right
semiparticle is constituted by three embedded semisheaves of rings:
\[ \theta ^{1-3}_{L,R}(t,r)_{ST}\subset
\theta ^{1-3}_{L,R}(t,r)_{MG}\subset \theta ^{1-3}_{L,R}(t,r)_{M}\]
such that the middle-ground and mass semisheaves of rings be generated
from the space-time semisheaf of rings.  This idea was initially
developed in order to get a bridge between quantum field theory and
general relativity.  Let us recall the pioneer work of L. Parker
\cite{Par} in this field.  Indeed, Sakharov \cite{Sak} suggested that
gravitation could be some quantized phenomenon due to the vacuum
energy.  The aim pursued in \cite{Pie1} was then to consider that:
\Be\item the two internal structures of an elementary semiparticle, i.e.
the space-time and middle-ground structures, could correspond to its
unobservable vacuum from which the mass structure could be generated. 
The middle-ground structure was then interpreted as being of
gravitational nature, which was promptly realized to be incorrect.

\item the equations of general relativity can be lightly modified
\cite{Pie1} by relating nicely the Ricci tensor to the matter stress
tensor in order to take into account the creation of matter from
gravitational energy.\Ee\Ee}\end{rem}
\vskip 11pt

\subsection{The strong interactions}

\begin{defi}[The space-time structure of a semibaryon]  { The strong
interactions are widely believed to be generated by a nonabelian
$SU(3)$ gauge theory of colored quarks and gluons which are permanently
confined in color singlet hadronic bound states: this is quantum
chromodynamics \cite{M-P}.

This theory is principally justified by the beautiful discovery that
nonabelian gauge theories are asymptotically free \cite{G-W},
\cite{Pol}, \cite{Wein1}, but unfortunately, quantum chromodynamics does
not give a simple qualitative and dynamical understanding of confinement
\cite{C-J-J-T-W}.

Now, in this algebraic quantum model, the strong interactions and the
nature of the confinement of the semiquarks result directly from the
space-time structure of the semibaryons.

Indeed, it was proved in proposition 1.4.3 that the algebraic time
structure of a semi\-ba\-ryon is given by:
\[ \theta ^{\rm Bar}_{R,L}(t) = \theta
^{*1}_{R,L}(t_c)\textstyle\bigoplus\limits^3_{i=1} \theta
^1_{I_{R,L_i}}(t_i)\]
where $\theta
^{*1}_{R,L}(t_c)$ is the core time structure of the semibaryon and where
$\theta ^1_{I_{R,L_i}}(t_i)$ is the time structure of a semiquark.}\end{defi}
\vskip 11pt

The algebraic space-time structure of a semibaryon is generated from the
semisheaf of rings $\theta ^{\rm Bar}_{R,L}(t)$ by $\gamma _{t_i\to
r_i}\circ E_i$ morphisms:
\[ \gamma _{t_i\to r_i}\circ E_i : \theta ^{\rm Bar}_{R,L}(t) \to
\theta ^{\rm Bar}_{R,L}(t,r)= \theta
^{*1}_{R,L}(t_c)\textstyle\bigoplus\limits^3_{i=1} \theta
^{1-3}_{R,L_i}(t_i,r_i)\]
according to proposition 1.4.4.

As in QCD (i.e. quantum chromodynamics), the color is 
related to a quark state \cite{Kok} and corresponds to one
of the indices ``~$i$~'' of the semiquark semimodule
$ \theta
^{1-3}_{R,L_i}(t_i,r_i)$~, $1\le 1\le 3$~.

We would thus have the equivalences: \bt[t]{lll}
$i=1$ & $\sim$ & red color,\\
$i=2$ & $\sim$ & blue color,\\
$i=3$ & $\sim$ & yellow color.\te
\vskip 11pt

Recall that the set of parameters
\[ G(\rho _B)_{t_c\to[t_1,t_2,t_3]}=\{ G_1(\rho _{B_1}),\cdots,
G_\mu (\rho _{B_\mu }),\cdots,G_q(\rho _{B_q})\}\;,\]
noted in abbreviated form $G(\rho _B)$ and defined by (see definition
1.4.5):\[
\phi ^*_{t_c;(n _B-\rho _B)_{R,L}}=G(\rho _B)_{t_c\to[t_1,t_2,t_3]}\cdot
\phi _{[t_1,t_2,t_3];\rho _{B_{R,L}}}\]
where
\Be\item $\phi ^*_{t_c;(n _B-\rho _B)_{R,L}}$ is the set of algebraic
Hecke characters related to the generation of the reduced semisheaf
$\theta ^{*1}_{R,L}(t_c)$ by Eisenstein homology,
\item $\phi _{[t_1,t_2,t_3];\rho _{B_{R,L}}}$ is the set of algebraic
Hecke characters related to the generation of the complementary
semisheaf
\[ \theta
^3_{I_{R,L}}(t_1,t_2,t_3)=\textstyle\bigoplus\limits^3_{i=1}\theta
^1_{I_{R,L}}(t_i)\;,\]
\Ee
leads to the definition of the strong constant of the strong
interaction:
\[\langle G(\rho _B)_{t_c\to[t_1,t_2,t_3]}\rangle =\left( \txt\sum\limits^{q_B}_{\mu
=1} G_\mu (\rho _{B_\mu })\right)\]
noted $G(\rho )$~.
\vskip 11pt

Now, we can state the proposition:
\vskip 11pt

\begin{propo} \Bena \item The confinement of the 3 semiquarks originates
naturally from the generation of the 3 semiquarks from the core time
semisheaf of rings of the semibaryon by the smooth endomorphism $E_t$~.
\vskip 11pt

\item The asymptotic freedom of the 3 semiquarks could result from a
complete transformation of the core time semisheaf of rings of the
semibaryon $\theta ^{*1}_{R,L}(t_c)$ into the complementary time
semisheaves of rings of the 3 semiquarks under the conditions that:
\Be\item $\theta ^{*1}_{R,L}(t_c)\simeq 0$~;
\item $G(\rho _B)_{t_c\to[t_1,t_2,t_3]}\simeq 0$~;
\item $(n _B-\rho _B)_{R,L}\to 0$~.\Ee\Ee\end{propo}
\vskip 11pt

\paragraph{Outline of proof:}  Asymptotic freedom, which is a
consequence of Bjorken scaling \cite{Bjo} at high momentum transfer
\cite{D-D-T}, occurs if the semiquarks are free, i.e. if $\theta
^{*1}_{R,L}(t_c)\simeq 0$~.

This is realized when the rank set $\rho _{B_{R,L}}$ of the
complementary semisheaf of rings of the 3 semiquarks $\theta
^3_{I_{R,L}}(t_1,t_2,t_3)$ is equal to the rank set of the core
time semisheaf of rings $\theta ^1_{R,L}(t_c)$~, i.e. if $(n _B-\rho
_B)_{R,L}\to 0$~.

And, from the definition of the set of parameters $G(\rho
_B)_{t_c\to[t_1,t_2,t_3]}$~, it appears that asymptotic freedom is
reached when $G(\rho _B)_{t_c\to[t_1,t_2,t_3]}\simeq 0$~.\epr
\vskip 11pt

\begin{defi} \ {\bf (Mass-operator of a semibaryon)\/}\ { Referring to the
space-time}  { structure of a semibaryon as given in definition 5.3.1, it is
immediate that the elliptic self-adjoint differential operator
corresponding to the mass operator of a right (resp. left)
semibaryon is
\nobeqn
\MM ^{\rm Bar}_{R,L}
&=& \{ \mp i\hbar s_0\ \frac\partial{\partial t_{c0}},\{ \pm i\hbar
G(\rho )s^R_0\ \frac\partial{\partial t_{0}},\pm i\ \frac\hbar{c_R}
G(\rho )s^R_x\ \frac\partial{\partial x},\cdots,
\pm i\ \frac\hbar{c_R}
G(\rho )s^R_z\ \frac\partial{\partial z}\},\\
\noalign{\vskip 11pt}
&&\{ \pm i\hbar
G(\rho )s^B_0\ \frac\partial{\partial t_0},\cdots,\pm i\ \frac\hbar{c_B}
G(\rho )s^B_z\ \frac\partial{\partial z}\},\\
\noalign{\vskip 11pt}
&&\{ \pm i\hbar
G(\rho )s^Y_0\ \frac\partial{\partial t_0},\cdots,\pm i\ \frac\hbar{c_Y}
G(\rho )s^Y_z\ \frac\partial{\partial z}\}\}\noeeqn
where
\Be\item the indices $R,B,Y$ refer to the colors;
\item $\{ s^R_i\}_{i=x,y,z}$ are the components of a $3D$ unit vector
referring to the spin of the red semiquark;
\item $G(\rho )$ is the strong constant defined in definition 5.3.1;
\item $c_{R,B,Y}$ is the abbreviated notation for the parameter $c_{t\to
r}(\rho _{q_{R,B,Y}})_{R,L}$ referring to the generation of the
$3D$ spatial semisheaf of rings of the semiquark $R$~, $B$ or $Y$ from its
corresponding $1D$ time semisheaf of rings.
\Ee}\end{defi}
\vskip 11pt

\begin{defi}\ {\bf (The space-time structure of a bisemibaryon)\/}\ { According
to}  { the axiom II 1.3.9 and definition 3.1.2, we have to consider
bisemibaryons whose ``~$ST$~'', ``~$MG$~'' or ``~$M$~'' structure is given
by the bisemisheaves\\  $\theta ^{\rm Bar}_R(t,r)_{ST,MG,M}\otimes 
\theta ^{\rm Bar}_L(t,r)_{ST,MG,M}$ characterized by the tensor products
between the  right semisheaf  $\theta ^{\rm
Bar}_R(t,r)_{ST,MG,M}$~, referring to a right semibaryon, and a
 left semisheaf   $\theta ^{\rm
Bar}_L(t,r)_{ST,MG,M}$ referring to a left semibaryon.  On the ``mass''
structure, we will have:
\nobeqn
\lefteqn{\theta ^{\rm Bar}_R(t,r)_{M}\otimes \theta ^{\rm
Bar}_L(t,r)_{M}}\\
\noalign{\vskip 6pt}
&=& (\theta ^{*1}_R(t_c)\oplus (\theta ^{1-3}_R(t_R,r_R)\oplus \theta
^{1-3}_R(t_B,r_B)\oplus \theta ^{1-3}_R(t_Y,r_Y))\\
\noalign{\vskip 6pt}
&&\txt\bigotimes\limits (\theta ^{*1}_L(t_c)\oplus (\theta ^{1-3}_L(t_R,r_R)\oplus
\theta ^{1-3}_L(t_B,r_B)\oplus \theta ^{1-3}_L(t_Y,r_Y)))\;.\noeeqn

Under the $S_L$-isomorphism, this baryonic bisemisheaf splits into
the following set of bisemisheaves:
\nobeqn
 \theta ^{\rm Bar}_R(t,r)_{M}\otimes \theta ^{\rm
Bar}_L(t,r)_{M} 
&\to & 
 (\theta ^{*1}_R(t_c)\otimes \theta ^{*1}_R(t_c))\\
\noalign{\vskip 6pt}
&&\textstyle\bigoplus\limits^3_{i=1}(\theta
^{1-3}_{R_i}(t_i,r_i)\otimes \theta
^{1-3}_{L_i}(t_i,r_i)) \textstyle\bigoplus\limits^3_{{i,j=1}\atop{i\neq j}}(\theta
^{1-3}_{R_i}(t_i,r_i)\otimes \theta
^{1-3}_{L_j}(t_j,r_j))\\
\noalign{\vskip 6pt}
&&\textstyle\bigoplus\limits^3_{i=1}(\theta
^{*1}_{R}(t_c)\otimes \theta
^{1-3}_{L_i}(t_i,r_i))
\textstyle\bigoplus\limits^3_{i=1}(\theta
^{1-3}_{R_i}(t_i,r_i)\otimes \theta
^{*1}_{L}(t_c))\noeeqn
where
\Be
\item the bisemisheaf $(\theta ^{*1}_R(t_c)\otimes \theta
^{*1}_L(t_c))$ refers to the core central time structure of the
bisemibaryon;
\vskip 11pt

\item the bisemisheaf $(\theta
^{1-3}_{R_i}(t_i,r_i)\otimes \theta
^{1-3}_{L_i}(t_i,r_i))$ refers to the $10D$-space-time structure of the
bisemiquark ``~$i$~'' (~$i=$ red, blue, yellow).

This bisemisheaf $(\theta
^{1-3}_{R_i}(t_i,r_i)\otimes \theta
^{1-3}_{L_i}(t_i,r_i))$ splits under the $S_L$-isomorphism into the
direct sum of the three bisemisheaves:
\begin{itemize}
\item 
the $4D$-diagonal space-time bisemisheaf $(\theta
^{1-3}_{R_i}(t_i,r_i)\otimes_D \theta
^{1-3}_{L_i}(t_i,r_i))$~,
\item
the $3D$ magnetic bisemisheaf
$\theta
^{3}_{R_i}(r_i)^{(m)}\otimes_m \theta
^{3}_{L_i}(r_i)^{(m)}$~, and
\item 
the $3D$ electric bisemisheaf $\theta
^{1}_{R_i}(t_i)^{(e)}\otimes_e \theta
^{3}_{L_i}(r_i)^{(e)}$ or $\theta
^{3}_{R_i}(r_i)^{(e)}\otimes_e \theta
^{1}_{L_i}(t_i)^{(e)}$~.\end{itemize}

Remark that the $4D$-space-time diagonal bisemisheaf $\theta
^{1-3}_{R_i}(t_i,r_i)\otimes_D \theta
^{1-3}_{L_i}(t_i,r_i)$ of the $i$-th bisemiquark is at the origin of
``biquanta'' $\widetilde M ^I_{R_i}(r_i)\otimes_D\widetilde M ^I_{L_i}(r_i)$ which
are generated from the $3D$ space orthogonal bisemisheaf $\theta
^3_{R_i}(r_i)\otimes_D \theta ^3_{L_i}(r_i)$ by a smooth biendomorphism
$E_R\otimes_D E_L$~.  

The electric bisemisheaf of a bisemiquark is at the origin of the
electric charge of this bisemiquark whose absolute value is $\left|\frac
13\right| e$ or $\left|\frac
23\right| e$~.  Indeed, following that the electric biquanta of the
electric bisemisheaf of a bisemiquark are invariant under an electric
subgroup of $SL(1,\rit)\times SL(3,\rit)$ (or of $SL(3,\rit)\times
SL(1,\rit)$~) (see definition 4.2.19) at one or two bigenerators, the
eigenbivalues (in ``~$e$~'' units) of the electric charge of a
bisemiquark will be $\left|\frac
13\right|  $ or $\left|\frac
23\right|  $~.
\vskip 11pt

\item the ``mixed'' direct sum $\textstyle\bigoplus\limits^3_{{i,j=1}\atop{i\neq j}}(\theta
^{1-3}_{R_i}(t_i,r_i)\otimes \theta
^{1-3}_{L_j}(t_j,r_j))$ refers to the bilinear interactions between the
right semisheaves $\theta ^{1-3}_{R_i} (t_i,r_i)$ of the right
semiquarks and the left semisheaves $\theta ^{1-3}_{L_j}(t_j,r_j)$ of the
left semiquarks.

The $(i-j)$-th interaction bisemisheaf $(\theta
^{1-3}_{R_i}(t_i,r_i)\otimes\theta ^{1-3}_{L_j}(t_j,r_j))$ between the
$i$-th right semiquark and the $j$-th left semiquark splits into
\Be \item a mixed diagonal $4D$- space-time bisemisheaf:
\[ (\theta ^{1-3}_{R_i}(t_i,r_i)\otimes_D\theta ^{1-3}_{L_j}(t_j,r_j))\]
composed of mixed $1D$ time biquanta and mixed $3D$ orthogonal space
biquanta which are biquanta of the gravitational field
between the $i$-th and the $j$-th bisemiquark,
\item a mixed $3D$  magnetic bisemisheaf
\[ (\theta ^3_{R_i}(r_i)^{(m)}\otimes_m \theta ^3_{L_j}(r_j)^{(m)})\]
composed of magnetic biquanta of the magnetic
field between the $i$-th and the $j$-th bisemiquark,
\item a mixed $3D$ electric bisemisheaf
\[ (\theta ^1_{R_i}(t_i)^{(e)}\otimes_e \theta ^3_{L_j}(r_j)^{(e)})
\quad \mbox{or} \quad (\theta ^3_{R_i}(r_i)^{(e)}\otimes_e \theta
^1_{L_j}(t_j)^{(e)})\] 
composed of electric biquanta of the electric field between the $i$-th and the $j$-th
bisemiquark.
\Ee
\vskip 11pt

\item the mixed direct sum of the bisemisheaves
$\bigoplus\limits^3_{i=1} (\theta ^{*1}_R(t_c) \otimes \theta
^{1-3}_{L_i}(t_i,r_i))$ and\\  $\bigoplus\limits^3_{i=1} (\theta
^{1-3}_{R_i}(t_i,r_i) \otimes \theta ^{*1}_{L}(t_c))$ refer respectively
to the bilinear interactions between the right central time semisheaf of
rings $\theta ^{*1}_R(t_c)$ of the right semibaryon and the left
semisheaves $\theta ^{1-3}_{L_i}(t_i,r_i)$ of the left semiquarks and to
the bilinear interactions between the right semisheaves $\theta
^{1-3}_{R_i}(t_i,r_i)$ of the right semiquarks and the left central time
semisheaf of rings $\theta ^{*1}_L(t_c)$ of the left semibaryon.

The $i$-th interaction bisemisheaf 
$(\theta ^{*1}_R(t_c)\otimes \theta ^{1-3}_{L_i}(t_i,r_i))$ splits into:
\Be \item 
a mixed $1D$ time bisemisheaf $(\theta ^{*1}_R(t_c)\otimes \theta
^1_{L_i}(t_i))$ referring to the interaction between the right central
time semisheaf of rings of the right semibaryon and the $i$-th left time
semisheaf of rings of the $i$-th left semiquark.

This bisemisheaf is composed of mixed $1D$ time biquanta $\widetilde M
^I_R(t_c)\otimes \widetilde M ^I_{L_i}(t_i)$ which are of gravitational nature
according to lemma 5.1.5.

\item a mixed $3D$  electric bisemisheaf
$ (\theta ^{*1}_{R}(t_c)^{(e)}\otimes_e \theta ^3_{L_i}(r_i)^{(e)})
$
composed of   electric biquanta $(\widetilde M ^I_R(t_c)^{(e)}\linebreak \otimes_e \widetilde M
^I_{L_i}(r_i)^{(e)})$
which must be of ``strong'' nature and responsible for a ``strong'' force
between the central core right semisheaf of the right semibaryon and
the $3D$ space semisheaf of the $i$-th left semiquark.

These strong electric biquanta $(\widetilde M ^I_R(t_c)^{(e)}\otimes_e \widetilde M
^I_{L_i}(r_i)^{(e)})$ are likely rather massive.

To each of the three electric strong bisemisheaves $(\theta^{*1}_R(t_c)^{(e)}\otimes_e \theta^3_{L_i}(r_i)^{(e)})$~, we can associate a ``blue'', ``yellow'' or ``red'' color in the sense that the localization of such a bisemisheaf on a bisemiquark gives to it the corresponding color.
\Ee
\vskip 11pt

We have a similar splitting of the $i$-th interaction bisemisheaf
$(\theta ^{1-3}_{R_i}(t_i,r_i)\otimes \theta ^{*1}_L(t_c))$~.\Ee}\end{defi}
\vskip 11pt

We are thus led to the following proposition:
\vskip 11pt

\begin{propo}  A right and a left semibaryon of a given bisemibaryon
interact by means of:
\Bena\item the electric charges and the magnetic moments of the 3
bisemiquarks;
\item a gravito-electro-magnetic field resulting from the bilinear
interactions between the right and the left semiquarks of different
bisemiquarks;
\item a strong gravitational and electric field resulting from the
bilinear interactions between the central core structures of the left
and right semibaryons and the right and left semiquarks.
\Ee\end{propo}
\vskip 11pt

\bpr $\!\!\!$The assertions of this proposition result from the developments
of definition \mbox{5.3.4.$\!$\square}
\vskip 11pt

\begin{coro}  In the lightest, stable (nonradioactive) nuclei, the number of protons is equal to the number of neutrons in such a way that the strong positive electric field of the up bisemiquarks equilibrates the strong negative electric field of the down bisemiquarks.  As a result, the strong force between up and down bisemiquarks is attractive.
\end{coro}
\vskip 11pt

\bpr According to proposition 3.1.6, the electric basis must be three-dimensional.  Consequently, for a given bisemiquark, the electric strong field must refer to the strong electric negative bisemisheaf $(\theta _R^{*1}(t_c)^{(e)}\otimes_e\theta ^3_{L_i}(r_i)^{(e)})$ or to the strong electric positive bisemisheaf $(\theta _{R_i}^{3}(r_i)^{(e)}\otimes_e\theta ^{*1}_{L}(t_c))$~.  

As the quark composition of the proton is $u,u,d$ and $u,d,d$ for the neutron, we see that the mixed $3D$ strong electric positive bisemisheaves 
$(\theta _{R_i}^{3}(r_i)^{(e)}\otimes_e\theta ^{*1}_{L}(t_c))$ of the right up semiquarks will compensate the mixed $3D$ strong electric negative bisemisheaves 
$(\theta _R^{*1}(t_c)^{(e)}\otimes_e\theta ^3_{L_i}(r_i)^{(e)})$ of the left down semiquarks: in other terms, the strong electric field between up and down bisemiquarks will be of attractive nature.\epr
\vskip 11pt

\begin{propo}  A set of bisemibaryons interact by means of:
\Bena
\item a gravito-electro-magnetic field resulting from the bilinear
interactions between the right and the left semiquarks belonging to
different bisemibaryons;
\item a strong gravitational and electric field resulting from the
bilinear interactions between the central core structures of the
semibaryons and the semiquarks belonging to different
bisemibaryons.\Ee\end{propo}
\vskip 11pt

\bpr According to definition 5.1.1, the general bisemisheaf of a set of
$N$ interacting bisemibaryons is given by:
\[ \Theta ^{\rm Bar}_{R_N}\otimes \Theta ^{\rm
Bar}_{L_N}=\textstyle\bigoplus\limits^N_{i=1} (\theta ^{\rm
Bar}_{R_i}\otimes \theta ^{\rm Bar}_{L_i}) 
\textstyle\bigoplus\limits^N_{{i,j=1}\atop{i\neq j}} (\theta ^{\rm
Bar}_{R_i}\otimes \theta ^{\rm Bar}_{L_j}) \;.\]
The $(i-j)$-th interaction bisemisheaf $(\theta ^{\rm
Bar}_{R_i}\otimes \theta ^{\rm Bar}_{L_j})$ decomposes under the
$S_L$-isomor\-phism into:
\nobeqn
 \theta ^{\rm Bar}_{R_i}\otimes \theta ^{\rm
Bar}_{L_j} 
&\to & 
 (\theta ^{*1}_{R_i}(t_{c_i})\otimes \theta ^{*1}_{L_j}(t_{c_j}))\\
\noalign{\vskip 6pt}
&&\textstyle\bigoplus\limits^3_{\alpha ,\beta =1}(\theta
^{1-3}_{R_{i\alpha }}(t_{i\alpha },r_{i\alpha })\otimes \theta
^{1-3}_{L_{j\beta }}(t_{j\beta },r_{j\beta }))
\textstyle\bigoplus\limits^3_{\alpha
=1}(\theta ^{*1}_{R_i}(t_{c_i})\otimes \theta
^{1-3}_{L_{j\alpha }}(t_{j\alpha },r_{j\alpha }))\\
\noalign{\vskip 6pt}
&&\textstyle\bigoplus\limits^3_{\alpha =1}(\theta
^{1-3}_{R_{i\alpha }}(t_{i\alpha },r_{i\alpha })\otimes \theta
^{*1}_{L_j}(t_{c_j}))\noeeqn
where
\Be\item the mixed direct sum $\bigoplus\limits^3_{\alpha ,\beta =1}(\theta
^{1-3}_{R_{i\alpha }}(t_{i\alpha },r_{i\alpha })\otimes \theta
^{1-3}_{L_{j\beta }}(t_{j\beta },r_{j\beta }))$ refers to the bilinear
interactions between the right semisheaves $
\theta
^{1-3}_{R_{i\alpha }}(t_{i\alpha },r_{i\alpha })$ of the right
semiquarks and the left semisheaves $
\theta ^{1-3}_{L_{j\beta }}(t_{j\beta },r_{j\beta }))$ of the left
semiquarks.

According to proposition 5.3.5, the $(i\alpha -j\beta )$-th interaction
bisemisheaf generates a gravito-electro-magnetic field.
\vskip 11pt

\item the mixed direct sum $\bigoplus\limits^3_{\alpha =1}(\theta
^{*1}_{R_i}(t_{c_i})\otimes \theta
^{1-3}_{L_{j\alpha }}(t_{j\alpha },r_{j\alpha }))$ refers to the
bilinear interactions between the central core right semisheaf $
\theta
^{*1}_{R_i}(t_{c_i})$ of the $i$-th right semibaryon and the left
semisheaves $
\theta ^{1-3}_{L_{j\alpha }}(t_{j\alpha },r_{j\alpha }))$ of the left
semiquarks of the $j$-th left semibaryon.

According to proposition 5.3.5, this $(i-j\alpha )$-th interaction
bisemisheaf generates a strong gravitational and electric field.\epr
\Ee
\vskip 11pt

\begin{rem}  { The classical charge parameter ``~$e$~'' is the
coupling constant modulating the connexion $A_{R,L}(t,r)$ tied up with
a right (resp. left) semilepton as envisaged in definition 5.2.1.  It would
then be natural to choose the parameter $\frac e3 G(\rho )$ \cite{Pie1} as
the coupling constant modulating the connexion $A_{q_{R,L}}(t,r)$ tied
up with a right (resp. left) semiquark.}\end{rem}
\vskip 11pt

\subsection{The decays of bisemiparticles}

\begin{defi}[Main decays of bisemiparticles] { In definition 2.4.2, it
was demonstrated that the second and the third families of elementary
right and left semiparticles are generated from the first (resp. the
second) family by a $SO(\cdot)\circ Vd(\cdot )$ morphism where:
\Be
\item $Vd(\cdot )$ denotes the versal deformation;
\item $SO(\cdot)$ is the spreading-out isomorphism.
\Ee
\vskip 11pt

As the second and the third families of elementary bisemiparticles are
unstable, they decay into lighter bisemiparticles, i.e. finally into
bisemiparticles of the first family.

The decays of bisemibaryons are of two types:
\Bena
\item leptonic decays which are of general form:
\[ A\to B+\ell+\nu _1\]
where\Be
\item $A$ and $B$ are bisemibaryons so that the bisemibaryon $A$ has a
bisemiquark composition of higher mass than that of the bisemibaryon
$B$~;
\item $\ell$ is a bisemilepton and $\nu _1$ is a bisemineutrino;
\Ee

\item nonleptonic decays which are of general form:
\[ A \to B+mes\]
where
\Be
\item the bisemibaryon $A$ has a bisemiquark composition of higher
mass than that of the bisemibaryon $B$~;
\item $mes$ denotes a meson having a bisemiquark composition $q_Rq_L$
such that the right semiquark $q_R$ has generally a different flavor
from the left semiquark $q_L$~.\Ee\Ee
\vskip 11pt

The decay of bisemileptons are of general type:
\[ \ell_a \to \ell_b +\nu_b +\o \nu \]
where the bisemilepton $\ell_a$ is of higher family than the
bisemilepton $\ell_b$ and where $\nu_b $ and $\o \nu $ are
bisemineutrinos.}\end{defi}
\vskip 11pt

\begin{propo}  The leptonic decay $A\to B+\ell+\nu _1$ of a
bisemibaryon $A$ results from the ``diagonal'' emission of a bisemilepton
``~$\ell$~'' by a bisemiquark $q_i$ of $A$ throughout the biendomorphism
$(E_R\otimes_D E_L)$ applied to the $10D$-space-time bisemisheaf
$(\theta ^{1-3}_{R_i}(t_i,r_i)\otimes \theta ^{1-3}_{L_i}(t_i,r_i))$ of
the bisemiquark $q_i$~.

As a consequence, the bisemiquark $q_i$ is transformed into a
bisemiquark $q'_i$ of lighter mass and a bisemineutrino $\nu _1$ is
emitted to take into account the bilinear interaction between the
bisemiquark $q'_i$ and the bisemilepton $\ell$~.

Summarizing, we have:
\[ q_i\to q'_i+\ell+\nu _1\;,\qquad q_i\in A\; ,\quad q'_i\in B\;.\]
\end{propo}
\vskip 11pt

\bpr Let $(\theta ^{1-3}_{R_i}(t_i,r_i)\otimes \theta
^{1-3}_{L_i}(t_i,r_i))$ be the $ST$~, $MG$ or $M$ bisemisheaf of the
$i$-th bisemiquark $q_i$ belonging to the bisemibaryon $A$~.  This
bisemiquark $q_i$ is supposed to be of the family $B$ or $C$ (see
definition 2.4.2).

Let then $(E_R\otimes_D E_L)$ be the diagonal smooth biendomorphism
applied to this bisemisheaf $(\theta ^{1-3}_{R_i}(t_i,r_i)\otimes \theta
^{1-3}_{L_i}(t_i,r_i))$~:
\nobeqn
\lefteqn{E_R\otimes_D E_L : \theta ^{1-3}_{R_i}(t_i,r_i)\otimes \theta
^{1-3}_{L_i}(t_i,r_i)}\\
\noalign{\vskip 6pt}
&\to & (\theta ^{1-3}_{R_{i'}}(t_{i'},r_{i'})\otimes \theta
^{1-3}_{L_{i'}}(t_{i'},r_{i'})) + (\theta ^{1-3}_{R_{1}}(t,r)\otimes 
\theta ^{1-3}_{L_{1}}(t,r))\\
\noalign{\vskip 6pt}
&&  +(\theta ^{1-3}_{R_{i'}}(t_{i'},r_{i'})\otimes \theta
^{1-3}_{L_{1}}(t,r)) + (\theta ^{1-3}_{R_{1}}(t,r)\otimes 
\theta ^{1-3}_{L_{i'}}(t_{i'},r_{i'}))\noeeqn
so that
\Be
\item  $(\theta ^{1-3}_{R_{i'}}(t_{i'},r_{i'})\otimes \theta
^{1-3}_{L_{i'}}(t_{i'},r_{i'}))$ is the bisemisheaf of the $i$-th bisemiquark
having decreased to a lighter family;
\item $(\theta ^{1-3}_{R_{1}}(t,r)\otimes 
\theta ^{1-3}_{L_{1}}(t,r))$ is the bisemisheaf of the generated bisemilepton
$\ell$ such that the left and right semisheaves $\theta
^{1-3}_{R_{\ell},L_\ell}(t,r)_{\rho _\ell}$ of the bisemilepton $\ell$ have
ranks
$\rho _\ell$ equal to the difference between the ranks $n_i$ and $(n
-\rho )_i$~, of the semisheaves 
$\theta ^{1-3}_{R_i,L_i}(t_i,r_i)_{n_i}$ and $\theta
^{1-3}_{R_{i'},L_{i'}}(t_{i'},r_{i'})_{(n -\rho  )_i}$ of the left and right
semiquarks $q_{R,L_i}$ and $q_{R,L_{i'}}$~.\\
We thus have that:
\[ \rho _\ell=n_i-(n -\rho )_{i}\;;\]

\item the leptonic bisemisheaf 
$(\theta ^{1-3}_{R_{\ell}}(t,r)\otimes 
\theta ^{1-3}_{L_{\ell}}(t,r))$ is disconnected from the $i'$-th bisemiquark
bisemisheaf
$(\theta ^{1-3}_{R_{i'}}(t_{i'},r_{i'})\otimes \theta
^{1-3}_{L_{i'}}(t_{i'},r_{i'}))$~;

\item the sum of the two bisemisheaves $(\theta ^{1-3}_{R_{i'}}(t_{i'},r_{i'})\otimes \theta
^{1-3}_{L_{1}}(t,r)) $ and \\ $ (\theta ^{1-3}_{R_{1}}(t,r)\otimes 
\theta ^{1-3}_{L_{i'}}(t_{i'},r_{i'}))$ is the interaction bisemisheaf between
the bisemilepton $\ell$ and the generated bisemiquark $q_{i'}$ and is allowed to
generate a new bisemifermion, under the circumstances a bisemineutrino $\nu
_\ell$~.\epr\Ee
\vskip 11pt

\begin{propo} The nonleptonic decay $A\to B +  mes$ of a bisemibaryon $A$ results
from the ``off-diagonal'' emission of a meson  ``~$mes$~'' by a bisemiquark $q_i$ of
$A$ throughout the nonorthogonal biendomorphism $(E_R\otimes_{m,e} E_L)$ (i.e. a
magnetic or electric biendomorphism) applied to the space-time bisemisheaf
$(\theta ^{1-3}_{R_{i}}(t_{i},r_{i})\otimes \theta
^{1-3}_{L_{i}}(t_{i},r_{i}))$ of the bisemiquark $q_i$~.

As a consequence, the bisemiquark $q_i$ is transformed into a bisemiquark $q_{i'}$
of a lighter mass.\end{propo}
\vskip 11pt

\bpr Let $(\theta ^{1-3}_{R_{i}}(t_{i},r_{i})\otimes \theta
^{1-3}_{L_{i}}(t_{i},r_{i}))$  be the $ST$~, $MG$ or $M$ bisemisheaf of the
$i$-th bisemiquark $q_i$ of the bisemibaryon $A$~.  

Let $(E_R\otimes_m E_L)$ be the ``magnetic'' smooth biendomorphism applied to the
$3D$ space bisemisheaf
$ \theta ^3_{R_i}(r_i)\otimes_D \theta ^3_{L_i}(r_i)\subset 
 \theta ^{1-3}_{R_i}(t_i,r_i)\otimes \theta ^{1-3}_{L_i}(t_i,r_i)$
so that the $3D$ space magnetic biquanta are emitted, i.e. are
disconnected from  $\theta ^3_{R_i}(r_i)\otimes_D \theta ^3_{L_i}(r_i)$~.

The set of emitted magnetic biquanta then generate the magnetic bisemisheaf 
$ (\theta ^3_{R_{mes}}\otimes_m \theta ^3_{L_{mes}})$ which can get a ``mass''
throughout the bimorphism  $(\gamma _{r\to t}\circ E)_R\otimes_m (\gamma _{r\to
t}\circ E)_L$~.  We then have that
\[ (\gamma _{r\to t}\circ E)_R\otimes_m (\gamma _{r\to
t}\circ E)_L :(\theta ^3_{R_{mes}}\otimes_m \theta ^3_{L_{mes}})
\to  (\theta ^{1-3}_{R_{mes}}\otimes_{\rm nonorth} \theta
^{1-3}_{L_{mes}})\]
where the nonorthogonal bisemisheaf
$(\theta ^{1-3}_{R_{mes}}\otimes_{\rm nonorth} \theta ^{1-3}_{L_{mes}})$
corresponds to the generated meson from the $i$-th bisemiquark $q_i$~.  This
bisemisheaf $(\theta ^{1-3}_{R_{mes}}\otimes_{\rm nonorth} \theta
^{1-3}_{L_{mes}})$ is characterized by a metric $g_{\alpha \beta }$ so that
\[ \left\{ \begin{array}{ll}
g_{\alpha  \beta }=0& \mbox{if\ } \alpha =\beta \;,\\
g_{\alpha  \beta }\neq 0& \mbox{if\ } \alpha \neq\beta \;.\end{array}\right.\]

Similarly, we can envisage an ``electric'' smooth biendomorphism $(E_R\otimes_e
E_L)$ applied to the bisemisheaf $(\theta ^{1-3}_{R_i}(t_i,r_i)\otimes 
\theta ^{1-3}_{L_i}(t_i,r_i))$ of the bisemiquark $q_i$ so that  electric biquanta are emitted, i.e. are disconnected from 
$(\theta ^{1-3}_{R_i}(t_i,r_i)\otimes 
\theta ^{1-3}_{L_i}(t_i,r_i))$~.

The set of emitted electric biquanta then generate the electric bisemisheaf  
$(\theta ^{1}_{R_{mes}}\otimes_e\theta ^{3}_{L_{mes}})$ which can get a mass
throughout the bimorphism\ $(\gamma  _{t\to r}\circ E)_R\otimes_e (\gamma_{r\to
t}\circ E)_L$~.  We then have that:
\[ (\gamma _{t\to r}\circ E)_R\otimes_e (\gamma _{r\to
t}\circ E)_L :(\theta ^1_{R_{mes}}\otimes_e \theta ^3_{L_{mes}})
\to  (\theta ^{1-3}_{R_{mes}}\otimes_{\rm nonorth} \theta
^{1-3}_{L_{mes}})\;.\]

Thus, in the case of a ``magnetic'' or an ``electric'' smooth biendomorphism, a
massive meson\linebreak $(\theta ^{1-3}_{R_{mes}}\otimes_{\rm nonorth} \theta
^{1-3}_{L_{mes}})$ can be generated from the bisemiquark $i$~.  \\
If this massive meson develops a morphism $(SO\circ Vd)$ on its time structure, it
will be endowed with an electric charge (see definition 2.4.1).  The consequence
of the generation of the meson ``~$mes''$~'' from the bisemiquark $q_i$ of
``~$A$~''  is the transformation of this bisemiquark into a bisemiquark $q_{i'}$ of
different flavor and with a mass lighter than this of the bisemiquark $q_i$~.\epr
\vskip 11pt

\begin{propo} The space-time structure of a meson is given by the nonorthogonal
space-time bisemisheaf $ST-MG-M$~: $(\theta ^{1-3}_{R_{mes}}\otimes_{\rm
nonorth} \theta ^{1-3}_{L_{mes}})_{ST-MG-M}$ characterized by a nonorthogonal
metric $g_{\alpha  \beta }\neq 0$ if $\alpha \neq \beta $~.\end{propo}
\vskip 11pt

\begin{propo}  The decay $\ell_a\to \ell _b +\nu _b+\o\nu $ of a bisemilepton
$\ell_a$ results from the diagonal emission of a bisemineutrino $\nu _b$
throughout the biendomorphism $(E_R\otimes_D E_L)$ applied to the space-time
bisemisheaf
$(\theta ^{1-3}_{R_{\ell_a}}(t,r)\otimes \theta
^{1-3}_{L_{\ell_a}}(t,r))$ of the bisemilepton $\ell_a$~.

As a consequence, the bisemilepton $\ell_a$ is transformed into a bisemilepton 
$\ell_b$ of lighter family than $\ell_a$ and a bisemineutrino $\o\nu $ of
different helicity from $\nu _b$ is emitted to take into account the bilinear
interaction between the bisemilepton $\ell_b$ and the bisemineutrino $\nu
_b$~.\end{propo}
\vskip 11pt

\bpr Let $(\theta ^{1-3}_{R_{\ell_a}}(t,r)\otimes \theta
^{1-3}_{L_{\ell_a}}(t,r))$ be the $ST$~, $MG$ or $M$ bisemisheaf of the
bisemilepton $\ell_a$~.  Let then $(E_R\otimes_D E_L)$ be the diagonal smooth
biendomorphism applied to this bisemisheaf:
\nobeqn
\lefteqn{(E_R\otimes_D E_L) : \theta ^{1-3}_{R_{\ell_a}}(t,r)\otimes \theta
^{1-3}_{L_{\ell_a}}(t,r)}\\
\noalign{\vskip 6pt}
&\to & (\theta ^{1-3}_{R_{\ell_b}}(t,r)\otimes \theta
^{1-3}_{L_{\ell_b}}(t,r)) + (\theta ^{1-3}_{R_{\nu _b}}(t,r)\otimes 
\theta ^{1-3}_{L_{\nu _b}}(t,r))\\
\noalign{\vskip 6pt}
&& + (\theta ^{1-3}_{R_{\ell_b}}(t,r)\otimes \theta
^{1-3}_{L_{\nu _b}}(t,r)) + (\theta ^{1-3}_{R_{\nu _b}}(t,r)\otimes 
\theta ^{1-3}_{L_{\ell_b}}(t,r))\noeeqn
where
\Be
\item $(\theta ^{1-3}_{R_{\ell_b}}(t,r)\otimes \theta
^{1-3}_{L_{\ell_b}}(t,r))$ refers to the bisemisheaf of the bisemilepton $\ell_b$
resulting from the decay of the bisemilepton $\ell_a$~;

\item $(\theta ^{1-3}_{R_{\nu _b}}(t,r)\otimes 
\theta ^{1-3}_{L_{\nu _b}}(t,r))$ is the bisemisheaf of the bisemineutrino $\nu
_b$ emitted by the bisemilepton $\ell_b$~;

\item $[(\theta ^{1-3}_{R_{\ell_b}}(t,r)\otimes \theta
^{1-3}_{L_{\nu _b}}(t,r)) + (\theta ^{1-3}_{R_{\nu _b}}(t,r)\otimes 
\theta ^{1-3}_{L_{\ell_b}}(t,r))]$ refers to the interaction between the
bisemilepton $\ell_b$ and the bisemineutrino $\nu _b$ and is allowed to generate a
new bisemineutrino, under the circumstances a bisemineutrino $\o\nu $ of which the
left semineutrino $\o\nu _L$ differs by its helicity (which is right) from the
helicity (left) of the left semineutrino $\nu _{b_L}$ of the bisemineutrino $\nu
_b$~.
\Ee

Thus, in the terminology, a bisemineutrino $\o\nu $ whose left semineutrino has
right helicity is the ``antineutrino'' of the bisemineutrino $\nu $ whose left
semineutrino has left helicity.
\epr
\vskip 11pt

\subsection{The EPR paradox}

Let us recall that the famous EPR paradox raises two kinds of questions 
\cite{E-P-R}, \cite{Bel}:
\Bena\item Does the wave function describe the objective reality of an elementary
particle?

\item How is it possible that two elementary particles, having interacted in the
past, can still interfere in the future, even instantaneously, although the
Hilbert space (representing the mathematical frame of quantum mechanics) only
deals with tensor products of one-particle Hilbert spaces, excluding interactions
between elementary particles?
\Ee
\vskip 11pt

We shall prove in the next proposition that this new algebraic quantum model gives
a response to the EPR paradox and that the two types of questions raised by this
paradox are in fact intimely interconnected.
\vskip 11pt

\begin{propo} \Bena\item The wave function of quantum mechanics, defined on the
linear Hilbert space $\Hs$~, is replaced in AQT by a wave ``bi''function referring
to the state of a bisemiparticle and defined on a bilinear Hilbert space $H^{\pm}$~.

\item  Two elementary bisemiparticles can interact:
\Be \item through the space by means of a
gravito-magneto-(electric) field;
\item through the time by means of a $1D$ time gravitational
field.\Ee\Ee\end{propo}
\vskip 11pt

\bpr \Bena \item Point 1) was already developed especially in definition 4.2.1. 
Thus, the linear wave function of linear quantum mechanics does not describe the
objective reality of an elementary (bisemi)particle.  Only, the wave ``bi''function
referring to a state of a bisemiparticle describes the objective reality of a
bisemiparticle.
\vskip 11pt

\item \Be\item Two bisemiparticles interact in an interval of time $dt$ through
the space by means of a gravito-electro-magnetic field according to proposition
5.1.6.  These two bisemiparticles can interact ``nonlocally'' through the internal time 
only by means of a $1D$ time gravitational field according to lemma 5.1.5.  The internal time can then be considered as an hidden variable.

\item the structure of the bilinear Hilbert space $H^{\pm}$ at two bisemiparticles
having interacted in the past makes possible their possible interaction in the
future at the condition that a gravito-electro-magnetic field might be generated
between these two bisemiparticles.\epr
\Ee\Ee
\vskip 11pt

It can be concluded that the description of the interferences between two bisemiparticles having
interacted in the past is only possible by the consideration of biobjects.
\vskip 11pt


\vfill
\noindent
Institut de Mathématique pure et appliquée, Université de Louvain, B-1348 Louvain-la-Neuve - Belgium
\\
E-mail: pierre@euler.math.ucl.ac.be

\end{document}